\documentclass[11pt]{article} 
\usepackage{natbib}
\usepackage{epsf}
\usepackage{graphicx}
\def\simlt{\mathrel{\rlap{\lower 3pt\hbox{$\sim$}}
        \raise 2.0pt\hbox{$<$}}}
\def\simgt{\mathrel{\rlap{\lower 3pt\hbox{$\sim$}}
        \raise 2.0pt\hbox{$>$}}}
\def\o3{[O~\small III\normalsize ]}
\def\ofig{[O~\small III] }
\def\therefore{%
       \leavevmode
       \lower0.2ex\hbox{$\circ$}%
       \kern-0.2em\raise0.7ex\hbox{$\circ$}%
       \kern-0.2em\lower0.2ex\hbox{$\circ$}%
       \thinspace}

\begin{document}

\begin{titlepage}

\vskip 0.5cm

\vskip 0.5cm
\ \ \\
\ \ \\
\ \ \\
\begin{center}
\Large
\bf{UNIVERSITY OF SOUTHAMPTON}
\large
\ \ \\
FACULTY OF SCIENCE\\
\ \ \\
School of Physics \& Astronomy\\
\ \ \\
\ \ \\
\ \ \\
\ \ \\
\ \ \\
\ \ \\
\ \ \\
\Large
\bf{Optical and infrared emission from discs, jets and nebulae associated with X-ray binaries}
\large
\ \ \\
\ \ \\
by\\
\ \ \\
\bf{David Russell}\\
\ \ \\
\ \ \\
\ \ \\
\ \ \\
\ \ \\
\ \ \\
\ \ \\
\ \ \\
\ \ \\
\large
Thesis for the degree of Doctor of Philosophy\\
\ \ \\
September 2007\\
\end{center}

\end{titlepage}

\normalsize

\begin{center}
\large
UNIVERSITY OF SOUTHAMPTON
\ \ \\
\underline{ABSTRACT}
\ \ \\
FACULTY OF SCIENCE
\ \ \\
SCHOOL OF PHYSICS \& ASTRONOMY
\ \ \\
\underline{Doctor of Philosophy}
\ \ \\
OPTICAL AND INFRARED EMISSION FROM DISCS, JETS AND NEBULAE ASSOCIATED WITH X-RAY BINARIES
\ \ \\
by David Michael Russell
\end{center}
\normalsize
X-ray binaries are binary star systems in which a compact object (a neutron star or a black hole) and a relatively normal star orbit a common centre of mass. Since the discovery of X-ray binaries with the first X-ray telescopes in the 1960s, astronomers have tried to understand how these bizarre objects behave, and why. Some change in X-ray luminosity by $10^8$ orders of magnitude on timescales of days to months due to an increased transfer of mass from the star towards the compact object. Many X-ray binaries are detected at all observable frequencies, from radio to $\gamma$-rays. It has been found that many different sources of emission, which peak at different frequencies, are present in X-ray binary spectra and together they produce the observed broadband spectrum. However, disentangling these components has proved challenging.
\ \ \\
$~~~~$Much of the work in this thesis concerns disentangling the components that occupy the optical and near-infrared (NIR) region of the spectrum of X-ray binaries; possibly the region in which the relative contributions of the different components are least certain. In particular one component, the synchrotron emission from jets of outflowing matter, is found in this work to contribute ubiquitously to the optical and NIR light of X-ray binaries with relatively faint stars. These results confirm that the jets are powerful and in some of this work, observations of the jets interacting with the surrounding matter are used to infer their power.
\ \ \\
$~~~~$An introduction to the subject matter is presented in Chapter 1.
Attention is given to the current thinking of the dominating optical and NIR emission processes, and to X-ray binaries that produce jets. In Chapters 2--6 I present a number of investigations into optical and infrared observations of X-ray binaries. Relations, patterns and general trends are discovered that have implications for not only the dominating emission processes but also for the physical conditions and general behaviour of the inflowing and outflowing matter. In Chapter 7 I summarise the results and discuss follow-up work that could further our understanding of these objects.

\newpage

\ \ \\
\ \ \\
\ \ \\
\ \ \\
\ \ \\
\ \ \\
\ \ \\
\ \ \\
\ \ \\
\ \ \\
\ \ \\
\ \ \\
\ \ \\
\ \ \\
\ \ \\
\begin{center}
To Mum

Your smiles will always be remembered

\end{center}

\newpage

\ \ \\
\ \ \\
\ \ \\
\ \ \\
\ \ \\
\ \ \\
\ \ \\
\ \ \\
\ \ \\
\ \ \\
\ \ \\
\ \ \\
\ \ \\
\ \ \\
\ \ \\
\begin{center}
In the whole history of this planet, in the 3--4 billion years it has

harboured life, it is only in the last few hundred years that a species

has begun to explore and to understand the universe in which it exists.
\end{center}

\newpage

\tableofcontents

\newpage

\listoftables

\listoffigures

\newpage

 (author's declaration here)

\newpage

\begin{center}
\Large
{\bf{Acknowledgements}}
\end{center}
\normalsize

Jagshemash! Thank you for read my thesis paper. I hope you like. I like a very much, is a good. Please to have visual pleasure. It would be big success! If it not success, I will be execute. High five! Dziekuje. Borat.
\emph{``I'd like to thank my family, my friends, my agent, my lawyer, my business manager, my publicist and passers-by who always wanted the best for me. Also all the nominees this evening and all the people again, that they've already thanked. My hormones are just too way out of control to be dealing with this."} (a collection of quotes from awards ceremonies by Sarah Jessica Parker, John Landau
and Catherine Zeta-Jones).

Let me begin traditionally -- by thanking all my office mates, past and present. Mo, Matthew, Mark, Martin and Matthew North, thanks to you I almost changed my name to Mdave to fit in. Mo if you weren't there being a fountain of knowledge I would've had to go up and ask Bobbie F all my trivial questions -- very much appreciated. I thank the Lamb for many hours of squash -- one day the Gecko will learn to climb the walls to return those awkward shots. Mark, well done for organising the postgrad gowns to be of West Ham colours -- I look forward to the ceremony. Carolyn -- thanks for your help in my first year when I knew nothing. Yang -- hopefully this will be the last time you move desks! To the Squid and the Jellyfish -- keep up the snooker, its been good. To Nessa G -- the office just hasn't been the same. The Poodle -- we were there first! Before youtube! Long live Google Videos.
\begin{eqnarray*}
  \rm productivity \propto morale \times pressure,~where~~morale \propto \frac{\rm social~life}{\rm pressure}
\end{eqnarray*} \begin{center}$\rm \therefore ~productivity \propto social~life$\end{center}

Therefore I'd like to thank the Koala and the Wheezel (whose average yearly publication record exceeds that of the Silverback) who from the beginning have always been up for a drink. I thank the Slug for providing encouragement when it comes to both working and drinking (hence enforcing both pressure and morale).
I thank the Possum for encouragement and help with work and job applications and for generally making sure people in the department talk to each other. To Ked -- thanks for the great conversations and the chocolate fountain. Mmm crisps dipped in chocolate.

I'd like to thank Lord Rayleigh (Noble Prize, 1904 for discovering argon) for supervising Sir J. J. Thompson, and J. J. Thompson (Noble Prize, 1906 for discovering the electron) for supervising Sir Edward Appleton, and Sir Edward Appleton (Noble Prize, 1947 for his work on the ionosphere which led to the development of radar) for supervising John Ratcliffe, and John Ratcliffe for supervising Sir Martin Ryle, and Sir Martin Ryle (Noble Prize, 1974 for his work on radio aperture synthesis which led to the discovery of pulsars) for supervising Prof. Antony Hewish, and Prof. Antony Hewish (Noble Prize, 1974 with Martin Ryle) for supervising Dame Jocelyn Bell Burnell, and Dame Jocelyn Bell Burnell for supervising Prof. Rob Fender, and Prof. Rob Fender for supervising me.~~~~~~~~~~~~~~~~~~~~~~~~...no pressure then!

I'd like to thank the Gibbon for dissuading me from taking up radio astronomy (at least using AIPS) on the day of my interview for PhD, and the Mosquito for definitely definitely not crashing his van of students on both the 2006 and 2007 Tenerife trips. To the Tapeworm -- you know the Poodle used to take the mick out of your straight line plots? He should see Mo's -- he's so happy when his data show a straight line!
Please can I also thank all of the staff in the Astro and ST(E)P groups (can you find your reference?) who I couldn't think anything humorous about and all my friends and co-workers on the lovely La Palma, because I forgot to make an acknowledgements section to my Masters thesis. I thank David G. Russell from Owego Free Academy for not publishing too much on ADS over the last few years, and Fraser Lewis for drowning me in Faulkes monitoring project data and for putting up with my e-mails. I thank Mischa Schirmer for extensive help with the usage of the excellent data reduction pipeline package \small THELI\normalsize. Thanks to Paul Callanan, Sera Markoff, Tariq Shahbaz, Stephane Corbel, John Miller, Rob Fender, Peter Jonker, Charles Bailyn and Mike Nowak for all showing some kind of vague interest in letting me work for them at some point (and to two of them for offering me a position). I also thank my examiners Stephane Corbel and Christian Knigge for joining me in the unique club of people who have actually read $all$ of my thesis (anyone else would be insane to do so).

Thank you Concert Band -- you doubled my number of friends at uni. To the Pfaffia -- I will never forget the ree-filled times. You're a bunch of insanely reeful people. Elaine you were a rock; I hope you find your planets -- have you looked behind the sofa? I thank Elena Gallo, Erik Kuulkers, Charles Bailyn and the RXTE teams for providing some data that were used in the thesis. I thank the anonymous MNRAS referees for remaining anonymous. This thesis uses observations made with the Isaac Newton Telescope, the Faulkes Telescope Project, the United Kingdom Infrared Telescope, the Very Large Telescope, the 2.2-m ESO/MPI Telescope, the Danish 1.54-m Telescope, the YALO Telescope and the Liverpool Telescope.

Finally, I'd like to thank God for being as cunning as a fox `what used to be professor of cunning at Oxford University, but has moved on, and is now working for the UN at the high commission of international cunning planning' (Baldrick, Middle Ages) and not revealing herself. I mean, if she just came along and said `Hey guys, look this is how I made the universe and this is how it works..' then I'd probably be out of a job. Oh and many thanks to my Dad, who is actually responsible for getting me into all of this in the first place.

\newpage

\begin{center}
\Large
{\bf{List of Definitions \& Abbreviations}}
\end{center}
\normalsize
\ \ \\
ADAF	........ advection-dominated accretion flow\\
AGN	........ active galactic nucleus/nuclei\\
AU	........ astronomical unit\\
BH	........ black hole candidate\\
BHXB	........ low-mass black hole X-ray binary\\
CCD	........ charge-coupled device\\
CO	........ compact object\\
CS	........ companion star\\
CV	........ cataclysmic variable\\
DDT	........ Director's Discretionary Time (observing proposal)\\
GRB	........ gamma-ray burst\\
HID	........ hardness--intensity diagram\\
HMXB	........ high-mass X-ray binary\\
IPHAS	........ INT Photometric H$\alpha$ Survey of the Northern Galactic Plane\\
IR	........ infrared\\
ISM	........ interstellar medium\\
JCMT	........ James Clerk Maxwell Telescope\\
LMXB	........ low-mass X-ray binary\\
LP	........ linear polarisation\\
MJD	........ Modified Julian Day\\
MNRAS	........ Monthly Notices of the Royal Astronomical Society\\
MSXP	........ millisecond X-ray pulsar\\
MTI	........ mass transfer instability\\
NIR	........ near-infrared\\
NS	........ neutron star\\
NSXB	........ low-mass neutron star X-ray binary\\
OIR	........ optical/NIR\\
PA	........ (polarisation) position angle\\
PSF	........ point spread function\\
QPO	........ quasi-periodic oscillation\\
S/N	........ signal-to-noise ratio\\
SED	........ spectral energy distribution\\
SMBH	........ supermassive black hole\\
SNR	........ supernova remnant\\
UKIRT	........ United Kingdom Infrared Telescope\\
ULX	........ ultra-luminous X-ray source\\
UV	........ ultraviolet\\
VLA	........ Very Large Array\\
VLT	........ Very Large Telescope\\
WFC	........ Wide Field Camera\\
XB	........ X-ray binary\\
XIN	........ X-ray ionised nebula\\
YSO	........ young stellar object\\

\newpage

\begin{center}
{\section{Introduction}}
\end{center}

Since the subject of this thesis is X-ray binaries (XBs), it is first necessary to introduce these objects. In the following sections I discuss what is known about XBs and why they are interesting. In particular I review the current understanding of the physical properties, emission properties and changing behaviour of low-mass X-ray binaries (LMXBs), which are the objects most of my work concerns. Following this I review the processes that can produce optical and near-infrared (NIR) emission that we can detect, and how this region of the spectrum in particular contains a large amount of information about LMXBs. In the subsequent Chapters I present my work, most of which is now published.
\\

\subsection{X-ray binaries}

\begin{center}
`\emph{So, the whooshy thing sometimes comes out from the spinny thing?}'

(My friend Katie sums up my work, 2004)
\end{center}

An X-ray binary is a binary star system in which one of the two stars is a compact object -- a black hole candidate (BH) or a neutron star (NS). The companion star is in most cases a normal main-sequence star but could also be e.g. a non-main-sequence supergiant or a white dwarf.\footnote{Much of the information in the Introduction comes from the chapters in two textbooks: `X-ray Binaries' \citep*{lewiet95} and `Compact Stellar X-Ray Sources' \citep{lewiva06} and the references therein. Other sources will be referenced individually.} When matter is transfered to the compact object the process usually produces high energy photons -- X-rays. This is because the compact object is small, and the luminosity radiated per unit area is proportional to the fourth power of the temperature: $L \propto R^2 T^4$, where $L$ is the luminosity, $R$ is the radius and $T$ is the temperature. A particle at a small radius will therefore be relatively hot, and will radiate at higher frequencies (and energies) than a particle at a larger radius. In general, the more matter that is captured in a given time, the brighter the X-rays, so inevitably these sources are discovered from X-ray outbursts. The highest luminosity XBs are some of the brightest X-ray sources in the sky. Several hundred XBs have been discovered in the Milky Way and some more in the Magellanic Clouds and nearby galaxies, in particular M31 (\citealt*{liuet01}; \citeyear{liuet06,liuet07}; \citealt{willet04}). It is estimated that there is a population of $10^7$--$10^9$ black holes in the galaxy which lie hidden because they are not accreting much matter \citep[either because they are in wide binaries where the star is not close enough for mass transfer or because they are isolated;][]{vand92}.

The luminosity of an XB can also change by many orders of magnitude in wavelength regimes from radio to $\gamma$-rays, over periods of time from minutes to decades (and probably centuries). The many different components of an XB, which can react to each other's physical and emission changes, result in a rich and complex phenomenology which is often challenging to interpret. However, the rewards of understanding these sources are also rich, as they are laboratories that have the potential to test theories of accretion\footnote{Accretion is the process that results when the compact object gravitationally attracts matter, and is explained in Section 1.1.2.}, jet formation, the existence of black holes and the general theory of relativity.

\subsubsection{The different flavours of X-ray binary}

Over the decades of X-ray binary observations many attempts have been made to separate the objects into different classes and sub-classes. Perhaps surprisingly, it is historically the mass of the companion star which defines the main two classes of XB, rather than the nature or mass of the compact object. There are many observational differences that were discovered early on between LMXBs, which contain a companion of mass $\simlt$ 2 M$_\odot$, and high-mass X-ray binaries (HMXBs), which have a more massive ($\simgt$ 2 M$_\odot$) companion. More recently, it has come to light that there are many areas of overlap in the properties and behaviour of LMXBs, HMXBs and their sub-categories.

The general properties and apparent behaviour of XBs are probably a natural consequence of the following parameters (which are usually constant to first order):

\begin{itemize}
\item the mass of the companion star
\item the spectral class of the companion star
\item the nature of the compact object (BH or NS)
\item the mass of the compact object
\item the orbital period of the system
\item the orbital separation (semi-major axis), which is defined by the masses and orbital period
\item the size of the accretion disc, which is related to the orbital separation
\item the orientation (orbital inclination to the line of sight)
\item the eccentricity of the orbit
\item the time-averaged mass accretion rate (inflow)
\item the time-averaged power of the jets (outflow)
\item the composition of the jets
\item the spin of the BH or NS
\item if NS -- the magnetic field of the NS
\item external influences, e.g. local interstellar medium (ISM) density and local photon field
\end{itemize}

\begin{figure}
\centering
\includegraphics[width=15cm,angle=0]{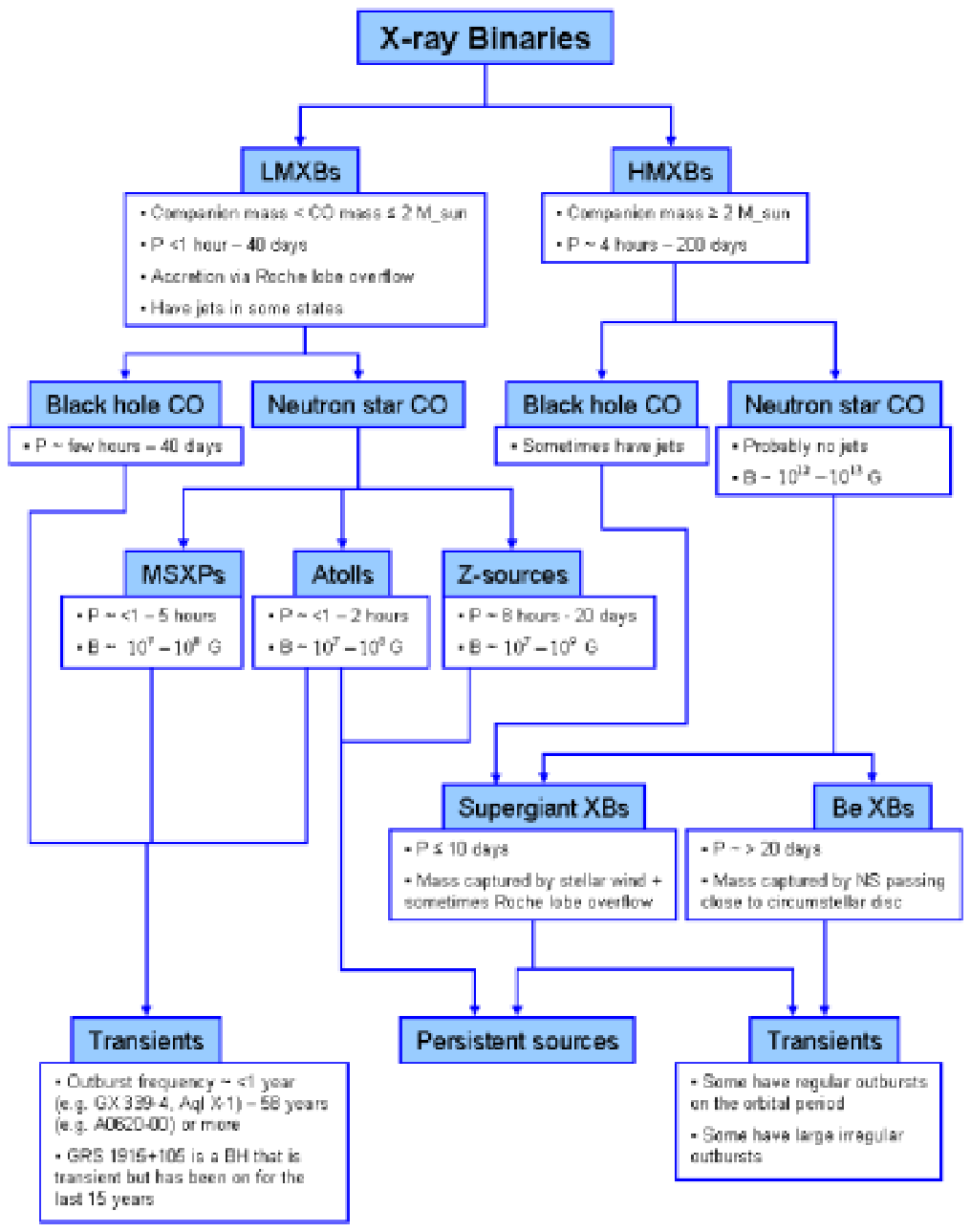}
\caption[The different flavours of XB]{A diagram showing the links between different flavours of XB.
References used to compile the diagram:
\cite{lewiet95};
\cite{lewiva06};
\cite{liuet07};
\cite{eachet76};
\cite{mass06};
\cite{hillet07}. Here, CO = compact object.}
\label{intro-XBflavours}
\end{figure}

Some combinations of the above parameters result in a system which dramatically changes luminosity and mass accretion rate in short periods of time. The systems which are persistently or transiently accreting matter are the ones which are detected and studied.
In Fig. \ref{intro-XBflavours} I present a diagram showing the different classes of observed XB as they have been defined historically, some of their basic properties and how they are related. In LMXBs the mass of the companion is generally less than that of the compact object. The region in which matter is gravitationally bound to a star in a binary system is called the Roche lobe; if matter from the companion crosses the Roche lobe in the direction of the compact object (see Section 1.1.2) it is gravitationally bound to the compact object. The orbital periods of LMXBs (empirically on scales of tens of minutes to tens of days) are usually less than those of HMXBs (a few hours to almost a year). Similarly the neutron star magnetic fields are larger for HMXBs ($10^{12}$--$10^{13}$ Gauss) than for LMXBs ($10^7$--$10^9$ G).

In LMXBs the nature of the compact object appears to define many aspects of the broadband emission behaviour. Low-mass black hole X-ray binaries (BHXBs) have different X-ray `states' (a description of the X-ray spectrum and timing properties; see Section 1.1.3) to low-mass neutron star X-ray binaries (NSXBs). Almost all BHXBs are transient -- usually having low accretion rates in a relatively stable `quiescent' state until at some point the mass build-up in the disc causes the system to become unstable and a dramatic increase of mass accretion in the inner regions causes a luminosity increase by up to $\sim$ eight or so orders of magnitude (the nearest exceptions are GRS 1915+105, which has been active for the last 15 years but was quiescent before then and LMC X--3 which some class as a HMXB; see discussion in Section 2.3). Some NSXBs are also transient but some are persistent. Transient LMXBs are traditionally referred to as `X-ray novae' or `soft X-ray transients'; here I will just call them transients as many do not go through a soft X-ray phase \citep[see][]{brocet04}. Outbursts typically last several months although some can be as short as minutes \citep[e.g. V4641 Sgr;][]{wijnva00} or as long as decades (as in the aforementioned GRS 1915+105). The frequency of outbursts depends on the particular source but can range from $\simlt 1$ a year (e.g. Aql X--1) to one every $\sim 58$ years \citep[e.g. A0620--00;][]{eachet76} or more.

There are three main classes of NSXB (discussed in Section 1.1.3): Z-sources, atolls and millisecond X-ray pulsars (MSXPs). The Z-sources have been active over the period of time X-ray telescopes and detectors have existed ($\simgt 30$ years) except one new source, XTE J1701--462 which `switched on' in 2006 \citep{remiet06} and at the time of writing is declining in luminosity via on atoll-like phase, possibly back to quiescence \citep{homaet07b}. Of the atolls, some are persistent (e.g. 4U 0614+09) and some are transient (e.g. XTE J2123--058). All MSXPs discovered so far are transient. The Z-sources generally have larger orbital periods than atolls and MSXPs \citep{liuet07}. It has recently been found that a source can behave at one time like a Z-source and at another time like an atoll \citep{homaet07b}, or at one time like an atoll and at another like a MSXP \citep{caseet07}, which implies they are in fact the same objects, but they vary in behaviour with accretion rate.

Unlike LMXBs, the main method of mass capture in HMXBs is usually not Roche lobe overflow. HMXBs with supergiant stars transfer matter from the star to the compact object via a strong stellar wind. These are one of the two main categories of HMXB, the other being Be XBs, in which a neutron star orbits a young Be (or Oe) star. Be stars have H$\alpha$ emission lines originating from a circumstellar disc around the star \citep[e.g.][]{zorebr91,grebet97}. Neutron stars in this latter class often have eccentric orbits and capture matter when they pass close to the circumstellar disc. These systems have the largest orbital periods and magnetic fields of all the XBs; $\sim 20$--200 days and $10^{12}$--$10^{13}$ G, respectively. The supergiant XBs have smaller periods than the Be XBs ($\simlt 10$ days), can contain a BH or NS compact object, and are sometimes transient and sometimes persistent.

In the following Sections I concentrate and elaborate on the LMXBs, which are the topic of most of my work.

\begin{figure}
\centering
\includegraphics[width=15cm,angle=0]{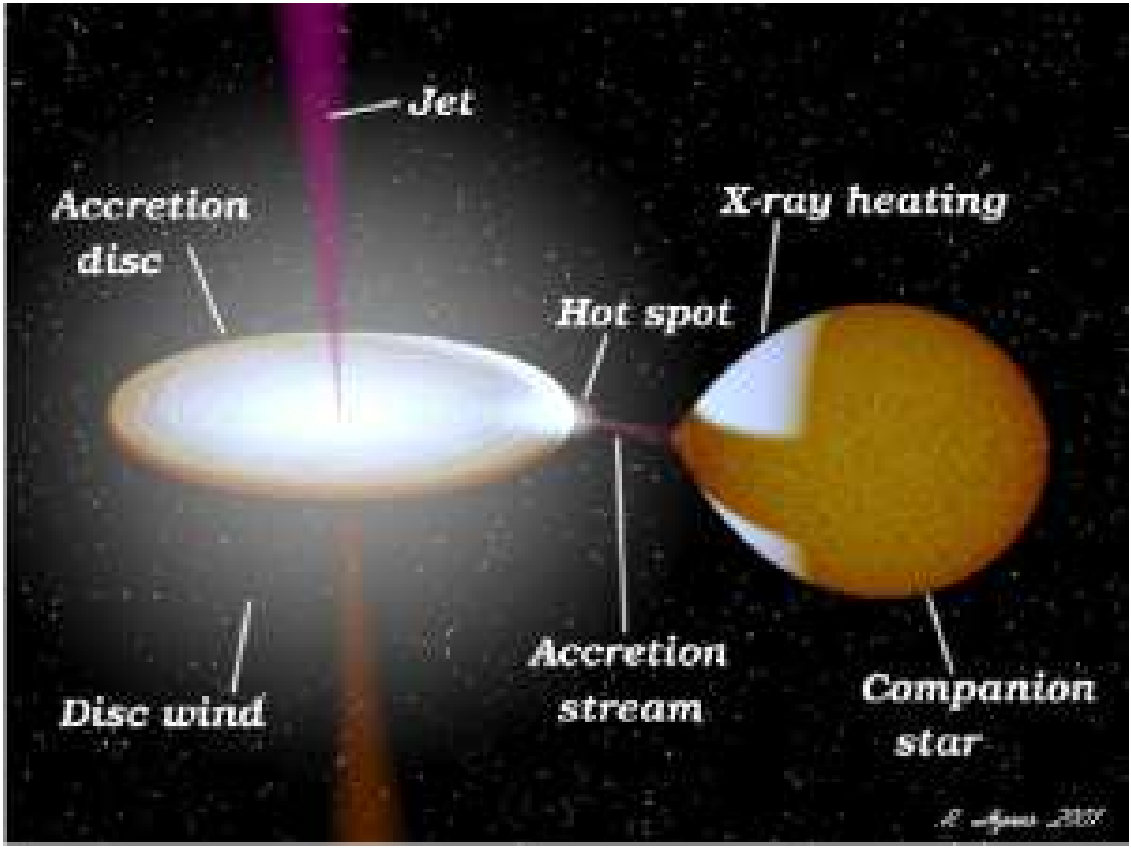}
\caption[The components of an LMXB]{Illustration of the different components of an LMXB (credit: Rob Hynes).}
\label{intro-hynes}
\end{figure}

\subsubsection{The physical properties of LMXBs and the process of accretion}

Essentially, the method we use to increase our understanding of LMXBs (and in fact most astronomical objects!) is to compare the behaviour of the emission (i.e. photons) we detect from the source to the emission we expect such a system to emit under the laws of physics. There is one and only one physical solution that can generally explain the emission behaviour of LMXBs -- the existence of a compact object -- either a black hole or a neutron star -- accreting matter from a companion star \citep[first suggested by][]{shkl67}.

Over decades of observing LMXBs, a general picture has emerged of their constituents. An illustration of an LMXB is shown in Fig. \ref{intro-hynes}. A star and compact object orbit their common centre of mass. The two objects are close enough (semimajor axis $\sim 0.01$--1 astronomical units; AU) for Roche lobe overflow to occur. If matter from the star passes through the inner Lagrange point L$_1$, it will be gravitationally bound to the compact object. In LMXBs, the surface of the star is close to L$_1$, which causes the star to stretch in that direction becoming non-spherical. The amount of light seen from the star varies with the amount of visible surface area; if the star passes in front of the compact object (inferior conjunction) it will appear more circular than when both objects lie on a plane of the sky (as in Fig. \ref{intro-hynes}). In the latter case the star will appear elongated and slightly larger. Systems with a high inclination (i.e. when the orbital plane is not parallel to the plane of the sky) display an `ellipsoidal'-type optical light curve due to this, first seen in transient LMXBs when they return to quiescence \citep[when the star dominates the optical regime; e.g.][]{mcclet83} and well-documented now in many systems.

When matter becomes gravitationally bound to the compact object, it has a net angular momentum and so forms a disc around the compact object. Matter moves inwards in a sequence of almost circular orbits in this `accretion disc', losing energy as angular momentum is transported outwards. The disc, probably in most cases is optically thick, and so radiates as a blackbody, the matter releasing energy and increasing in temperature as it moves towards the compact object. The spectrum of the disc is in the form of a multi-temperature blackbody due to its radial temperature gradient. $Viscosity$ is the process responsible for energy loss and angular momentum transportation -- how this process works is still under debate. Atomic and molecular viscosity can only account for a very small amount of the angular momentum transport; magnetic fields probably play a very significant role and a number of models have been put forward, in particular the magnetorotational instability (MRI) model \citep[e.g.][]{balbha91}.

Finally the matter reaches the inner disc radius and continues towards the compact object in a hot accretion flow of debatable structure \citep*[e.g.][]{naraet97,blanbe99,ferret06,petret06} which under some conditions forms a collimated jet which escapes the system. These jets predominantly reveal themselves as radio counterparts emitting synchrotron radiation. In addition, a `corona' of hot electrons and positrons surrounds the compact object and the disc is likely to give off a disc wind.

For an LMXB to form, a massive star must have been present in the binary to eventually form the compact object after a supernova explosion. The longer-lived companion star must end up in a close orbit, probably due to losing angular momentum while spiralling inwards in the common envelope of the massive star before the explosion. The compact object is formed during the supernova.

The above process is accretion as envisaged for LMXBs -- many other types of astronomical object undergo accretion:

\begin{itemize}
\item Active galactic nuclei (AGN) -- supermassive black holes (SMBHs) that accrete from the surrounding matter at the centre of a galaxy (see Section 1.1.5)
\item Cataclysmic variables (CVs) -- similar systems to XBs but with a white dwarf compact object -- these are more numerous than XBs but they cannot probe the dense matter and the extreme-gravity environments of XBs
\item Young stellar objects (YSOs) -- pre-main-sequence stars collecting mass from a protoplanetary disc
\item Ultra-luminous X-ray sources (ULXs) -- which may either harbour intermediate-mass black holes or are XBs that emit non-isotropically
\end{itemize}

Accretion onto a black hole is the most efficient means known in the universe of converting mass into energy; the fastest spinning Kerr BHs can radiate up to $\sim 40$\% of the rest-mass energy of the matter. On the other hand, accretion onto non-rotating BHs radiates about 6\% of the rest-mass energy \citep{thor74,lipa00}; for comparison, nuclear fusion converts $\sim 1$\% of the rest mass into energy. The gravitational potential energy released through accretion is proportional to $M / R$ where $M$ is the mass of the accreting object. $R$, the radius, increases linearly with black hole mass (if the spin is a constant) so the radiative efficiency $\epsilon$ of a SMBH is the same of that of a stellar-mass BH: $\epsilon \propto M / R \propto M / M \propto 1$.

\subsubsection{X-ray states and the classes of NSXB}

In BHXBs, properties of the emission in all wavebands are often related to changes in the X-ray spectrum.  The two main X-ray spectral states are the \emph{hard} (or \emph{low/hard}) state, which is characterised by a hard power-law spectrum and strong variability, and the \emph{soft} (or \emph{high/soft}; \emph{thermal--dominant}) state, where a thermal spectrum dominates with a power-law contribution. The low luminosity `quiescent' state is likely to be an extension to the hard state \citep*{nara96,esinet97,mcclet03,fendet03,fendet04,gallet06} but currently this is not universally accepted. In general I treat quiescence as an extension to the hard state in this work but also show how some results differ when this is not assumed.

The most striking observation of X-ray state-dependency at other wavelengths comes from the radio regime. The radio counterpart of Cyg X--1 appeared when a transition was made to a harder X-ray spectrum \citep{tanaet72}, and it has since been shown that this strong correlation between X-ray spectral shape and strength of radio emission is ubiquitous \citep[e.g.][]{fendet99a}. The shape of the X-ray spectrum of a BHXB is also strongly correlated with its timing properties, with strong variability generally correlated with bright radio and hard X-ray emission. The X-ray power spectrum (the power as a function of frequency in a time series; the square of the Fourier transform) varies with X-ray state and consists of broad structures (`$noise$') and narrow structures (\emph{quasi-periodic oscillations}; QPOs).

It was observed from early on that the different spectral states occur
at different fractions of the Eddington luminosity; $L_{\rm Edd} = 1.25\times 10^{38}(M / M_\odot )$ erg s$^{-1}$ (see e.g. \citealt{nowa95} for a review), the maximum isotropic luminosity a source can emit without the outward radiation pressure becoming greater than the inward gravitational force. The hard state, which exhibits strong
radio emission, usually occurs at lower luminosities than the
soft state, which has low amplitude variability and no radio emission.
Observations which fit neither of these states show similar
phenomenology over a rather wide range of luminosities; they have
typically been called intermediate states when at lower luminosities
and very high states when at high luminosities, but their spectral and
variability properties are sometimes comparable and sometimes varied. For a detailed review of BHXB X-ray states see \cite*{homabe05} and for the most recent effort at putting together a unified picture of spectral states incorporating jets, see \cite*{fendet04} \citep[see also][]{homaet01,mcclet06}.

\begin{figure}[ht]
\centering
\includegraphics[width=8cm,angle=270]{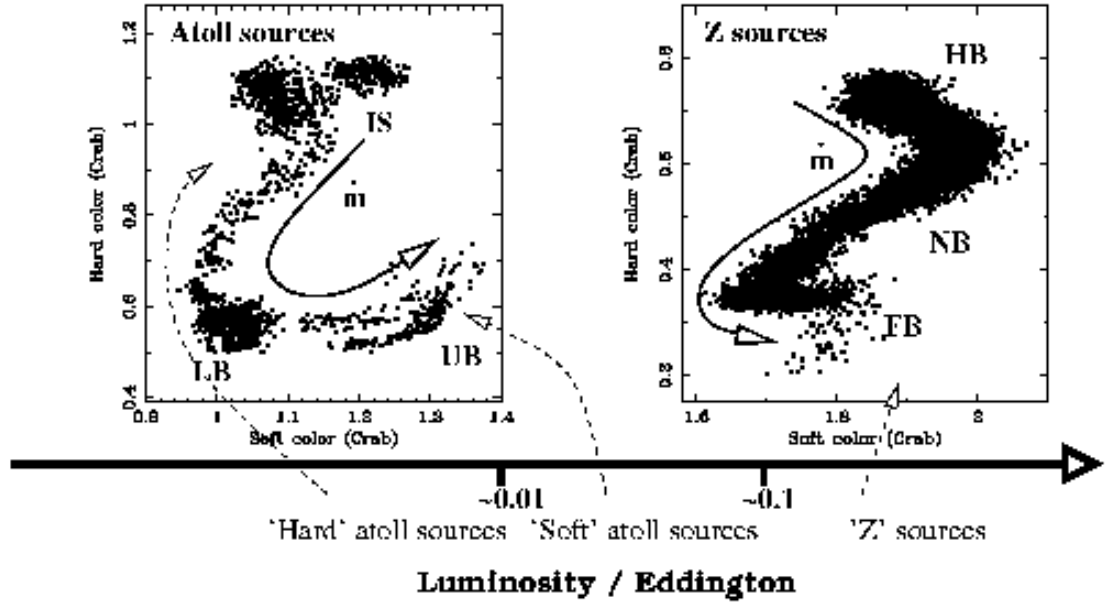}
\caption[X-ray colour--colour diagrams for NSXBs]{NSXBs are traditionally classed by the tracks they trace out in X-ray colour--colour diagrams (see text). This summary is from \cite{miglfe06}. The `$\dot{m}$' arrows indicate the expected increase in mass accretion rate along the tracks. FB = flaring branch; HB = horizontal branch; IS = island state; LB = lower banana; NB = normal branch; UB = upper banana.}
\label{intro-NSstates}
\end{figure}

NSXBs exhibit many of the same X-ray spectral and timing properties as BHXBs. However, luminosity correlates with X-ray state only in some cases, and other (e.g. timing) properties appear to be related to their flux ratios derived from four X-ray bands (in the $\sim 2$--16 keV range) -- their position in the X-ray colour-colour diagram \citep{hasiva89}. Indeed, the tracks that NSXBs trace out in the colour--colour diagram or hardness--intensity diagram (HID) are usually `atoll'-type tracks or `Z'-type tracks, and most NSXBs can be classed as an atoll source or a Z-source (see Fig. \ref{intro-NSstates}). In atolls, the X-ray states consist of the \emph{island state} (harder spectrum) and the \emph{banana state} (softer) and in Z-sources the states are the \emph{horizontal branch} (hardest), the \emph{normal branch} and the \emph{flaring branch} (softest). Weaker NSXBs (mainly MSXPs) generally have a hard spectrum throughout their outbursts and reach similar peak luminosities to transient BHXBs that do not make a transition to a softer state (see Section 1.1.4). NSXBs exhibit many of the same timing properties as BHXBs but there are differences \citep{vand06}; in fact structures like QPOs seem to be common to accreting objects in general \citep*{papala93,pretet06,zhanet07}.

\subsubsection{Patterns in outburst behaviour}

A picture has begun to develop in which the BHXB X-ray state transitions are hysteretic, with
the transition from hard state to soft state occurring at higher
luminosities than the transition from soft state to hard state
\citep*[see e.g.][]{miyaet95,nowaet02,barret02,smitet02,maccco03}.  At
sufficiently low luminosities, it appears that only hard states
can exist, but the brightest hard states are just as bright as the
brightest soft states \citep{homabe05}.
Transient NSXBs can also exhibit this hysteretical behaviour (namely the atolls) \citep{maitba04}.

\begin{figure}[ht]
\centering
\includegraphics[width=10cm,angle=0]{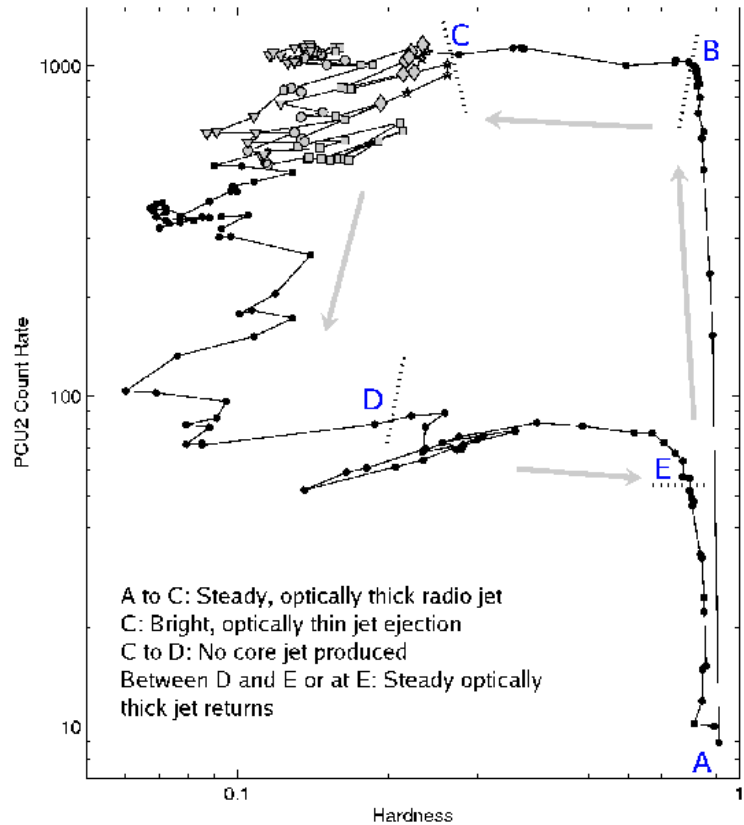}
\caption[X-ray hardness--intensity diagram for BHXBs]{X-ray HID for a `typical' XB outburst that enters the soft state \citep[adapted from][]{bellet05}. The ordinate axis is proportional to the flux (data from the $RXTE$ PCA), and hardness (abscissa) is defined as the 6.3--10.5 keV count rates divided by the 3.8--6.3 keV count rates.}
\label{intro-turtle}
\end{figure}

\begin{figure}[ht]
\centering
\includegraphics[width=15cm,angle=0]{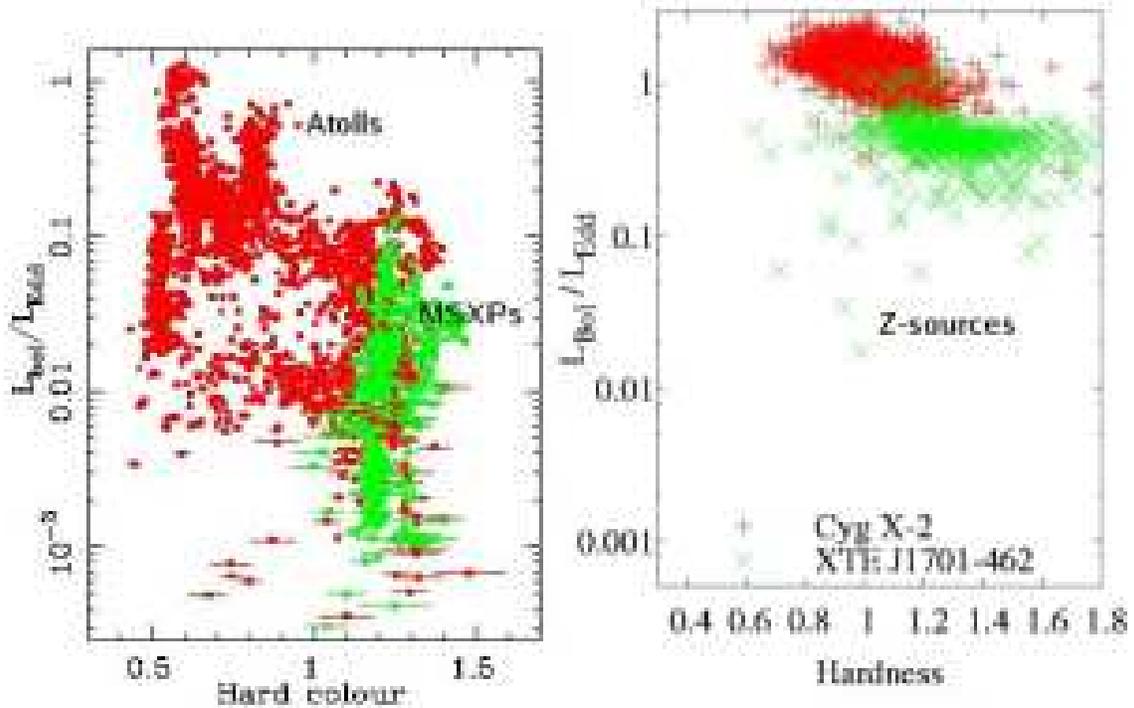}
\caption[X-ray hardness--intensity diagram for NSXBs]{X-ray HIDs for NSXBs, to compare to BHXBs (Fig. \ref{intro-turtle}). Left panel: Atolls (red) and MSXPs (green) \citep*[adapted from Fig. 1 of][]{gladet07}. The hard colour refers to the 9.7--16 keV band / 6.4--9.7 keV band. Right panel: Z-sources Cyg X--2 (red) and XTE J1701--462 (the only transient Z-source to date; green). The data is from the $RXTE$ ASM; the hardness is the 5--12 keV / 3--5 keV. The distances and hydrogen column densities are listed in Table \ref{tab-19NSXBs-1}.}
\label{intro-NSs-HID3}
\end{figure}

A well-established relationship exists between X-ray luminosity and radio luminosity in BHXBs in the hard state \citep*{corbet00}; $L_{\rm radio} \propto L_{\rm X}^{0.7}$ (\citealt*{corbet03}; \citealt*{gallet03}), where self-absorbed, compact steady jets are the origin of the radio emission, which is optically thick. It has been suggested that the correlation requires the hard state to be a radiatively inefficient flow, with $L_{\rm X} \propto \dot{m}^2$ \citep[approximately as predicted by advection-dominated accretion flow (ADAF) models; e.g.][]{narayi95,maha97}, and with the radio emission providing a good tracer of $\dot{m}$, the mass accretion rate \citep*{kordet06}. Both the rising hard state and the falling hard state connect to a radiatively efficient flow in the soft state, where $L_{\rm X} \propto \dot{m}$ \citep{shaksu73}, albeit at different luminosities.
A picture has emerged \citep*{fendet04} of a universal X-ray--radio pattern in transient BHXB outbursts (Fig. \ref{intro-turtle}) in which a typical outburst follows a specific path in an X-ray HID and the radio behaviour is dependent on the position in the diagram. `Hardness' here refers to the X-ray colour, as in the colour--colour plots for NSXBs (see Section 1.1.3). During an outburst, the source remains in the hard state as the X-ray and radio (optically thick compact steady jet) luminosity rises (from positions A to B in Fig. \ref{intro-turtle}). The X-ray spectrum softens at B and the steady radio jet persists until C where a bright radio flare is seen. Here (termed the `jet line'), the radio spectrum becomes optically thin, and the bright flare corresponds to the launching of a (or a number of) discrete plasma ejection(s). These ejections may have higher bulk Lorentz factors than the steady, hard state jets \citep{fendet04} and so the former may plough into the latter, producing optically thin emission from the collision.

From C to D the radio jet (core) is quenched and the X-ray spectrum is soft, however radio emission may be visible from the aforementioned discrete jet ejections. The source transits back into the hard state at E and the steady self-absorbed jet returns either before (Fender et al., in preparation) or at \citep{kaleet05} E. There is no bright optically thin radio flare on the soft-to-hard state transition. It has been thought that in the soft state, the inner radius of the accretion disc is closer to the compact object than in the hard state \citep*[e.g.][]{mcclet95}. More recently it has been suggested that the inner accretion disc may not change dramatically during state transitions \citep[e.g.][]{fronet01a,millet06,rykoet07} although in most cases is likely to be truncated at low luminosities \citep{mcclet95,esinet01,fronet01b,fronet03,chatet03,mcclet03,yuanet05,yuanet07}.

The different flavours of NSXB can be imagined to occupy various regions of the HID that transient BHXBs trace out (Fig. \ref{intro-NSs-HID3}). The Z-sources, which nearly always appear persistent, occupy the most luminous region of the HID (right panel of Fig. \ref{intro-NSs-HID3}). Since they produce transient, discrete jet ejections \citep[e.g.][]{miglfe06} they occupy the `jet line' region (around C in Fig. \ref{intro-turtle}) and are similar to GRS 1915+105, a (currently) persistent BHXB which continuously cycles back and forth through the jet line \citep{fendet04}. The recent transient Z-source XTE J1701--462 interestingly behaved more like an atoll on its outburst decline \citep{homaet07b}. The MSXPs appear to be analogous to the faint BHXBs that remain in the hard state during outburst (e.g. XTE J1118+480) whereas the atolls are softer at their highest luminosities than at their lowest (Fig. \ref{intro-NSs-HID3}; left panel). Both MSXPs and atolls have been associated with a steady radio counterpart (analogous to the BHXB hard state jet) but little is known of NSXB jets in general since they are intrinsically fainter than BHXB jets and are less-well sampled \citep[e.g.][]{miglfe06}. The spectral index of the hard state jet of NSXBs has not been measured so it is uncertain whether it is optically thick or thin emission. See \cite{maitba04} for an example of outburst X-ray hysteresis in one NSXB, Aql X--1.

\subsubsection{Black holes of all shapes and sizes}

The three characteristics of a BH are its mass, angular momentum (spin) and electric charge. Black holes that occur naturally have approximately zero charge as the matter they accumulate is not electrically charged, nor are they likely to be born intrinsically charged. A BH with zero spin is described by the Schwarzschild metric whereas a spinning BH is described by the Kerr metric and has an ergosphere (the region in which all matter and even light is forced to rotate with the BH) whose surface is an oblate spheroid. The radius of the event horizon (and the ergosphere) is larger for more massive BHs. The spin can therefore be imagined as its `shape' (the shape of the ergosphere) and the mass, its `size'.

The size of a BH, and hence its dynamical timescales, scales linearly with its mass. AGN contain a SMBH of mass $10^5$--$10^{10}M_\odot$ which is accreting at the centre of a galaxy and unlike their smaller BHXB cousins which orbit with a companion star, the most massive SMBHs can swallow whole stars \citep*[e.g.][]{rees84,oste89,gezaet06,bellet06}. They are, it appears, scaled up versions of stellar-mass BHs; they produce jets several orders of magnitude more powerful than those of XBs. Quasars are very powerful AGN at high redshifts; X-ray binaries with jets are sometimes termed microquasars \citep{miraet92}. Observational relations that exist in XBs can be extended to include AGN (e.g. \citealt*{merlet03,falcet04,donegi05,mchaet06a},b) and it has recently been shown \citep*[e.g.][]{maccet03,kordje06,summet07} that it may be possible to unify the different classes of AGN in terms of the states of XBs and their orientation. However, there are some differences between the environments of stellar-mass and SMBHs. For example, AGN have broad-absorption line regions -- powerful emission line-driven outflows \citep[e.g.][]{nortet06} and BHXBs of course have a companion star whose presence affects many of its properties.

\subsubsection{One of the most dynamically changing systems in the universe?}

XBs vary in luminosity in all wavebands from radio to $\gamma$-ray by several orders of magnitude on timescales accessible to us. Few astronomical objects are as dynamical on these timescales. Gamma-ray bursts (GRBs) including their afterglows are one \citep[e.g.][]{berget05}, but extremely swift multiwavelength follow-up is required to track their light curves. Other interacting binaries such as CVs are another -- CVs are more abundant than XBs but are generally fainter at very high and very low frequencies \citep[e.g.][as also are supernovae]{pattra85}. There are also oddball binary stars which emit at radio through to X-rays and are variable \citep[e.g.][]{whitet04} but they generally do not span the luminosity range of XBs. The luminosity of SMBHs must vary by many orders of magnitude (sometimes they are AGN and sometimes they are relatively inactive) but in our lifetime we see just a snapshot of their `outburst' light curve (see also Section 1.1.5).

The dynamical availability of these sources only makes them interesting if their behaviour tells us something useful. Luckily, it does. General relativity can be tested in the strong gravitational field near compact objects, and XBs could reveal the existence of event horizons, i.e. black holes (by e.g. the differences between NSXBs and BHXBs), or the nature of the matter in neutron stars. The study of XBs and their populations also constrains theories of various astrophysical processes including stellar evolution (the compact object is the endpoint of massive stars), the composition of the ISM (e.g. the progenitor of the compact object enriches the ISM with heavy elements from the supernova), black hole and neutron star formation and evolution, and jet formation mechanisms, to name a few. The process of accretion in XBs can be compared to other accreting objects: CVs, AGN, GRBs, YSOs. Finally, stellar mass BHs can, as shown recently, tell us a wealth of information about SMBHs (see Section 1.1.5). Of course, the work in this thesis comes nowhere near answering any fundamental questions, but does contribute to our understanding of the behaviour of these interesting objects.

\subsection{Optical/NIR emission from X-ray binaries}

Historically, the main uses of identifying and monitoring the optical or NIR counterpart of an XB are to establish the geometry of the accretion flow, in particular the mass of the compact object and nature of the companion star \citep*{whitet95}. The best localisation of an XB is also traditionally obtained from radio or optical follow-up of an X-ray detection. It has now become clear that studying the optical/NIR (OIR) counterparts of XBs can uncover a varied wealth of information about these sources. For example, the detection of optical emission lines can tell us about the geometry and workings of the accretion disc \citep*[e.g.][]{marset94} or (very rarely) the outflow \citep{marg84,blunbo05} and absorption lines can tell us about the companion star and level of interstellar dust extinction towards the source \citep[e.g.][]{wagnet91,casaet93}.

Most of the OIR light from LMXBs comes from the continuum, which is (in outburst) usually blue and generally thought to arise in the outer accretion disc as the result of X-ray reprocessing \citep*{cunn76,vrtiet90,vanpet95}, a process whereby the surface of the disc absorbs high energy ultraviolet (UV) or X-ray photons and is heated, then re-emitted as thermal radiation at lower energies. Indeed, timing and spectral analysis in many cases has led to this conclusion \citep*[e.g.][]{wagnet91,callet95,obriet02,hyneet02a,hyne05}. However, reprocessed X-rays are often misleadingly \emph{assumed} to dominate the OIR light in LMXBs. Some observations (of BHXBs in particular but also sometimes of NSXBs) point towards alternative physical processes contributing (and sometimes dominating) the OIR emission. It is important to quantify these contributions to constrain the broadband spectrum. In addition, many analyses assume the emission is dominated by a specific mechanism. For example, optical colours have been used (i) in quiescence to uncover the spectral type of the companion, sometimes assuming that the light is coming solely from the companion, and (ii) in outburst to measure the temperature of the accretion disc, usually assuming the light is dominated by X-ray reprocessing.

\emph{The aim of a large proportion of this thesis is to constrain the OIR contributions of the different emission processes at a given time for a given LMXB. This is motivated by the recent realisation that jet emission may be a significant contributor in this waveband.}

\subsubsection{Data availability}

\begin{figure}
\centering
\includegraphics[width=10cm,angle=270]{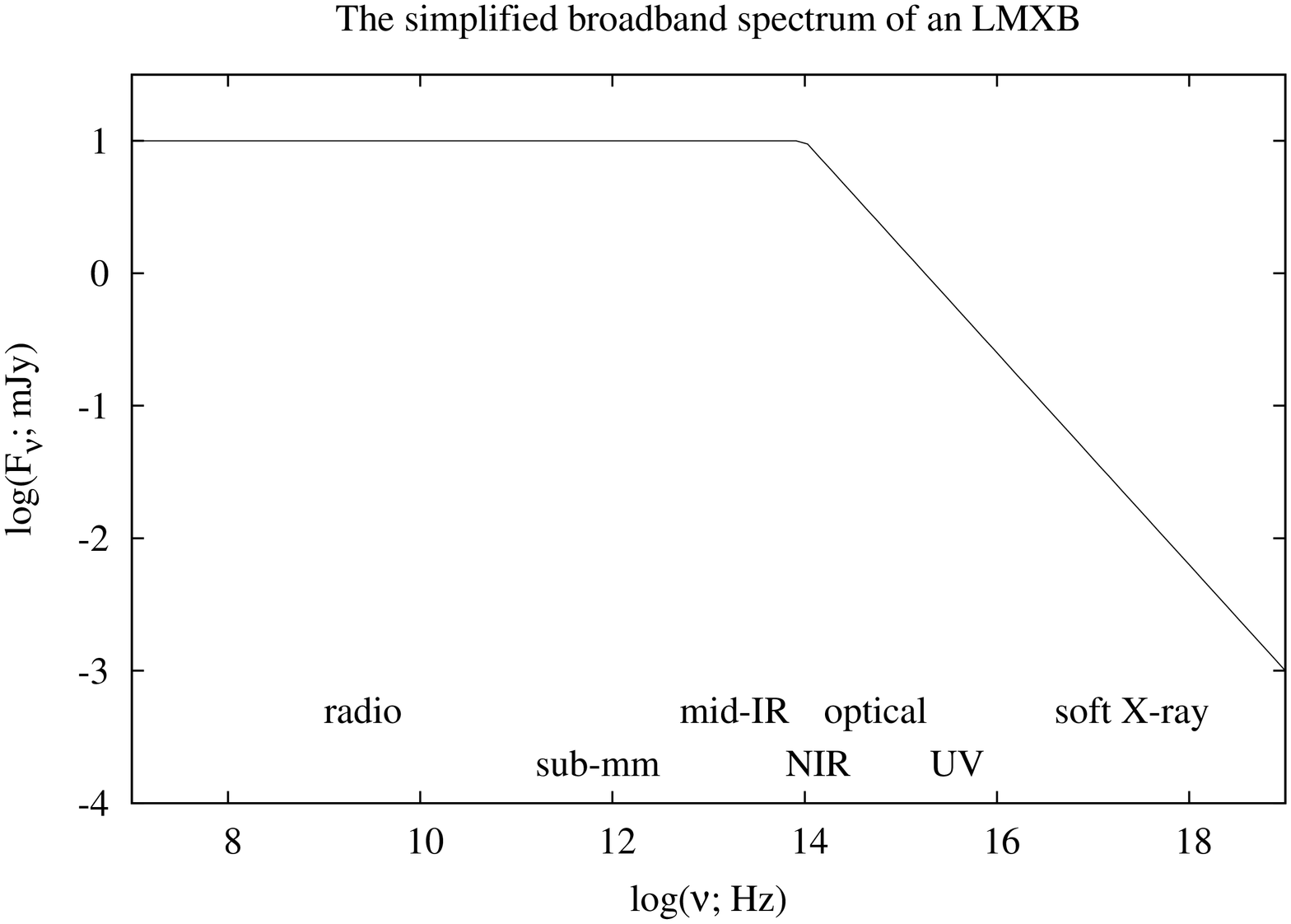}
\caption[The simplified broadband spectrum of an LMXB]{The simplified SED is adapted from \cite{market01}.}
\label{whyoptical-1}
\end{figure}

\begin{figure}
\centering
\includegraphics[width=10cm,angle=270]{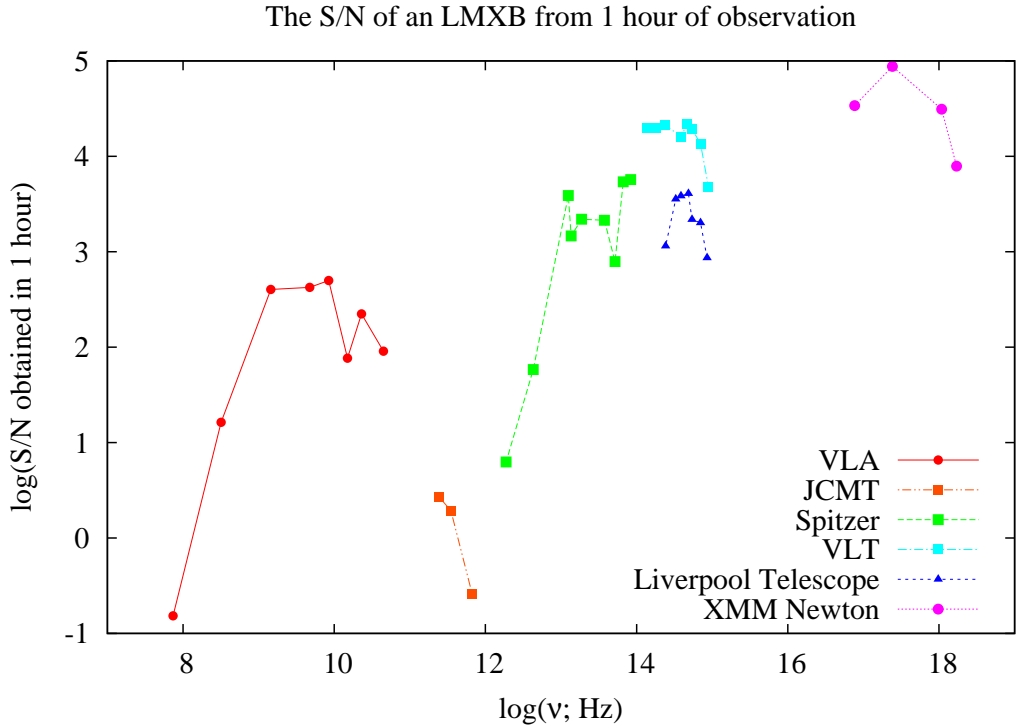}
\caption[The S/N in one hour of observation of an LMXB]{The S/N in one hour of observation of an LMXB using various telescopes. Telescopes: VLA = Very Large Array; JCMT = James Clerk Maxwell Telescope; Spitzer = Spitzer Space Telescope; VLT = Very Large Telescope; The Liverpool Telescope; XMM-Newton.}
\label{whyoptical-3}
\end{figure}

It is clear that OIR studies of LMXBs can reveal important informations about these objects. Conveniently, optical and infrared (IR) telescopes are the most abundant. Many galactic LMXBs can be detected with a high level of confidence in only a few minutes on relatively small optical telescopes.
In Fig. \ref{whyoptical-1} I show a very simplified broadband spectral energy distribution (SED) of a typical LMXB in outburst in the hard state. It is based on an observed broadband SED with unprecedented wavelength coverage from radio to X-ray \citep*[from][]{market01}. The flux density is plotted against frequency, both in logarithmic scales. The flat ($\alpha = 0$, where $F_{\nu} \propto \nu^{\alpha}$) part of the spectrum is normalised to 10 mJy and the non-flat part has $\alpha = -0.8$ \citep[approximating to the SED in][]{market01}.

The signal-to-noise ratio (S/N) can be used to measure the level of confidence in a detection of a source. Using the various exposure time calculators for a number of telescopes, it is possible to estimate what the S/N of an LMXB with the above SED would be for each telescope, in a given amount of integration time. The S/N obtained in one hour of on-source integration for a number of telescopes is shown in Fig. \ref{whyoptical-3}. Exposure time calculators were used with the optimum setup for each waveband/instrument/filter under average (or default) external conditions and are given in Table \ref{whyoptical-table}. Data from \cite{hasiet01} are used for XMM Newton. Fig. \ref{whyoptical-3} shows that an LMXB can be detected with a S/N of almost $10^5$ using XMM Newton at soft X-ray energies and $> 2\times 10^4$ with the Very Large Telescope (VLT; 8 metre mirror) in the optical and NIR.  Even the smaller Liverpool Telescope (2 metre mirror) can achieve a high ($> 10^3$) S/N in one hour. These calculations do not take into account (i) the time needed for calibrations and overheads, which vary between telescopes, and (ii) interstellar dust extinction -- sources that are obscured will be harder to detect in the optical and NIR -- but radio telescopes are unaffected.

\begin{table}
\begin{center}
\caption[Telescopes and instruments used for Fig. \ref{whyoptical-3}]{Telescopes and instruments used for Fig. \ref{whyoptical-3}.}
\label{whyoptical-table}
\small
\begin{tabular}{|l|lr|lr|}
\hline
Telescope&Waveband name&log $\nu$&Waveband name&log $\nu$\\
\hline
VLA&4 (73--74.5 MHz)&7.87&P (300--340 MHz)&8.50\\
VLA&L (1.24--1.70 GHz)&9.16&C (4.5--5.0 GHz)&9.68\\
VLA&X (8.1--8.8 GHz)&9.93&U (14.6--15.3 GHz)&10.18\\
VLA&K (22.0--24.0 GHz)&10.36&Q (40.0--50.0 GHz)&10.65\\
JCMT&A$_3$ (211--272 GHz)&11.38&HARP (325--375 GHz)&11.54\\
JCMT&W$_{\rm D}$ (630--710 GHz)&11.83&&\\
Spitzer&IRAC: 3.6 $\mu$m&13.92&IRAC: 4.5 $\mu$m&13.82\\
Spitzer&IRAC: 5.8 $\mu$m&13.71&IRAC: 8.0 $\mu$m&13.57\\
Spitzer&IRS PUI: 16 $\mu$m&13.27&IRS PUI: 22 $\mu$m&13.14\\
Spitzer&MIPS: 24 $\mu$m&13.10&MIPS: 70 $\mu$m&12.63\\
Spitzer&MIPS: 160 $\mu$m&12.27&&\\
VLT&ISAAC: Ks&14.13&ISAAC: H&14.26\\
VLT&ISAAC: J&14.38&FORS1: I&14.59\\
VLT&FORS1: R&14.66&FORS1: V&14.73\\
VLT&FORS1: B&14.84&FORS1: U&14.95\\
Liverpool&SupIRCam: H&14.26&SupIRCam: J&14.38\\
Liverpool&RATCam: z'&14.52&RATCam: i'&14.59\\
Liverpool&RATCam: r'&14.69&RATCam: V&14.73\\
Liverpool&RATCam: B&14.84&RATCam: u'&14.93\\
XMM Newton&EPIC: 0.2--0.5 keV&16.88&EPIC: 0.5--2.0 keV&17.38\\
XMM Newton&EPIC: 2-10 keV&18.03&EPIC: 5-10 keV&18.23\\
\hline
\end{tabular}
\end{center}
\normalsize
Log $\nu$ is the central frequency in log space. The S/N is calculated from the VLA Sensitivity web page, the Heterodyne Integration TimE Calculator for the JCMT, the Sensitivity -- Performance Estimation Tool (SENS-PET) for Spitzer, the FORS1 Exposure Time Calculator and the  ISAAC Exposure Time Calculator for the VLT, the Liverpool Telescope Exposure Calculator and \cite{hasiet01} for XMM Newton.
\normalsize
\end{table}

It seems that optical and infrared telescopes are not only the most abundant, but can achieve a higher S/N in a given time than telescopes at most other wavebands. It is therefore important to take advantage of this data availability; to observe details in their behaviour that are probably not observable at other frequencies. The wavebands with the highest S/N from the data collected are OIR and X-ray, so correlations between these two bands can be explored over a greater luminosity range than say, radio--X-ray correlations (see Chapter 2).

\subsubsection{The nature of the OIR emission}
	
The origin of the emission from LMXBs is known in some wavebands and not so well established in others. Radio emission is produced by the synchrotron process in collimated outflows/jets \citep*[see e.g.][]{hjelet95,fend06}, whereas X-ray emission could originate directly from the hot inner accretion disc, from a Comptonising corona, from an advective flow, or from the compact jets \citep*[all of which have been successfully modelled;][]{pout98,naraet98,mccoet00,market01,brocet04,market05,rogeet06}.  The OIR is perhaps the waveband for which the relative contributions of the different emission processes are least certain.

Various techniques have been adopted to infer the physical emission mechanisms responsible for the OIR light from X-ray binaries.  Numerous processes may contribute to the OIR emission depending on the flux and spectral state of the source at the time.  This is evident from the complex variety of spectral, timing and luminosity properties observed between sources and between states for an individual source.  In HMXBs, the OIR light is largely dominated by the massive companion star in the system \citep*{vandet72} with occasional additional contributions, for example from the reprocessing of X-rays.  For NSXBs, there is strong evidence for a central X-ray source illuminating a disc that reprocesses the light to OIR wavelengths \citep*[see e.g.][]{mcclet79}.

OIR emission from BHXBs has been extensively studied in outburst and quiescence (for a review see \citealt{vanpet95}; see also \citealt*{chenet97} and \citealt{charco06}).  The techniques employed in the literature to attribute an emission process to the behaviour of a BHXB are numerous.  Here I discuss the most cited emission mechanisms that can contribute to the OIR light, and the common techniques used to infer their presence.  BHXBs are discussed first, followed by any differences that exist in NSXB systems. The findings so far are discussed for BHXBs in the three most stable spectral/luminosity states: the low-luminosity hard state (i.e. $\sim$ quiescence; $L_{\rm X}\simlt 10^{-5} L_{\rm Edd}$), the high-luminosity hard state (i.e. in outburst; $L_{\rm X}\simgt 10^{-4} L_{\rm Edd}$) and the soft state (also in outburst; usually at $L_{\rm X}\simgt 10^{-2}L_{\rm Edd}$).  `Optical' and `NIR' emission is here classed as that seen in the $BVRI$ ($\sim 4400-7900$\AA) and $JHK$ ($\sim1.25-2.22\mu m$) wavebands, respectively.
\newline\newline
\textbf{The reprocessing of X-rays in the outer accretion disc or on the companion star surface:}\newline
Reprocessing is apparent in many accreting systems by the signature of correlated OIR and X-ray behaviour, whereby the OIR lags the X-rays on light-travel timescales \citep*[e.g.][on the order of seconds for XBs; this may not be a diagnostic of X-ray reprocessing however; see below]{obriet02}.  The surface of the disc (or companion star) is being illuminated by X-rays, which are absorbed by the surface and then reprocessed to OIR wavelengths. Models predict the OIR to lie in the part of the blackbody spectrum where the spectral index $\alpha$ (where $F_{\nu}\propto \nu^{\alpha}$) is positive; some observations indicate the Rayleigh Jeans tail region of the blackbody, where $\alpha = 2$, and some have a lower spectral index \citep[e.g.][]{hyne05}.  X-ray reprocessing is associated with very strong emission lines which arise from atoms recombining after becoming ionised by the high energy photons. Double-peaked emission lines have been observed in the optical \citep*{marset94,orosba95,oroset02} and NIR \citep[e.g.][]{shahet99} in quiescence, and at higher luminosities in both the hard state \citep[e.g.][]{shraet94} and in the soft state (\citealt*{casaet91}; \citeyear{casaet99}), which signify the red- and blue-shifted components of the rotating accretion disc.

Optical SEDs of some sources are fit well by models with a contribution from X-ray reprocessing in the hard and soft states \citep*[e.g.][]{hyneet02a,hyne05} and \cite{chenet97} found that the properties of most transient outbursts in BHXBs and NSXBs are consistent with the disc thermal instability model \citep*{minewh89}, where X-ray reprocessing dominates the optical regime \citep{kingri98}.  X-ray reprocessing on the surface of the companion star heats the side facing the compact object, producing periodic optical variations \citep[e.g.][]{basket74}.  BHXBs, which are LMXBs, mostly have small mass ratios \citep{rittet03}; hence emission from reprocessing on the companion star is generally weak \citep{vanpet95}.
\newline\newline
\textbf{Intrinsic thermal emission from the viscously heated outer accretion disc:}\newline
In optically thick accretion disc models, the viscously heated disc is expected to contribute significant light in the optical, through UV to X-ray wavelengths \citep*{shaksu73,franet02}.  The accretion discs of BHXBs are generally larger than those of NSXBs and may be more luminous.  The spectrum is that of a multi-temperature blackbody with a positive OIR spectral index, similar to that of the X-ray heated disc.  However, the different temperature distributions that arise from viscous and irradiative heating result in spectra with different spectral indices in the UV \citep[e.g.][]{hyne05}.  Changes in the accretion rate affect the OIR before the X-rays since the OIR and X-ray disc emission originate in the outer and inner parts of the accretion disc, respectively.

\begin{figure}
\centering
\includegraphics[height=10cm,angle=0]{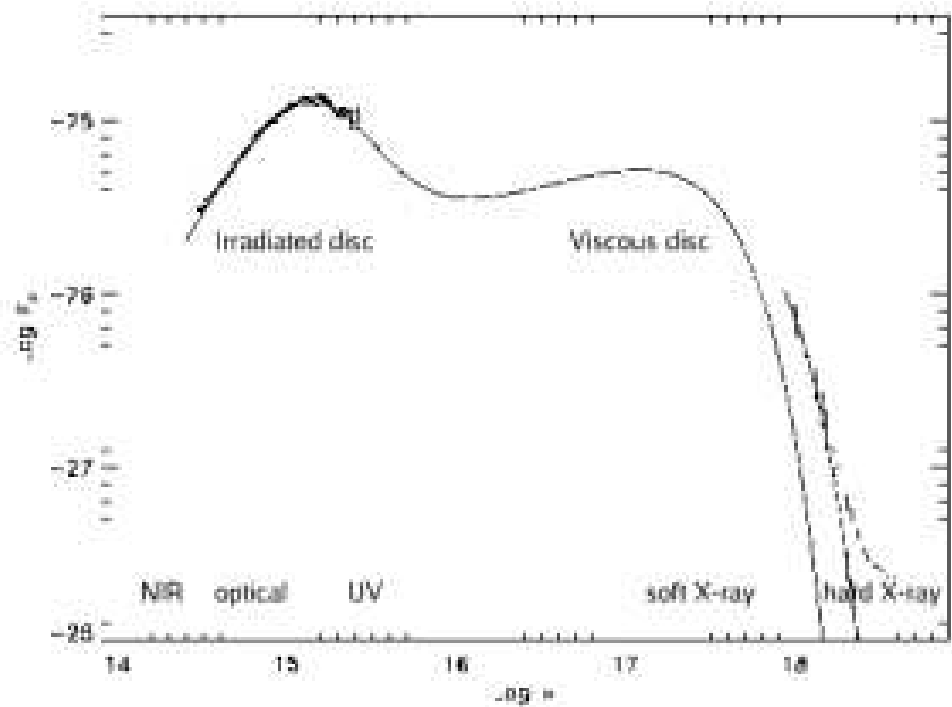}
\caption[The model SED of a multi-temperature XB accretion disc]{The IR to X-ray model SED of a multi-temperature XB accretion disc, fit to NIR/optical/UV/X-ray data from a BHXB \citep[from][]{hyneet02a}.}
\label{intro-hynes2}
\end{figure}

In one BHXB, A0620--00, changes in the NIR light in the soft state was seen to precede the optical, which in turn preceded the X-rays on the viscous timescale (typically days--weeks for BHXBs); an indication of intrinsic disc emission \citep{kuul98,esinet00}.  Similar optical--X-ray \citep*[e.g.][]{soriet99,brocet01a,brocet01b} and NIR--X-ray \citep*[e.g.][]{homaet05a} behaviour has been noted in other BHXBs in the hard and soft states.  The presence of superhumps (periodic variations whose periods are slightly longer than the orbital period; thought to originate in the hot spot; see Fig. \ref{intro-hynes}) during outburst in the hard state \citep*[e.g.][]{katoet95} and behaviour consistent with the Mass Transfer Instability \citep*[MTI;][]{hameet90} model (e.g. \citealt*{callet95,kuul98} and references therein) have also been interpreted as evidence for emission from the viscously heated disc.

A model \citep{hyneet02a} of the broadband SED of an X-ray plus viscously heated disc is shown in Fig. \ref{intro-hynes2}. In this case, the OIR is dominated by the Rayleigh Jeans tail of the blackbody from the X-ray reprocessing component.
\newline\newline
\textbf{Intrinsic thermal emission from the companion star:}\newline
In HMXBs, the characteristic blackbody with stellar absorption lines provides evidence for the domination of the companion star at OIR wavelengths \citep[e.g.][]{trevet80}.  The presence of an ellipsoidal period with the characteristic light curve of a gravitationally distorted companion also implies thermal emission from the star; normal unperturbed stars only vary by very small amplitudes in comparison \citep[e.g.][]{lin77,simpet06}.  These features are sometimes observed in LMXBs in quiescence.  They have been seen in BHXBs in the optical \citep{oke77,mcclet83,bail92,callet92,casaet95,oroset96,garcet96,soriwu99} and the NIR \citep*{shahet99,fronro01,greeet01,geliha03,mikoet05}.  The companion may play a significant role in outburst for BHXBs in which the star is comparatively large \citep*[e.g. GRO J1655--40;][]{hyneet98}.  Large increases or decreases in OIR luminosity cannot be explained by thermal stellar emission from the companion.
\newline\newline
\textbf{Synchrotron emission from a conical jet:}\newline
Optically thick synchrotron emission from partially self-absorbed jets is observed at radio wavelengths in BHXBs as a flat/inverted power-law ($\alpha\geq$ 0) spectrum, when the source is in the hard X-ray state \citep{fend01}.  Models show that a simple `isothermal' conical jet can produce a flat spectrum, where the size scale of the emission region is linearly proportional to the wavelength \citep{blanko79,kais05}.  The jet is optically thick at radio frequencies to relatively large distances (light hours) from the launch region. Evidence has been mounting only in the last 5 or so years, that this flat spectrum may extend to the OIR regime or beyond from spectral \citep{hanet92,fend01,corbet01,brocet02,market03,chatet03,brocet04} and timing \citep{kanbet01,uemuet04} studies.  If the spectrum extends to the OIR, the jet must be self-similar over the same range in physical size, i.e. $\geq$ 5 orders of magnitude \citep{fend01}.  NIR flares in GRS 1915+105 were found to be associated with radio (i.e. jet) ejections \citep*{fendet97,miraro98,eikeet98} and NIR jets have been ambiguously resolved in this source \citep*{samset96}.

The optically thick synchrotron jet spectrum is predicted to break to an optically thin spectrum ($\alpha\sim -$0.6) at a `turnover' at shorter wavelengths \citep{blanko79}.  This turnover is predicted to lie in the IR waveband according to some models (\citeauthor{market03} \citeyear{market01,market03}), which has been tentatively confirmed in observations of one source: GX 339--4 \citep*{corbfe02}.  The position of this turnover is essential in estimating the power in the jets, as the power is dominated by the higher energy photons.  \cite*{nowaet05} have also analysed GX 339--4 and claim that the position of the turnover from optically thick to optically thin emission may decrease in wavelength with decreasing luminosity, as expected in some models.  The NIR spectra of some BHXBs in the hard state in outburst can be fit by a negative power law and suggest an optically thin synchrotron emission origin (e.g. 4U 1543--47; \citealt*{buxtet04,kaleet05} and XTE J1118+480; \citealt{hyneet06b}).

Recently, \cite{homaet05a} extensively studied an outburst of GX 339--4 at OIR wavelengths, and revealed strong evidence that the NIR is dominated by optically thin synchrotron emission from the jet when the source is in the hard state.  A strong drop of the $H$-band flux was observed as the jet was quenched in the soft state, similar to the quenching of the radio emission seen in many BHXBs in the soft state (\citealt*{tanaet72,fendet99a,gallet03}; and perhaps in AGN, see \citealt*{maccet03}).  In XTE J1550--564 and 4U 1543--47, \cite*{jainet01b} and \cite*{buxtet04} (respectively) noted a drop and then rise in the OIR flux (moreso in the NIR than the optical) in transition from and to the hard state respectively, supporting the notion of a jet origin to the flux in the hard state.

Other characteristics of emission from the compact jet include fast variability and a featureless continuum with no emission lines.  Variability and flaring activity (seconds--hours) and a flat/slightly negative OIR spectral index (after subtraction of the companion star spectrum) observed from BHXBs in quiescence \citep{hyneha99,zuriet03,shahet03a,hyneet03c,shahet04,zuriet04} and at higher luminosities in the hard state \citep*{motcet82,bartet94,uedaet02,hyneet03b,hyneet06b} has sometimes been attributed to a non-thermal origin.  In addition, \cite*{malzet04} showed that OIR emission lagging X-rays on timescales $\simlt$ seconds (a behaviour normally attributed to X-ray reprocessing) can be explained by OIR jet synchrotron emission fed by an X-ray corona.  Indeed, many components of XBs can be variable; even in HMXBs the circumstellar discs around Be stars can be extremely variable \citep{schuet07}.  In some BHXBs, correlations between the emission line and continuum intensities during flaring events in quiescence can constrain their origin \citep[e.g.][]{hyneet02c}.  Models also predict a high level of linear polarisation for optically thin synchrotron emission -- this may be the definitive test of optically thin synchrotron emission in these systems (see Section 1.2.4 for further discussion).
\newline\newline
\textbf{Additional emission processes:}\newline
Behaviour that is not consistent with intrinsic disc or reprocessed emission has in the past been attributed to e.g. magnetic loop reconnection in the outer disc \citep*[e.g.][]{zuriet03} or emission from a magnetically dominated compact corona \citep*[e.g.][]{merlet00}, although this corona may be indistinguishable from the base of the compact jet.  Variability and flaring have alternatively been interpreted as emission from an advective region \citep[e.g.][]{shahet03a}.  In addition, Doppler images of Balmer lines sometimes show a stream from the companion hitting the accretion disc at the `hot spot' \citep{marset94,shahet04,torret04}; a further origin of OIR light. In quiescent LMXBs, it is possible that a circumbinary disc that surrounds the whole system can be detected in the IR \citep{munoma06} but recent evidence suggests the observations are explained by the jet \citep{gallet07}. Circumbinary discs are thought to exist around some accreting binaries, for example CVs \citep[e.g.][]{hoaret07}.
\newline\newline
\textbf{Neutron star X-ray binaries:}\newline
In NSXBs, many spectral and timing studies have established the presence of an accretion disc reprocessing X-ray photons to optical wavelengths \citep*[e.g.][]{mcclet79,lawret83,konget00,mcgoet03,hyneet06a}. In quiescence, the companion star can come to dominate the OIR emission \citep*[e.g.][]{thoret78,chevet89,shahet93}, as is the case in HMXBs and many BHXBs. X-ray reprocessing is generally thought to dominate the OIR emission of non-quiescent NSXBs (see \citealt*{vanpet95} for a review; see also \citealt*{chenet97,charco06}). In these systems, matter is accreted onto a neutron star but, like the BHXBs, some of the matter and energy can be released from the system through jets.  Evidence for jets associated with NSXBs date back more than a decade \citep*{stewet93,bradet99,fomaet01} but not until recently has the evidence emerged from any waveband other than the radio. \cite*{callet02} found that IR $K$-band flaring in the Z-source NSXB GX 17+2 could not be explained by an X-ray driven wind or reprocessed X-rays, but shared many properties with the radio (i.e. jet) variability previously seen in the same source. It is worth noting that the NIR counterpart of GX 13+1 is also largely variable \citep{charna92}. Recently, \cite*{miglet06} for the first time spectrally detected optically thin synchrotron emission from the compact jets of an atoll NSXB 4U 0614+09 in the mid-IR.

Recently, an anomalous transient NIR excess has been observed in a number of MSXPs at high luminosities, which is equivocal in nature. The source most studied is SAX J1808.4--3658, for which \cite*{wanget01} found a NIR flux almost one order of magnitude too bright to originate from X-ray heating. The NIR flux density was comparable to a radio detection of 0.8 mJy (with a flat 2.5--8.6 GHz spectrum) seen one week after the NIR excess. \cite*{greeet06} also reported an $I$-band excess in a different outburst of the same source, which they attributed to synchrotron emission.  In addition, a variable $I$- and $R$-band excess in XTE J0929--314 seen on the same day as a radio detection \citep*{gileet05}, and a transient NIR excess in XTE J1814--338 \citep*{krauet05} and IGR J00291+5934 \citep{torret07} were all interpreted as synchrotron emission from the steady jets in the systems. The NIR excess appears to be ubiquitously absent at lower luminosities.

Steady, partially self-absorbed jets probably exist in low-magnetic field ($B \simlt 10^{\sim 11}$ G) NSXBs in hard X-ray states \citep*{miglfe06,mass06}. These include atolls in the `island' state, Z-sources in the `horizontal branch' and possibly the `normal branch' and transients at low accretion rates ($L_{\rm X}\simlt 0.1 L_{\rm Edd}$) such as MSXPs.  \cite{mass06} argues on theoretical grounds that the existence of jets in NSXBs depends on the magnetic field and mass accretion rate, and the conditions required for jet ejection are probably fulfilled for most of the NSXBs (Z-sources, atolls and MSXPs) with known magnetic field strengths, further supporting the existence of jets in theses systems.

\subsubsection{Towards a unified model for the OIR behaviour}

Power-law correlations between OIR and X-ray luminosities are naturally expected from a number of emission processes. \cite{vanpet94} showed that the optical luminosity of an X-ray reprocessing accretion disc varies as $L_{\rm OPT}\propto T^2 \propto L_{\rm X}^{0.5}a$, where $T$ is the temperature and $a$ is the orbital separation of the system, and that this correlation has been observed in a selection of LMXBs. $L_{\rm OIR}$--$L_{\rm X}$ correlations are also expected when the OIR originates in the viscously heated disc as both X-ray and OIR are linked through the mass accretion rate (see Section 2.4.2).

In addition, OIR--X-ray correlations can be predicted if the OIR emission originates in the jets. Models of steady, compact jets demonstrate that the total jet power is related to the radio luminosity as $L_{\rm radio}\propto L_{\rm jet}^{1.4}$ \citep{blanko79,falcbi96,market01,heinsu03}. It was shown that the jet power is linearly proportional to the mass accretion rate in NSXBs and BHXBs in the hard state \citep*{falcbi96,miglfe06,kordet06} and the X-ray luminosity scales as $L_{\rm X}\propto$ \.m and $L_{\rm X}\propto$ \.m$^{\sim 2}$ for radiatively efficient and inefficient objects, respectively \citep[e.g.][Section 1.1.4]{shaksu73,narayi95,maha97,kordet06}. The accretion in hard state BHXBs is found to be \emph{radiatively inefficient} (the majority of the liberated gravitational potential is carried in the flow and not radiated locally), where jet-dominated states can exist, whereas in NSXBs, the accretion is \emph{radiatively efficient}, and jet-dominated states are unlikely to exist \citep[see also][]{fendet03}. We therefore have:

\smallskip BHXBs: $L_{\rm radio}\propto L_{\rm jet}^{1.4}\propto $ \.m$^{1.4}\propto L_{\rm X}^{0.7}$

\smallskip NSXBs: $L_{\rm radio}\propto L_{\rm jet}^{1.4}\propto $ \.m$^{1.4}\propto L_{\rm X}^{1.4}$
\newline\newline
The correlation for BHXBs has been observed \citep*{corbet03,gallet03} and very recently, \cite*{miglfe06} have applied this technique to NSXBs and found $L_{\rm radio}\propto L_{\rm x}^{\ge1.4}$; which is also consistent with the above NSXB model. If the optically thick jet spectrum is indeed flat from the radio regime to OIR, we can expect the following correlations:

\smallskip BHXBs: $L_{\rm OIR}\propto L_{\rm radio}\propto L_{\rm X}^{0.7}$

\smallskip NSXBs: $L_{\rm OIR}\propto L_{\rm radio}\propto L_{\rm X}^{1.4}$
\newline\newline
\cite{homaet05a} discovered a correlation between the quasi-simultaneous NIR (which was shown to originate in the jets) and X-ray fluxes for GX 339--4 in the hard state, with a slope $F_{\rm NIR}\propto F_{\rm X}^{0.53\pm 0.02}$ (3--100 keV). To date, no other sources have been tested for jet OIR emission using OIR--X-ray correlations.

It is now becoming clear that this profitable but simple technique of analysing the dependence of OIR and X-ray luminosities over many orders of magnitude, may prove fruitful for the understanding of the emission mechanisms involved. The unification of jet--X-ray state activity is now underway; a steady jet exists in the hard state, which is accelerated as the X-ray spectrum softens, and is finally quenched as it passes the `jet line' into the soft state \citep[][see Section 1.1.4]{fendet04}. A unification (if one exists) of the origins of OIR light from BHXBs and NSXBs in different spectral and luminosity states is desired to understand the behaviour of these systems. Furthermore, a measure of the level of OIR emission from jets may be used to constrain jet power estimates.

\subsubsection{Polarised OIR emission}

A further dimension of information is available if we include polarimetry. Polarimetric capabilities are generally possible at radio, OIR and UV regimes but not in X- or $\gamma$-rays \citep[although see][]{willba05}. OIR polarimetric studies can provide information about the physical conditions of LMXBs and inner accretion flow. Most radiation from LMXBs is expected to be unpolarised, for example thermal blackbody radiation from the accretion disc or companion star. The scattering of unpolarised photons could result in a small degree of net polarisation in certain geometries \citep[e.g.][]{dola84}. There is one emission mechanism known to be present in LMXBs that intrinsically produces polarised light -- synchrotron emission (coherent emission from e.g. radio pulsars can be highly polarised but has not been observed in LMXBs). It has been known for decades that optically thin synchrotron radiation can produce a high level (tens of percent) of linear polarisation if the magnetic field structure is ordered \citep[e.g.][]{west59,bjorbl82}.

Linear polarisation (LP) of BHXBs is measured at radio frequencies at a level of $\sim 1$--3\% in a number of sources in the hard state, and up to $\sim 30$\% during transient radio events associated with X-ray state transitions and jet ejections (for a review see \citealt{fend06}). During transient jet ejections the synchrotron spectrum is optically thin, with a negative spectral index $\alpha$. For optically thin synchrotron emission, a strong LP signal is expected, of order 70\% $\times$ $f$, where $f$ can be considered to crudely parameterise the degree of ordering of the large scale magnetic field \citep{rybili79,bjorbl82}. The high polarisation levels measured in the radio from these optically thin ejections have indicated that $f$ may be as large as 0.5, and have provided clues towards the fundamental jet and magnetic field properties \citep{fendet99b,hannet00,gallet04,brocet07}.

\begin{figure}
\centering
\includegraphics[width=14cm,angle=0]{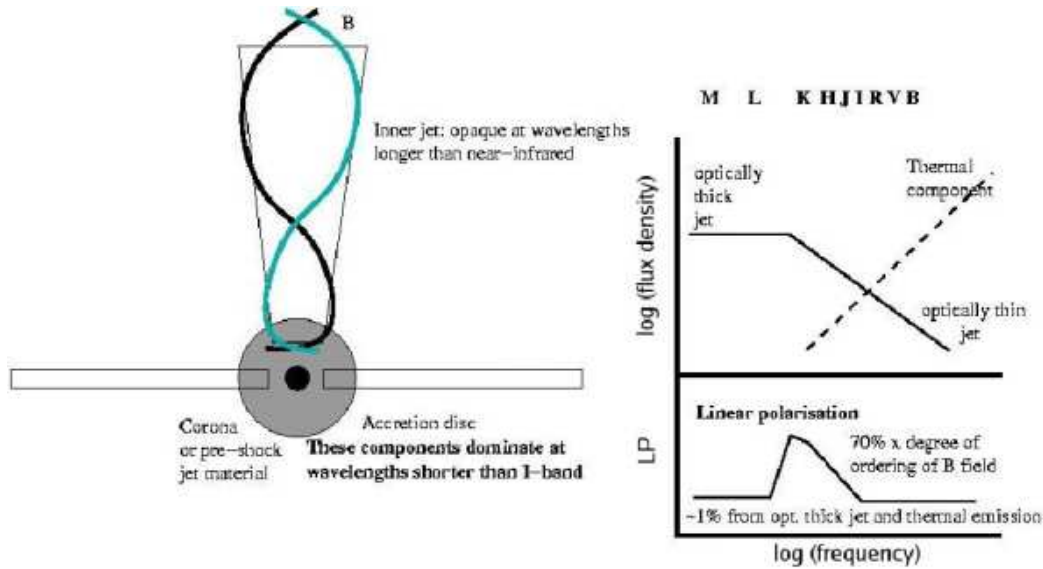}
\caption[Schematic of the cross section of the inner disc/jet region, and the expected polarisation as a function of frequency]{Schematic of the cross section of the inner disc/jet region (left) and the expected LP as a function of frequency (right). Figure courtesy of Rob Fender.}
\label{intro-pol}
\end{figure}

In the hard state which has a steady, compact jet, the higher frequency photons from the jet are emitted in its inner regions (in the absence of shocks downstream), closest to the compact object (see the left panel of Fig. \ref{intro-pol}). In the optical regime the reprocessed light from the X-ray illuminated accretion disc should be no more than $\sim 1$\% polarised, much like the optically thick jet. In the NIR, a strong polarised signal from the optically thin synchrotron emission is expected. The suggested expected LP as a function of frequency is presented in Fig. \ref{intro-pol} (right panel). It is currently uncertain whether the jet component dominates the NIR at low luminosities (near quiescence); in most systems the companion star comes to dominate \citep[e.g.][]{charco06} but in some it does not \citep*[e.g. GX 339--4;][]{shahet01}. If the value of $f$ is high, a polarised signal from the jet component should be detectable. Higher levels of LP ($\sim 5$\% in Cyg X--1 and up to $\sim 20$\% in SS 433), which are variable, have been detected in the UV \citep{woliet96,dolaet97} and result from a combination of Thomson scattering  (whereby charged particles scatter the light) and Rayleigh scattering (scattering by particles smaller than the wavelength of the light).

Radio polarimetry has shown that in some cases, the magnetic field is fairly ordered in the optically thin transient jet ejections. When the radio spectrum is optically thick, LP is detected at a level of a few percent; lower than from the optically thin spectrum. Consequently, if a high level of LP is observed from the hard state IR optically thin spectrum, the magnetic field at the base of the \emph{steady} hard state jet must also be ordered. Therefore IR LP is key to understanding the conditions of the inner regions of the steady jet flow, in particular the level of ordering of the magnetic field. In addition, the higher frequency IR photons do not suffer from Faraday rotation, which can confuse radio results.

In the optical regime of LMXBs, just two sources (A0620--00 and GRO J1655--40) possess intrinsic LP to my knowledge \citep{dolata89,glioet98}. The LP varies as a function of orbital phase and is likely caused by the scattering of intrinsically unpolarised thermal emission \citep[e.g.][]{dola84}. No intrinsic LP has been detected from optical observations of NSXBs, except for tentatively in Aql X--1 \citep{charet80}.

In 2006, \citeauthor{dubuch06} were the first to report IR polarimetric observations of LMXBs. They found no evidence for intrinsic LP in H1743--322 in outburst or GRO J1655--40 in quiescence, but did find significant (at the 2.5$\sigma$ level) LP (which is probably intrinsic) in XTE J1550--564 during a weak X-ray outburst. No polarised standard star was observed so the authors were unable to calibrate the polarisation position angle (PA), however if the calibration correction is small then PA $\sim 10^\circ$, which is perpendicular to the known jet axis \citep{corbet02}. The polarisation PA is a measure of the electric vector, which for optically thin synchrotron emission, is perpendicular to the magnetic field vector. Therefore in XTE J1550--564 the magnetic field may be parallel to the jet. Very recently, \cite{shahet07} performed IR spectropolarimetry of three LMXBs and found two of them (the NSXBs Sco X--1 and Cyg X--2) to be intrinsically polarised, with an increasing LP at lower frequencies. They interpret this as the first detection of the polarised inner regions of the compact jets.

It is interesting to note that polarimetric observations of jets from AGN (which are resolved and are of course orders of magnitude larger and more powerful than X-ray binary jets) have revealed a strong link between the local magnetic field and the dynamics of the jet. LP levels of $>20$\% are observed in the optical and radio, confirming the emission is synchrotron, and the levels and position angles are often correlated with intensity and morphology \citep[for overviews see][]{saiksa88,perlet06}. In addition, NIR flares from the black hole at the centre of our Galaxy, Sgr A$^\ast$, are also highly polarised ($\sim 10$--20\%) and may originate in its jets \citep{eckaet06,meyeet06}.

\subsubsection{Extended OIR emission associated with X-ray binaries}

\begin{figure}
\centering
\includegraphics[width=14cm,angle=0]{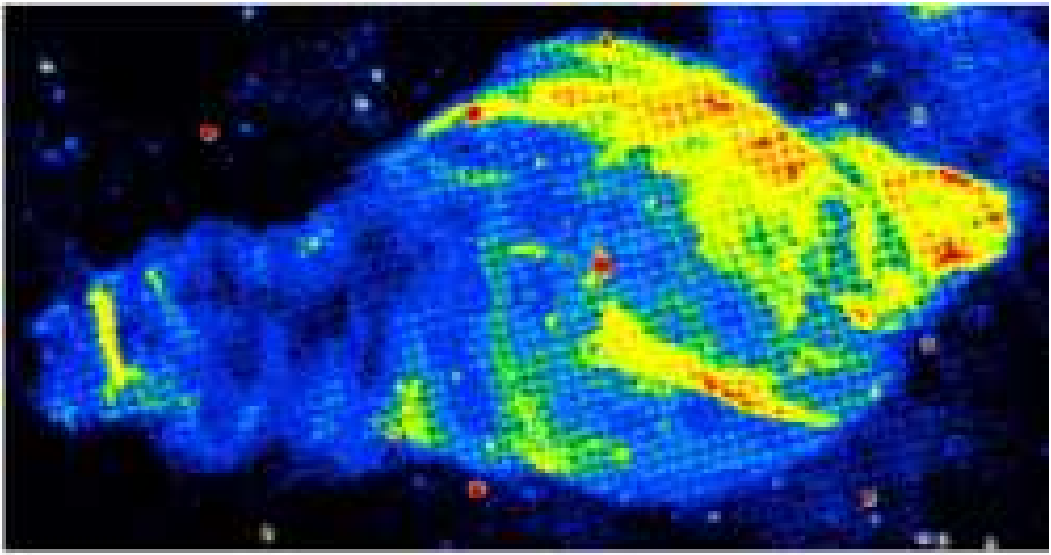}
\caption[The W50 nebula surrounding SS 433]{Combined radio pseudocolour representation of the W50 nebula surrounding SS 433 \citep[from][]{dubnet98}. The approximately circular structure is the supernova remnant from the explosion of the progenitor; the resolved radio jets of SS 433 are thought to be interacting with this nebula causing it to appear elongated to the east and west.}
\label{intro-ss433}
\end{figure}

In addition to direct light produced within the XB system, there are ways in which XBs can indirectly produce emission \emph{outside} the system. The progenitor of the compact object is responsible for the supernova which results in a shell of expelled matter; the supernova remnant (SNR), and some XBs have been found to be associated with these large, extended structures \citep[e.g. SS 433 and the W50 nebula;][see Fig. \ref{intro-ss433}]{spen79,dubnet98}.

XBs themselves release both energy and matter into the surrounding ISM via photons, jets, disc winds, stellar winds and high-energy particles (cosmic rays). The disruption and interaction of these sources of energy with the local environment could cause detectable emission. High energy photons from the X-ray source can ionise the surrounding gas and if the local density is high enough, an X-ray-ionised nebula can be seen \citep[e.g. LMC X--1;][]{pakuan86} in optical line emission. XB jets plough into the local ISM and if again the local density is high, shock waves are formed which can be detected in the radio and optical \citep{gallet05} in front of radio lobes \citep[e.g.][]{miraet92}.

The power carried by the steady, hard state jets when averaged over their lifetime stands as one of the key questions in our understanding of BHXB accretion, and possibly for black hole accretion on all mass scales. In recent years it has become apparent that a significant fraction of the liberated accretion power is ejected as jets in hard state BHXBs \citep[e.g.][]{gallet03}, and it now seems likely that in hard X-ray states the accretion flow is `jet-dominated' \citep*{fendet03}, with more power in the jets than in X-rays. Measuring as accurately as possible the power of the compact jet is key to understanding both the overall physics of the accretion process and the matter and energy input from BHXBs into the ISM. The jets may be the strongest ISM-energising sources produced by XBs.

While the presence of the hard state jets is commonly inferred from the flat radio-through-IR spectral component, radio observations of the nearby \citep*[d = 2.1$\pm$0.1 kpc;][]{masset95} high-mass BHXB Cygnus X--1 have directly resolved a collimated jet $\sim$30 AU in length \citep{stiret01}, whilst in the hard state. A transient radio jet has also been observed of length $\sim$140 AU in this source, that was launched during a period of X-ray state transitions \citep{fendet06}. Attempts at estimating the jet power content from core radio luminosities of the hard state jets are riddled with assumptions about its spectrum and radiative efficiency, the latter of which is poorly constrained \citep[e.g.][]{ogleet00,fend01,homaet05a,hein06}. The radiative efficiency, although uncertain is estimated to be low: typically $\sim 5$ percent \citep{ogleet00,fend01}. The jet power estimated in this way is highly sensitive to the location of the high-frequency break of the flat (spectral index $\alpha \approx 0$) optically thick part of the jet spectrum, as the radiative power is dominated by the higher energy photons \citep{blanko79}.

The jet power may alternatively be constrained by analysing its interaction with the surrounding medium, without requiring prior knowledge of the jet spectrum and radiative efficiency. Radio lobes associated with jets from AGN are commonly used as accurate calorimeters of the \emph{power $\times$ lifetime} product of the jets \citep{burb59}, a method only very recently applied to jets from stellar mass black holes. Radio lobes have been identified and associated with an increasing number of BHXBs \citep{miraet92,rodret92,corbet02} and a couple of NSXBs \citep*{fomaet01,tudoet06}.

\begin{figure}
\centering
\includegraphics[width=10cm,angle=270]{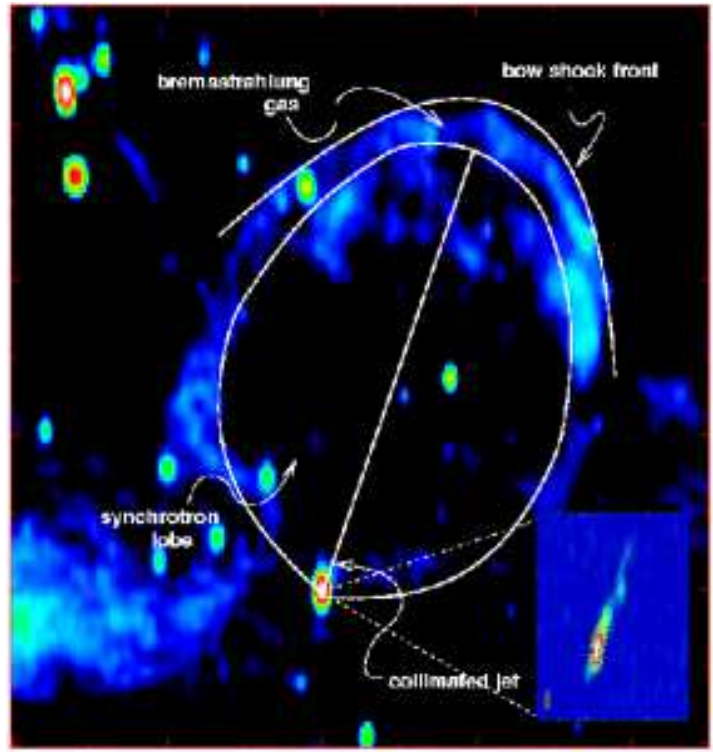}
\caption[Radio image of the jet-powered nebula of Cyg X--1]{Radio image of the jet-powered nebula of Cyg X--1 \citep[from][]{gallet05}.}
\label{intro-gallo}
\end{figure}

Recently, deep radio observations of the field of Cyg X--1 resulted in the discovery of a shell-like structure which is aligned with the aforementioned resolved radio jet \citep{gallet05}. This \emph{jet-blown nebula} with a diameter of $\sim 5$ pc is interpreted as the result of a strong shock that develops at the location where the collimated jet impacts on the ambient ISM (Fig. \ref{intro-gallo}). The nebula was subsequently observed at optical wavelengths by the Isaac Newton Telescope (INT) Wide Field Camera (WFC) and is clearly visible in the H$\alpha$ (6568 \AA) filter, and coincident with the radio shell (see Chapter 6). The flux density $F_\nu \geq 0.02$ mJy arcsec$^{-2}$ is $\geq 20$ times the measured radio flux density, inconsistent with optically thin synchrotron emission between the two spectral regimes and instead favouring a thermal plasma with H$\alpha$ line emission \citep{gallet05}.

Models of X-ray binary jet--ISM interactions predict a shell of shocked compressed ISM emitting bremsstrahlung radiation containing a bubble of relativistic plasma emitting synchrotron radiation  \citep{kaiset04}. As with any interstellar shock wave where the preshock gas is at least partially ionised by the approaching radiation field, line emission from recombination of the shocked gas is also expected \citep[e.g.][]{cox72}. The shell, which is essentially a radiative shock wave, consists of a bow shock front where the gas is in transition to a higher temperature due to collisions and plasma instabilities \citep[e.g.][]{mckeho80}. The post-shocked gas then enters the optically thin `cooling region', where the overall ionisation level rises then falls as the gas radiatively cools to $\sim 10^4$ K \citep[e.g.][]{coxra85}. Shock waves are commonly observed in SNRs, and shocks created from outflows exist in Herbig-Haro objects, where emission lines are produced in the shock wave created by bipolar flows from pre-main-sequence stars impacting the ISM \citep[e.g.][]{schw83}. Shock fronts associated with AGN jets interacting with the intra-cluster gas \citep[which is much hotter than the ISM; e.g.][]{fabi94,edge01,popeet06} are also seen at X-ray energies \citep*{cariet94,krafet03,formet05,nulset05,wilset06,goodet07}. In addition, IR sources found symmetric about the position of the XB GRS 1915+105 \citep{chatet01} may be jet--ISM impact sites \citep{kaiset04}. The IR does not suffer from interstellar dust extinction as much as the optical, so the continuum bremsstrahlung spectrum could be detected more easily in this regime.

By modelling the shell emission of the Cyg X--1 nebula as originating from radiatively shocked gas \citep*{castet75,kaisal97,heinet98}, the total power carried by the compact jet was estimated to be $\sim 9 \times 10^{35}\leq P_{\rm Jet} \leq 10^{37}$ erg s$^{-1}$ which, when taking into account the power of the counter jet, is equal to $\sim $0.06--1 times the bolometric X-ray luminosity of Cyg X--1 \citep{gallet05}. These calculations have led to estimates of the matter content of the jet \citep{hein06}; a similar technique to those applied to the jets of AGN, where their matter content are inferred from a combination of information from the core and lobes \citep*[e.g.][]{celofa93,dunnet06}.

These jet power calculations are highly sensitive to the velocity of the gas in the shock front. From temperature constraints and from the non-detection of an X-ray counterpart, this was estimated to be $20 \leq v_{\rm s} \leq 360$ km s$^{-1}$. Optical emission line ratios of shock-heated gas can constrain its parameters, including the velocity of the shock \citep[e.g.][]{oste89}.

It seems that two methods can be used to constrain the power of the compact jets using OIR emission. Firstly, OIR core emission can constrain the position of the flat spectrum high-frequency break and hence the broadband spectrum of the jet. Secondly, indirect OIR emission (in particular optical emission line ratios and continuum IR imaging) from jet-powered shock waves can constrain the velocity or the shock and hence the time-averaged jet power.

\subsection{Brief summary of projects}

In the following Chapters, the projects I have carried out are described. The general aim of the projects is to help answer some of the outstanding questions detailed in Chapter 1. Chapters 2--5 are related to the nature of the core OIR emission from LMXBs. In Chapter 2, correlation analysis is used to constrain the dominating OIR emission mechanisms in BHXBs and NSXBs; the results have implications for the spectrum and power of the jets. These are further constrained in Chapter 3 by analysing OIR SEDs. In Chapter 4 the emphasis is on the discovery of a jet--X-ray hysteresis; the data also reveal the dominating NIR emission processes. Linear polarimetry is used to probe the local conditions of the inner jets in Chapter 5; again the results have implications for the OIR emission.

Chapter 6 describes the analysis of optical imaging of the XB jet-powered nebula of Cyg X--1. The velocity of the shocked gas powered by the jet of Cyg X--1 is estimated, which is used to infer the time-averaged jet power.
Finally, a summary of the results and possible future projects are given in Chapter 7; by including literature by other authors a general picture seems to emerge.

\newpage

\begin{center}
{\section{Relations between OIR and X-ray emission}}
\end{center}

\subsection{Introduction}

The OIR region of the spectra of low-mass X-ray binaries appears to lie at the intersection of a variety of different emission processes. In this Chapter I present a comprehensive collection of quasi-simultaneous OIR--X-ray observations of 43 XBs mostly taken from the literature, with a few new observations. Correlation analysis between the OIR and X-ray luminosities is used in conjunction with radio luminosities to estimate the contributions of various emission processes in these sources, as a function of X-ray state and luminosity. The results are used to test the predictions mentioned in Section 1.2.3. Relations between X-ray luminosity and the shape of the OIR spectrum are also examined for NSXBs (for a detailed study of the OIR spectrum in both black hole and neutron star systems see Chapter 3). These results have been published in two papers; Russell et al. (2006; hereafter \citeauthor{paper1}) concentrates on the BHXBs and Russell, Fender \& Jonker (2007; hereafter \citeauthor{paper4}) concentrates on the NSXBs.

In this Chapter, the data collection and the majority of the text, presentation and analysis is my own work, and the co-authors of the papers made various contributions. The new data (Sections 2.2.3 and 2.2.4) were reduced by myself (Peter Jonker made the Danish 1.54-m Telescope observations in Chile). The idea for the project came from my supervisor Rob Fender, and ideas from Peter Jonker (in particular bringing to my attention the NIR excess seen in a number of MSXPs), myself and Rob led to the second publication (\citeauthor{paper4}). All of the co-authors (Rob Fender, Peter Jonker, Rob Hynes, Catherine Brocksopp, Jeroen Homan and Michelle Buxton) contributed fully-reduced observational data, plus Erik Kuulkers who kindly contributed data of one source A0620--00 and Elena Gallo who provided radio and X-ray data from \cite{gallet03}. Rob Hynes contributed the theoretical models for OIR--X-ray correlations in viscously heated discs. All of the co-authors gave valuable comments and suggested changes which included a substantial re-ordering of \citeauthor{paper1} (which is partially reflected in this Chapter). Use of the word `we' is adopted in this Chapter when referring to work not done by myself alone.

\subsection{Methodology}

\subsubsection{Data Collection}

For this work, I have collected a wealth of OIR and X-ray data from a large number of BHXBs, low-magnetic field NSXBs (atolls, Z-sources and MSXPs) and HMXBs in order to find relations that may help determine the processes responsible for the OIR light in these systems. I apply the technique of testing the dependency of OIR luminosity with X-ray luminosity for the three types of XB and between different X-ray states in BHXBs, and attempt to identify the dominant luminosity-dependent emission mechanisms.

A literature search for quasi-simultaneous (no more than $\sim$ 1 day between observations; for sources with outbursts $\leq$10 days in length we only use data with separations of $\leq$0.1 times the outburst length, since a few outbursts are very short, e.g. V4641 Sgr) X-ray and OIR fluxes from BHXBs, NSXBS and HMXBs was conducted. Quasi-simultaneity was achieved in many cases with use of the \emph{RXTE} ASM X-ray daily monitoring. I used non-simultaneous OIR--X-ray luminosities only in quiescence for some sources, and for these we have included errors that encompass all observed values of the quiescent flux in one of the two wavebands. Where possible, tabulated fluxes or magnitudes were noted. In some cases I obtained data directly from the authors. I also made use of the \emph{DEXTER} applet provided by \emph{NASA ADS} to extract data from light curves where the data themselves were unattainable. For each source, the best estimates of its distance, optical extinction $A_{\rm V}$ and HI absorption column $N_{\rm H}$ were sought. The X-ray state of each BHXB was also noted for all data collected (but we do not discriminate between different X-ray states for NSXBs or HMXBs). The properties of each source are listed in Tables \ref{tab-16BHXBs1} \& \ref{tab-16BHXBs2} (BHXBs), \ref{tab-19NSXBs-1} \& \ref{tab-19NSXBs-2} (NSXBs), and \ref{tab-9HMXBs1} \& \ref{tab-9HMXBs2} (HMXBs).

\begin{table*}
\begin{center}
\caption[Properties and data collected for the 16 BHXBs (1)]{Properties and data collected for the 16 BHXBs (1).}
\label{tab-16BHXBs1}
\small
\begin{tabular}{lllll}
\hline
Source name      	     &Distance             &Period                  &$M_{\rm BH}$ / $M_\odot$&$M_{\rm cs}$ / $M_\odot$\\
= alternative name           &/ kpc (ref)          &/ hrs (ref)             &(ref)                   &(ref)                   \\
(I)                          &(II)                 &(III)                   &(IV)                    &(V)                     \\
\hline
M31 r2--70                   &784$\pm$30 (1)       &192$^{+290}_{-120}$ (12)&--                      &--                      \\
GRO J0422+32 = V518 Per      &2.49$\pm$0.30 (2)    &5.09 (5)                &3.97$\pm$0.95 (15)      &0.46$\pm$0.31 (15)      \\
LMC X--3                     &50$\pm$10 (3, 4)     &$\sim$40.8 (13)         &$\sim$9--10 (16)        &$\sim$4--8 (16)         \\
A0620--00    = V616 Mon      &1.2$\pm$0.4 (5)      &7.75 (5)                &11.0$\pm$1.9 (17)       &0.74$\pm$0.13 (17)      \\
XTE J1118+480 = KV UMa       &1.71$\pm$0.05 (6)    &4.08 (5)                &6.8$\pm$0.4 (15)        &0.28$\pm$0.05 (15)      \\
GRS 1124--68 = GU Mus        &5.5$\pm$1.0 (5)      &10.4 (5)                &6.0$^{+1.5}_{-1.0}$ (18)&0.80$\pm$0.11 (18)      \\
GS 1354--64  = BW Cir        &$\geq$27 (7; we      &61.1 (7)                &$>$7.83$\pm$0.50    &$>$1.02$\pm$0.06            \\
                             &use 33$\pm$6)        &                        &(15)                    &(15)                    \\
4U 1543--47 = IL Lup         &7.5$\pm$0.5 (5)      &26.8 (5)                &9.4$\pm$1.0 (15)        &2.45$\pm$0.15 (15)      \\
XTE J1550--564 = V381 Nor    &5.3$\pm$2.3 (5)      &37.0 (5)                &10.6$\pm$1.0 (19)       &1.30$\pm$0.43 (19)      \\
GRO J1655--40                &3.2$\pm$0.2 (5)      &62.9 (5)                &7.02$\pm$0.22 (20)      &2.35$\pm$0.14 (20)      \\
= Nova Sco 1994              &                     &                        &                        &                        \\
GX 339--4 = V821 Ara         &8$^{+7.0}_{-1.0}$ (8)&42.1 (5)                &$\sim$5.8 (21)          &$\sim$0.52 (21)         \\
GRO J1719--24                &2.4$\pm$0.4 (9)      &14.7 (14)               &$\sim$4.9 (9)           &$\sim$1.6 (9)           \\
= Nova Oph 1993              &                     &                        &                        &                        \\
XTE J1720--318               &8$^{+7}_{-5}$ (10)   &--                      &--                      &--                      \\
= INTEGRAL1 51               &                     &                        &                        &                        \\
XTE J1859+226 = V406 Vul     &6.3$\pm$1.7 (5)      &9.17 (5)                &5--12 (22)              &0.68--1.12 (22)         \\
GRS 1915+105 = V1487 Aql     &9.0$\pm$3.0 (11)     &816 (5)                 &14.0$\pm$4.4 (15)       &0.81$\pm$0.53 (15)      \\
V404 Cyg = GS 2023+338       &4.0$^{+2.0}_{-1.2}$ (5)&155.28 (5)            &10.0$\pm$2.0 (15)       &0.65$\pm$0.25 (15)      \\
\hline
\end{tabular}
\end{center}
\normalsize
Columns give:
(I) source names;
(II) distance estimate;
(III) orbital period of the system;
(IV) mass of the black hole in solar units;
(V) mass of the companion star in solar units.
References for tables \ref{tab-16BHXBs1} to \ref{tab-9HMXBs2}:
(1) \cite{stanga98};
(2) \cite{geliha03};
(3) \cite{boydet00};
(4) \cite{kova00};
(5) \cite{jonket04};
(6) \cite{chatet03};
(7) \cite{casaet04};
(8) \cite{zdziet04};
(9) \cite{maseet96};
(10) \cite{cadoet04};
(11) \cite{chapco04};
(12) \cite{willet05};
(13) \cite{hutcet03};
(14) \cite{liuet01};
(15) \cite{rittet03};
(16) \cite{cowlet83};
(17) \cite{geliet01};
(18) \cite{esinet00};
(19) \cite{oroset02};
(20) \cite{hyneet98};
(21) \cite{hyneet03a};
(22) \cite{hyneet02a};
(23) \cite{shraet97};
(24) \cite{soriet01};
(25) \cite{wilmet01};
(26) \cite{wanget05};
(27) \cite{konget02};
(28) \cite{ebiset94};
(29) \cite{kitaet90};
(30) \cite{zdziet98};
(31) \cite{tana93};
(32) \cite{hyne05};
(33) \cite{mcclet95};
(34) \cite{torret04};
(35) \cite{oroset96};
(36) \cite{oroset98};
(37) \cite{greeet01};
(38) \cite{shahet01};
(39) \cite{zuriet02};
(40) \cite{casaet93};
(continued over).
\normalsize
\end{table*}

\begin{table*}
\begin{center}
\caption[Properties and data collected for the 16 BHXBs (2)]{Properties and data collected for the 16 BHXBs (2).}
\label{tab-16BHXBs2}
\small
\begin{tabular}{lllll}
\hline
Source            &$A_{\rm V}$, $N_{\rm H}$ / 10$^{21} cm^{-2}$  &$q_{\rm cs}$                   &$\Delta t$ /&Fluxes -- data    \\
                  &(refs)                                        &(band, ref)                    &days        &references        \\
(I)               &(II) 	                                 &(III) 		         &(IV)        &(V)               \\
\hline
M31 r2--70        &1.0$\pm$0.3, 1.8$\pm$0.5 (12)                 &--                             &2.0         &12                \\
GRO J0422+32      &0.74$\pm$0.09, 1.6$\pm$0.4 (2, 23)            &0.76$^{+0.04}_{-0.46}$ (R, 2)  &0.5         &2, 41, 42         \\
LMC X--3          &0.19$\pm$0.03, 0.32$^{+0.31}_{-0.07}$ (24--26)&--                             &1.0         &43, 44            \\
A0620--00         &1.17$\pm$0.08, 2.4$^{+1.1}_{-1.0}$ (5 ,27)    &0.58$^{+0.25}_{-0.22}$ (V, 33) &1.0         &33, 45            \\
XTE J1118+480     &0.053$^{+0.027}_{-0.016}$, 0.11$\pm$0.04 (6)  &0.55$^{+0.15}_{-0.28}$ (R, 34) &1.0         &6, 41--43, 46--49 \\
GRS 1124--68      &0.9$\pm$0.1, 1.58$^{+0.42}_{-0.58}$ (5, 28)   &0.55$\pm$0.05 (B--V, 35)       &--          &28, 50--52        \\
GS 1354--64       &2.60$\pm$0.31, 37.2$^{+14}_{-7}$ (7, 29)      &--                             &1.0         &43, 53            \\
4U 1543--47       &1.55$\pm$0.15, 4.3$\pm$0.2 (5)                &0.68$^{+ 0.11}_{-0.07}$ (R, 36)&1.0         &43, 54, 55        \\
XTE J1550--564    &2.5$\pm$0.6, 8.7$\pm$2.1 (5)                  &0.7$\pm$0.1 (V, 19)            &1.0         &19, 56, 57        \\
GRO J1655--40     &3.7$\pm$0.3, 6.66$\pm$0.57 (5, 20)            &$\sim$1.0 (B--K, 37)           &1.0         &20, 43, 58--60    \\
GX 339--4         &3.9$\pm$0.5, 6.0$^{+0.9}_{-1.7}$ (5, 30)      &$\leq$0.3 (B--K, 38)           &1.0         &43, 60--64        \\
GRO J1719--24     &2.8$\pm$0.6, 4.0$^{+0.0}_{-2.6}$ (9, 31, 32)  &--                             &0.5         &41                \\
XTE J1720--318    &6.9$\pm$0.1, 12.4$\pm$0.2 (10$^{\ast}$)       &--                             &1.0         &43, 65            \\
XTE J1859+226     &1.80$\pm$0.37, 8$\pm$2 (5)                    &0.59$\pm$0.04 (R, 39)          &1.0         &22, 39, 43, 66, 67\\
GRS 1915+105      &19.6$\pm$1.7, 35$\pm$3 (11)                   &--                             &1.0         &68                \\
V404 Cyg          &3.65$\pm$0.35, 6.98$\pm$0.76 (5, 27)          &0.87$\pm$0.03 (R, 40)          &0.5         &69--71            \\
\hline
\end{tabular}
\end{center}
\normalsize
Columns give:
(I) source name;
(II) interstellar reddening in $V$-band, and interstellar HI absorption column ($^{\ast}A_{\rm V}$ is estimated here from the relation $N_{\rm H} = 1.79 \times 10^{21} cm^{-2}A_{\rm V}$; \citealt{predet95});
(III) the companion star OIR luminosity contribution in quiescence;
(IV) The maximum time separation, $\Delta t$, between the OIR and X-ray observations defined as quasi-simultaneous;
(V) References for the quasi-simultaneous OIR and X-ray fluxes collected.
References for tables \ref{tab-16BHXBs1} to \ref{tab-9HMXBs2} (continued):
(41) \cite{brocet04};
(42) \cite{garcet01};
(43) \emph{RXTE} ASM;
(44) \cite{brocet01a};
(45) \cite{kuul98};
(46) \cite{mcclet03};
(47) \cite{kiziet05};
(48) \cite{uemuet00};
(49) \cite{hyneet05};
(50) \cite{kinget96};
(51) \cite{dellet98};
(52) \cite{sutaet02};
(53) \cite{brocet01b};
(54) \cite{buxtet04};
(55) \cite{kaleet05};
(56) \cite{hameet03};
(57) \cite{jainet01b};
(58) \cite{markwet05};
(59) \cite{torret05};
(60) \cite{chatet02};
(61) \cite{corbfe02};
(62) \cite{homaet05a};
(63) \cite{kuulet04};
(64) \cite{israet04};
(65) \cite{nagaet03};
(66) \cite{tomset03a};
(67) \cite{haswet00};
(68) \cite{fendpo00};
(69) \cite{hyneet04};
(70) \cite{zycket99};
(71) \cite{hanet92};
(72) \cite{torret07};
(73) \cite{branet92};
(74) \cite{gallet02};
(75) \cite{rutlet02};
(76) \cite{chevet89};
(77) \cite{kennet06};
(78) \cite{wachma96};
(79) \cite{intzet01};
(continued over).
\normalsize
\end{table*}

\begin{table*}
\begin{center}
\caption[Properties and data collected for the 19 NSXBs (1)]{Properties and data collected for the 19 NSXBs (1).}
\label{tab-19NSXBs-1}
\small
\begin{tabular}{llllll}
\hline
Source name      	     &Type &Distance                 &Period                  &$M_{\rm NS}$ / $M_\odot$&$M_{\rm cs}$ / $M_\odot$   \\
= alternative name           &     &/ kpc (ref)              &/ hrs (ref)             &(ref)                   &(ref)			   \\
(I)                          &(II) &(III)                    &(IV)                    &(V)                     &(VI)			   \\
\hline
IGR J00291+5934              &MSXP &2.8$\pm$1.0 (72)         &2.457 (85)              &1.4 (89)                &0.039--0.16 (89) 	   \\
LMC X--2 = 4U 0520--72       &Z    &50$\pm$10 (3, 4)         &8.16 (15)               &$\sim 1.4$ (90)         &$\sim 1.2$ (90) 	   \\
4U 0614+09 = V1055 Ori       &Atoll&3.0$^{+0.0}_{-2.5}$ (73) &0.25--0.33              &1.4 (91)                &$\leq$1.9 (91)             \\
                             &     &                         &(86)                    &                        &($\sim$1.45)               \\
XTE J0929--314               &MSXP &$>5$ (74)                &0.726 (15)              &--                      &$\sim 0.008$ (74)	   \\
= INTREF 390                 &     &($\sim$8$\pm$3)          &                        &                        &                           \\
CXOU 132619.7--472910.8      &?    &5.0$\pm$0.5 (75)         &--                      &--                      &--			   \\
Cen X--4 = V822 Cen          &Atoll&1.2$\pm$0.3 (76)         &15.1 (87)               &1.3$\pm$0.8 (15)        &0.31$\pm$0.27 (96)	   \\
Cir X--1 = BR Cir            &Z    &9.2$\pm$1.4 (5)          &398 (5)                 &--                      &4$\pm$1 (97)		   \\
4U 1608--52 = QX Nor         &Atoll&3.3$\pm$0.5 (5)          &$\sim$12.9 (5)          &$\sim$1.4 (92)          &$\sim$0.32 (92) 	   \\
Sco X--1 = V818 Sco          &Z    &2.8$\pm$0.3 (5)          &18.9 (5)                &1.4 (93)                &0.42 (93)		   \\
XTE J1701--462               &Z    &8.5$\pm$8.0 (77)         &--                      &--                      &--			   \\
GX 349+2 = Sco X--2          &Z    &9.25$\pm$0.75 (78)       &--                      &--                      &--			   \\
SAX J1808.4--3658            &MSXP &2.5$\pm$0.1 (79)         &2.0 (5)                 &$\geq$1.7 (94)          &0.05--0.10 (98)            \\
= XTE J1808--369             &     &                         &                        &($\sim$1.7)             &                           \\
XTE J1814--338               &MSXP &8.0$\pm$1.6 (80)         &4.27 (5)                &--                      &$\sim 0.5$ (99)            \\
GX 13+1 = 4U 1811--17        &Z    &7$\pm$1 (81)             &--                      &--                      &--			   \\
GX 17+2 = 4U 1813--14        &Z    &8.0$\pm$2.4 (82)         &--                      &--                      &--			   \\
HETE J1900.1--2455           &MSXP &5$\pm$1 (83, 84)         &1.39 (88)               &--                      &0.016--0.07 (88)           \\
Aql X--1 = V1333 Aql         &Atoll&5.15$\pm$0.75 (5)        &19.0 (5)                &$\sim$1.4 (95)          &$\sim$0.6 (95)             \\
XTE J2123--058 = LZ Aqr      &Atoll&18.4$\pm$2.7 (5)         &5.96 (5)                &1.3 (15)                &0.60 (15)                  \\
Cyg X--2 = V1341 Cyg         &Z    &13.4$\pm$2.0 (5)         &236.2 (5)               &1.78 (15)               &0.60 (15)                  \\
\hline
\end{tabular}
\end{center}
\normalsize
Columns give:
(I) source names;
(II) X-ray classification (Z = Z-source);
(III) distance estimate;
(IV) orbital period of the system;
(V) mass of the neutron star in solar units (assumed to be $\sim 1.4 M_\odot$ if unconstrained);
(VI) mass of the companion star in solar units.
References for tables \ref{tab-16BHXBs1} to \ref{tab-9HMXBs2} (continued):
(80) \cite{stroet03};
(81) \cite{bandet99};
(82) \cite{kuulet02};
(83) \cite{kawasu05};
(84) \cite{gallet05b};
(85) \cite{shawet05};
(86) \cite{neleet04};
(87) \cite{campet04a};
(88) \cite{kaaret06};
(89) \cite{gallma05a};
(90) \cite{cramet90};
(91) \cite{vanset00};
(92) \cite{wachet02};
(93) \cite{steeet02};
(94) \cite{campet02};
(95) \cite{welset00};
(96) \cite{torret02};
(97) \cite{johnet99};
(98) \cite{wanget01};
(99) \cite{krauet05};
(100) \cite{bonnet89};
(continued over).
\normalsize
\end{table*}

\begin{table*}
\begin{center}
\caption[Properties and data collected for the 19 NSXBs (2)]{Properties and data collected for the 19 NSXBs (2).}
\label{tab-19NSXBs-2}
\small
\begin{tabular}{lllll}
\hline
Source            &$A_{\rm V}$, $N_{\rm H}$ / 10$^{21} cm^{-2}$  &$q_{\rm cs}$                   &$\Delta t$ /&Fluxes -- data    \\
                  &(refs)                                        &(band, ref)                    &days        &references        \\
(I)               &(II) 	                                 &(III) 		         &(IV)        &(V)               \\
\hline
IGR J00291+5934   &2.5$\pm$0.3, 4.64$\pm$0.58 (72)               &--                             &1.0         &72, 125           \\
LMC X--2          &0.15, 0.91$\pm$0.07 (100, 101)                &--                             &0.5         &43, 100, 126      \\
4U 0614+09        &1.41$\pm$0.17, 2.99$\pm$0.01 (86, 101)        &--                             &0.5         &126, 127          \\
XTE J0929--314    &0.42$\pm$0.10, 0.76$\pm$0.24 (102, 103)       &--                             &1.0         &43, 102           \\
CXOU 132619.7     &0.34$\pm$0.03, 0.9$\pm$0.1 (75, 104)          &0.8$\pm$0.1 (B, 88)            &--          &75, 104           \\
Cen X--4          &0.31$\pm$0.16, 0.55$\pm$0.16 (105, 106)       &0.75$\pm$0.05 (R),             &0.5         &87, 121,          \\
                  &                                              &0.70$\pm$0.05 (V,94)           &            &128--130          \\
Cir X--1          &10.5$\pm$1.5, 19$\pm$3 (107)                  &--                             &0.2         &43, 131--133      \\
4U 1608--52       &4.65$^{+3.25}_{-0.18}$, 15$\pm$5 (14, 92, 108)&--                             &1.0         &43, 92, 134, 135  \\
Sco X--1          &0.70$\pm$0.23, 1.25$\pm$0.41 (14$^{\dagger}$) &--                             &1.0         &43, 136           \\
XTE J1701--462    &9$\pm$4, 9$\pm$5 (77, 109)$^{\ast}$           &--                             &0.5         &43, 137           \\
GX 349+2          &5$\pm$1, 7.7$\pm$1.0 (78, 110)                &--                             &1.0         &43, 78, 126       \\
SAX J1808.4--3658 &0.68$^{+0.37}_{-0.15}$, 0.11$\pm$0.03 (98, 111)&0$^{+0.3}_{-0.0}$ (V--I, 122) &1.0         &43, 98, 138--142  \\
XTE J1814--338    &0.71$\pm$0.10, 1.63$\pm$0.21 (99)             &--                             &1.0         &43, 99            \\
GX 13+1           &15.3$\pm$2.3, -- (112)                        &--                             &1.0         &112               \\
GX 17+2           &12.5$\pm$1.5, 15$\pm$2 (113, 114)             &--                             &1.0         &43, 143           \\
HETE J1900.1      &0.89$\pm$0.22, 1.6$\pm$0.4 (115)$^{\ast}$     &--                             &2.0         &43, 115, 144,145  \\
Aql X--1          &1.55$\pm$0.31, 4.0$^{+3.8}_{-3.2}$ (116, 117) &--                             &0.5         &43, 146--153      \\
XTE J2123--058    &0.37$\pm$0.15, 0.66$\pm$0.27 (118)$^{\dagger}$&0.77 (R, 123,124)              &--          &124, 154          \\
Cyg X--2          &1.24$\pm$0.22, 1.9$\pm$0.5 (119, 120)         &--                             &1.0         &155, 156          \\
\hline
\end{tabular}
\end{center}
\normalsize
Columns give:
(I) source name;
(II) interstellar reddening in $V$-band, and interstellar HI absorption column ($^{\ast}A_{\rm V}$ and $^{\dagger}N_{\rm H}$ are estimated here from the relation $N_{\rm H} = 1.79 \times 10^{21} cm^{-2}A_{\rm V}$; \citealt{predet95});
(III) the companion star OIR luminosity contribution in quiescence;
(IV) The maximum time separation, $\Delta t$, between the OIR and X-ray observations defined as quasi-simultaneous;
(V) References for the quasi-simultaneous OIR and X-ray fluxes collected.
References for tables \ref{tab-16BHXBs1} to \ref{tab-9HMXBs2} (continued):
(101) \cite{schu99};
(102) \cite{gileet05};
(103) \cite{juetet03};
(104) \cite{hagget04};
(105) \cite{blaiet84};
(106) \cite{rutlet01};
(107) \cite{jonket07};
(108) \cite{grinli78};
(109) \cite{prodet06};
(110) \cite{iariet04};
(111) \cite{campet05};
(112) \cite{charna92};
(113) \cite{deutet99};
(114) \cite{vrtiet91a};
(115) \cite{steeet05b};
(116) \cite{chevet99};
(117) \cite{campet03};
(continued over).
\normalsize
\end{table*}

\begin{table*}
\begin{center}
\caption[Properties and data collected for the 9 HMXBs (1)]{Properties and data collected for the 9 HMXBs (1).}
\label{tab-9HMXBs1}
\small
\begin{tabular}{llll}
\hline
Source name&Type&Distance &$A_{\rm V}$, $N_{\rm H}$ / 10$^{21} cm^{-2}$\\
= alternative name &&/ kpc (ref)& (refs)\\
(I)&(II)&(III)&(IV)\\
\hline
SMC X--3 &NS &58.1$\pm$5.6 (157) &1.5$\pm$0.7, 2.9$\pm$1.4 (164)\\
CI Cam = XTE J0421+560 &unknown&5$^{+3}_{-4}$ (158)&3.2$\pm$1.2, 5$\pm$2 (165, 168)\\
LMC X--4 &NS &50$\pm$10 (3, 4) &0.31$\pm$0.06, 0.55$\pm$0.10 (166)$^{\ast}$\\
A0535+26 = HDE 245770 &NS &2$^{+0.4}_{-0.7}$ (159) &2.3$\pm$0.5, 11.8$\pm$1.5 (159, 169)\\
GX 301--2 = 4U 1223--62 &NS &5.3$\pm$0.1 (160) &5.9$\pm$0.6, 20$\pm$10 (160, 170)\\
V4641 Sgr = SAX 1819.3--2525&BH &9.6$\pm$2.4 (5)&1.0$\pm$0.3, 2.3$\pm$0.1 (5, 171)\\
KS 1947+300 = GRO J1948+32 &NS &10$\pm$2 (161) &3.38$\pm$0.16, 34$\pm$30 (161)\\
Cyg X--1 = HD 226868 &BH &2.1$\pm$0.1 (162) &2.95$\pm$0.21, 6.21$\pm$0.22 (167, 172)\\
Cyg X--3 = V1521 Cyg &unknown&10.3$\pm$2.3 (163) &20$\pm$5, 85$\pm$1 (163, 173)\\
\hline
\end{tabular}
\end{center}
\normalsize
Columns give:
(I) source names;
(II) Compact object (BH = black hole, NS = neutron star);
(III) distance estimate;
(IV) interstellar reddening in $V$-band, and interstellar HI absorption column ($^{\ast}A_{\rm V}$ is estimated here from the relation $N_{\rm H} = 1.79 \times 10^{21} cm^{-2}A_{\rm V}$; \citealt{predet95});
References for tables \ref{tab-16BHXBs1} to \ref{tab-9HMXBs2} (continued):
(118) \cite{hyneet01};
(119) \cite{mcclet84};
(120) \cite{costet05};
(121) \cite{shahet93};
(122) \cite{burdet03};
(123) \cite{casaet02};
(124) \cite{shahet03b};
(125) \cite{steeet04};
(126) \citeauthor{paper4};
(127) \cite{machet90};
(128) \cite{kaluet80};
(129) \cite{caniet80};
(130) \cite{vanpet80};
(131) \cite{shir98};
(132) \cite{glas78};
(133) \cite{mone92};
(134) \cite{corbet98};
(135) \cite{wach97};
(136) \cite{mcnaet03};
(137) \cite{maitba06};
(138) \cite{campst04};
(139) \cite{homeet01};
(140) \cite{gilfet98};
(141) \cite{campet04b};
(142) \cite{greeet06};
(143) \cite{callet02};
(144) \cite{fox05};
(145) \cite{steeet05a};
(146) \cite{charet80};
(147) \cite{jainet99};
(148) \cite{jainet00};
(149) \cite{maitet03};
(150) \cite{maitet04a};
(151) \cite{maitet04b};
(152) \cite{maitet05};
(153) \cite{bailet06};
(154) \cite{tomset04};
(155) \cite{hasiet90};
(156) \cite{vanpet90};
(157) \cite{cole98};
(158) \cite{miodet04};
(159) \cite{steeet98};
(160) \cite{kapeet95};
(161) \cite{neguet03};
(162) \cite{masset95};
(163) \cite{vanket96};
(164) \cite{lequet92};
(165) \cite{hyneet02b};
(166) \cite{naiket03};
(167) \cite{wuet82};
(168) \cite{orlaet00};
(169) \cite{orlaet04};
(170) \cite{mukhet04};
(171) \cite{dicket90};
(172) \cite{schuet02};
(173) \cite{staret03};
(174) \cite{claret78};
(175) \cite{cowlet04};
(176) \cite{claret00};
(177) \cite{heemet89};
(178) \cite{katoet99};
(179) \cite{buxtet05b};
(180) \cite{bochet98};
(181) \cite{brocet99b};
(182) \cite{brocet99a};
(183) \cite{kochet02}.
\normalsize
\end{table*}

\begin{table*}
\begin{center}
\caption[Properties and data collected for the 9 HMXBs (2)]{Properties and data collected for the 9 HMXBs (2).}
\label{tab-9HMXBs2}
\small
\begin{tabular}{lll}
\hline
Source&$\Delta t$ / days&Fluxes - data references\\
(I)&(II)&(III)\\
\hline
SMC X--3 &--&174, 175\\
CI Cam &1.0&43,168, 176\\
LMC X--4 &--&43, 177\\
A0535+26 &1.0&169\\
GX 301--2 &--&160, 170\\
V4641 Sgr &0.2&43, 178, 179\\
KS 1947+300 &1.0&43, 161\\
Cyg X--1 &1.0&43, 180--182\\
Cyg X--3 &1.0&43, 183\\
\hline
\end{tabular}
\end{center}
\normalsize
Columns give:
(I) source name;
(II) The maximum time separation, $\Delta$t, between the OIR and X-ray observations defined as quasi-simultaneous;
(III) References for the quasi-simultaneous OIR and X-ray fluxes collected.
\normalsize
\end{table*}

\subsubsection{Luminosity Calculation}

The X-ray unabsorbed 2--10 keV flux was calculated for all X-ray data\footnote{This energy range was adopted to be consistent with that used for the radio--X-ray correlations of \citeauthor{gallet03} (2003; 2--11 keV) and \citeauthor{miglfe06} (2005; 2--10 keV).}. I made note of the X-ray state of each BHXB on each observation, as defined by the analysis by authors in the literature. I was unable to apply strict definitions to the spectral states due to the differing nature (e.g. X-ray energy ranges and variability analysis) of each data set. I have therefore used the judgement of the authors to determine the spectral states; a method likely to be much more accurate than any conditions that could be imposed by us. A power-law with a spectral index $\alpha=-$0.6 (photon index $\Gamma=1.6$) where $F_{\nu}\propto \nu^{\alpha}$ is assumed when the source is in the hard state (and for all NSXBs), and a blackbody at a temperature of 1 keV for soft state data. These models for the X-ray spectrum are the same as those adopted by \cite{gallet03}, and altering the values in the models to other reasonable approximations does not significantly change the estimated X-ray luminosities. For the BHXB GRS 1915+105 I use data from the radio-bright plateau state, which is approximately analogous to the hard state \citep[e.g.][]{fendpo00}. The \emph{NASA} tool \emph{Web-PIMMS} was used to convert from instrument X-ray counts per unit time (e.g. day-averaged $RXTE$ ASM counts s$^{-1}$). \cite{brocet04} also provide a table of approximate instrument counts--flux conversion factors used in this work.

For the OIR luminosities (or monochromatic luminosities; $L_{\rm \nu, OIR}$), data were collected from the optical $B$-band at 440 $nm$ to the NIR $K$-band at 2220 $nm$. OIR absorbed fluxes were de-reddened using the best-known value of the extinction $A_{\rm V}$ to each source and the dependence of extinction with wavelength given by \cite*{cardet89}. For OIR data at fluxes low enough for the companion star to significantly contribute (i.e. $quiescence$ for most low-mass XBs), the data were discarded if the fractional contribution of the companion had not been estimated in the OIR band of the data (adopting the quoted contributions given in the papers from which the data were acquired). The wavelength-dependent fractional contribution of the companion (column 3 of Tables \ref{tab-16BHXBs2} and \ref{tab-19NSXBs-2}) was subtracted from the low-flux OIR data in order to acquire the flux from all other emission processes in the system. This contribution is not well constrained in many cases due to the uncertain spectral type of the companion star \citep*[e.g.][]{haswet02}. For the Z-sources, OIR data were only included when the OIR fluxes were significantly higher than the lowest measured for each source in each waveband, to ensure minimal contamination from the companion (which has an unknown contribution in nearly all Z-sources). We have therefore propagated the errors associated with this into the errors of the OIR luminosities for all quiescent data. In sources for which the companion is comparatively bright \citep*[e.g. GRO J1655--40;][]{hyneet98}, its contribution has been subtracted in outburst in addition to quiescence. For The HMXBs, which are expected to be dominated by the massive companion, the light of the companion was not subtracted.

The intrinsic (de-reddened) OIR and X-ray luminosities were then calculated given the best-known estimate of the distance to each source (Tables \ref{tab-16BHXBs1}, \ref{tab-19NSXBs-1} and \ref{tab-9HMXBs1}). I adopted the approximation $L_{\rm OIR}\approx \nu F_{\nu,OIR}$ to estimate the OIR luminosity (we are approximating the spectral range of each filter to the central wavelength of its waveband). The errors associated with the luminosities are propagated from the errors quoted in the original data. Where no errors are quoted, I apply a conservative error of 30\%. Errors associated with estimates of the distance, extinction $A_{\rm V}$ and NI absorption column $N_{\rm H}$ were sought (Tables \ref{tab-16BHXBs1}--\ref{tab-9HMXBs1}). Where these limits are not directly quoted in the reference, I used the most conservative estimates implied from the text. These errors were not propagated into the errors associated with the derived luminosities because the resulting plots would be dominated by error bars, however one error bar for each data set is shown, representing the average total systematic 1$\sigma$ errors associated with each luminosity data point.

For the NSXBs, a wealth of OIR data were collected which had two or more quasi-simultaneous OIR wavebands. In these cases it was possible to calculate the spectral slope $\alpha$ of the OIR continuum. For some NSXBs however, OIR--X-ray quasi-simultaneity was achieved but there was only one OIR waveband available so OIR colours were not obtained in this way for these sources. In addition to the data collected from the literature, we obtained OIR photometry of three NSXBs using two telescopes; the observations and data reduction are discussed in the following subsections.

\subsubsection{Observations with the Danish 1.54-m Telescope}

\begin{figure*}
\centering
\includegraphics[width=14cm,angle=0]{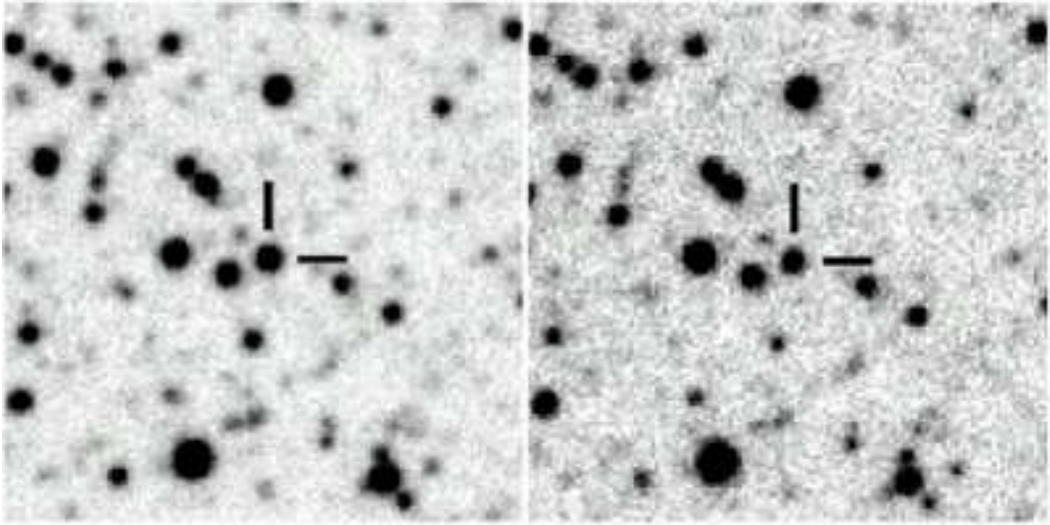}
\caption[LMC X--2 finding chart]{High resolution optical finding chart for LMC X--2 in $V$-band (left panel) and $I$-band (right panel) \citep[for a lower resolution $B$-band finding chart see][]{johnet79}. North is to the top and east is to the left. The images are $\sim 0.8\times 0.8$ arcmins and were taken on 2001-07-27 with DFOSC on the Danish 1.54-m Telescope.}
\label{NSs-LMCX2}
\end{figure*}

$VRIZ$-band imaging of two Z-sources, GX 349+2 and LMC X--2, were taken in July 2001 using the camera on the Danish Faint Object Spectrograph and Camera (\small DFOSC). \normalsize Table \ref{tab-NSs-Danish-obs} lists the observations used in this work. De-biasing and flat fielding was performed with \small IRAF \normalsize and aperture photometry of the targets and the standard star LTT 7987 ($V$ = 12.23 mag; $R$ = 12.29; $I$ = 12.37; $Z$ = 12.50; \citealt{hamuet92}; \citeyear{hamuet94}) was achieved using \small PHOTOM\normalsize. GX 349+2 was detected with a significance of $>8\sigma$ in all exposures (most were $>50\sigma$ detections). For LMC X--2 I aligned and combined the six exposures in each filter on MJD 52115 to achieve a higher signal-to-noise ratio (S/N). No combined images were created from the observations of this source on MJD 52117 as the S/N was sufficiently high for photometry. LMC X--2 was detected at a significance level of $>4\sigma$ in the $Z$-band images and $>20\sigma$ in all other images used in this work. In Fig. \ref{NSs-LMCX2} I present the finder charts for LMC X--2.

The measured fluxes were accounted for airmass-dependent atmospheric extinction \citep[the atmosphere absorbs and scatters some optical light; e.g.][]{hayela75,dahlet06} according to \cite{burket95}. I flux-calibrated the data using LTT 7987. For GX 349+2 the flux was measured of three stars in the field of view with known $V$ and $R$-band magnitudes \citep[stars 2, 5 and 7 listed in][]{wachma96}. Their magnitudes differ from those previously reported by $\sim 0.04$ mag. The resulting fluxes obtained for GX 349+2 and LMC X--2 were de-reddened to account for interstellar extinction using the values of $A_{\rm V}$ listed in Table \ref{tab-19NSXBs-2}.

\begin{table*}
\begin{center}
\caption[Log of the Danish 1.54-m Telescope observations]{Log of the Danish 1.54-m Telescope observations.}
\label{tab-NSs-Danish-obs}
\small
\begin{tabular}{lllrrrrllll}
\hline
MJD&Target&Expo-&\multicolumn{4}{c}{Integration time}&\multicolumn{4}{c}{Apparent magnitudes}\\
   &      &sures  &\multicolumn{4}{c}{/ exp.}&\multicolumn{4}{c}{(not de-reddened)}\\
   &      &/filter&\multicolumn{1}{c}{V}&\multicolumn{1}{c}{R}&\multicolumn{1}{c}{i}&\multicolumn{1}{c}{z}&\multicolumn{1}{c}{V}&\multicolumn{1}{c}{R}&\multicolumn{1}{c}{i}&\multicolumn{1}{c}{z}\\
\hline
52114.2&GX 349+2&9&300&200&120&120&18.54(8)&17.53(9)&16.59(5)&16.14(10)\\
52115.0&GX 349+2&4&300&200&120&120&18.44(12)&17.44(9)&16.52(5)&16.12(5)\\
52115.3&LTT 7987&1&2  &2  &4  &6  &&&&\\
52115.4&LMC X--2&6&120&120&180&240&19.19(7)&19.01(7)&18.91(7)&19.12(14)\\
52117.4&LMC X--2&2&120&120&180&300&18.47(7)&18.43(7)&18.45(8)&18.57(26)\\
\hline
\end{tabular}
\normalsize
\end{center}
MJD 52114.0 corresponds to 2001-07-24.0 UT. The filters used were Bessel $V$, Bessel $R$, Gunn $i$ and Gunn $z$. LTT 7987 is the standard star used for flux calibration. For GX 349+2 the range of magnitudes measured are tabulated, whereas for LMC X--2 I tabulate the magnitudes from the combined images.
\end{table*}

\subsubsection{UKIRT Observations of 4U 0614+09}

NIR imaging of 4U 0614+09 was obtained with the 3.8 m United Kingdom Infrared Telescope (UKIRT) on 2002 February 14 (MJD 52319.3), using UFTI, the UKIRT Fast Track Imager \citep*{rochet03}. Jittered observations of 4U 0614+09 were made in the $J$, $H$ and $K$ filters, with 9$\times$30 s exposures in both $J$ and $H$, and 9$\times$60 s exposures in $K$. The infrared standard star FS 120 ($J$ = 11.335 mag; $H$ = 10.852; $K$ = 10.612) was also observed for a total of 50 s, 25 s and 25 s in $J$, $H$ and $K$, respectively. The airmass of the standard and 4U 0614+09 were very similar: 1.004--1.036.  The `JITTER-SELF-FLAT' data reduction recipe was used, which created a flat field from the sequence of 9 jittered object frames and a dark frame. After dark subtraction and flat fielding, a mosaic was generated from the 9 object frames.

Photometry was carried out using \small IRAF\normalsize. 4U 0614+09 was detected with a significance of 7.1$\sigma$ in $J$, 15.3$\sigma$ in $H$ and 11.5$\sigma$ in $K$. Flux calibration was achieved using FS 120, yielding the following de-reddened ($A_{\rm V}=1.41$) flux densities for 4U 0614+09: $F_{\rm \nu,J}=0.145 \pm 0.037$ mJy; $F_{\rm \nu,H}=0.139 \pm 0.020$ mJy; $F_{\rm \nu,K}=0.111 \pm 0.022$ mJy (the apparent reddened magnitudes are $J=18.12$; $H=17.50$; $K=16.38$).

\subsection{Results}

\begin{figure}
\centering
\includegraphics[height=21.5cm,angle=180]{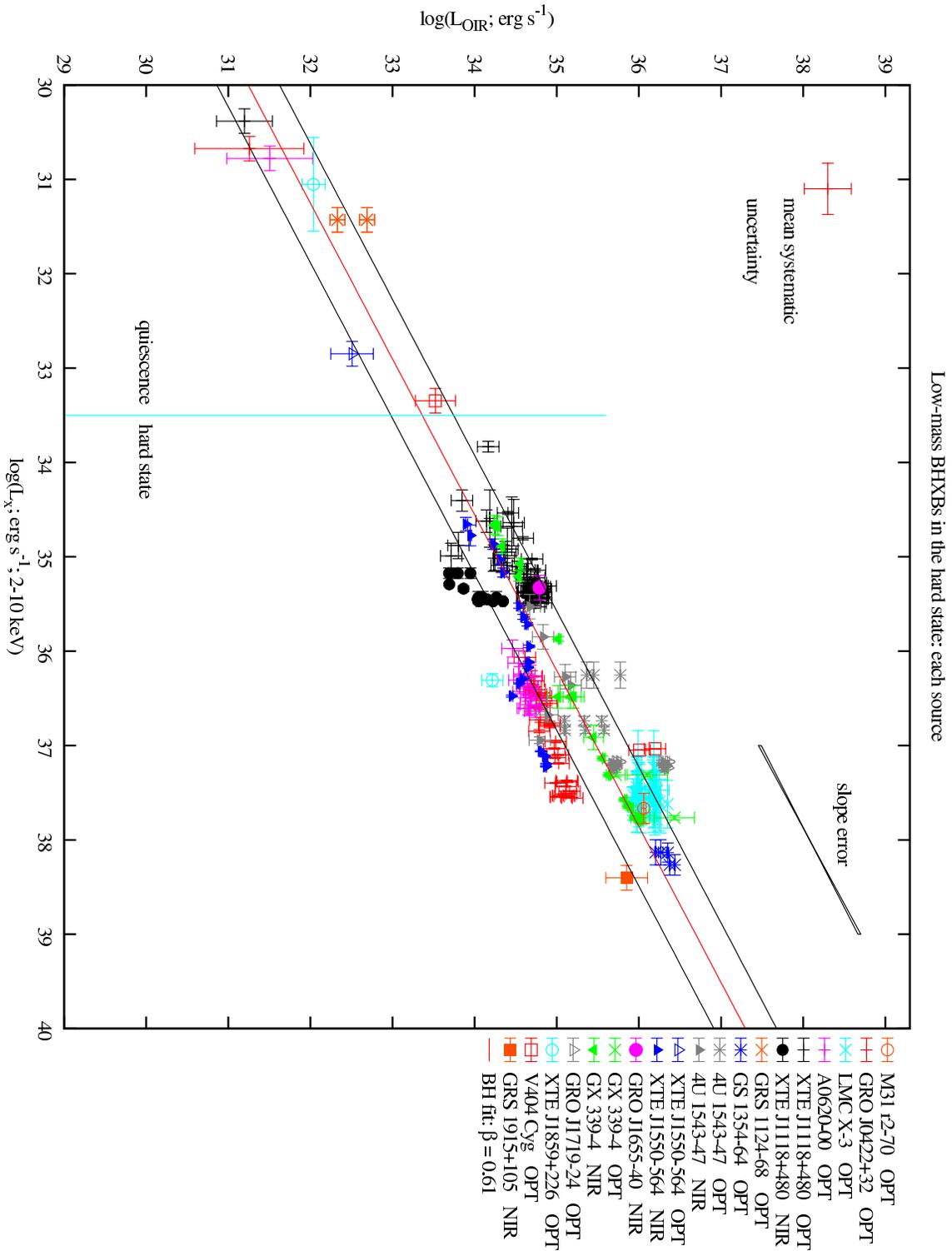}
\caption[X-ray versus OIR luminosities for BHXBs in the hard state (each source)]{OPT and NIR refer to the optical ($BVRI$) and NIR ($JHK$) wavebands, respectively. As a rule, filled symbols represent NIR data and unfilled symbols represent optical data. The mean 1$\sigma$ errors associated with each value (which include errors on distance and reddening) are shown top left. Quasi-simultaneous X-ray versus OIR luminosities for BHXBs in the hard X-ray state are shown, with each source indicated by a different symbol. The best power-law fit is $L_{\rm OIR} = 10^{13.1} L_{\rm X}^{0.605\pm0.018}$. The black lines represent the 1$\sigma$ uncertainties in the normalisation of the fit, which is 0.38 dex in $L_{\rm OIR}$ (slope error top right).}
\label{corr-bhxbs1}
\end{figure}

\begin{figure*}
\centering
\includegraphics[height=21.5cm,angle=180]{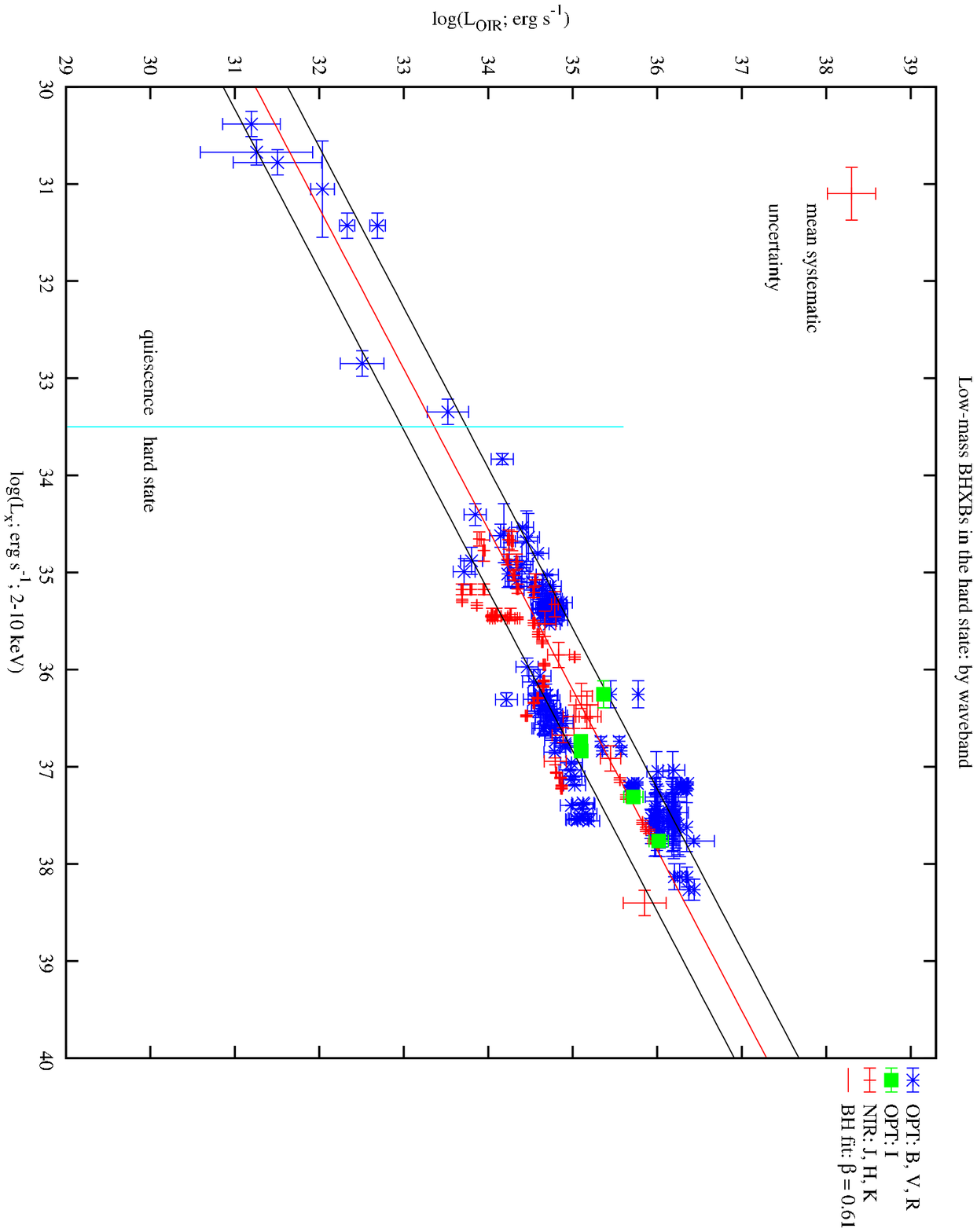}
\caption[X-ray versus OIR luminosities for BHXBs in the hard state (by waveband)]{The same as Fig. \ref{corr-bhxbs1}: BHXBs in the hard state, but with the data split into three groups: $BVR$, $I$ and $JHK$ wavebands.}
\label{corr-bhxbs2}
\end{figure*}

\begin{figure*}
\centering
\includegraphics[height=21.5cm,angle=180]{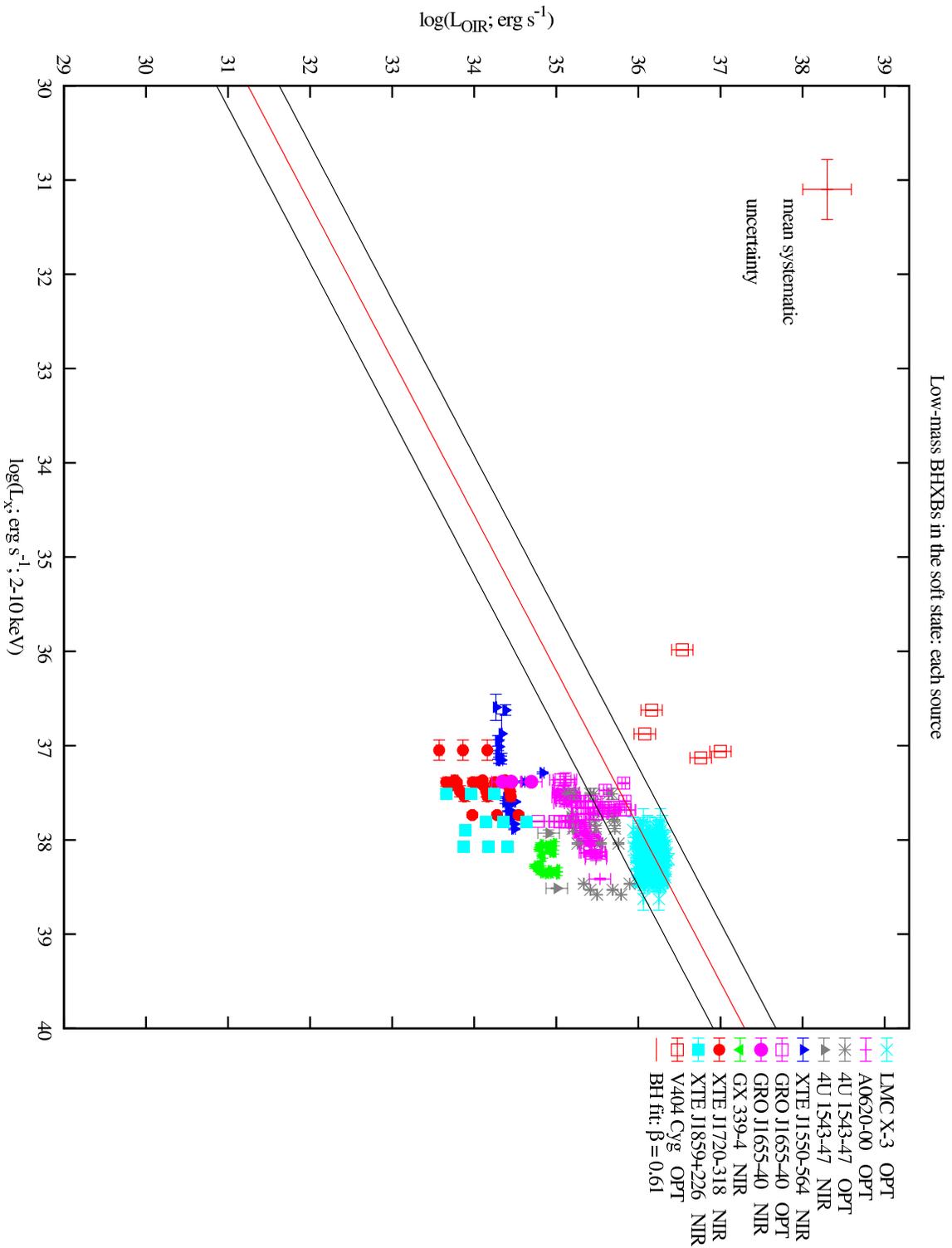}
\caption[X-ray versus OIR luminosities for BHXBs in the soft state (each source)]{As of Fig. \ref{corr-bhxbs1} but for BHXBs in the soft state (the BHXB hard state correlation is shown for comparison).}
\label{corr-bhxbs3}
\end{figure*}

\begin{figure*}
\centering
\includegraphics[height=21.5cm,angle=180]{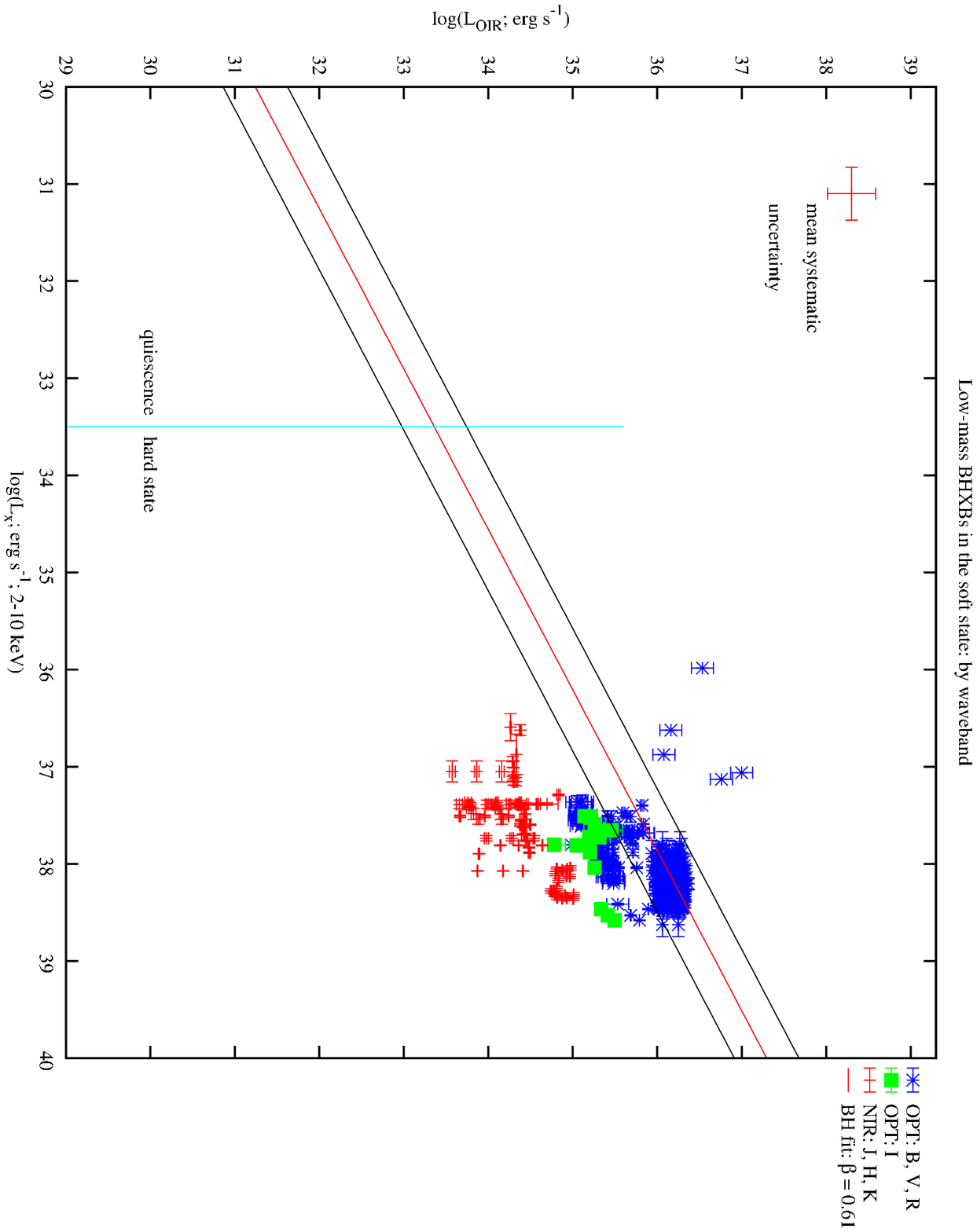}
\caption[X-ray versus OIR luminosities for BHXBs in the soft state (by waveband)]{The same as Fig. \ref{corr-bhxbs3}: BHXBs in the soft state, but with the data split into three groups: $BVR$, $I$ and $JHK$ wavebands.}
\label{corr-bhxbs4}
\end{figure*}

\begin{figure*}
\includegraphics[height=21.5cm,angle=180]{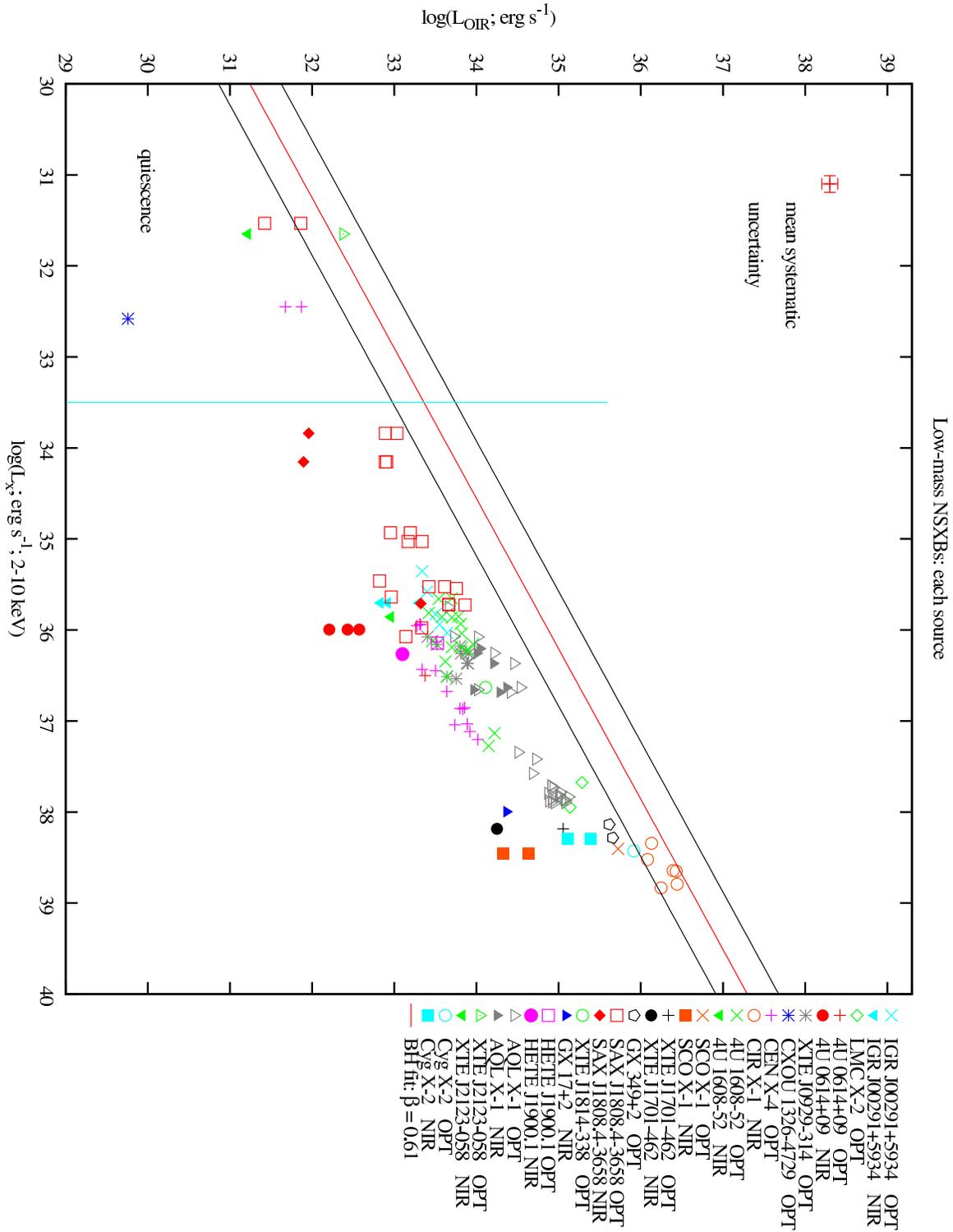}
\caption[X-ray versus OIR luminosities for NSXBs (each source)]{OPT and NIR refer to the optical ($BVRI$) and NIR ($JHK$) wavebands, respectively. Quasi-simultaneous X-ray versus OIR luminosities for all NSXBs are shown, with each source indicated by a different symbol. The BHXB hard state correlation (red line with black 1$\sigma$ uncertainties) is shown for comparison.}
\label{corr-nsxbs1}
\end{figure*}

\begin{figure*}
\includegraphics[height=21.5cm,angle=180]{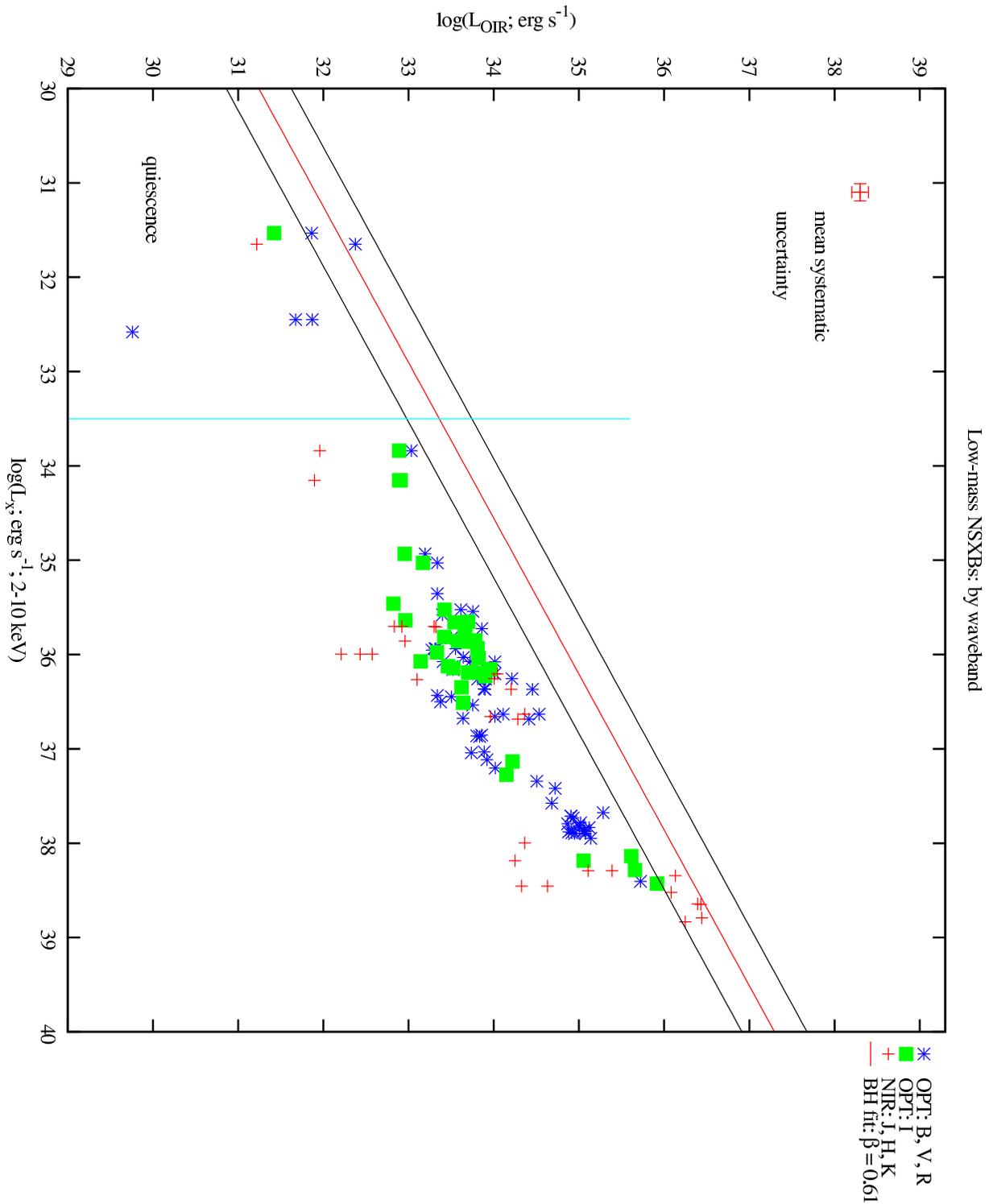}
\caption[X-ray versus OIR luminosities for NSXBs (by waveband)]{The same as Fig. \ref{corr-nsxbs1}: all NSXBs, but with the data split into three groups: $BVR$, $I$ and $JHK$ wavebands.}
\label{corr-nsxbs2}
\end{figure*}

\begin{figure*}
\includegraphics[height=21.5cm,angle=180]{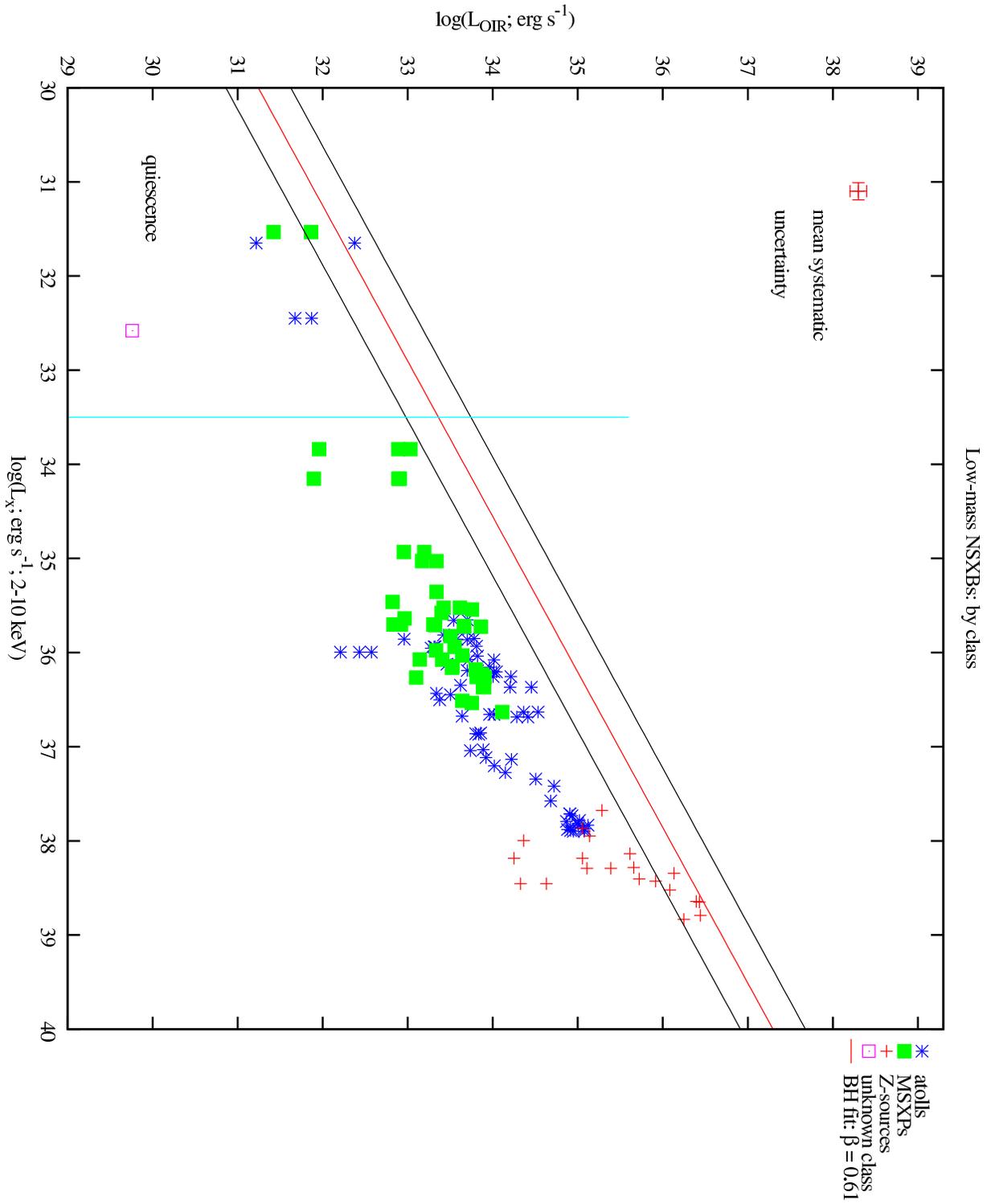}
\caption[X-ray versus OIR luminosities for NSXBs (by source class)]{The same as Fig. \ref{corr-nsxbs1}: all NSXBs, but with the data split into the classes of NSXB.}
\label{corr-nsxbs3}
\end{figure*}

\begin{figure*}
\includegraphics[height=21.5cm,angle=180]{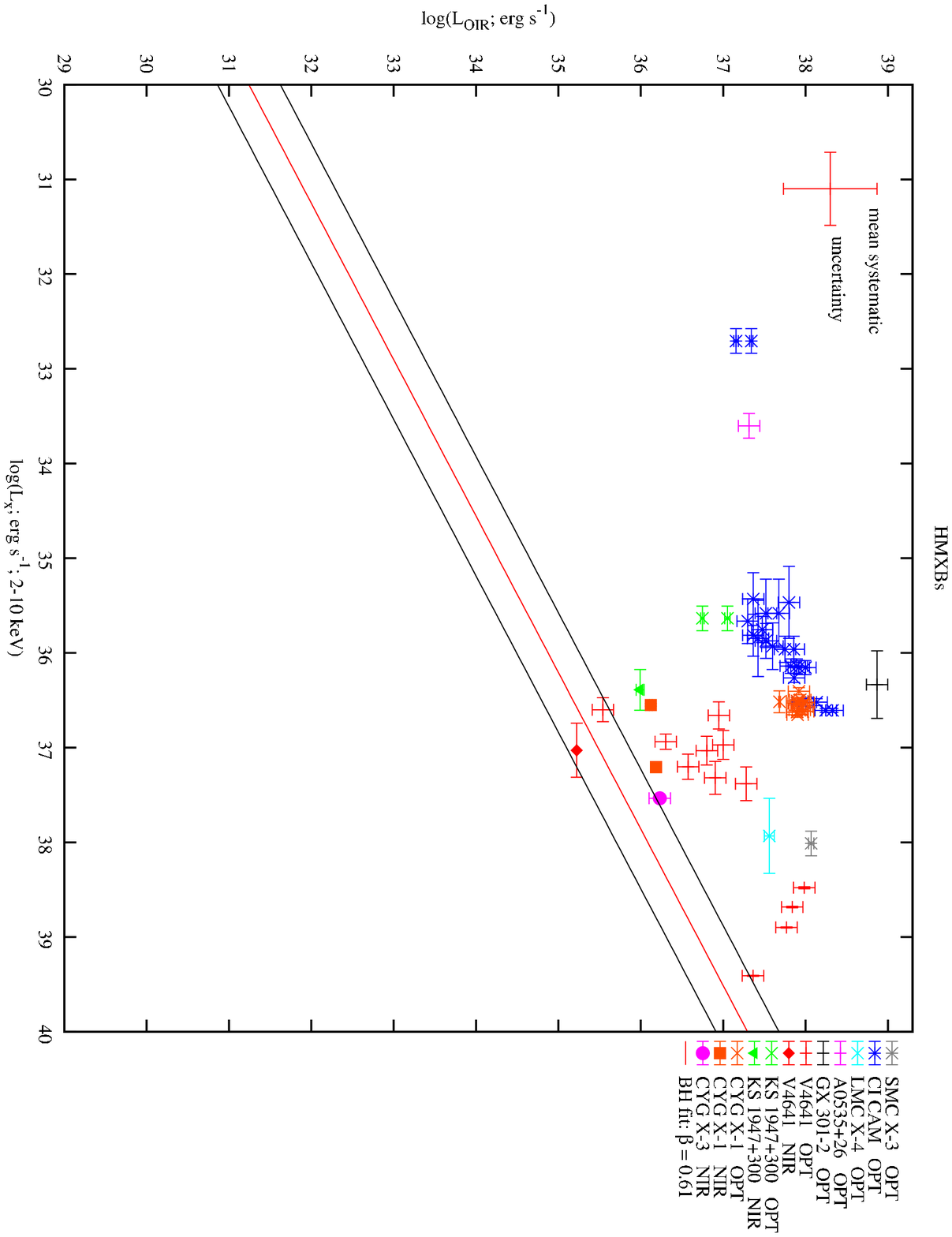}
\caption[X-ray versus OIR luminosities for HMXBs]{As of Fig. \ref{corr-nsxbs1} but for HMXBs (black hole and neutron star XBs).}
\label{corr-hmxbs}
\end{figure*}

Quasi-simultaneous OIR and X-ray luminosities are plotted of 15 BHXBs in the hard state (Figs. \ref{corr-bhxbs1} and \ref{corr-bhxbs2}), 9 BHXBs in the soft state (Figs. \ref{corr-bhxbs3} and \ref{corr-bhxbs4}), 18 NSXBs (Figs. \ref{corr-nsxbs1}--\ref{corr-nsxbs3}) and 9 HMXBs (Fig. \ref{corr-hmxbs}). We have classed LMC X--3 as a low-mass XB (BHXB) because \cite{brocet01a} found from 6 years of observations on this source that its optical emission is dominated by long-term variations rather than the bright companion star in the system, due to its persistent nature (and we are using data from their paper).

For BHXBs in the hard state, a strong $L_{\rm OIR}$--$L_{\rm X}$ correlation exists over 8 orders of magnitude in X-ray luminosity. The slope of the global correlation is $\beta$=0.605$\pm$0.018, where $L_{OIR} \propto L_{\rm X}^{\beta}$ (I do not take into account the $L_{\rm OIR}$ and $L_{\rm X}$ error bars in calculating $\beta$). The package \emph{gnuplot} is used to calculate correlations; equal weighting is applied to each data point. The relations will be biased towards sources with more data points, but I argue that all data points are equally important because a maximum of one data point is used per day per waveband per source, and the sources tend to vary on timescales $\geq$ days. The data from all 15 individual sources lie close to this correlation but deviations in the slopes and normalisations of individual sources are present. For example, there is a strong correlation for the outburst of GRO J0422+32 but its slope is shallower; $\beta$=0.33 (however including the quiescent point, $\beta$=0.54) and for XTE J1118+480, $\beta$=0.58 from the optical data (including quiescence) whereas $\beta$=1.36 for the NIR data (although this does not cover much range in luminosity and does not include any quiescent data). There are no obvious differences in the correlation for optical data compared to NIR data (Fig. \ref{corr-bhxbs2}).

In the soft state, the optical data ($BVRI$-bands) lie close to the hard state correlation for BHXBs (Fig. \ref{corr-bhxbs4}), and the NIR data ($JHK$-bands) lie below the correlation. For the NSXBs (Fig. \ref{corr-nsxbs1}), an OIR--X-ray correlation is also found but it is not a single power-law; $\beta$ appears to be larger at high $L_{\rm X}$ compared with at low $L_{\rm X}$. The upturn (steeper slope) is dominated by the Z-sources and the shallower slope is dominated by the MSXPs below $L_{\rm X}\sim 10^{37}$ erg s$^{-1}$ (there is little data of atolls below this luminosity except in quiescence; Fig. \ref{corr-nsxbs3}). In addition most of the NSXB data lie below the BHXB data, i.e. at a given $L_{\rm X}$, the OIR luminosity of the BHXBs in the hard state is around an order of magnitude or so higher than that of a NSXB. As with the hard state BHXB correlation, there is no strong relation between slope/normalisation and waveband (Fig. \ref{corr-nsxbs2}).

If we neglect the data from sources in quiescence ($L_{\rm X}<10^{33.5}$ erg s$^{-1}$), the BHXB hard state fit is $L_{\rm OIR} = 10^{13.2} L_{\rm X}^{0.60\pm0.02}$. This is very similar to the fit with quiescent data included. The similarity of the fits strengthens the case for the quiescent state being an extension to the hard state. For the analysis in the following subsections, we use the fits with quiescent data included, as they are similar enough either way. In addition, it is also interesting that the fit to the BHXB quiescent data alone has a slope $L_{OIR} \propto L_{\rm X}^{0.67\pm 0.14}$.

From Fig. \ref{corr-hmxbs} it is clear that the OIR luminosity of HMXBs is typically orders of magnitude larger than that of LMXBs, and does not appear to correlate with $L_{\rm X}$. This is consistent with the high-mass companion star dominating the OIR emission. The large range in OIR luminosities between sources is likely to be due to the differing masses and spectral types of the companions. Since this interpretation agrees with the evidence in the literature, I will not discuss the HMXBs in further detail (although they are mentioned in Section 2.5.3).

There are clear advantages and disadvantages of these OIR--X-ray compilations compared to the $L_{\rm radio}$--$L_{\rm X}$ approach of \cite{gallet03} and \cite{miglfe06}. The hard state OIR--X-ray correlation in BHXBs includes data from many sources in low-luminosity \emph{quiescent} states (7 sources with $10^{30}<L_{\rm X}<10^{33.5}$ erg s$^{-1}$), which was not possible for the radio--X-ray correlation due to radio detector limits \citep[see Section 1.2.1; although this has recently been achieved with long radio integration times;][]{gallet06}. The OIR samples also include sources in the LMC and M31, which are too distant (and hence faint) to observe at radio wavelengths. However, unlike the radio--X-ray comparisons, HMXBs cannot be included in these OIR--X-ray plots as the OIR emission is dominated by the companion. In addition, two BHXBs (for example) situated in the galactic plane (1E 1740.7--2942 and GRS 1758--258) were included in the radio--X-ray correlation but are not observable at OIR wavelengths due to the high levels of extinction towards the sources.

In Section 2.4 we attempt to interpret the relations found between OIR and X-ray luminosities in terms of the most likely dominant emission processes. In Section 2.5 we discuss additional patterns, applications and implications of the empirical relations. The results and interpretations are summarised in Section 2.6.

\subsection{Interpretation \& Discussion}

\subsubsection{Jet Suppression in the Soft State in BHXBs}

\begin{figure*}
\includegraphics[height=17cm,angle=270]{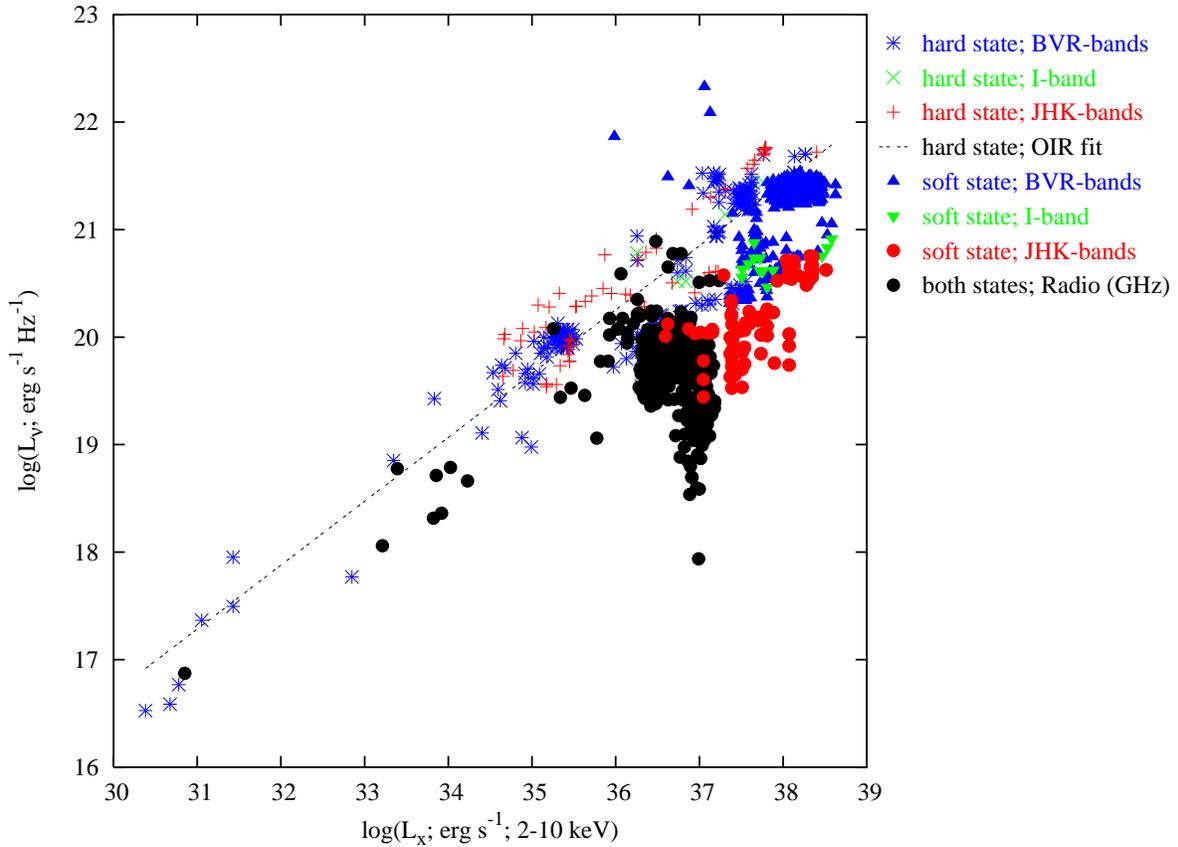}
\caption[X-ray luminosity versus OIR and radio monochromatic luminosities for BHXBs]{X-ray 2--10 keV luminosity versus OIR and radio monochromatic luminosities for BHXBs.}
\label{corr-radio}
\end{figure*}

The significant drop in some of the OIR data in the soft state for BHXBs compared to the hard state (Figs. \ref{corr-bhxbs3} and \ref{corr-bhxbs4}) is not due to changes in the X-ray luminosity during state transition because although the X-ray spectrum significantly changes during transition, the bolometric (and 2--10 keV) X-ray luminosity does not \citep[e.g.][]{zhanet97,kordet06}. The jet component at radio wavelengths is known to decrease/increase in transition to/from the soft X-ray state, due to the quenching of the jets in the soft state \citep[e.g.][]{gallet03}. In Fig. \ref{corr-radio}, the OIR data from BHXBs are split into three wavebands: $BVR$-bands, $I$-band and $JHK$-bands (as in Figs. \ref{corr-bhxbs2} and \ref{corr-bhxbs4}), the monochromatic OIR luminosity ($L_{\nu}$; i.e. flux density scaled for distance) is plotted against $L_{\rm X}$, overplotted with the radio data $L_{\rm \nu ,radio}$--$L_{\rm X}$ for BHXBs from \citeauthor{gallet03} (2003; 2006). It is clear that the normalisations in $L_{\nu}$ (as well as the power-law slopes) for the radio--X-ray and OIR--X-ray hard state correlations are similar to one another for BHXBs: at a given $L_{\rm X}$, the radio and OIR monochromatic luminosities are $\sim$ equal, implying a flat spectrum from radio to OIR wavelengths for all BHXBs in the hard state. I return to the hard state interpretation in Section 2.4.2.

Fig. \ref{corr-radio} also shows a clear suppression of all the $JHK$ data, and a little, if any of the $BVR$ in the soft state. The $I$-band data appear to sit in the centre of the two groups. All 9 BHXBs with soft state data are consistent with this behaviour, suggesting it is ubiquitous in BHXBs. We interpret this as the NIR wavebands being quenched as the jet is switched off, as is observed at radio wavelengths. According to this plot, the NIR appears to be quenched at a higher X-ray luminosity than the radio, but this effect is due to an upper $L_{\rm X}$ limit adopted by \cite{gallet03} when compiling their radio--X-ray data.

The optical wavebands in the soft state lie close to the OIR hard state correlation. Most of the optical data lie below the centre of the hard state correlation, with the exception of V404 Cyg, whose optical luminosity is enhanced in the soft state. There is still a debate as to whether V404 Cyg entered the soft state, and the data here were taken close to the supposed state transition. The $I$-band appears to be the ``pivot'' point, as already shown by \cite{corbfe02} and \cite{homaet05a} to be where the continuum of the optically thin jet meets that of (possibly) the disc. The NIR quenching in the soft state implies that this waveband is dominated by the jets in luminous hard states, just before/after transition to/from the soft state. The optical data are not quenched, suggesting a different process is dominating at these wavelengths.

An alternative interpretation could be that the disc dominates the OIR in both the hard and soft states, but changes temperature during state transition, shifting the blackbody from (e.g.) the OIR in the hard state, to the optical--UV in the soft state. This would have the effect of reducing the NIR in the soft state but maintaining the optical, as is observed. We argue that this is not the case because there is evidence in many BHXBs for two spectral components of OIR emission, the redder of which is quenched in the soft state \citep[Chapter 3;][]{jainet01b,buxtet04,homaet05a}.

The jets should contribute negligible OIR flux in the soft state, so the OIR contribution of the jets at high luminosity in the hard state can be estimated from the level of soft state quenching. The mean offset of the OIR soft state data from the hard state fit is 0.30$\pm$0.32 dex, 0.71$\pm$0.21 dex and 1.10$\pm$0.26 dex in $L_{\rm\nu,OIR}$ for the $BVR$, $I$ and $JHK$-bands, respectively. This corresponds to a respective fractional jet component of 50$^{+26}_{-50}$, 81$^{+7}_{-13}$ and 92$\pm$5 percent in the $BVR$, $I$ and $JHK$-bands. It is clear that the spectrum of the jet extends, with spectral index $\alpha\sim 0$, from the radio regime to the NIR in BHXBs. The position of the turnover from optically thick to optically thin emission must be close to the NIR waveband for the radio--NIR spectrum to appear flat \citep[unless the optically thick spectrum is highly inverted, which is not seen in the radio spectrum most of the time; see e.g.][]{fend01}. The OIR spectrum in the hard state is flatter than in the soft state, as is expected if the jet component is present in the former and not in the latter; this is explored further in Chapter 3.

There are no clear OIR--X-ray relations in the soft state alone; the data do not span a large enough region in $L_{\rm X}$. In Chapter 4, OIR--X-ray relations in the soft state are observed in individual sources which indicate the viscously-heated disc dominates the soft state emission (see Chapter 4 for analysis and discussion).

\subsubsection{Modelling OIR--X-ray Relations}

We now attempt to interpret the empirical OIR--X-ray relations in terms the three most cited OIR emission processes: X-ray reprocessing in the disc, the viscously heated disc and jet emission.
For X-ray reprocessing in the disc, we adopt the theoretical model between optical and X-ray luminosities of \cite{vanpet94}
but also take into account a dependency of the relation on system mass that was neglected by \citeauthor{vanpet94}, but is needed here to compare between NSXBs and BHXBs. From Kepler's third law and equation 5 of \citeauthor{vanpet94} (and the discussion that follows), we have:
\begin{eqnarray*}
  L_{\rm OPT}\propto L_{\rm X}^{1/2}a \propto L_{\rm X}^{1/2}(M_{\rm BH,NS}+M_{\rm cs})^{1/3}P^{2/3}
\end{eqnarray*}

\begin{figure}
\centering
\includegraphics[height=15.5cm,angle=270]{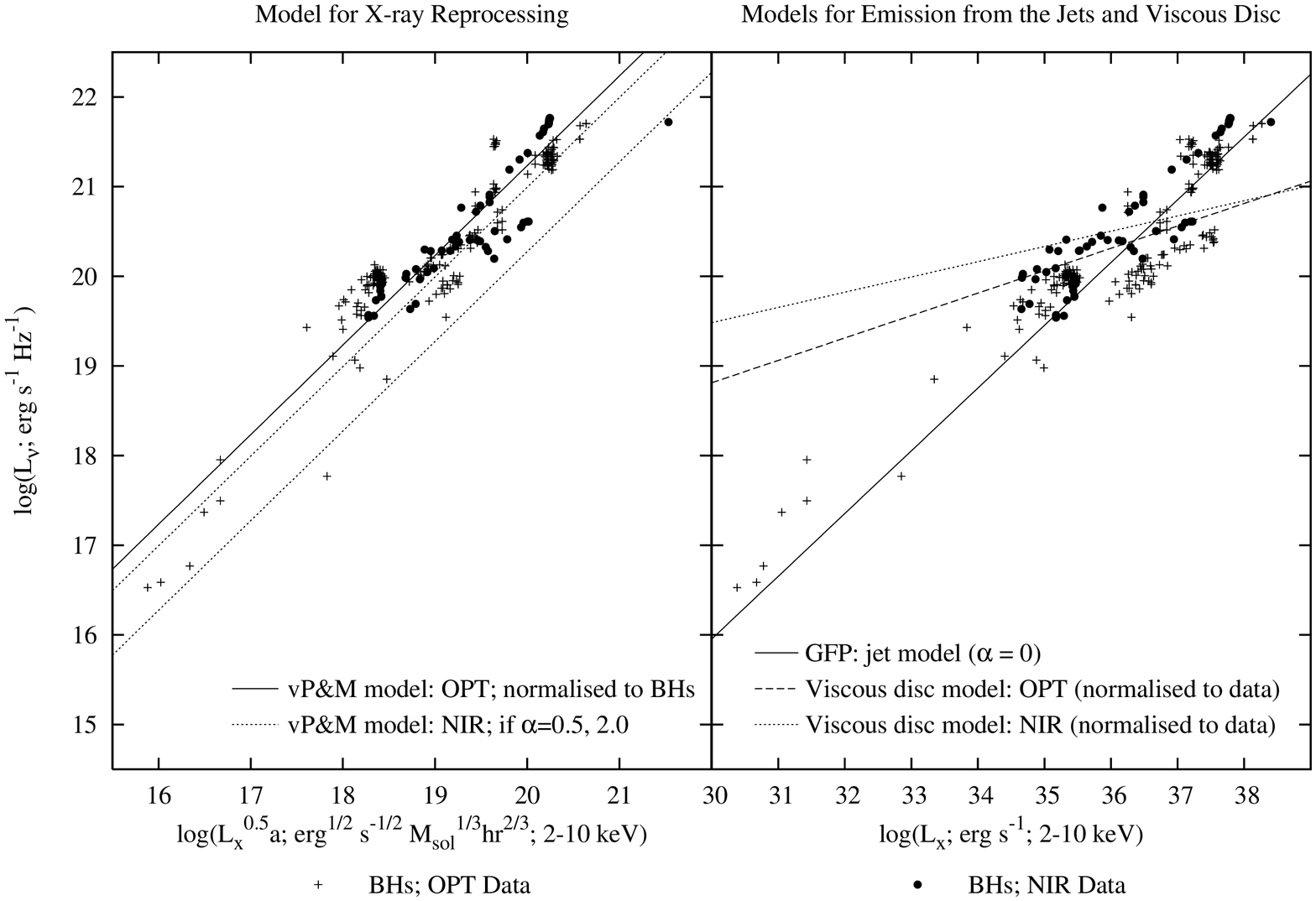}
\caption[X-ray--OIR models for BHXBs]{X-ray luminosity (or $L_{\rm X}^{1/2}a$ in the left panel) versus monochromatic OIR luminosity for BHXBs in the hard X-ray state, with models for X-ray reprocessing (left), and jet and viscous disc (right) components. The X-ray reprocessing model is derived from vP\&M \citep{vanpet94} and the jet model uses the relation from GFP \citep{gallet03}. The X-ray reprocessing and viscous disc models are fit to the data whereas the jet model is normalised assuming a flat spectrum from radio to OIR ($\alpha$ = 0).}
\label{corr-models}
\end{figure}

\begin{figure}
\centering
\includegraphics[width=12cm,angle=0]{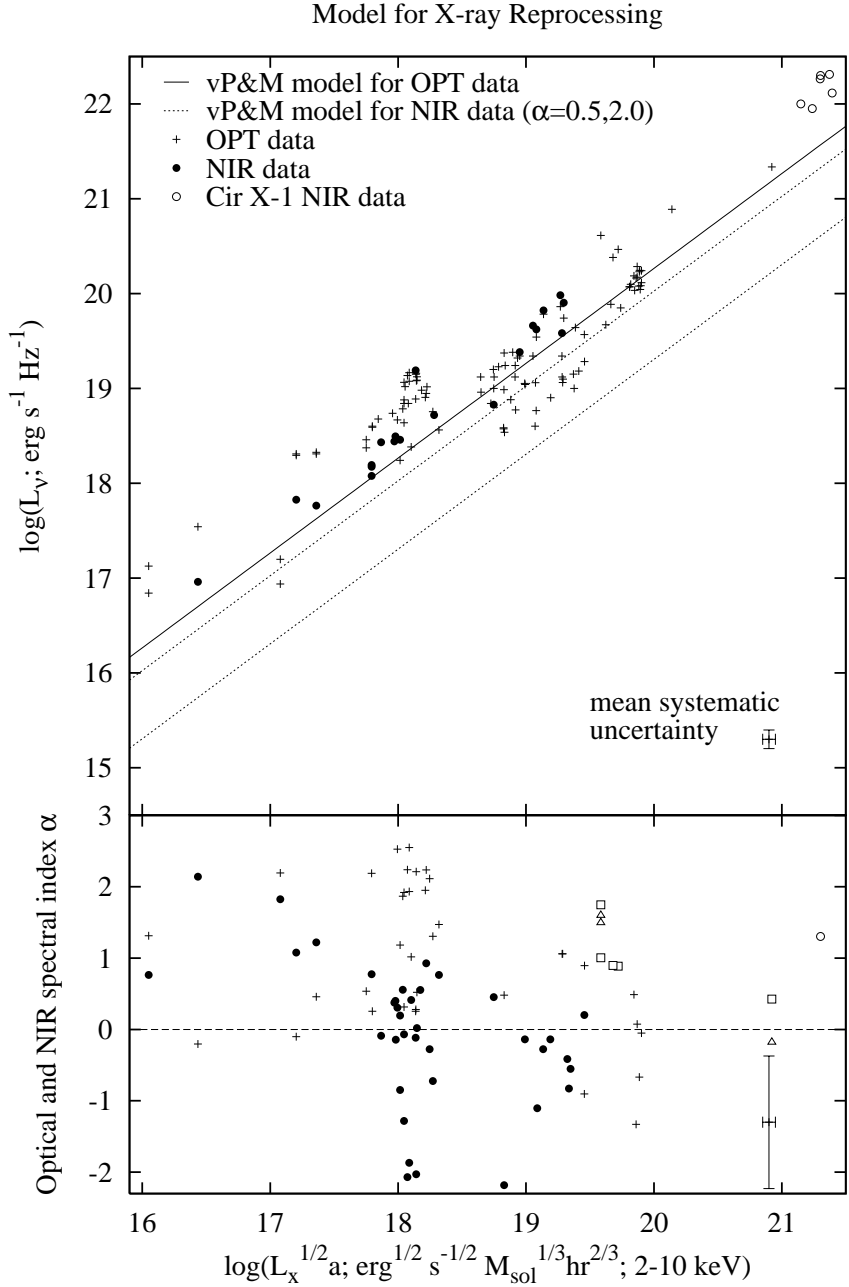}
\caption[X-ray--OIR model for X-ray reprocessing, and optical/NIR spectral indices for NSXBs]{Quasi-simultaneous $L_{\rm X}^{1/2}a$ (where $a$ is the orbital separation) versus OIR monochromatic luminosity (upper panel) and optical and NIR spectral index (lower panel) for NSXBs. In the upper panel, the model for optical emission from X-ray reprocessing (solid line) is derived from vP\&M \citep{vanpet94} and normalised to the optical data of \citeauthor{paper1}. The dotted lines show the expected relation for X-ray reprocessing from NIR data, assuming a NIR--optical spectral index of $\alpha = 0.5$ (upper dotted line) and $\alpha = 2.0$ (lower dotted line). In the lower panel, the Z-sources have different symbols to the atolls and MSXPs (see Fig. \ref{NSs-jet-model} for legend). Data for Cir X--1 are denoted by open circles; the orbital separation may not be accurate for this source. The mean error on the spectral index is very likely overestimated in the lower panel (see text).}
\label{NSs-XR-model}
\end{figure}

In the left panel of Fig. \ref{corr-models} we plot the monochromatic luminosity (optical and NIR) versus $L_{\rm X}^{1/2}a$ for each BHXB data point in the hard state (adopting the values of $P$, $M_{\rm BH,NS}$ and $M_{\rm cs}$ for each source from Table \ref{tab-16BHXBs1}). In the upper panel of Fig. \ref{NSs-XR-model} the same is plotted for each NSXB data point, using the system parameters from Table \ref{tab-19NSXBs-1}. The solid line in these panels represents the power-law fit to the optical hard state BHXB data and optical NSXB data, respectively, fixing the slope at unity ($L_{\rm OPT}\propto L_{\rm X}^{1/2}a$). We see that the X-ray reprocessing model describes the slope of the data well for both types of source. We expect a lower normalisation for the NIR data due to the shape of the OIR spectrum, which may have a spectral index $0.5\leq\alpha\leq2.0$ \citep[a conservative range based on theoretical and empirical results of X-ray reprocessing; see e.g.][]{hyne05}. The upper and lower dotted lines indicate the expected correlations of the NIR data (approximating the optical to the $V$-band centred at $550nm$ and the NIR to the $H$-band at $1660 nm$) if $\alpha =0.5$ and $2.0$, respectively. However, the NIR data lie above these expected correlations for emission from X-ray reprocessing in both the BHXB and NSXB systems.

For NSXBs with observationally unconstrained neutron star masses, we assume $M_{\rm NS}\approx 1.4M_\odot$. This is the approximate mass of all known neutron stars, although there are exceptions where the mass is likely to be up to $\approx 2M_\odot$ \citep[e.g.][]{campet02,cornet07}. We do not include data in Fig. \ref{NSs-XR-model} from systems in which the orbital period or the companion mass is unconstrained (see Table \ref{tab-19NSXBs-1}). We note that data from Cir X--1 in Fig. \ref{NSs-XR-model} (open circles in the top right corner of the figure) may not be representative because its orbit is eccentric and so the orbital separation cannot be accurately inferred using this method.

\begin{figure}
\centering
\includegraphics[width=12cm,angle=0]{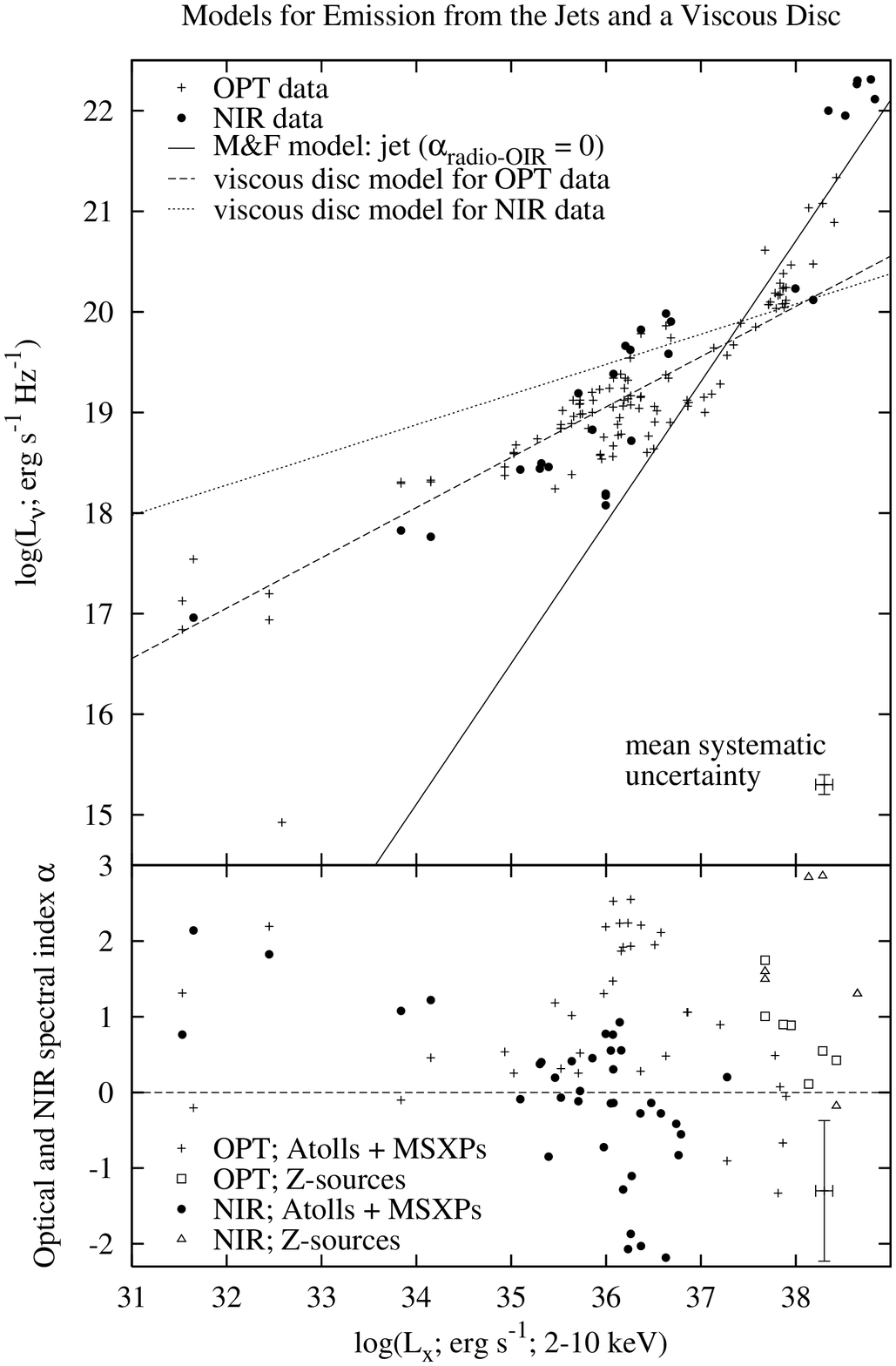}
\caption[X-ray--OIR model for jets and viscous disc, and optical/NIR spectral indices for NSXBs]{Quasi-simultaneous X-ray luminosity versus OIR monochromatic luminosity (upper panel) and optical and NIR spectral index (lower panel) for NSXBs. In the upper panel, the jet model for NSXBs assumes a flat spectrum from radio to OIR and adopts the relation $L_{\rm radio}\propto L_{\rm X}^{1.4}$ \citep[M\&F:][]{miglfe06}. The models for optical and NIR emission from a viscously heated disc are also shown, normalised to the optical and NIR data, respectively. The mean error on the spectral index is very likely overestimated in the lower panel (see text).}
\label{NSs-jet-model}
\end{figure}

For the jets, we take the models described in Section 1.2.3 where the spectrum is flat (optically thick) from radio to OIR, and normalise them using the empirical $L_{\rm radio}$--$L_{\rm X}$ relations found by \cite{gallet03} and \cite{miglfe06} for BHXBs and NSXBs, respectively. In the right panel of Fig. \ref{corr-models} we plot $L_{\rm \nu ,OIR}$ versus $L_{\rm X}$ for hard state BHXBs and in the upper panel of Fig. \ref{NSs-jet-model} the same is plotted for NSXBs. The same relation is expected between optical and NIR jet emission when plotting monochromatic luminosity because we are assuming the jet spectrum is flat; $\alpha \sim 0$.

We find that the model for OIR emission from the jet in BHXBs can approximately describe the optical and NIR data of the BHXBs (the data are close to the solid line in the right panel of Fig. \ref{corr-models}). The model for OIR emission from jets in NSXBs lies close to the NSXB data at high $L_{\rm X}$ but cannot describe the data at low $L_{\rm X}$ ($\simlt 10^{36}$ erg s$^{-1}$; Fig. \ref{NSs-jet-model}, upper panel). However, since all the optical NSXB data also lie close to the expected relation for X-ray reprocessing (Fig. \ref{NSs-XR-model}, upper panel), X-ray reprocessing may be responsible for all the OIR emission instead of the jet. Only the data above $L_{\rm X}\simgt 10^{36}$ erg s$^{-1}$ could arise due to the presence of a jet (Fig. \ref{NSs-jet-model}).

The Z-sources, which tend to have much longer orbital periods (and hence larger orbital separations) than atolls and MSXPs, dominate the highest X-ray luminosities. These sources spend most of their time in a soft X-ray state and in fact have radio luminosities lower than predicted by the NSXB hard state radio--X-ray relation if the radio emission originates in the jet \citep{miglfe06}. We would therefore expect the OIR emission from the jets in Z-sources to also be lower than the jet model in Fig. \ref{NSs-jet-model} unless the radio--OIR jet spectrum is inverted (positive). In fact from Fig. 3 of \cite{miglfe06} we see that at $L_{\rm X} \sim 10^{38}$ erg s$^{-1}$, $L_{\rm radio} \sim 10^{30}$ erg s$^{-1}$ and so from Fig. \ref{NSs-jet-model} here, the radio--OIR spectral index for Z-sources\footnote{At $L_{\rm X} \sim 10^{38}$ erg s$^{-1}$ we can calculate the spectral index between radio and OIR for Z-sources since we know the radio and OIR luminosities.} is $\alpha \sim 0.2$. With this information alone, the jets can only dominate the OIR of Z-sources if the radio--OIR spectrum is inverted.

The models for OIR emission from a viscously heated disc are described as follows. For the simplest viscously heated steady-state disc, there are two limiting regimes \citep{franet02}. For $h\nu \ll kT$, we expect $L_{\rm OIR}\propto$ \.m$^{1/4}$. This is simply the Rayleigh-Jeans limit and will only apply well into the IR. For $h\nu \gg kT$, the relationship is steeper, $L_{\rm OIR}\propto$ \.m$^{2/3}$. For typical disc edge temperatures of 8,000--12,000\,K, expected power-law slopes ($L_{\rm OIR}\propto$ \.m$^{\gamma}$), are calculated to vary from $\gamma\sim0.3$ in the $K$-band, to $\gamma\sim0.5$ in $V$, and $\gamma\sim0.6$ in the UV. Using the calculations linking $L_{\rm X}$ and \.m in Section 1.2.3, this corresponds to expected correlations of the form $L_{\rm OIR}\propto L_{\rm X}^{\beta}$, where 0.15$\le\beta\le$0.25 for BHXBs and 0.30$\le\beta\le$0.50 for NSXBs.

For the BHXBs, the viscous disc models clearly cannot describe the data; the correlation is too steep. Focusing on the NSXBs (Fig. \ref{NSs-jet-model}, upper panel), we see that the model for emission from a viscously heated disc (which requires a relation $L_{\rm OIR}\propto L_{\rm X}^{0.5}$ for optical data and $L_{\rm OIR}\propto L_{\rm X}^{0.3}$ for NIR) can describe the optical data below $L_{\rm X}\simlt 10^{37.5}$ erg s$^{-1}$. Above this $L_{\rm X}$ the OIR is more luminous than expected from this relation. The viscous disc model is not consistent with the NIR data. We can therefore rule out a viscous disc origin to the OIR emission, at least at $L_{\rm X}\simgt 10^{37.5}$ erg s$^{-1}$.

\begin{table*}
\begin{center}
\caption[Parameters for the BHXB hard state models: theory versus observations]{Parameters for the BHXB hard state models: theory versus observations.}
\label{corr-HSmodels}
\small
\begin{tabular}{lccccc}
\hline
Sample&Model&$\beta_{\rm model}$&$\beta_{\rm data}$&$\mid \beta_{\rm data}$-$\beta_{\rm model}\mid$&$\frac{n_{\rm data}}{n_{\rm model}}^\ast$\\
\hline
\multicolumn{6}{c}{X-ray reprocessing model}\\
\vspace{2mm}
BHs; OPT&$L_{\rm \nu} = n L_{\rm X}^{\beta}a$&0.5&0.55$\pm$0.03&0.05$\pm$0.03&9.3$\pm$0.4\\
\vspace{2mm}
BHs; NIR&$L_{\rm \nu}=(\frac{\nu_{\rm NIR}}{\nu_{\rm OPT}})^\alpha n L_{\rm X}^{\beta}a$&0.5&0.56$\pm$0.03&0.06$\pm$0.03&15.5--81.3$^\dagger$\\
\hline
\multicolumn{6}{c}{Viscous disc model}\\
\vspace{2mm}
BHs; OPT&$L_{\rm \nu} = L_{\rm X}^{\beta}$&0.25&0.59$\pm$0.02&0.34$\pm$0.02&--\\
\vspace{2mm}
BHs; NIR&$L_{\rm \nu} = L_{\rm X}^{\beta}$&0.17&0.61$\pm$0.04&0.44$\pm$0.04&--\\
\hline
\multicolumn{6}{c}{Jet model}\\
\vspace{2mm}
BHs; OPT&$L_{\rm \nu} = n L_{\rm X}^{\beta}$&0.7&0.59$\pm$0.02&0.11$\pm$0.02&1.05$\pm$0.07\\
\vspace{2mm}
BHs; NIR&$L_{\rm \nu} = n L_{\rm X}^{\beta}$&0.7&0.61$\pm$0.04&0.09$\pm$0.04&1.78$\pm$0.16\\
\hline
\end{tabular}
\end{center}
\normalsize
When measuring $\beta_{data}$ for a sample, $n$ (the normalisation) is also a free parameter (i.e. this is a fit to the data) whereas for measuring $n_{data}$, $\beta$ is fixed at the value of the model $\beta_{model}$, and $n$ is a free parameter (i.e. this is measuring the mean normalisation offset of the data from the model). $^\ast$ for the X-ray reprocessing model, $n$ is defined by the fit to the optical NSXB data used in \citeauthor{paper1}; $^\dagger$ the range corresponds to an OIR spectral index $0.5\leq\alpha\leq 2.0$.
\end{table*}

A summary of the results of fitting these models to the observed BHXB data is provided in Table \ref{corr-HSmodels}. There is no global power-law correlation found in NSXBs (the relation steepens at higher $L_{\rm X}$) so we I do not include NSXBs in this table. The following Section (2.4.3) discusses the models for NSXBs with the additional information of how the optical and NIR spectral index changes with $L_{\rm X}$. It is clear that the slopes $\beta$ of the observed relations for BHXBs can be explained by the X-ray reprocessing model or the jet model, but not by the viscous disc model (Homan et al. 2005 also ruled out a viscous disc origin to the OIR emission in the hard state of GX 339--4 because the mass accretion rate inferred from the luminosity is much higher than expected).

The normalisation $n$ of the BHXB data is closer to the jet model than the reprocessing model. Although this is consistent with the constraints derived from Section 2.4.1 (the jets are contributing $\sim90$ percent of the NIR luminosity here), an optical excess of $\sim$1 order of magnitude (compared to NSXBs) from the reprocessing model is not expected. The excess is unlikely to be fully explained by the jets because of the lack of optical quenching in the soft state (Section 2.4.1). In addition, the optical spectrum of most BHXBs is inconsistent with jet emission dominating (Chapter 3).
Instead we suggest that OIR emission from reprocessing is enhanced for BHXBs at a given $L_{\rm X}$ due to the localisation of the source of X-rays. For example, the X-ray emitting region in BHXBs may have a larger scale height than in NSXBs and will therefore illuminate the disc more readily \citep[e.g.][]{minifa04}. This may account for the high value of $n$ for the optical BHXB data, but still struggles to explain the even higher $n$ for the NIR BHXB data, which strongly suggests the NIR is indeed jet-dominated. On the other hand, some of the X-ray emission could originate in an outflow \citep[e.g.][]{market01,market05} which may be relativistic and therefore beamed, which would result in less X-rays illuminating the accretion disc. In this sense our results suggest the X-rays are not significantly beamed if X-ray reprocessing dominates the OIR.

We note that there are deviations of individual sources from the correlations which may be caused by distance and reddening errors, or by differing emission process contributions due to the range of orbital separations or slight differences in the slope of the radio--OIR jet spectrum between sources (or other system parameters not considered). For example, XTE J1118+480 has a small disc and is known to produce significant optical jet emission \citep*[e.g.][]{malzet04,hyneet06b} whereas V404 Cyg possesses a large disc and is dominated by X-ray reprocessing in the disc \citep[e.g.][]{wagnet91}. Other OIR emission mechanisms that could contribute, for example disc OIR emission due to magnetic reconnection, have not been modelled here and could also contribute to the scatter in Fig. \ref{corr-models}. These processes cannot be ruled out, but are unlikely to easily explain the observed correlations.

\begin{figure}
\centering
\includegraphics[width=6cm,angle=270]{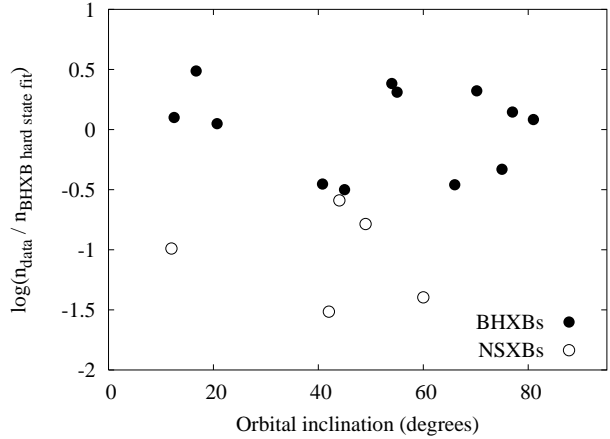}
\caption[Orbital inclination effects on X-ray--OIR relations]{Mean normalisation $n$ of each source, fixing the slope at $L_{\rm OIR}\propto L_{\rm X}^{0.6}$, versus orbital inclination.}
\label{corr-inclin}
\end{figure}

In addition, the orbital inclination may affect the level of OIR emission, in particular from X-ray reprocessing in the disc. To explore this, I have plotted in Fig. \ref{corr-inclin} the average normalisation $n$ of the data for each source against the known estimated orbital inclination, fixing the slope at $L_{\rm OIR}\propto L_{\rm X}^{0.6}$ for each source. The best estimate of the inclination of each system (where known) was obtained from \cite{rittet03}, \cite{trevet88}, \cite*{fomaet01}, \cite{wanget01} and \cite{falaet05}. Note that for the NSXBs I use the data from \citeauthor{paper1}; a smaller sample than that of \citeauthor{paper4}, but a global OIR--X-ray relation was found for NSXBs in \citeauthor{paper1} and so one can measure the $n$ more easily than from the data in \citeauthor{paper4}. We find no evidence for there being a direct relation between $L_{\rm OIR}$ and the orbital inclination $i$, suggesting the effect of inclination is subtle.

\subsubsection{OIR Spectral Index -- X-ray Relations in NSXBs}

In Chapter 3 the OIR spectrum of BHXBs and NSXBs is discussed as a function of OIR luminosity. For BHXBs this can be directly related to X-ray luminosity since $L_{\rm OIR}$ and $L_{\rm X}$ are well correlated with a single power-law in BHXBs. In this Section we explore the direct relationship between the optical/NIR spectral index and $L_{\rm X}$ in NSXBs. This analysis is also motivated by the apparent NIR-excess recently seen in MSXPs (see Section 1.2.2; \citeauthor{paper4}).

The lower panels of Figs. \ref{NSs-XR-model} and \ref{NSs-jet-model} describe the dependence of the shape of the OIR spectrum of NSXBs with $L_{\rm X}^{1/2}a$ and $L_{\rm X}$, respectively. The spectral index $\alpha$ is calculated for all NSXB OIR data where two or more optical or NIR data points are quasi-simultaneous. The spectral index of the optical (using data in the $B$, $V$ or $R$-bands) and NIR ($R$, $I$, $J$, $H$ or $K$-bands) data are shown separately. We define the $R$-band as the break between optical and NIR here because it is often the $R-I$ colours that indicate the NIR excess (and for many of these observations $JHK$ data were not obtained). For consistency (and to separate the two wavebands used to calculate $\alpha$) the two lowest frequency NIR bandpasses in each OIR SED are used to calculate $\alpha_{\rm NIR}$ and the two highest frequency optical bands are used to calculate $\alpha_{\rm OPT}$. Data from the Z-sources are shown as different symbols to the atolls/MSXPs because the spectral index of the former group appears to behave differently to that of the latter (see below).

As the plots show, there is a relatively large mean systematic uncertainty here in measurements of the spectral index $\alpha$. These are very likely (unavoidable) substantial overestimates, for the following reason. For the data gathered, errors are quoted for each magnitude or flux measurement, but there is no knowledge of if these errors are related between two wavebands. In most cases they are likely to be correlated for consecutive data taken with the same telescope but with different filters. The errors associated with the spectral index as derived from the colour (e.g. $V$-$I$) will therefore be less than this error as estimated by error propagation from the magnitude or flux errors. For example, if the fluxes in $V$ and $I$ have quoted errors of 30\%, the spectral index $\alpha = 0.0 \pm 0.6$ since the two wavebands are close together in log(frequency). This estimate is therefore a huge overestimate if the flux errors are correlated.

It is evident from the lower panel of Fig. \ref{NSs-jet-model} that there is a relation between $L_{\rm X}$ and $\alpha_{\rm NIR}$ for MSXPs and atolls: the NIR spectrum becomes redder at higher luminosities. Quantitatively, $\alpha_{\rm NIR}$ becomes negative when $L_{\rm X}\simgt 10^{36}$ erg s$^{-1}$ and $\alpha_{\rm NIR}$ is positive for all data below $L_{\rm X}\approx 10^{35}$ erg s$^{-1}$.
This is opposite to the expected relation if the origin of the emission is the disc blackbody, where one would expect a bluer (hotter) spectrum at higher luminosities.

The only process expected to produce an OIR spectrum of index $\alpha < 0$ at high luminosities in these systems is optically thin synchrotron. It is therefore intriguing that $\alpha_{\rm NIR} < 0$ for atolls and MSXPs when $L_{\rm X}\simgt 10^{36}$ ergs s$^{-1}$; the X-ray luminosity range in which the jet could play a role (Fig. \ref{NSs-jet-model}, upper panel). Since there are just five NIR data points in the lower panel of Fig. \ref{NSs-jet-model} below $L_{\rm X} = 10^{35}$ erg s$^{-1}$, we perform a Kolmogorov-Smirnov (K-S) test to quantify the significance of the apparent $\alpha_{\rm NIR}$--$L_{\rm X}$ relation for the atolls/MSXPs. We use the `Numerical recipes in \small FORTRAN\normalsize ' \citep{preset92} routine `kstwo' which is the K-S test for two data sets, to determine if the values of $\alpha_{\rm NIR}$ differ significantly below and above $L_{\rm X} = 10^{35}$ erg s$^{-1}$. The maximum difference between the cumulative distributions is $D = 0.91$ with a corresponding probability of $P = 5.0\times 10^{-4}$; i.e. the probability that the NIR spectral index of the data below $L_{\rm X} = 10^{35}$ erg s$^{-1}$ belongs to the same population as the data above $L_{\rm X} = 10^{35}$ erg s$^{-1}$ is 0.05 percent.

In addition, we have carried out a Spearman's Rank correlation test on the $\alpha_{\rm NIR}$--$log~L_{\rm X}$ and $\alpha_{\rm OPT}$--$log~L_{\rm X}$ data of atolls/MSXPs (Table \ref{tab-NSs-spearman}). An anti-correlation between $\alpha_{\rm NIR}$ and $log~L_{\rm X}$ is found at the 3.8$\sigma$ confidence level, supporting the above K-S test results. We do not find a significant relation between $\alpha_{\rm OPT}$ and $log~L_{\rm X}$ except when we impose a somewhat arbitrary X-ray luminosity cut: for data above $L_{\rm X}> 10^{36}$ erg s$^{-1}$ there is an anti-correlation at the 3.5$\sigma$ confidence level. This again could be due to the jet contribution dominating at these highest luminosities. The confidence of this result should be taken with caution as it could be dominated by the group of data with the highest $\alpha_{\rm OPT}$ values which happen to lie just above the $L_{\rm X}= 10^{36}$ erg s$^{-1}$ cut.

\begin{table}
\begin{center}
\caption[Results of the Spearman's Rank correlation]{Results of the Spearman's Rank correlation between $\alpha$ and $log~L_{\rm X}$ for atolls/MSXPs.}
\label{tab-NSs-spearman}
\begin{tabular}{llll}
\hline
OIR data&Range in $L_{\rm X}$&Correlation&Significance\\
used    &(erg s$^{-1}$)      &coefficient&            \\
\hline
$\alpha_{\rm NIR}$&all                 &$r_{\rm s}=-0.63$&3.8$\sigma$\\
$\alpha_{\rm OPT}$&all                 &$r_{\rm s}=-0.14$&0.8$\sigma$\\
$\alpha_{\rm OPT}$&$L_{\rm X}> 10^{36}$&$r_{\rm s}=-0.77$&3.5$\sigma$\\
\hline
\end{tabular}
\normalsize
\end{center}
\end{table}

The optical spectral index $\alpha_{\rm OPT}$ is generally positive for atolls and MSXPs (Fig. \ref{NSs-jet-model} lower panel), but decreases at $L_{\rm X} \simgt 10^{37}$ erg s$^{-1}$. We would expect $\alpha_{\rm OPT}$ to become negative at a higher X-ray luminosity than $\alpha_{\rm NIR}$ if the origin of the redder emission component is the jets. As the X-ray luminosity is increased, the optically thin synchrotron jet component will dominate over X-ray reprocessing in the NIR bands before the optical as the jet component has a negative spectral index. At $L_{\rm X} > 10^{37}$ erg s$^{-1}$, $\alpha_{\rm OPT}$ is negative in 63 percent of the optical data of atolls/MSXPs. Below $L_{\rm X} = 10^{37}$ erg s$^{-1}$ this is 7 percent. These few data in the latter group with $\alpha_{\rm OPT} < 0$ are at low $L_{\rm X}$ ($< 10^{34}$ erg s$^{-1}$) and may be due to cooler accretion discs (possibly like the BHXBs; \citeauthor{paper1}). However, the mean uncertainty in the values of $\alpha$ are fairly large, so we can only make conclusions from general trends and not individual data points.

Almost all of the spectra of the Z-sources are blue ($\alpha > 0$; Fig. \ref{NSs-jet-model}, lower panel). This supports the suggestion in Section 2.4.2 that X-ray reprocessing dominates the OIR of the Z-sources due to their larger accretion discs and lower radio jet luminosities (at a given $L_{\rm X}$). The Z-sources cannot be dominated by optically thin synchrotron emission as this requires $\alpha < 0$. Since the radio--OIR spectrum of Z-sources is $\alpha \approx 0.2$ (Section 2.4.2), the optically thick part of the jet spectrum could dominate the OIR if $\alpha_{\rm OIR} \approx 0.2$ also (although this requires the optically thick/thin break to be in the optical regime or bluer). $\alpha_{\rm OPT,NIR} > 0.2$ is observed for most of the data from the Z-sources (Fig. \ref{NSs-jet-model} lower panel), which implies this is not the case, however we cannot rule out an optically thick jet origin to a few of the OIR Z-source data (those with lower spectral indices).

The lower panel of Fig. \ref{NSs-XR-model} shows how the optical and NIR spectral index changes with $L_{\rm X}^{1/2}a$. The Z-source data in this panel have the largest values of $L_{\rm X}^{1/2}a$. The expected level of OIR emission from reprocessing in the disc in these systems is larger than that of the atolls/MSXPs due to their larger accretion discs and higher X-ray luminosities, further supporting the X-ray reprocessing scenario.

A direct measurement of the jet radio--NIR spectral index for the atolls/MSXPs can be made, using the NIR data which are dominated by the jets. In the luminosity range $10^{36} < L_{\rm X} < 10^{37}$ erg s$^{-1}$, the NIR data with a negative spectral index are on average 0.70 dex more luminous than expected from the jet model \citep[which assumes a flat radio--NIR spectrum; we know the radio luminosity at this $L_{\rm X}$ from][]{miglfe06}. Some low level contribution from the disc could only partly explain this excess. The corresponding radio--NIR spectral index in that range of $L_{\rm X}$ is $\alpha \approx 0.16$ (0.70 dex in luminosity divided by 4.5 dex in frequency between radio and NIR). The optically thick to optically thin break frequency in NSXBs is thought to be in the mid-IR \citep{miglet06}, making the optically thick radio--mid-IR jet spectrum more inverted; $\alpha_{\rm thick} \geq 0.2$.

\subsection{Applications of the Correlations}

The existence of the OIR--X-ray relations leads to a number of intriguing tools and uses for quasi-simultaneous multi-wavelength data.

\subsubsection{The Rise and Fall of a BHXB Outburst}

During a transient outburst typical of BHXBs, the source will either remain in the hard state for the entire outburst \citep[e.g. XTE J1118+480; GRO J1719--24; see also][]{brocet04} or transit into the soft (or intermediate or very high) state, before returning to the hard state and declining in luminosity. A hysteresis effect has been identified in BHXB transients, and at least one NSXB transient: the hard to soft state transition always occurs at a higher X-ray luminosity than the transition back to the hard state (\citealt*{maccco03,maccet03,yuet03}; \citealt{fendet04,homabe05}).

\begin{figure}[h]
\centering
\includegraphics[width=10.5cm,angle=0]{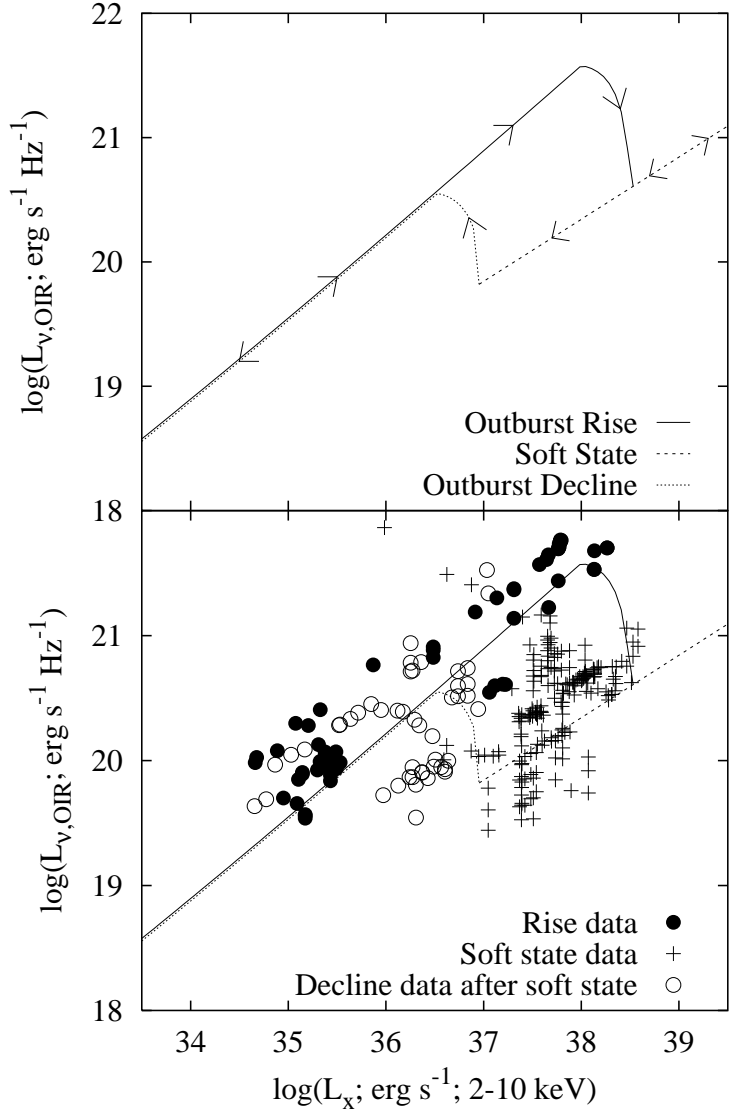}
\caption[The rise and fall of a BHXB outburst]{Top panel: Schematic of the expected $L_{\rm OIR}$--$L_{\rm X}$ behaviour for a BHXB outburst that enters the soft state. Lower panel: As the top panel but with OIR data from the rise, soft state and decline (for sources that entered the soft state) stages of outbursts.}
\label{corr-minird}
\end{figure}

Here, we present a prediction of the OIR behaviour of an outburst that results from this hysteresis effect and the models in Section 3.2. The following sequence of events should occur for a BHXB outburst that enters the soft state:

\begin{itemize}
\item $L_{\rm OIR}$ and $L_{\rm X}$ increase in the hard state rise of the outburst, with jets and reprocessing contributing to the OIR.
\item The source enters the soft state, quenching the jet component. $L_{\rm OIR}$ drops but $L_{\rm X}$ is maintained.
\item $L_{\rm OIR}$ and $L_{\rm X}$ decrease before transition back to the hard state, when the jet component returns.
\item $L_{\rm OIR}$ and $L_{\rm X}$ continue to decline, with jets and X-ray reprocessing contributing towards $L_{\rm OIR}$.
\end{itemize}

This sequence is illustrated in the top panel of Fig. \ref{corr-minird}. In this schematic, we fix the hard-to-soft state transition at $L_{\rm X}\sim 10^{38}$ erg $s^{-1}$ and the soft-to-hard, at $L_{\rm X}\sim 10^{37}$ erg $s^{-1}$. To test our hysteresis prediction, we split the hard state BHXB data into data from the rise of outbursts and data from the declines. We define rise and decline as before and after the peak X-ray (2--10 keV) luminosity of the outburst, respectively. We do not include data on the decline for sources that remained in the hard state throughout the outburst, as this is not what our prediction is testing. In the lower panel of Fig. \ref{corr-minird} we plot these and the soft state data. Data from persistent sources (LMC X--3 and GRS 1915+105) and from sources in or near quiescence are not included. We find that the prediction is consistent with the data, with some inevitable scatter from errors as described in Section 2.4.2. The loop is larger in the NIR data, as is expected because the jets contribute more in the NIR than in the optical regime. Other reasons for any deviations from the expected models are also discussed in Section 2.4.2.

I return to this subject in Chapter 4, where the above hysteresis is indeed observed with the predicted hard state slopes. However there is a twist: the NIR luminosity at a given $L_{\rm X}$ is higher on the decline than on the rise of an outburst (at least in one source).

\subsubsection{The Mass Accretion Rate}

In Section 1.2.3 I show how $L_{\rm X}$ and \.m are thought to be linked for BHXBs and NSXBs. Here, we can use the empirical $L_{\rm OIR}$--$L_{\rm X}$ correlations to link the OIR luminosity to the mass accretion rate. For the NSXBs this can be compared to observed OIR luminosities only when (a) $L_{\rm X}$ is low and the jet component is not dominant, and (b) when the NSXB is in a hard state (see above).

\smallskip BHXBs: $L_{\rm OIR}\propto L_{\rm X}^{0.6}\propto$ \.m$^{1.2}$

\smallskip NSXBs: $L_{\rm OIR}\propto L_{\rm X}^{0.6}\propto$ \.m$^{0.6}$
\newline\newline
Essentially, $L_{\rm OIR}$ for BHXBs and NSXBs respond to $L_{\rm X}$ in the same way, but not to \.m, since $L_{\rm X}$ varies with \.m differently for BHXBs and NSXBs. From equations (1) and (7) of \cite{kordet06} we can estimate accretion rates directly from $L_{\rm OIR}$ in the hard state:

\smallskip BHXBs: $L_{\rm OIR}/erg$ $s^{-1}\approx 5.3\times 10^{13}$ (\.m/$g$ $s^{-1}$)$^{1.2}$

\smallskip Or: \.m/$g$ $s^{-1}\approx 3.7\times 10^{-12}$($L_{\rm OIR}/erg$ $s^{-1}$)$^{0.8}$

\smallskip NSXBs: $L_{\rm OIR}/erg$ $s^{-1}\approx 3.2\times 10^{23}$ (\.m/$g$ $s^{-1}$)$^{0.6}$

\smallskip Or: \.m/$g$ $s^{-1}\approx 6.7\times 10^{-40}$($L_{\rm OIR}/erg$ $s^{-1}$)$^{1.7}$
\newline\newline
Given the level of the scatter in the correlations, we expect these calculations to be accurate to $\sim$ one order of magnitude. The OIR luminosity is all that is required to estimate \.m for hard state objects, however \.m would be more accurately measured from $L_{\rm X}$, where most of the energy is usually released. In addition, it is possible to estimate the X-ray, OIR and radio luminosities in the hard state quasi-simultaneously, given the value of just one because they are all linked through correlations.

\subsubsection{The Parameters of an X-ray Binary}

\begin{figure}
\centering
\includegraphics[width=10.5cm,angle=270]{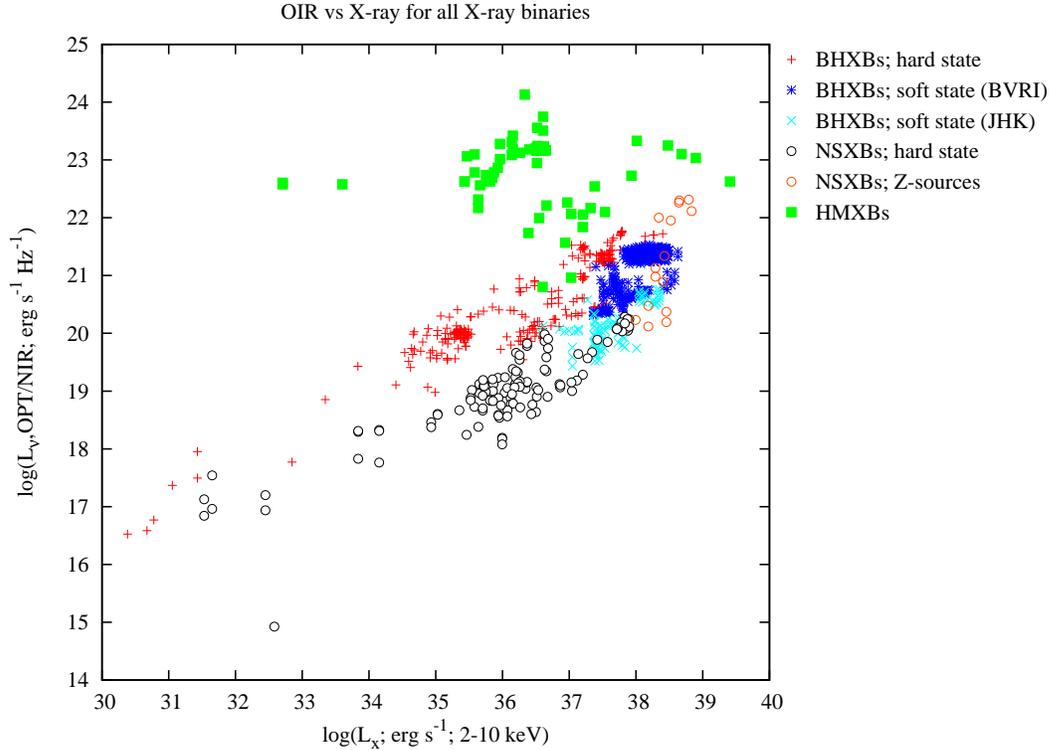}
\caption[OIR--X-ray plot for all data collected]{$L_{\rm \nu,OIR}$ versus $L_{\rm X}$ for all data collected. This diagram could be used to identify the nature of a new X-ray binary (e.g. transients in M31, where the distance and extinction are known).}
\label{corr-everything}
\end{figure}

Quasi-simultaneous OIR and X-ray luminosities can constrain the nature of the compact object (BHXB or NSXB), the mass of the companion (HMXB or LMXB) and the distance and reddening to an X-ray binary. This is possible because the data from BHXBs (in the hard and soft states), NSXBs and HMXBs lie in different areas of the $L_{\rm X}$--$L_{\rm OIR}$ diagram, with some areas of overlap (Fig. \ref{corr-everything}). If the distance and reddening towards a source are known (not necessarily at a high level of accuracy), its quasi-simultaneous OIR and X-ray fluxes could reveal the source to be any one of the above types of XB. In addition, if the nature of the compact object is known (BH or NS), but the distance and/or reddening is not, the fluxes can constrain these parameters. I stress that the total errors associated with the data (top left error bars in Figs. \ref{corr-bhxbs1}--\ref{corr-hmxbs}) need to be considered to define the areas of overlap.

Current techniques used to infer the nature of the compact object in XBs include X-ray timing analysis (e.g. thermonuclear instabilities on accreting neutron stars produce Type I X-Ray bursts), the X-ray spectrum \citep[the well-known X-ray states of BHXBs and tracks in the colour--colour diagrams of NSXBs; e.g.][]{mcclet06} and optical timing analysis in quiescence (the orbital period and radial velocity amplitude constrain the mass function). This new tool has the power to constrain the nature of the compact object requiring only $L_{\rm OIR}$, $L_{\rm X}$ and, only at high luminosities, the X-ray state of the source at the time of observations.

This tool may have many applications. X-ray all-sky monitors such as the \emph{RXTE ASM} are continuously discovering new XBs which are subsequently identified at optical wavelengths. In addition, campaigns are underway to find extragalactic XBs, many of which have optical counterparts discovered with the $HST$ \citep[e.g.][]{willet05}. It would also be interesting to see where ULXs and SMBHs \cite*[e.g. the X-ray and NIR flares seen from Sagittarius A*; e.g.][]{yuanet04} lie in the $L_{\rm OIR}$--$L_{\rm X}$ diagram with the inclusion of a mass term, and whether this could be used to constrain the BH mass in these systems.

\subsection{Summary}

I have collected a wealth of OIR and X-ray fluxes from over 40 XBs (including new observations of three sources) in order to identify the mechanisms responsible for the OIR emission using correlation analysis between the two regimes. A strong correlation between quasi-simultaneous OIR and X-ray luminosities has been discovered for BHXBs in the hard state: $L_{\rm OIR}\propto L_{\rm X}^{0.61\pm0.02}$. This correlation holds over 8 orders of magnitude in $L_{\rm X}$ and includes data from BHXBs in quiescence and at large distances (LMC and M31), which were \citep[until recently; see][]{gallet06} unattainable for radio--X-ray correlations in XBs. All the NIR (and some of the optical) BHXB luminosities are suppressed in the soft state; a behaviour indicative of synchrotron emission from the jets at high luminosities in the hard state. Comparing the hard state OIR data to radio data of \cite{gallet03}, we find that the radio--OIR jet spectrum in BHXBs is $\sim$ flat ($F_{\nu}=$ constant) at a given $L_{\rm X}$ \citep[see also][]{fend01}.

A $L_{\rm OIR}$--$L_{\rm X}$ relation is found for NSXBs but it cannot be described by a single power-law. Below $L_{\rm X}\sim 10^{38}$ erg s$^{-1}$, a NSXB is typically $\sim$20 times fainter in OIR than a BHXB, at a given $L_{\rm X}$. The average radio--OIR spectrum for NSXBs is $\alpha\approx +0.2$ at least at high luminosities when the radio jet is detected. In atolls and MSXPs, the observed spectral index of the continuum becomes red ($\alpha < 0$) above $L_{\rm X} \approx 10^{36}$ erg s$^{-1}$ and $L_{\rm X} \approx 10^{37}$ erg s$^{-1}$ for the NIR and optical continuum, respectively. The Z-sources (which occupy the highest luminosities) have a blue ($\alpha > 0$) OIR spectrum.

By comparing the observed OIR--X-ray relations with those expected from models of a number of emission processes, one is able to constrain the mean OIR contributions of these processes for XBs. Table \ref{corr-emproc1} summarises the results. We find from the level of soft state quenching in BHXBs that the jets are contributing $\sim$90 percent of the NIR emission at high luminosities in the hard state. The optical BHXB data could have a jet contribution between zero and 76 percent. In BHXBs, ambiguity arises from the fact that the slope of the expected OIR--X-ray relations from the jets and X-ray reprocessing are essentially indistinguishable. Emission from the viscously heated disc may contribute at low luminosities in BHXBs, but cannot account for the observed correlations. The OIR light from soft state BHXBs could originate from a combination of X-ray reprocessing (as the OIR--X-ray relations suggest) and the viscous disc \citep[e.g.][ see also Chapter 4]{homaet05a}. 

The dominating OIR emission processes in NSXBs vary with X-ray luminosity and between source types (atolls/MSXPs and Z-sources). Models predict that X-ray reprocessing in the accretion disc should dominate the OIR at low luminosities and the jets, if present, should dominate at high luminosities for NSXBs with relatively small accretion discs. The observed spectral index of the OIR continuum and the OIR--X-ray relations indicate this is the case in atolls and MSXPs: the jets dominate the NIR and optical emission above $L_{\rm X} \approx 10^{36}$ erg s$^{-1}$ and $L_{\rm X} \approx 10^{37}$ erg s$^{-1}$, respectively. Below these luminosities X-ray reprocessing dominates, although we cannot rule out a viscously heated disc origin to the optical emission. The average radio--NIR spectral index of the actual jets in atolls/MSXPs (at high luminosities when they are detected in the NIR) is slightly inverted: $\alpha \approx 0.16$. In the Z-sources, which have larger discs, we find that X-ray reprocessing is responsible for all OIR emission, and the radio--OIR jet spectrum has to be $\alpha \leq 0.2$ (otherwise the jet spectrum would dominate over the disc). However, the optically thick part of the jet spectrum could dominate in a few cases.

In general, the exact contributions of the emission processes are likely to be sensitive to many individual parameters, such as the size of the accretion disc and the shape of the jet spectrum (e.g. the OIR luminosity of the jets is very sensitive to the slope of the optically thick jet spectrum).

\begin{table}
\small
\caption[The OIR emission processes that can describe the empirical OIR--X-ray relations]{The OIR emission processes that can describe the empirical OIR--X-ray relations (and the OIR spectral index--$L_{\rm X}$ relations for NSXBs).}
\label{corr-emproc1}
\small
\begin{tabular}{lcccc}
\hline
Sample&X-ray&Jet emission&Viscous disc&Intrinsic\\
&reprocessing&&&companion\\
\hline
BHXBs; OPT; hard state&$\surd$&$\surd$&$\times$&$\times$\\
BHXBs; NIR; hard state&$\times$&$\surd$&$\times$&$\times$\\
BHXBs; OIR; soft state&$\surd$&$\times$&$\surd$&$\times$\\
Atolls/MSXPs; OPT; $\simgt 10^{37}$ erg s$^{-1}$&$\times$&$\surd$&$\times$&$\times$\\
Atolls/MSXPs; NIR; $\simgt 10^{36}$ erg s$^{-1}$&$\times$&$\surd$&$\times$&$\times$\\
Atolls/MSXPs; OPT; $\simlt 10^{37}$ erg s$^{-1}$&$\surd$&$\times$&$\surd$&$\times$\\
Atolls/MSXPs; NIR; $\simlt 10^{36}$ erg s$^{-1}$&$\surd$&$\times$&$\times$&$\times$\\
Z-sources; OIR&$\surd$&$\times$&$\times$&$\times$\\
HMXBs; OIR; all&$\times$&$\times$&$\times$&$\surd$\\
\hline
\end{tabular}
\normalsize
\end{table}

I have presented a prediction of the $L_{\rm OIR}$--$L_{\rm X}$ behaviour of a BHXB outburst that enters the soft state, where the peak $L_{\rm OIR}$ in the hard state rise is greater than in the hard state decline (the well known hysteretical behaviour). The data obtained in this Chapter appear to agree with the prediction, with some scatter (see Chapter 4 for a slight revision of this behaviour motivated by empirical hysteresis). It is also possible to estimate the X-ray, OIR and radio luminosity and the mass accretion rate in NSXBs and hard state BHXBs quasi-simultaneously, from observations of just one of these wavebands, since they are all linked through correlations. This is currently possible due to daily monitoring of the X-ray and OIR fluxes being done for some sources by e.g. the \emph{RXTE ASM} and by ground-based telescopes. In addition, we have discovered a potentially powerful tool to complement current techniques, that can constrain the nature of the compact object, the mass of the companion and the distance/reddening towards an XB, given only the quasi-simultaneous X-ray and OIR luminosities. Data from BHXBs (in the hard and soft states), NSXBs and HMXBs lie in different areas of the $L_{\rm X}$--$L_{\rm OIR}$ diagram, with small areas of overlap. The tool is most useful for e.g. faint sources with poor timing analysis or at large distances, i.e. extragalactic XBs and new transients. Finally, future $L_{\rm OIR}$--$L_{\rm X}$ correlation analysis could be used to constrain the emission process contributions in the intermediate and very high X-ray states of BHXBs and in ULXs and SMBHs.

\newpage

\begin{center}
{\section{OIR spectral energy distributions}}
\end{center}

\subsection{Introduction}

In Chapter 2 correlation analysis was used to constrain the dominating OIR emission processes in X-ray binaries. In this Chapter I present a collection of OIR Spectral Energy Distributions (SEDs) which are used to further constrain the emission processes, this time from the shape of the spectrum of the continuum. In Chapter 2 it was shown that a number of separate processes could explain most of the relations between OIR, X-ray and radio luminosities in BHXBs and NSXBs. However, these processes are expected to have different spectral slopes, and so SEDs can help to reveal the ones which dominate at a given luminosity. The results of this Chapter are presented and discussed in \citeauthor{paper1} (focusing on the BHXBs) and \citeauthor{paper4} (focusing on the NSXBs).

As with Chapter 2, most of the data collection, presentation and analysis is my own work. The co-authors of the papers provided valuable comments and suggested changes. Data were used that were provided by the co-authors (Rob Fender, Peter Jonker, Rob Hynes, Catherine Brocksopp, Jeroen Homan and Michelle Buxton) and also Erik Kuulkers. Use of the word `we' is adopted in this Chapter when referring to work not done by myself alone.

\subsection{Methodology}

I conducted a literature search for OIR Spectral Energy Distributions (SEDs) of BHXBs and NSXBs. Fluxes were used when two or more OIR wavebands were quasi-simultaneous (preferably during the same run using the same instrument, but often within one day using different optical and NIR telescopes). No X-ray fluxes were required for the SEDs, however for the BHXBs the X-ray state of the source on the date was noted. No distinctions were made in the data from NSXBs except to note the different source type; atoll, Z-source or MSXP. Where the companion star significantly contributes to the emission (i.e. in quiescence), the estimated wavelength-dependent contribution of the companion was subtracted (adopting the quoted contributions given in the papers from which the data were acquired). The lowest flux data from the Z-sources were not used as the companion contribution is largely unknown.

Fluxes and magnitudes were de-reddened and converted to intrinsic monochromatic luminosities in the same manner as described in Sections 2.2.1 and 2.2.2, adopting the estimates of source distance and interstellar extinction given in Tables \ref{tab-16BHXBs1}--\ref{tab-19NSXBs-2}. The data reduction procedure for observations obtained for the first time by us (the sources GX 349+2, LMC X--2 and 4U 0614+09) are described in Sections 2.2.3 and 2.2.4.

\subsection{Results}

\begin{figure*}
\includegraphics[height=20cm,angle=0]{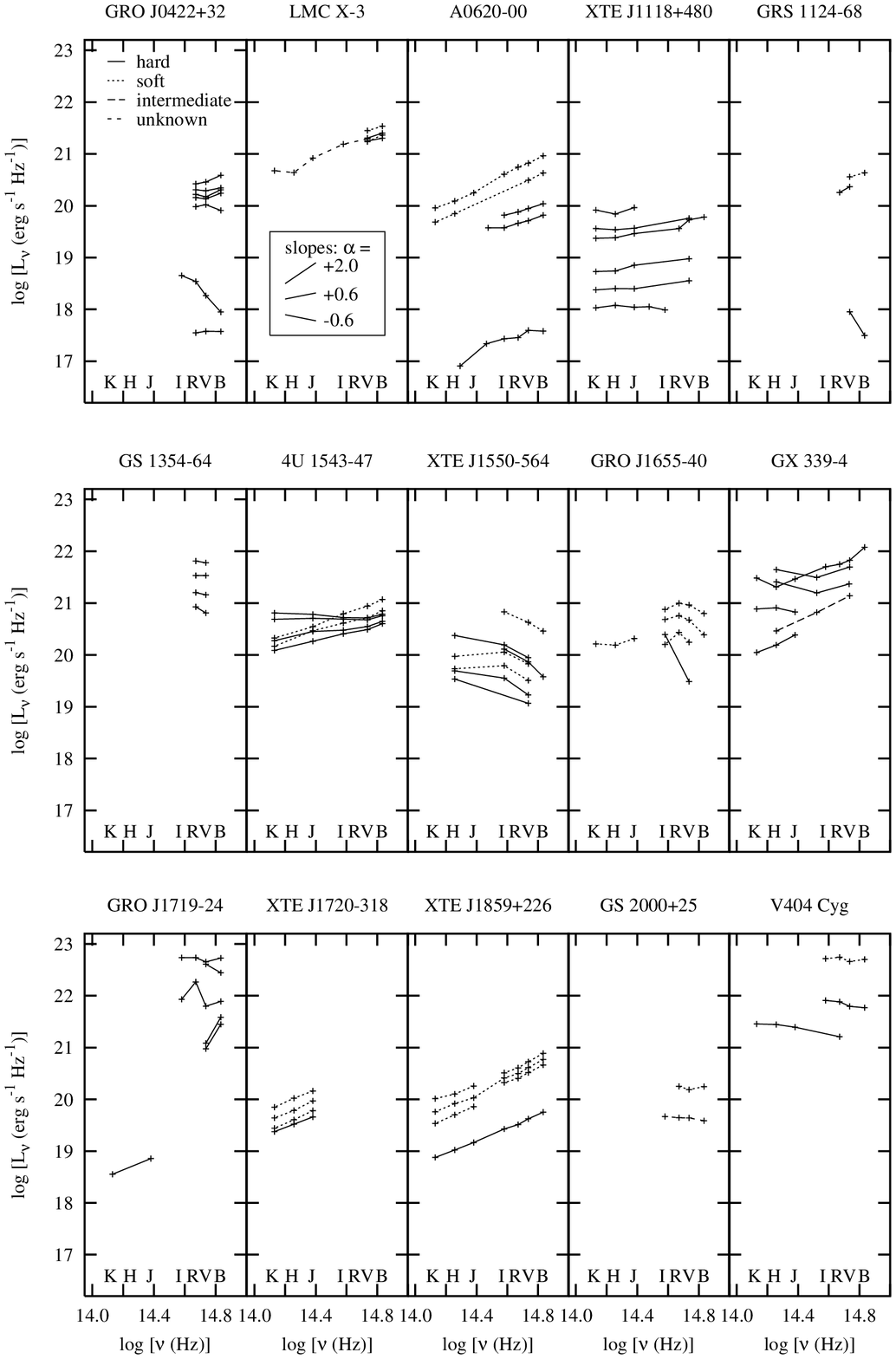}
\caption[OIR SEDs of BHXBs]{Spectral Energy Distributions (SEDs) of 15 BHXBs. The key in the top left panel corresponds to the X-ray state of the source on the date of observation. A key to the slope of the continuum, i.e. the spectral index $\alpha$ (where $L_{\nu} \propto \nu^{\alpha}$), is indicated in the second panel on the top row. The dates of observations and references for the data are listed in Table \ref{BH-SEDs1}.}
\label{corr-seds}
\end{figure*}

\begin{table}
\begin{center}
\caption[Dates and references for the data in Fig. \ref{corr-seds} (1)]{Dates and references for the data in Fig. \ref{corr-seds} (1).}
\label{BH-SEDs1}
\small
\begin{tabular}{llll}
\hline
Source&X-ray&Dates (MJD)&References\\
      &state&&\\
\hline
GRO J0422+32 &hard&48866, 48870, 49006--7, 49039, 49209, 49590&1 -- 4\\
LMC X--3 &hard&50151, 50683&\\
         &soft&50324, 51038&\\
         &unknown&46804    &5 -- 7\\
A0620--00 &hard&42792, 43097--100, 42859&\\
          &soft&42650--2, 42703 &8 -- 11\\
XTE J1118+480 &hard&51652, 51975, 53385-6, 53393, 53409, 53412&12 -- 13$^1$\\
GRS 1124--68 &hard&48453, 48622&\\
             &soft&48367-71    &14 -- 16\\
GS 1354--64 &hard&50778, 50782, 50851, 50888&17\\
4U 1543--47 &hard&52486, 52490, 52495, 52501&\\
            &soft&52454, 52469&18\\
XTE J1550--564&hard&51260, 51630, 51652, 51717&\\
              &soft&51210, 51660, 51682&19 -- 20\\
GRO J1655--40 &hard&53422 &\\
              &soft&50217, 50254, 50286, 50648&21 -- 23\\
\hline
\end{tabular}
\normalsize
\end{center}
$^1$Some of this data is previously unpublished and was obtained with the Liverpool Telescope and the United Kingdom Infrared Telescope. See Brocksopp et al. (in preparation) for the data reduction recipe used. References for Table \ref{BH-SEDs1} and \ref{BH-SEDs2}:
(1) \cite{bartet94};
(2) \cite*{goraet96};
(3) \cite*{castet97};
(4) \cite*{hyneha99};
(5) \cite{brocet01a};
(6) \cite*{trevet87};
(7) \cite*{trevet88};
(8) \cite*{robeet76};
(9) \cite*{okeet77};
(10) \cite{oke77};
(11) \cite*{kleiet76};
(12) \cite{chatet03};
(13) \citeauthor{paper1} (see Brocksopp et al. in preparation);
(14) \cite{kinget96};
(15) \cite{dellet98};
(16) \cite{bail92};
(17) \cite{brocet01b};
(18) \cite{buxtet04};
(19) \cite{jainet01a};
(20) \cite{jainet01b};
(21) \cite*{buxtet05a};
(22) \cite{hyneet98};
(23) \cite{chatet02};
(24) \cite{corbfe02};
(25) \cite{homaet05a};
(26) \cite{sekiwy93};
(27) \cite{allejo93};
(28) \cite{allegi93};
(29) \cite{brocet04};
(30) \cite{nagaet03};
(31) \cite{haswet00};
(32) \cite{hyneet02a};
(33) \cite{charet81};
(34) \cite*{chevil90};
(35) \cite*{szkoet89};
(36) \cite*{gehret89};
(37) \cite{casaet91};
(38) \cite{hanet92}.
\end{table}

\begin{table}
\begin{center}
\caption[Dates and references for the data in Fig. \ref{corr-seds} (2)]{Dates and references for the data in Fig. \ref{corr-seds} (2).}
\label{BH-SEDs2}
\small
\begin{tabular}{llll}
\hline
Source&X-ray&Dates (MJD)&References\\
      &state&&\\
\hline
GX 339--4 &hard&44748--59, 49539, 50648, 52368, 52382&\\
          &intermediate&52406&23 -- 25\\
GRO J1719--24 &hard&49096, 49101, 49108, 49266, 49280, 49539&23, 26 -- 29\\
XTE J1720--318&hard&52782 &\\
              &soft&52659, 52685, 52713&30\\
XTE J1859+226 &hard&51605--8 &\\
              &soft&51465, 51469, 51478, 51488, 51501&31, 32\\
GS 2000+25$^1$&soft&47360 &\\
              &unknown&47482&33, 34\\
V404 Cyg &hard&47684, 47718--27 &\\
         &soft&47678&35 -- 38\\
\hline
\end{tabular}
\normalsize
\end{center}
$^1$ This source (alternative name: QZ Vul) is not tabulated in Tables \ref{tab-16BHXBs1} and \ref{tab-16BHXBs2}; its distance and reddening are 2.7$\pm$0.7 kpc and $Av\sim$ 3.5, respectively \citep{jonket04}. See Table \ref{BH-SEDs1} for references.
\end{table}

OIR SEDs were collected of 15 BHXBs in a range of luminosities and X-ray states and are presented in Fig. \ref{corr-seds}. The dates of observations and references are given in Table \ref{BH-SEDs1}--\ref{BH-SEDs2}. The SEDs of 17 NSXBs are shown in Fig. \ref{NSs-SEDs} and the references for the data are listed in Table \ref{tab-19NSXBs-2}. A separate table is provided for the BHXBs because the references of much of the data are different from those for the BHXB data used in Chapter 2. The references for the NSXB data are all in Table \ref{tab-19NSXBs-2} because the SEDs of NSXBs are already used in Chapter 2 to estimate the optical and NIR spectral indices. Unlike Chapter 2, all of the quasi-simultaneous OIR wavebands are shown in the SEDs in this Chapter (Fig. \ref{NSs-SEDs}) whereas only a few wavebands were used to estimate the spectral indices in Chapter 2 (see Section 2.4.3). In the following subsection we attempt to interpret the OIR SEDs in terms of the dominating emission processes.

\subsection{Interpretation \& Discussion}

\subsubsection{The BHXB SEDs}

Although no clear patterns are visible from the BHXB SEDs in Fig. \ref{corr-seds} on first inspection, closer analysis reveals support for many of the conclusions made in Chapter 2. Optically thin synchrotron emission is expected to produce an OIR spectral index $\alpha<$ 0, and this is seen in part of the SEDs of 10 out of 14 BHXBs in the hard state. In the soft state only 3 out of 10 BHXBs are observed to have $\alpha<$ 0, suggesting a synchrotron component playing a larger role in the hard state than in the soft \citep[$\alpha$ in some sources could be dominated by uncertainties in $A_{\rm V}$; e.g. the SEDs of XTE J1550--564 are red even in the soft state, which is likely due to an underestimated extinction; see][]{jonket04}. In comparison, $\alpha>$ 0 is seen in the SEDs of 10/14 BHXBs in the hard state and 10/10 in the soft state. These spectra are likely to have a thermal origin and agree with recent analysis of optical/UV SEDs of 6 BHXBs \citep{hyne05}. The SEDs generally appear redder in the hard state than in the soft, as is observed by the NIR suppression in the soft state (Section 2.4.1).

A clear suppression of the NIR, and not the optical bands in the soft state is visible in the SEDs of GX 339--4, XTE J1550--564 and 4U 1543--47 \citep{homaet05a,jainet01b,buxtet04}. In contrast, the NIR spectral index of XTE J1720--318 and XTE J1859+226 appear not to change between the hard and soft states. The SEDs show no substantial evidence for the turnover in the jet spectrum from optically thick ($\alpha\sim$ 0) to optically thin ($\alpha\sim-$0.6) synchrotron emission, as we may expect to see in the hard state \citep[except in GX 339--4; see][]{corbfe02}. This is consistent with the turnover lying redward of the $K$-band. In some systems, $\alpha$ is more negative at low luminosities. This effect could be the result of (a) a cooler accretion disc, or (b) the jets contributing more than the disc at low luminosities. Since the former process cannot explain the steeply negative SEDs of a few BHXBs at low luminosities, we suspect both processes may play a role.

\begin{figure*}
\includegraphics[height=20cm,angle=0]{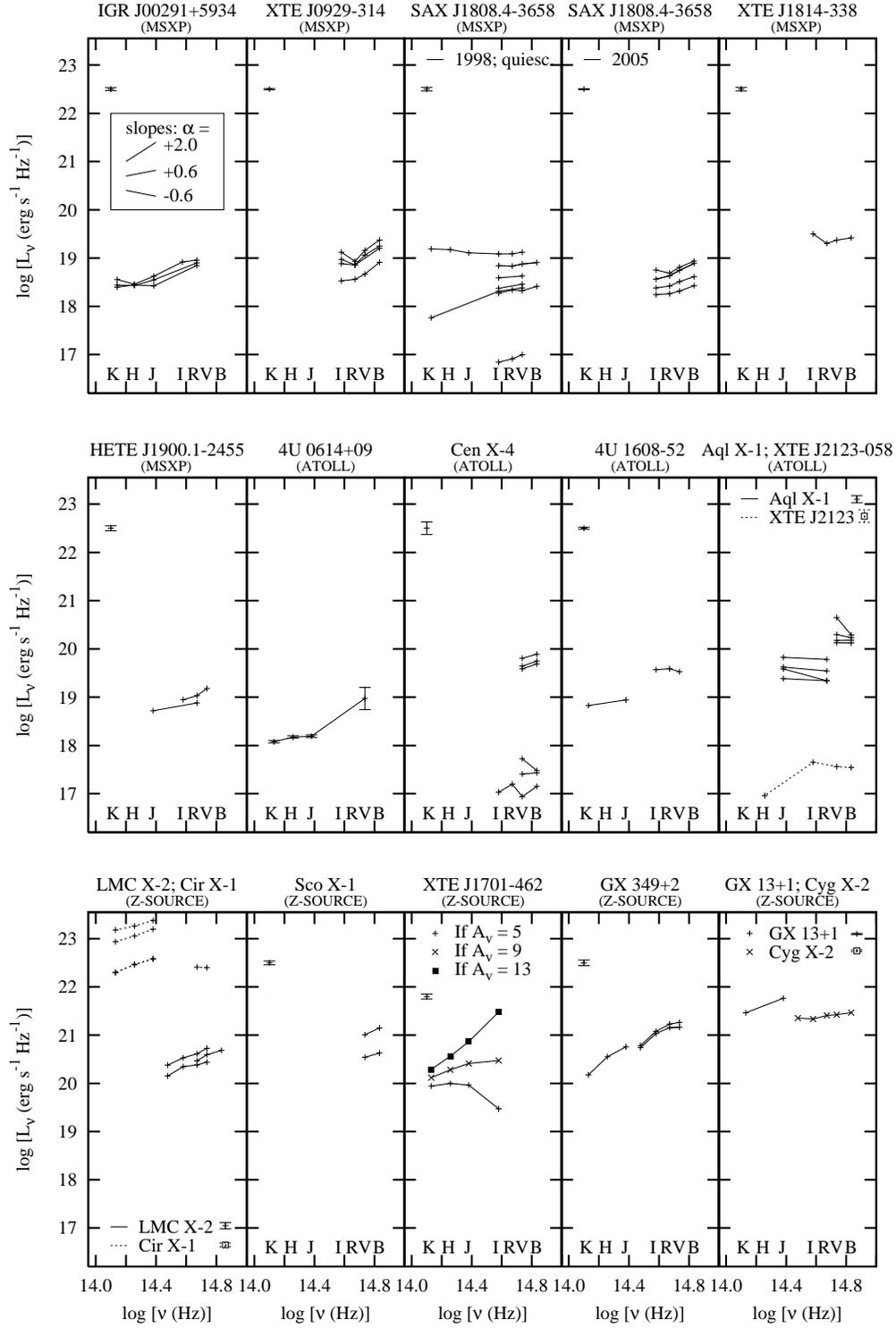}
\caption[OIR SEDs of NSXBs]{Spectral Energy Distributions of 17 NSXBs. The key in the top left panel corresponds to the slope of the continuum, i.e. the spectral index $\alpha$ (where $L_{\nu} \propto \nu^{\alpha}$). MSXP, ATOLL and Z-SOURCE refer to millisecond X-ray pulsars, atoll sources and Z-sources, respectively. The references for the data are listed in Table \ref{tab-19NSXBs-2}. The mean systematic error for the data of each source is indicated in each panel (in the top left corner in most cases).}
\label{NSs-SEDs}
\end{figure*}

Finally, the hard state NIR excess seen in GX 339--4 (\citealt{corbfe02,homaet05a}; interpreted as where the optically thin jet spectrum meets the blue thermal spectrum) is here also seen in LMC X--3 (tentatively), 4U 1543--47, XTE J1550--564 and V404 Cyg. In these sources, we interpret the NIR excess as originating in the optically thin part of the jet spectrum.

\subsubsection{The NSXB SEDs}

We group the panels in Fig. \ref{NSs-SEDs} into the different types of NSXB: MSXPs in the upper panels, then atolls, then Z-sources in the lower panels. One can visually see a NIR excess ($\alpha < 0$) joining a blue ($\alpha > 0$) optical spectrum in the SEDs of four of the five MSXPs (the `V'-shape). This is indicative of two separate spectral components and, like the BHXBs, is interpreted as the optically thin jet spectrum meeting the thermal spectrum of the accretion disc. The NIR excess disappears at low luminosities in a total of four outbursts of three MSXPs; no NIR excess is seen below $L_{\rm \nu,OIR} < 10^{18}$ erg s$^{-1}$ Hz$^{-1}$ in any MSXP or in fact any NSXB. This behaviour was suggested by the $\alpha$--$L_{\rm X}$ plots in Section 2.4.3 but here the two spectral slopes are identified in the same spectra of the same source. Radio emission has been detected briefly from MSXPs at the peak of their outbursts \citep*[e.g.][]{gaenet99}; this is consistent with the jet being brightest at their highest X-ray luminosities, as the NIR jet suggests.

Of the five atolls in Fig. \ref{NSs-SEDs}, one (Aql X--1; the atoll with the most data and highest luminosity) has a negative spectral index; more-so at high luminosity. There is little NIR coverage of atolls in the literature (at least at high luminosities) and from the data we have, the OIR SEDs are positive in all sources except Aql X--1, consistent with the disc dominating the OIR. The optical spectral index for Aql X--1 is very negative at the highest luminosities, suggesting all of the OIR was dominated by the jets during the peak of the bright 1978 outburst of this source. The SED of 4U 0614+09 flattens (becomes redder) in the NIR; this is known to be where the disc spectrum meets that of the jet \citep{miglet06}.

There is no evidence for a jet ($\alpha < 0$) component contributing to the OIR spectra of any of the seven Z-sources in Fig. \ref{NSs-SEDs}. The jets should be OIR-bright in Z-sources but because of their larger discs, X-ray reprocessing dominates. For the new transient Z-source XTE J1701--462 \citep{homaet07a}, the optical extinction is not well constrained and we show three SEDs in Fig. \ref{NSs-SEDs} of the same data, using different values of $A_{\rm V}$ to illustrate the possible range in spectral indices. The spectral indices in a few observations are also consistent with an optically thick jet which extends from the radio regime (e.g. the shallower spectral index of Cyg X--2). This would require the optically thick--optically thin turnover to be at a higher frequency than the optical, contradictory to its position for at least one atoll \citep{miglet06}.

\subsubsection{The Power of the Jets}

The evidence in this Chapter confirms the presence of a flat or slightly inverted jet spectrum from radio to NIR wavelengths in some NSXBs (atolls and MSXPs at least) and in hard state BHXBs, which breaks to an optically thin spectrum redward of the OIR regime. This turnover is close to the $K$-band in BHXBs (see Section 2.4.1). The turnover may be further into the NIR for NSXBs \citep{miglet06} but if the radio--NIR spectrum is flat (as is shown here), this would require an inverted ($\alpha > 0$) optically thick radio--mid-IR jet spectrum. The power of the jets is sensitive to the position of the turnover since it is dominated by the higher energy photons. However, the most recent calculations of the jet power \citep[e.g.][]{gallet05,heingr05,miglfe06} already assume the jet spectrum extends to the IR (for the BHXBs at least), and extra calculations are beyond the scope of this work.

\subsection{Summary}

Quasi-simultaneous OIR data have been collected for 32 LMXBs: 15 black hole systems, five MSXPs, five atolls and seven Z-sources. The optical and NIR spectral slopes (parameterised by the spectral index $\alpha$) are compared between object types and as a function of luminosity. In Chapter 2 it was not possible to differentiate between different emission processes which are predicted to have similar OIR--X-ray relations, but this has been possible in this Chapter for emission processes which are expected to differ in spectral slope.

\begin{table}
\caption[The OIR emission processes that can describe the OIR SEDs]{The OIR emission processes that can describe the OIR SEDs.}
\label{seds-emproc1}
\small
\begin{tabular}{lcccc}
\hline
Sample&X-ray       &Jet emission&Jet emission&Viscous\\
      &reprocessing&(opt. thin) &(opt. thick)&disc   \\
\hline
BHXBs; OPT; hard state&$\surd$&$\times$&$\times$&$\surd$\\
BHXBs; NIR; hard state&some&$\surd$&$\times$&some\\
BHXBs; OIR; soft state&$\surd$&$\times$&$\times$&$\surd$\\
Atolls/MSXPs; OPT; highest $L_{\rm \nu,OPT}$&$\times$&$\surd$&$\times$&$\times$\\
Atolls/MSXPs; NIR; high $L_{\rm \nu,NIR}$&$\times$&$\surd$&$\times$&$\times$\\
Atolls/MSXPs; OPT; low/moderate $L_{\rm \nu,OPT}$&$\surd$&$\times$&$\times$&$\surd$\\
Atolls/MSXPs; NIR; low $L_{\rm \nu,NIR}$&$\surd$&$\times$&$\times$&$\surd$\\
Z-sources; OIR&$\surd$&$\times$&some&$\surd$\\
\hline
\end{tabular}
\normalsize
\end{table}

In Table \ref{seds-emproc1} I summarise the emission processes that can explain the OIR SEDs of BHXBs and NSXBs. The majority of the OIR SEDs of BHXBs are thermal, with $\alpha > 0$. This is consistent with X-ray reprocessing in the disc dominating the emission, largely agreeing with the literature. The NIR region of the BHXB SEDs are almost all non-thermal ($\alpha < 0$) in the hard state, indicating optically thin synchrotron emission from the jets; a separate spectral component which is seen to join the disc component in some sources, around the $I$-band. This emission is in fact found to be suppressed in the soft state, which is expected as the jet component is quenched in both this regime and the radio.

The emission in the soft state of BHXBs could originate from a combination of X-ray reprocessing (as the OIR--X-ray relations suggest; see Chapter 2) and the viscous disc \citep[see Chapter 4 and e.g.][]{homaet05a}. The SEDs of many BHXBs are redder at low luminosities, which seems to be due to both a cooler disc blackbody and possibly a higher fractional contribution from the jets. BHXBs may be jet-dominated at low luminosities (for more on jet-dominated states, see \citealt*{fend01,falcet04}; \citealt{fendet04,gallet05,kordet06}). The optically thick jet spectrum of BHXBs must extend to near the $K$-band for (a) the radio spectrum to be flat \citep{fend01}, (b) the radio--to--OIR spectrum to appear flat (Section 2.4.1), and (c) the NIR spectrum to be optically thin (this Chapter).

We find that thermal emission due to X-ray reprocessing can explain all the NSXB data except at high luminosities for some NSXBs, namely the atolls and MSXPs. Optically thin synchrotron emission from the jets (with an observed OIR spectral index of $\alpha < 0$) dominate the NIR light at high luminosities and the optical at the peak of the bright outbursts in these systems. For NSXB Z-sources, the OIR observations can be explained by X-ray reprocessing alone, although synchrotron emission may make a low level contribution to the NIR. The optically thick jet spectrum could dominate the OIR in one or two cases if the spectral break is at a higher frequency than currently thought. The turnover in the optically thick jet spectrum of NSXBs must be redward of the $K$-band for the NIR to be optically thin (as seen in the atolls/MSXPs). The radio spectral index of NSXBs is fairly unconstrained \citep[e.g.][]{miglfe06} and could be more inverted than that of BHXBs because (a) the radio--to--OIR spectral index is slightly inverted (Section 2.4.3) and (b) the turnover is in the mid-IR in at least one atoll \citep{miglet06}.

\newpage

\begin{center}
{\section{NIR -- X-ray hysteresis}}
\end{center}

\subsection{Introduction}

In Chapter 2 a global correlation between X-ray and OIR luminosities was found for BHXBs in the hard state. This led onto a prediction (Section 2.5.1) of the track a BHXB should trace out during a transient hard state--soft state--hard state outburst. Here, outbursts from individual BHXBs are examined and a new hysteresis effect is found: the NIR emission appears to be weaker in the hard state rise of an outburst than the hard state decline of an outburst at a given X-ray luminosity. The hysteresis could be the result of a number of different effects, each of which are discussed in detail. In addition, correlations are found between X-ray and NIR luminosities in the soft state in two sources. In the same manner as Chapter 2, correlation analysis is used to constrain the emission mechanisms which dominate in the soft state.

This project led to a publication in the Monthly Notices of the Royal Astronomical Society ($MNRAS$; Russell, Maccarone, K\"ording \& Homan 2007; \citeauthor{paper5}). Thomas Maccarone led the initial motivation of this project; to test whether there are parallel tracks in jet power versus X-ray power in X-ray binaries in the hard state, and provided much of the discussion and introduction. Elmar K\"ording contributed to the discussion, in particular the dependency of the mass accretion rate on a number of system variables. Jeroen Homan provided most of the data and suggested some changes to the manuscript. The literature search, methodology and results were provided by myself and I also contributed to all sections and compiled, re-arranged and edited the manuscript. Since in this Chapter there are many discussions involving the dimensionless viscosity parameter $\alpha$ (the symbol alpha is used extensively for this in the literature), I adopt the symbol $\gamma$ for the spectral index ($F_{\nu}\propto \nu^{\gamma}$) as opposed to $\alpha$ (which is used in other Chapters), in this Chapter.

\subsection{The data}

I searched the literature for BHXBs with well sampled quasi-simultaneous ($\Delta t \leq 1$ day) NIR--X-ray or radio--X-ray data on the rise and/or decline of a transient outburst (where NIR refers to $J$, $H$ or $K$-band). Data from three sources were found: 4U 1543--47, XTE J1550--564 and GX 339--4. XTE J1550--564 is the only source for which both rise and decay data of the same outburst were available. Apparent fluxes were converted to intrinsic luminosities adopting the best known estimates of the distance to the sources, the extinction $A_{\rm V}$ and neutral hydrogen column density $N_{\rm H}$. In Table \ref{tab-hyst-sourcedata} these values and the outbursts, telescopes used and references of the data are listed. $A_{\rm V}=5$ is adopted for XTE J1550--564 as opposed to $A_{\rm V}=2.5$, which is adopted in Chapters 2 and 3. A value of $A_{\rm V}\sim 5$ is inferred from the hydrogen column density measured from X-ray observations of this source \citep*{tomset01,tomset03b,kaaret03}. A value of $A_{\rm V}\sim 2.5$ produces a red optical spectrum ($\gamma < 0$ where $F_{\nu}\propto \nu^{\gamma}$; Chapter 3), which would indicate a much cooler disc than is expected, whereas with $A_{\rm V}\sim 5$ the spectrum is blue ($\gamma > 0$), as is expected for the Rayleigh-Jeans side of the disc blackbody spectrum. While uncertainties in the extinction will affect the relative normalisations of the $L_{\rm NIR}$--$L_{\rm X}$ relations for different sources, they will not affect the slopes of the relations for the individual sources. At the time of carrying out the projects of Chapters 2 and 3, the lower value of $A_{\rm V}$ was the most cited for this source and so that was the one we adopted; but the above more careful analysis indicates the larger value is more accurate, so it is adopted here in this later project.

To calculate the X-ray unabsorbed 2--10 keV luminosity, a power-law of spectral index $=-$0.6 (photon index $\Gamma=1.6$) is assumed when the source is in the hard state and a blackbody at a temperature of 1 keV when the source is in the soft state (these are the same as in the previous Chapters). The resulting luminosities do not change significantly if similar reasonable approximations to these values are adopted. \emph{RXTE} ASM X-ray counts were converted to flux units using the \emph{NASA} tool \emph{Web-PIMMS}. Since data close to state transitions are used, the dates of these transitions are adopted from those found by the analyses of the authors in the literature. Inaccurate background subtraction could severely affect the X-ray luminosity estimates at low luminosities. For 4U 1543--47, \cite{kaleet05} have accounted for the Galactic X-ray background \citep[e.g.][]{lahaet93,branet02,pageet06}. The X-ray data used here for XTE J1550--564 are more than two orders of magnitude above the quiescent level so no background subtraction is necessary. The X-ray background contamination for GX 339--4 is uncertain \citep{nowaet02}, but for a reasonable choice of background fluxes the slope of the resulting NIR--X-ray correlation (see Section 4.3) does not change.

\begin{figure}
\centering
\includegraphics[width=13.5cm,angle=0]{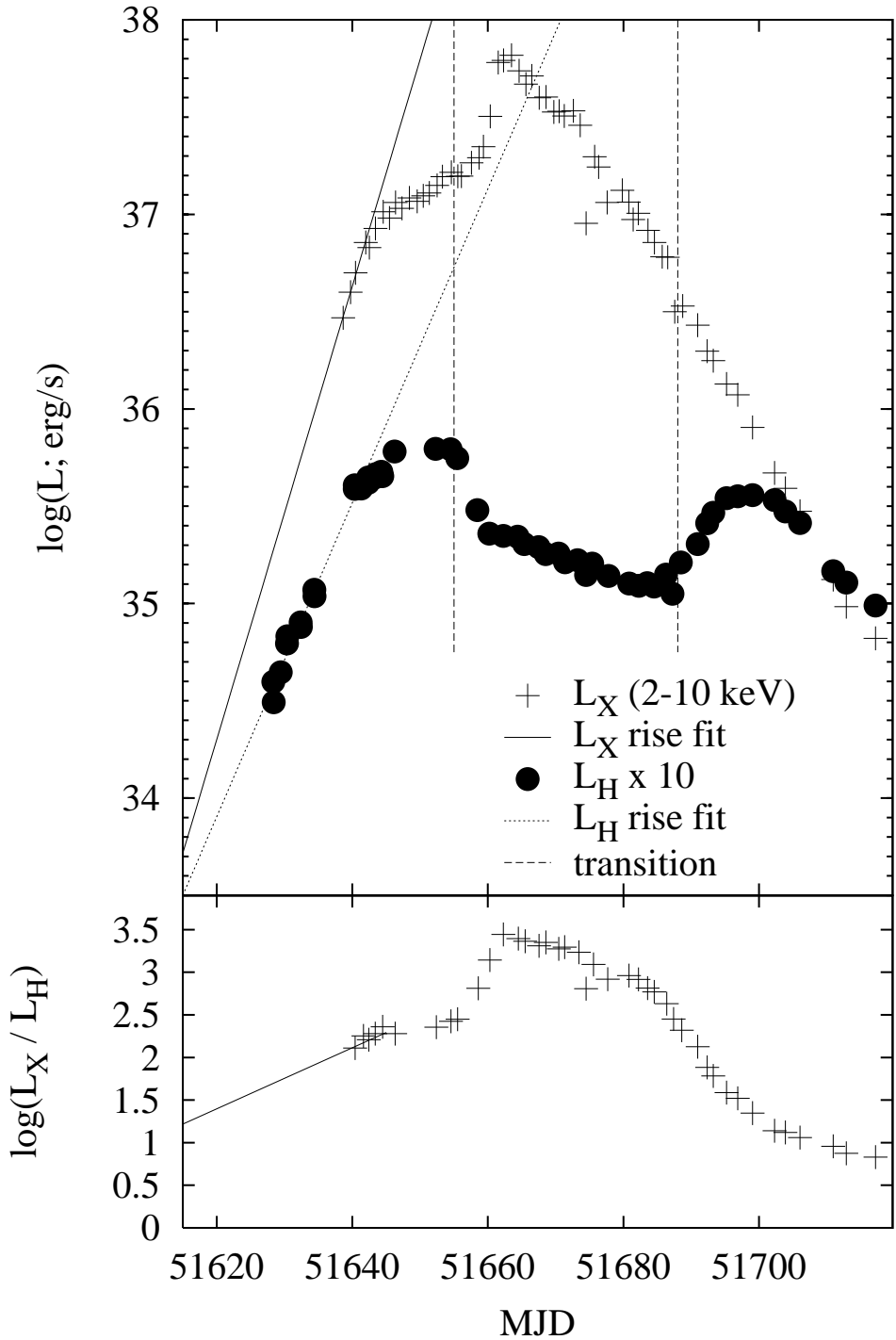}
\caption[Light curve of the 2000 outburst of XTE J1550--564]{The NIR and X-ray light curves of the 2000 outburst of XTE J1550--564 (upper panel). The lines are fit to the initial rise data before the break in the rise behaviour of both light curves (see Fig. \ref{hyst-J1550lc-X} for errors on these fits). All X-ray data are from the \emph{RXTE} ASM except in the hard state decline, which are from the \emph{RXTE} PCA/HEXTE. The ratio of X-ray to NIR fluxes are shown in the lower panel.}
\label{hyst-J1550lc}
\end{figure}

\begin{figure}
\centering
\includegraphics[width=10.7cm,angle=270]{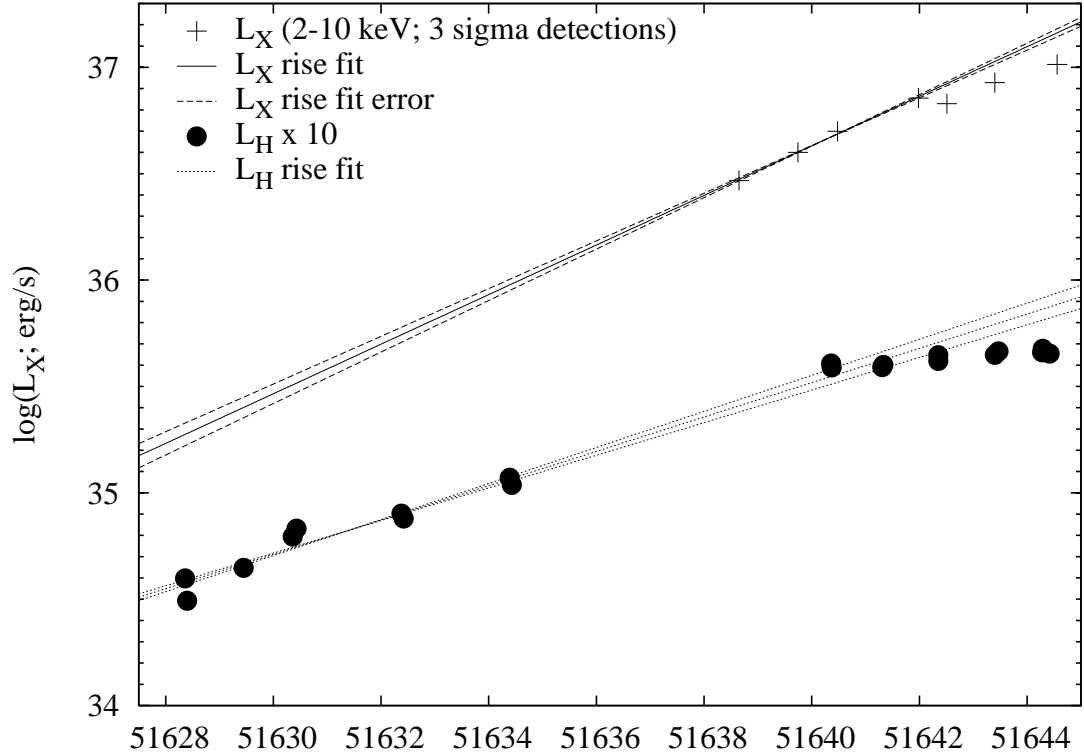}
\caption[Light curve of the 2000 outburst of XTE J1550--564: zoom-in on the rise fit]{A zoom-in on the fits to the X-ray and NIR light curve rise in Fig. \ref{hyst-J1550lc}.}
\label{hyst-J1550lc-X}
\end{figure}

\begin{table*}
\begin{center}
\caption[The sources used in Chapter 4 and data collected]{The sources used and data collected to calculate their luminosities.}
\label{tab-hyst-sourcedata}
\small
\begin{tabular}{lllllll}
\hline
Source &Outburst&Telescopes&Distance&$A_{\rm V}$&$N_{\rm H}$ /    &Flux\\
       &        &          &/ kpc   &           &10$^{21} cm^{-2}$&refs\\
\hline
4U 1543--47 &2002; decline&YALO 1.0 m, \emph{RXTE} &7.5$\pm$0.5 &1.55$\pm$0.15&4.3$\pm$0.2&1, 2\\
\vspace{1mm}
&&PCA/HEXTE/ASM&&&&\\
XTE  &2000; rise &YALO 1.0 m, \emph{RXTE} &5.3$\pm$2.3&5.0&8.7$\pm$2.1&3, 4\\
\vspace{1mm}
J1550--564&+ decline&PCA/HEXTE/ASM&&&&\\
GX 339--4 &2002; rise&YALO 1.0 m, &8.0$^{+7.0}_{-1.0}$&3.9$\pm$0.5&6$^{+0.9}_{-1.7}$&5\\
&&\emph{RXTE} PCA/ASM&&&&\\
GX 339--4 &1999; decline&ATCA$^1$, \emph{RXTE} PCA& & & &6\\
\hline
\end{tabular}
\normalsize
\end{center}
\normalsize
$^1$Australia Telescope Compact Array. The distances and extinction estimates are from the references in Tables \ref{tab-16BHXBs1} and \ref{tab-16BHXBs2}, except $A_{\rm V}=5$ for XTE J1550--564 \citep[][ see text]{tomset01,tomset03b,kaaret03}. References for the fluxes used:
(1) \cite*{buxtet04};
(2) \cite*{kaleet05};
(3) \cite*{jainet01b};
(4) \cite*{millet01};
(5) \cite*{homaet05a};
(6) \cite*{nowaet05}.
\normalsize
\end{table*}

For NIR data near quiescence (or close to the lowest NIR fluxes observed from the source) I only include data which are $\geq 0.5$ mag above the measured quiescent flux level. This quiescent level is also subtracted from all NIR data so that only the outburst flux remains (the NIR in quiescence is likely dominated by the companion star). The well-sampled NIR quiescent level of XTE J1550--564 is brighter by $\sim 0.25$ magnitudes after the 2000 outburst than prior to it \citep[Fig 1a of][]{jainet01b}. This enhancement could be due to thermal emission from the heated accretion disc, or jet emission.

It is possible that the star or outer regions of the accretion disc are re-emitting X-ray radiation which was absorbed during the outburst.  The donor star in XTE J1550--564 \citep[which has spectral type G8 to K4;][]{oroset02} occupies around 10 percent of the solid angle seen by the X-ray source and should absorb $\sim 70$ percent of the radiation incident on it \citep{basksu73}.  Since the outburst gave off $\sim 3\times10^{43}$ ergs and the source brightened by about $10^{33}$ ergs s$^{-1}$ in the infrared, it could maintain this brightening for $\sim 70$ years if all the energy was lost in this infrared waveband. The brightening was present in at least the optical $V$-band too, but the brightening of the broadband optical--infrared spectrum could still be sustained for more than one year, longer than the duration of the quiescent light curve used here. In addition, \cite{oroset02} found that the quiescent optical flux level of XTE J1550--564 was lower in May 2001 than before \emph{and} after the 2000 outburst. The (not-dereddened) base levels of $H=16.25$ have been subtracted from the data on the hard state rise and $H=16.0$ subtracted from the hard state decline data. Long-term variations in infrared quiescent levels have been observed in GRO J0422+32, A0620--00 and Aql X--1 (\citealt{reynet07}; C. Bailyn, private communication).

NIR intrinsic \citep*[de-reddened using the extinction $A_{\rm V}$ listed in Table \ref{tab-hyst-sourcedata}, according to][]{cardet89} luminosities were calculated adopting the approximation $L_{\rm NIR}\approx \nu F_{\nu,NIR}$ (i.e. approximating the spectral range of each filter to the central wavelength of its waveband; the same method as \citeauthor{paper1}).

\subsection{Results}

The X-ray and NIR light curves of the 2000 outburst of XTE J1550--564 are plotted in Fig. \ref{hyst-J1550lc} \citep[upper panel; see also][]{jainet01b}. In both X-ray and NIR light curves there is a break in the relation between luminosity and time on the outburst rise. One explanation for this could be that the point at which the luminosity versus time curve flattens corresponds to the start of the hard/intermediate state, and not to the bona fide hard state.  However, this traditionally occurs when the X-ray spectrum begins to soften, which happened about six days after the X-ray break (the first \emph{transition} dashed line in Fig. \ref{hyst-J1550lc}; see also \citealt{millet01}). We note also that the flattening in the infrared appears to happen a few
days before the flattening in the X-rays, and the optical ($V$-band) precedes the NIR ($H$-band) also by $\sim 3$ days \citep{jainet01b}.  One may expect these delays to reveal a relation between wavelength and delay time, but the order between the wavebands is incorrect for this to be the case; the optical change is followed by the infrared and then X-ray. The reasons for these delays between wavebands are not clear. \cite{uemuet00} found that the X-rays lagged the optical rise of an outburst of XTE J1118+480 by about 10 days, and interpreted it as an `outside-in' outburst; the optical emission originating from the viscously-heated (as opposed to X-ray heated) accretion disc. In the case here of XTE J1550--564, we would not expect the NIR light to lag the optical, as is observed, if the viscous disc is the origin of the emission.

We have fit exponentials in luminosity versus time for the initial rise data before the break to a shallower slope in both light curves (Fig. \ref{hyst-J1550lc} upper panel; Fig. \ref{hyst-J1550lc-X}). These are required to obtain the $L_{\rm NIR}$--$L_{\rm X}$ relation before the break, as there is little quasi-simultaneous data on the rise but clear luminosity--time relations exist. For the X-ray light curve we have made a fit (with errors since we are extrapolating) to the first four \emph{RXTE} ASM data points (all have $>3\sigma$ significance). The break appears just after these data in the X-rays and all four points lie very close to the fit. All $H$-band data before the gap at MJD 51634--40 were used for the $L_{\rm NIR}$--time fit.

\begin{figure}
\centering
\includegraphics[width=12cm]{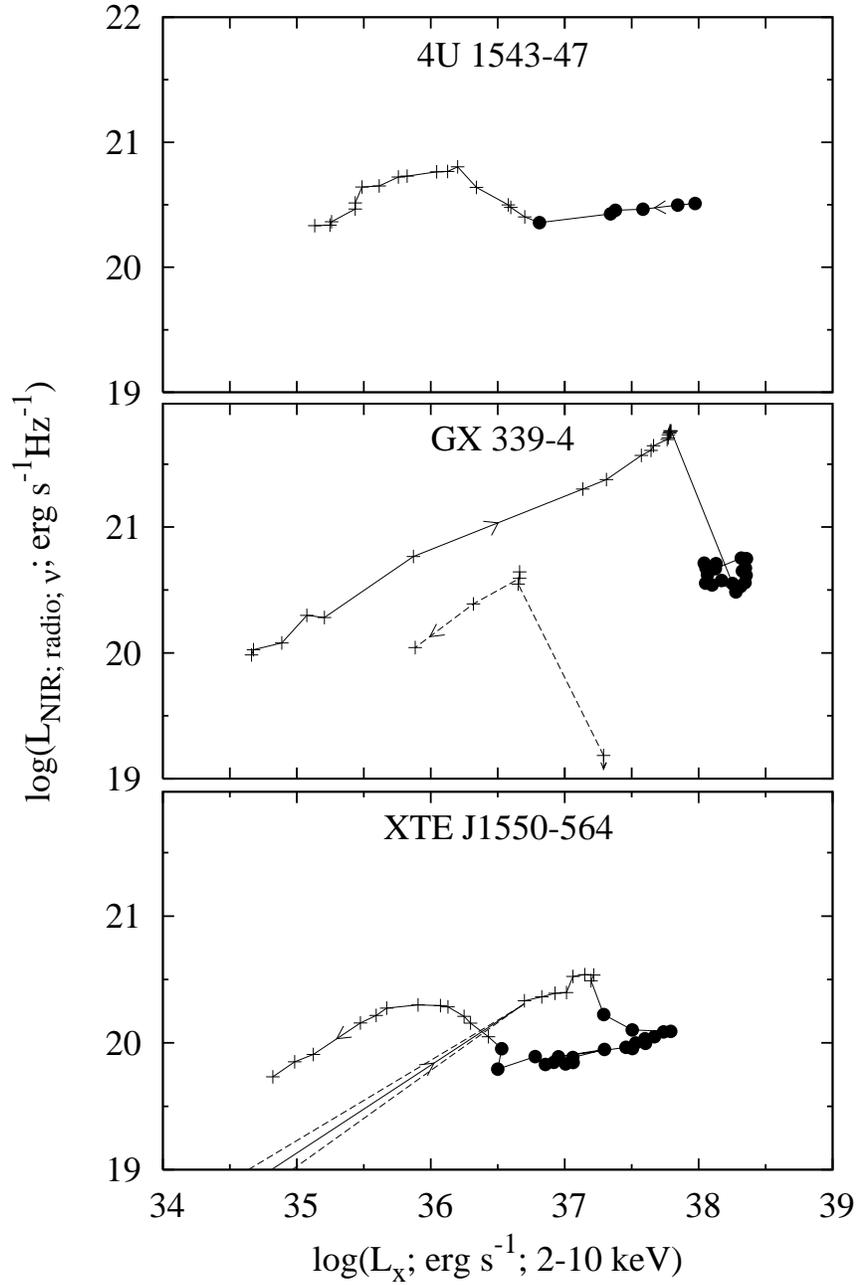}
\caption[NIR/radio--X-ray rise, soft state and decline]{X-ray luminosity versus quasi-simultaneous monochromatic NIR/radio luminosity for the three sources. Crosses are hard state data and filled diamonds represent data in the soft state when the jet is quenched. The NIR data are $J$-band for 4U 1543--47 and $H$-band for XTE J1550--564 and GX 339--4. The data on the decline for GX 339--4 are radio 8.64 GHz from a different outburst, not NIR (and is the only radio data we use).}
\label{hyst-3panels}
\end{figure}

In Fig. \ref{hyst-3panels} we plot the quasi-simultaneous NIR (or radio) and X-ray luminosities for each source. The luminosity--time fits to the rise of the XTE J1550--564 outburst in NIR and X-ray (Figs. \ref{hyst-J1550lc} and \ref{hyst-J1550lc-X}) are used to infer the $L_{\rm NIR}$--$L_{\rm X}$ relation on the rise for this source before the data are sampled well (propagating the errors from both fits). This fit has a slope $L_{\rm NIR}\propto L_{\rm X}^{\beta}$ where $\beta=0.69 \pm 0.03$. Interestingly, fitting the data after the aforementioned NIR and X-ray breaks (and before the state transition) in the XTE J1550--564 light curve yields the same slope: $\beta=0.7$.

For GX 339--4 we have NIR rise data and radio decline data from different outbursts \citep{homaet05a,nowaet05}. We therefore cannot compare the normalisations of the rise and decline in this source even if we assume the two outbursts were similar hysteretically; assumptions are required of the radio-to-NIR spectral index. For 4U 1543--47 we have outburst decline data only, but there is an increase in the NIR flux when the source enters the hard state, as is expected from jet emission (\citealt{homaet05a}; \citeauthor{paper1}).

\begin{table}
\begin{center}
\caption[Measured slopes of the NIR--X-ray (or radio--X-ray) relations]{Measured slopes of the $L_{\rm NIR}$--$L_{\rm X}$ (or $L_{\rm radio}$--$L_{\rm X}$) relations. $L_{\rm NIR}\propto L_{\rm X}^{\beta}$.}
\label{tab-hyst-slopes}
\small
\begin{tabular}{lll}
\hline
Source &Data&Slope $\beta$\\
\hline
4U 1543--47&soft state&0.13$\pm$0.01\\
4U 1543--47&hard state decline&0.5--0.8\\
GX 339--4 &hard state rise&0.55$\pm$0.01\\
GX 339--4 &hard state decline$^2$&0.70$\pm$0.06\\
XTE J1550--564&hard state rise&0.69$\pm$0.03\\
XTE J1550--564&soft state$^1$&0.23$\pm$0.02\\
XTE J1550--564&hard state decline&0.63$\pm$0.02\\
\hline
\end{tabular}
\normalsize
\end{center}
\normalsize
$^1$All soft state data are used for this fit except those above the apparent relation at the beginning and end of the soft state phase (likely to be during state transition). $^2$The slope here is measured from the five radio data points.
\normalsize
\end{table}

One sees from Fig. \ref{hyst-3panels} that the NIR is enhanced at a given X-ray luminosity on the decline compared to the rise for the only source with both rise and decline data from the same outburst, XTE J1550--564 (based on the extrapolation of the X-ray data on the rise, as explained above). The level of enhancement is $\sim 0.7$ dex in NIR luminosity, or a factor of $\sim 5$. The NIR luminosity at a given $L_{\rm X}$ (i.e. the normalisation) also differs between sources; this may reflect distance and/or interstellar absorption uncertainties.

\begin{figure}
\centering
\includegraphics[height=12cm,angle=270]{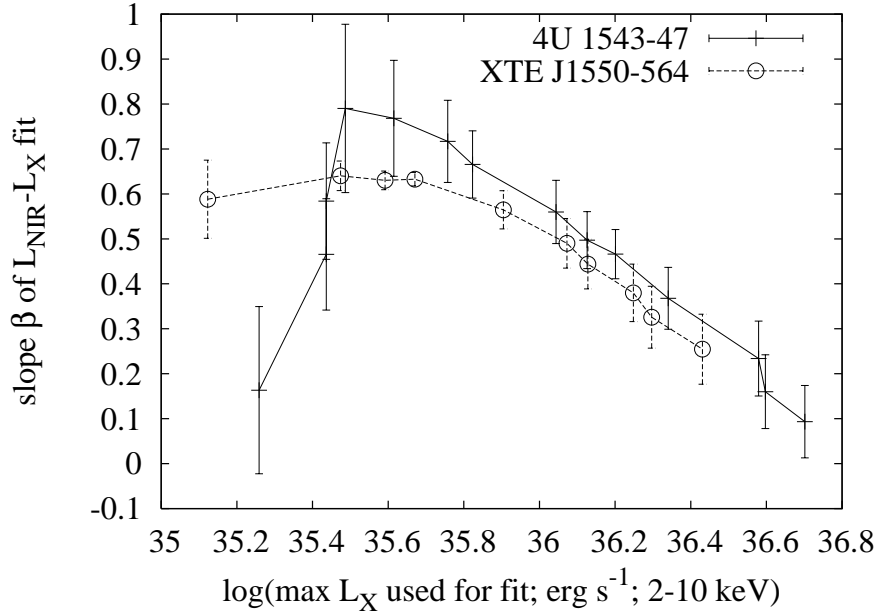}
\caption[The slope of the NIR--X-ray relation for the decline data of 4U 1543--47 and XTE J1550--564]{The slope $\beta$ of the $L_{\rm NIR}$--$L_{\rm X}$ relation for the decline data of 4U 1543--47 and XTE J1550--564, as a function of the maximum $L_{\rm X}$ of the data used to infer the fit.}
\label{hyst-slope-vs-Lx}
\end{figure}

A striking observation from Fig. \ref{hyst-3panels} is the similarity of the slope of the $L_{\rm NIR}$--$L_{\rm X}$ (or $L_{\rm radio}$--$L_{\rm X}$) relation in the hard state rise and decline between sources. In Table \ref{tab-hyst-slopes} we tabulate the slopes measured for each source. The slope of the relation for 4U 1543--47 and XTE J1550--564 on their declines depends on which hard state data are used, since there is an initial NIR rise after the transition to the hard state. The inferred slope $\beta$ is plotted as a function of the range of data used to measure the fit, in Fig. \ref{hyst-slope-vs-Lx}. We use all the data at low X-ray luminosity and impose a cut at a higher X-ray luminosity: $L_{\rm X}^{\rm max}$. $\beta$ is plotted against log($L_{\rm X}^{\rm max}$) in Fig. \ref{hyst-slope-vs-Lx}. There appears to be a smooth relation between $\beta$ and $L_{\rm X}^{\rm max}$; the slope becomes shallower when we include the data at high $L_{\rm X}$ because of the initial NIR rise as the source declines in $L_{\rm X}$. The cut required to measure the $L_{\rm NIR}$--$L_{\rm X}$ relation once the two wavebands are positively correlated should be at a value of $L_{\rm X}$ after the NIR begins to drop. At this cut we have $\beta \sim 0.5$--0.8 and $\beta = 0.63\pm0.02$ for 4U 1543--47 and XTE J1550--564, respectively. From Table \ref{tab-hyst-slopes} we can see that generally, the slopes in the hard state ($\beta = 0.5$--0.7) are in agreement with the radio--X-ray and optical/NIR--X-ray correlations found from these and other sources (\citealt{corbet03,gallet03,gallet06}; \citeauthor{paper1}).

In addition, Fig. \ref{hyst-3panels} shows that all the NIR and radio data we use are suppressed in the soft state, indicating its most likely origin in the hard state is the jet, which is expected to be quenched. In addition, the optical--NIR colours are red ($\gamma < 0$) in all three sources in the hard state, indicating optically thin emission \citep{jainet01b,buxtet04,homaet05a}. The optical and NIR luminosity from the viscously-heated accretion disc should vary relatively smoothly across the state transition, as should X-ray reprocessing in the disc unless it largely depends on the hard spectral component (see Chapter 2 for a more detailed discussion of these and other sources). The emission may also not be smooth if it is correlated with $\alpha$, the dimensionless viscosity parameter, and if $\alpha$ changes across the transition (see Section 4.4.2). Since we see a stronger soft state quenching in the NIR than in the optical wavebands in our sources \citep{jainet01b,buxtet04,homaet05a}, the longer wavelengths have much larger ratios of jet to disc contributions (this is expected since the spectral index of the jet spectrum is less than that of the disc).

In Table \ref{tab-hyst-slopes} we also state the slopes of the $L_{\rm NIR}$--$L_{\rm X}$ relations in the soft state for 4U 1543--47 and XTE J1550--564. The theoretically expected slope if the origin of the NIR emission is X-ray reprocessing in the disc is $\beta = 0.5$ \citep{vanpet94}. The shallower slopes observed are close to those expected if the origin of the NIR emission is the viscously heated disc: $\beta = 0.15$ (\citeauthor{paper1}). Since we are using a finite X-ray energy range (2--10 keV) as opposed to the bolometric X-ray luminosity, $\beta$ may be inaccurate because the fraction of the bolometric X-ray luminosity in this range will change with $L_{\rm X}$ in the soft state. However, this fraction would have to change by a factor of several during the $\sim$ one dex drop in $L_{\rm X}$ for $\beta$ to be inaccurate by a factor of $\sim 3$--4 (as is required if $\beta \sim 0.5$ for X-ray reprocessing). We can therefore conclude from correlation analysis that the viscously heated disc is likely to dominate the NIR in the soft state in these two sources. \cite{homaet05a} also found this to be the dominating NIR emission process for GX 339--4 in the soft state (from the same data that we use here) since changes in the NIR preceded those in the X-ray by several weeks.

It is unlikely that disc emission contaminating the infrared luminosity plays a substantial role in the observed hysteresis effect in XTE J1550--564 in the hard state, because the separation between the rise and decline parallel tracks is a factor of about five in $L_{\rm NIR}$; the disc component would have to be dominating the NIR luminosity in the rising hard state for this to be the explanation of the effect. It would then be quite surprising for the colours to change so dramatically at the state transition and would have to be for some other reason than the quenching of the jet. In the decaying
hard state of XTE J1550--564, a rise is seen in $I$ and $V$ as well as in $H$ at the
time of the state transition, indicating again that the jet
power-to-disc power ratio is larger for the decaying hard state than
for the rising hard state \citep[see the discussion of the optical `reflare' in][]{jainet01b}.

\subsection{Discussion}

In this Section we concentrate mainly on interpreting the hysteresis effect. Nearly all models for jet production suggest that the kinetic power
supplied to the relativistic jet will be linearly proportional to the
mass accretion rate \citep[e.g.][]{falcbi95,meie01}. The observed monochromatic luminosity of the synchrotron self-absorbed part of the jet spectrum will be proportional
to the kinetic luminosity of the jet to the 1.4 power \citep[e.g.][]{blanko79,falcbi95,heinsu03}.  As a
result, infrared and longer wavelength fluxes may be expected to
be excellent tracers of the mass accretion rate, and would be much
less sensitive to changes in the accretion efficiency than at X-ray
wavelengths \citep{kordet06}. However, in at least one neutron star X-ray binary 4U 0614+09, the NIR jet contribution is optically thin with the break between optically thick and optically thin emission lying further into the mid-infrared \citep{miglet06}. For BHXBs, it has been shown that this break lies close to the NIR (\citealt{corbfe02}; \citeauthor{paper1}) but may change with luminosity \citep{nowaet05}.

For a jet being powered from an advection
dominated accretion flow \citep[ADAF;][]{narayi95}, it has been suggested
that the relation between $\dot{m}$ and the jet's kinetic luminosity
can also depend on the dimensionless viscosity parameter $\alpha$, and
on some fudge factors related to the fact that the standard ADAF treatment of
rotational velocities and azimuthal magnetic field strengths are only
approximations \citep{meie01}.  This assumes that an ADAF produces the X-ray emission. 
We then find that:
\begin{equation}
P_{\rm J} \propto \dot{m}\alpha^{-1}f(g,j)
\label{hyst-eqn1}
\end{equation}
where $P_{\rm J}$ is the total power injected into the jet, $\dot{m}$ is the
mass accretion rate through the disc, and $f(g,j)$ is a correction
term which is a function of the black hole spin $j$ and the fudge factors which are rolled into $g$.

In the following subsections we use these arguments and discuss possible explanations for the observed hysteresis effect.

\subsubsection{Hysteresis traces the radiative efficiency of the accretion flow?}

\begin{figure}
\centering
\includegraphics[width=13.5cm,angle=0]{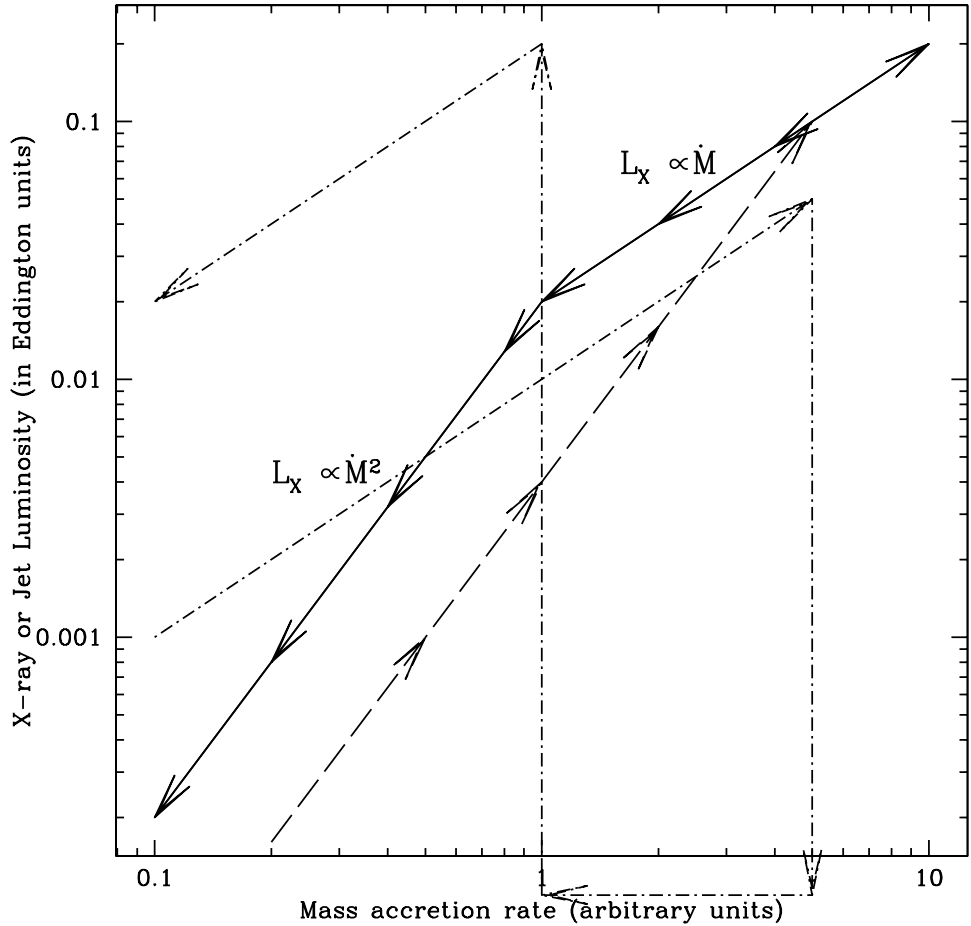}
\caption[A schematic of the X-ray luminosity versus accretion rate]{A schematic diagram of the X-ray luminosity versus accretion
rate.  The units of the accretion rate are such that, in the
`normal' hard state, at the state transition \emph{back to} the hard state, the X-ray
luminosity is 2\% of the Eddington luminosity and the mass accretion
rate is unity.  The solid line corresponds to the decaying hard
state, and to sources like Cygnus X-1 which show no state transition/luminosity hysteresis
effects \citep{smitet02}, while the dashed line corresponds to the rising portion of
hysteretic light curves.  The arrows trace a typical outburst's track
in this diagram.  The dot-dashed line shows the evolution of the jet
kinetic power, as inferred from the observed hysteresis effect.  It drops below the $y$-axis in the high soft state to
indicate that the jet is `quenched' in this state.}
\label{hyst-Lx-vs-mdot}
\end{figure}

Let us first consider the simplest possible interpretation of the
parallel tracks in XTE J1550--564 -- that they represent variations in the
radiative efficiency in the rising and decaying hard states, with
the mass accretion rate being traced perfectly by the infrared flux.
Here, we effectively assume that the variations in $\alpha$ and the
fudge factors are unimportant, as are variations in the radiative
efficiency of the jet itself for producing infrared emission.  Since
both the rising hard state and the decaying hard state are
observed to follow the $L_{\rm radio} \propto L_{\rm X}^{\sim 0.7}$ relation (\citealt{gallet03}; 2006), they are both expected to have $L_{\rm X} \propto \dot{m}^2$ (a radiatively inefficient flow).  At
the same time, they are both expected to join to the same soft
state solution, which is believed to be a radiatively efficient ($L_{\rm X} \propto \dot{m}$)
geometrically thin accretion disc \citep{shaksu73}.  XTE J1550--564 was said to be in an intermediate state between the rise and decline of the 2000 outburst based on timing properties and a power law component in the X-ray spectrum. However, the blackbody flux was always dominant over the power law flux \citep{millet03}, and the radio/infrared jet was quenched so for purposes of this work we can treat it as a soft state from its jet properties \citep[see e.g.][]{fendet04} in the context of our following analysis. A
schematic version of the expected track followed by a hysteretic
source is plotted in Fig. \ref{hyst-Lx-vs-mdot}.  We then see that the scenario where
$\dot{m}$ is well traced in the hard state by the infrared luminosity
and in the soft state by the X-ray luminosity is immediately
invalidated as this would require the infrared luminosity to be lower at a given $L_{\rm X}$ on the decline compared to the rise -- the opposite to what is observed. To achieve what is observed, we instead require the jet (NIR) luminosity to evolve as the dot-dashed lines in Fig. \ref{hyst-Lx-vs-mdot}.

\subsubsection{Hysteresis traces $\alpha$?}

Next, let us consider the possibility that $\dot{m}$ is instead sensitive to variations in $\alpha$ ($P_{\rm J}$ and $\alpha$ are linked through equation \ref{hyst-eqn1}).  This is a natural explanation for the different
state transition luminosities in the rising and falling hard
states; it is found by \cite*{esinet97}, for
example, that the state transition luminosity should be 1.3$\alpha^2$,
while it is found by \cite{zdzi98} that the state transition
luminosity should scale approximately as $\propto \alpha^{7/5}$.
In either of these cases, then, one
would qualitatively expect that increasing $\alpha$ would increase the
state transition luminosity.  A similar result, albeit without an
analytic approximation for the size of the state transition
luminosity, is found by \cite*{meyeet05}, who
showed that hysteresis may result from the different efficiencies of
hard photons and soft photons in evaporating mass from the thin disc
into the corona.

Some of the other hysteresis properties of outbursts seem to be
qualitatively consistent with the idea that $\alpha$ variations play
at least some role.  A larger mass density at the onset of an outburst
may lead to a higher $\alpha$ when the instability is triggered.  If
this is the case, then it will naturally follow that both the peak
luminosity of the outburst, and the luminosity at which the
hard-to-soft transition occurs, will be correlated with one another
through the connection to $\alpha$, as has been shown \citep*{yuet04}.

The radiative efficiency of the hot disc will depend on which $L_{\rm NIR}$--$L_{\rm X}$ track
one chooses, and hence on $\alpha$.  In general, the X-ray luminosity
is given by $L_{\rm X}=\epsilon \dot{m}$, where $\epsilon$ is the radiative
efficiency.  For a solution where $L_{\rm X}\propto\dot{m}^2$, then
$\epsilon\propto\dot{m}$.  More specifically, when the two hard state
tracks must join onto the same radiatively efficient soft state track,
then $\epsilon\propto\frac{\dot{m}}{\dot{m_{\rm cr}}}$, where
$\dot{m_{\rm cr}}$ is the mass accretion rate at the critical state transition luminosity (which has one value for the rise track and a different one for the decline). Let us now take a general form for the $\alpha$ dependence of this value, $\dot{m_{\rm cr}}=\alpha^c$.  The radiative
efficiency of the hot disc will then scale as $\alpha^{-c}$.  As a
result, $\dot{m}$ at a given X-ray luminosity will scale as
$\alpha^{-c/2}$, since $L_{\rm X} \propto \dot{m}^2$ in the hard state.

The $\alpha$ variations will thus not lead to parallel tracks in $L_{\rm NIR}$
versus $L_{\rm X}$ if the state transition luminosity scales as
$\alpha^2$.  The $\alpha$ dependence of the jet power extraction
efficiency will cancel out exactly with the $\alpha$ dependence of the
X-ray radiative luminosity.  The estimate of the state transition
luminosity given by \cite{esinet97} can therefore be valid only if
the fudge factors vary hysteretically or if the radiative efficiency
of the jet varies hysteretically.  The dependence of the transition
luminosity on $\alpha$ must be weaker than $\alpha^2$ in order for
$\alpha$ variations to be solely responsible for the parallel tracks in $L_{\rm X}$
versus $L_{\rm NIR}$.

The estimate of the state transition luminosity put forth by \cite{zdzi98} does have a weaker than $\alpha^2$ dependence for the state
transition luminosity: $\alpha^{7/5}$.  This leads to a relation where
$L_{\rm X}\propto L_{\rm jet} \alpha^{3/5}$, a closer match to
the observations, but still problematic quantitatively.  For a
given $L_{\rm NIR}$, the X-ray luminosity is $\sim 10$ times higher in the rising
hard state than the decaying one.  The difference
between the X-ray luminosity at the hard-to-soft transition compared to the
soft-to-hard transition is a factor of about five.  If the observations
are to be interpreted in terms of variations in $\alpha$ only, then
either the state transition luminosity must scale slightly weaker than
linearly with $\alpha$, or the jet power must scale more strongly than
linearly with $\alpha$ (e.g. if the fudge factors in the \citealt{meie01}
relation scale with $\alpha$ in some way).

\subsubsection{Hysteresis traces jet behaviour?}

The other alternative is that the spectral shape or the radiative efficiency of the jet may change
hysteretically.  A test of whether the jet power, or the jet spectrum is the most
important factor may come from correlated high resolution timing in
the X-rays and OIR bands.  It has previously been shown
that a negatively lagged anti-correlation of
optical emission relative to X-ray emission \citep{kanbet01,hyneet03b} can be explained as a result of the jet taking away
the bulk of the kinetic luminosity from the accretion flow \citep{malzet04}.  If the hysteresis effect we see is the
result of a different amount of kinetic power being taken away by the
jet at the same accretion disc $\dot{m}$, then there should be a signature in the magnitude of the
negatively lagged anti-correlation.  We note that observing the
spectral changes directly would be preferable, but this is quite
difficult due to the convergence around the $I$-band of the optical
components from the accretion disc, the donor star, and the jet in
most systems (e.g. \citealt{homaet05a}; \citeauthor{paper1}).

A completely independent emission mechanism does exist for explaining
the X-ray emission in X-ray binaries -- jet emission (e.g. \citealt{market01}; 2005).  In this scenario, the observed parallel tracks would form a jet--jet hysteresis as opposed to a jet--inflow (e.g. jet--corona) hysteresis. In the pure
jet-synchrotron model \citep{market01}, the NIR-to-X-ray ratio can
change at a given $L_{\rm NIR}$ only if either the spectral slope changes,
or the break frequency where the jet becomes optically thin changes
(see e.g. \citealt{nowaet05} for a discussion of attempts to test the
model based on observed spectral shapes).  In the more recent, more
complicated scenarios, where the corona forms the `base of the jet'
\citep{market05} there are additional parameters which can
change this flux ratio, by changing the fraction of the power that
comes from Compton scattering at the base of the jet.  Given that the
physics of the underlying accretion flow feeding the jet is
considerably less developed in these models than for ADAF and related
models, there are no quantitative determinations of what should be
the state transition luminosities or how the resulting hysteresis in the state
transitions should behave.  It thus seems premature to
comment on whether the observations presented in this paper have any
implications for or against the viability of jet X-ray models, but we
note that we here provide an additional constraint for such
models.

\subsubsection{Consequences of the interpretation}

These findings indicate, on both observational and
theoretical grounds, that there is reason to expect substantial
scatter in relations between radio/NIR and X-ray luminosity especially if
some of the data points are taken during the rising parts of
outbursts.  At the present time, it seems that most of the scatter in
the relations of Gallo et al. (2003; 2006) and \citeauthor{paper1} are due to the
uncertainties in the measurements of the masses and distances of the
different systems used to make the correlations, however two recently discovered sources could be exceptions (\citealt{cadoet07,rodret07}; see also \citealt{gall07}).  As more data is
collected for some of the sources (especially GX 339-4), it may become
clear that the relation is multiple-valued, and the hysteresis effects
will need to be considered.

\subsection{Conclusions}

Quasi-simultaneous NIR and X-ray data have been collected from BHXBs with well-sampled outburst rises and/or declines. Three sources were found satisfying this criteria. The data confirm the intrinsic correlation between NIR (which we show comes from the jets) and X-ray luminosity; $L_{\rm NIR}\propto L_{\rm X}^{0.5-0.7}$ found in the hard state (\citealt{homaet05a}; \citeauthor{paper1}). In the soft state, there is evidence for the NIR emission to be dominated by the viscously heated (as opposed to X-ray heated) accretion disc in all three sources, with $L_{\rm NIR}\propto L_{\rm X}^{0.1-0.2}$ observed in two of the three. The emission processes that can explain the NIR--X-ray relations (and the X-ray state-dependent OIR colours) in this Chapter are shown in Table \ref{hyst-emproc}.

The normalisation of the hard state correlation can differ with time for a single source. The infrared flux in the 2000 outburst of XTE J1550--564 shows a hysteresis effect, in that the rising hard state
has a lower infrared flux at a given X-ray flux than the decaying hard
state.  Several theoretical explanations for this effect seem
plausible, and have direct implications for the accretion process, jet power and the dominating X-ray emission processes.  We have ruled out the possibility that the hysteresis is directly caused solely by different radiative efficiencies in the inner accretion flow on the rise and decline. The effect may be due to changes in the viscosity parameter $\alpha$ and/or shifts in the spectrum/radiative efficiency of the jet.

\begin{table}
\begin{center}
\small
\caption[The NIR emission processes that can describe the NIR--X-ray relations and OIR colours in Chapter 4]{The NIR emission processes that can describe the NIR--X-ray relations and OIR colours of the data in this Chapter.}
\label{hyst-emproc}
\small
\begin{tabular}{lcccc}
\hline
Sample&X-ray       &Jet emission&Jet emission&Viscous\\
      &reprocessing&(opt. thin) &(opt. thick)&disc   \\
\hline
BHXBs; NIR; hard state&$\times$&$\surd$&$\times$&$\times$\\
BHXBs; NIR; soft state&$\times$&$\times$&$\times$&$\surd$\\
\hline
\end{tabular}
\normalsize
\end{center}
\end{table}

Hopefully, these results will help motivate future
observational searches for more examples of jet-disc hysteresis
effects and more theoretical work which might help to understand
whether the disc viscosity parameter $\alpha$ should vary as a function of time over X-ray binary outbursts.

Little OIR data are published from the rise of X-ray transient outbursts since this requires a monitoring campaign; it has been shown that the OIR counterpart begins to rise before the X-ray detection at least in some cases \citep[e.g.][]{jain01,jainet01b}. Consequently, the behaviour of the OIR rise is not understood as well as the declines. I, with Fraser Lewis, Rob Fender and Paul Roche am currently carrying out an optical monitoring project of a dozen or so quiescent LMXBs using the 2.0 m Faulkes Telescope North and South \citep{lewiet06} in order to spot and track LMXB outbursts and to alert the community, before they are detected at X-ray energies. As of September 2007, two outbursts have been detected, of Aql X--1 -- a source which has frequent outbursts and one which is also being monitored (more regularly) by the Yale team \citep{maitba07}. With the Yale team we prepared an Astronomer's Telegram \citep{maitet07} to encourage the community to make multiwavelength follow-up observations of the bright September 2007 outburst. Most of the sources we are monitoring are not being covered by other teams, and have less frequent outbursts. In a future work we plan to publish our results, including variability analysis of a number of sources in quiescence.

\newpage

\begin{center}
{\section{NIR Polarimetry}}
\end{center}

\subsection{Introduction}

The results of Chapters 2, 3 and 4 indicate that the NIR (and sometimes optical) emission of LMXBs, in particular the black hole systems, is dominated by optically thin synchrotron emission from the jets. This component is identified by its characteristic spectral shape and its relations with other wavebands. However, similar spectral indices and correlations can sometimes be reproduced by other components -- a low-temperature blackbody (accretion disc or companion star) peaks in the IR and therefore has a negative OIR spectral index (this would only be seen at low luminosities), and OIR emission from X-ray reprocessing has a similar dependency on X-ray luminosity to OIR jet emission (Chapter 2). There is one characteristic of optically thin synchrotron emission which is predicted to be very different from the other OIR components in an LMXB -- strongly polarised emission (Chapter 1).

A strong detection of linearly polarised OIR light would be a diagnostic of optically thin synchrotron emission from a region whose magnetic field is to some extent ordered. A non-detection would not rule out optically thin synchrotron emission, but constrain its magnetic field structure if present. Not only is polarisation a powerful tool to constrain the contribution of the jet at these frequencies, it can also probe the local conditions of the jets, namely the degree of ordering and orientation of the magnetic field. Here, linear polarimetric observations of eight XBs have been obtained using the 8 m VLT and the 3.8 m United Kingdom Infrared Telescope (UKIRT) in order to constrain the origin of the emission and the level of ordering of the jet magnetic field. Seven of the eight targets are observed in NIR filters and the other, in optical filters.
This work was completed by myself and Rob Fender only; any occurrences of `we' refer to these persons. At the time of writing, this work is submitted for publication (\citeauthor{paper6}).

\subsection{Observations}

In an effort to identify the polarised signature of jets in the infrared, the best suited sources and telescopes were selected. The evidence for optically thin emission from the jet in the NIR (and hence our best chance of detecting high levels of LP) comes mainly from BHXBs in outburst in the hard state \citep[e.g.][]{homaet05a,hyneet06b} and low-magnetic field NSXBs that are active (\citealt{miglet06}; Chapters 2 and 3). However in these earlier Chapters it was proposed that the jet may also contribute or dominate the NIR flux in some quiescent BHXBs, but probably not in quiescent NSXBs. Consequently, we obtained observations of both outbursting and quiescent sources.

\subsubsection{Target selection}

\begin{table*}
\caption[Log of polarimetry observations]{Log of polarimetry observations.}
\label{polar-obslog}
\small
\begin{tabular}{llllllllll}
\hline
Source&Date&MJD&Source  &\multicolumn{3}{c}{Exposure times}&Airmass\\
      &    &   &activity&\multicolumn{3}{c}{      (sec)               }&       \\
\hline
      &	         &\multicolumn{2}{l}{\emph{VLT + FORS1}}   &	          $V$&$R$&$I$                            &       \\
GX 339--4    &22 Aug 2005&53604.1&L&264&264&264&1.19--1.26\\
GX 339--4    &24 Aug 2005&53606.0&L&264&264&264&1.14--1.18\\
GX 339--4    &31 Aug 2005&53613.0&L&264&264&264&1.13--1.17\\
GX 339--4    &12 Sep 2005&53625.0&L&264&264&264&1.16--1.23\\
GX 339--4    &19 Sep 2005&53632.0&L&264&264&264&1.20--1.28\\
Hiltner 652  &19 Sep 2005&53632.0&PS&2&2&2&1.03--1.04\\
\hline
      &	         &\multicolumn{2}{l}{\emph{VLT + ISAAC}}   &	          $J$&$H$&$Ks$                           &       \\
GRO J1655--40&14 Oct 2005&53657.0&H&&56&&1.52\\
GRO J1655--40&28 Oct 2005&53671.0&H&32&56&80&2.05--2.13\\
WD 2359--434 &28 Oct 2005&53671.0&US&14&14&14&1.16--1.17\\
\hline
      &	         &\multicolumn{2}{l}{\emph{UKIRT + UIST}}  &	          $J$&$H$&$K$                            &       \\
XTE J1118+480&15 Feb 2006&53781.5&Q&2880&&1440&1.13--1.25\\
Sco X--1     &15 Feb 2006&53781.6&P&3$\times$32&3$\times$28&3$\times$28&1.27--1.48\\
Sco X--1     &16 Feb 2006&53782.6&P&96&84&84&1.37--1.52\\
WD 1344+106  &17 Feb 2006&53783.5&US&720&&&1.49--1.58\\
XTE J1118+480&17 Feb 2006&53783.5&Q&2880&&&1.13--1.15\\
Sco X--1     &07 Mar 2006&53801.6&P&3$\times$32&3$\times$28&3$\times$28&1.24--1.27\\
SS 433       &07 Mar 2006&53801.6&P&24&&&1.49--1.53\\
HD 38563C    &08 Mar 2006&53802.3&PS&47&&&1.42--1.48\\
Sco X--1     &08 Mar 2006&53802.6&P&96&84&84&1.37--1.52\\
WD 1615-154  &08 Mar 2006&53802.6&US&720&&&1.23--1.24\\
SS 433       &19 Aug 2006&53966.5&P&24&&&1.79--1.84\\
GRO J0422+32 &19 Aug 2006&53966.6&Q&2160&&2160&1.17--1.30\\
Aql X--1     &20 Aug 2006&53967.3&F&720&&720&1.06--1.10\\
Aql X--1     &08 Oct 2006&54016.4&Q&&&1680&1.51--1.88\\
4U 0614+09   &08 Oct 2006&54016.7&LP&&&480&1.02\\
Aql X--1     &11 Oct 2006&54019.2&Q&2160&2160&&1.07--1.21\\
\hline
\end{tabular}
\normalsize
\\
MJD = Modified Julian Day; Exposure times are total on-source integration time. Source activity: F = fading from mini outburst; H = hard state; L = low luminosity state; LP = low, persistent state; P = persistent; PS = polarised standard; Q = quiescent; US = unpolarised standard.
\normalsize
\end{table*}

GX 339--4 undergoes quasi-regular outbursts of varying length and peak luminosity. The infrared jet of GX 339--4 is one of the most studied, and the NIR spectrum in the hard state is consistent with optically thin synchrotron emission \citep{corbfe02,homaet05a}. This spectral component is observed to join to a thermal spectrum in the optical, with the two components producing half of the flux each around the $I$-band. We were allocated VLT imaging polarimetry with FORS1 (optical) for these observations (ESO Programme ID 076.D-0497). The 2005 outburst of GRO J1655--40 was studied at X-ray, optical and radio wavelengths \citep[e.g.][]{shapet07}. On the outburst decline the source entered the hard state \citep{homaet05b} so NIR imaging polarimetry was obtained with a Director's Discretionary Time (DDT) proposal with ISAAC (ESO Programme ID 275.D-5062).

In addition to outbursting sources, a number of quiescent and persistent sources were selected for NIR polarimetry. Target selection of quiescent systems was based on known NIR magnitudes and the companion star contribution, where measured. For example, the companion in the A0620--00 system is known to contribute $\sim 75$\% of the $H$-band quiescent flux \citep*{shahet99}. Even if a polarised jet provides the remaining 25\% (as opposed to the disc), the level of LP would be dampened by a factor of 4 due to the unpolarised light from the star. Most NIR-bright BHXBs are dominated by their companion stars in quiescence \citep[e.g. V404 Cyg;][]{casaet93} but there are some good candidates (e.g. XTE J1118+480 and in outburst, GRS 1915+105). Three NSXBs are also good candidates to identify a NIR jet component. Optically thin synchrotron emission appears to join to the thermal disc spectrum in the NIR in the NSXB 4U 0614+09 \citep{miglet06}. Sco X--1 is the brightest ($J = 11.9$) of the Z-sources, a class of NSXB whose jet spectrum may significantly contribute to the NIR. Aql X--1 is very active (it has about one outburst per year) and has a NIR spectrum in outburst consistent with a jet origin (Chapter 3).

The observations used in this Chapter are listed in Table \ref{polar-obslog}. For all three instruments (on two telescopes) imaging polarimetry is obtained using a Wollaston prism inserted in the optical path. The light is split into simultaneous, perpendicularly polarised ordinary and extra-ordinary beams (o- and e-beams). The waveplate was rotated to four angles: 0$^\circ$, 22.5$^\circ$, 45$^\circ$ and 67.5$^\circ$; this achieves a higher accuracy of LP and PA than from just two angles 0$^\circ$ and 45$^\circ$. The flux of the source was then measured in the o- and e-beams for each prism angle ($F_0^o$, $F_0^e$, $F_{22}^o$, $F_{22}^e$, etc.) using aperture photometry in \small IRAF \normalsize and LP and PA were calculated using the `ratio' method from the Stokes parameters $q$ and $u$ thus:
\begin{eqnarray}
\label{polar-eqn1}
  R_Q^2 = \frac{(F_0^e/F_0^o)}{(F_{45}^e/F_{45}^o)};~~R_U^2 = \frac{(F_{22}^e/F_{22}^o)}{(F_{67}^e/F_{67}^o)}\\
\label{polar-eqn2}
  q = \frac{R_Q-1}{R_Q+1};~~u = \frac{R_U-1}{R_U+1}\\
  LP = (q^2 + u^2)^{0.5}\\
  PA / deg = 0.5 tan^{-1}(u/q)
\end{eqnarray}

Flux calibration was possible for most of the targets. The total flux of each source was estimated from the sum of the o- and e-beam fluxes at each angle. Magnitudes were converted to de-reddened flux densities by accounting for interstellar extinction. As with the earlier Chapters, the method of \cite*{cardet89} was adopted, and the known values of $A_{\rm V}$ were used for each source (Table \ref{polar-Av}).

\begin{table}
\caption[The interstellar extinction towards each source used for polarimetry]{Values of interstellar extinction towards each source.}
\label{polar-Av}
\begin{center}
\small
\begin{tabular}{lll}
\hline
Source&$A_{\rm V}$&Reference\\
\hline
GRO J0422+32 &0.74$\pm$0.09&\cite{geliha03}\\
4U 0614+09   &1.41$\pm$0.17&\cite{neleet04}\\
XTE J1118+480&0.053$^{+0.027}_{-0.016}$&\cite{chatet03}\\
Sco X--1     &0.70$\pm$0.23&\cite{vrtiet91b}\\
GRO J1655--40&3.7$\pm$0.3&\cite{hyneet98}\\
GX 339--4    &3.9$\pm$0.5&\cite{jonket04}\\
Aql X--1     &1.55$\pm$0.31&\cite{chevet99}\\
SS 433       &6.5$\pm$0.5&\cite{fukuet97}\\
\hline
\end{tabular}
\small
\normalsize
\end{center}
\end{table}

\subsubsection{ESO VLT observations and reduction}

The VLT FORS1 Service Mode programme 076.D-0497 was carried out between 22 August and 19 September 2005. The target GX 339--4 was observed on five nights in Bessel $V$, $R$ and $I$ filters, and the polarised standard star Hiltner 652, on one night (see Table \ref{polar-obslog} for observations and Table \ref{polar-vlt} for the results). The seeing ranged from 0.48 to 2.09 arcsec, with a mean of $\sim 1.2$ arcsec. The conditions were clear on all nights and photometric on three of the five. Using \small IRAF\normalsize, the science images were de-biased and flat-fielded and aperture photometry was performed on GX 339--4 and three stars in the FOV. The point spread function (PSF) of GX 339--4 was found to be blended with two close stars \citep[B and C in][]{shahet01}. We therefore could not measure small differences in the flux of GX 339--4 alone; we used an aperture which encompassed the PSFs of the contaminating stars. GX 339--4 contributes $\sim 60$\% of the total flux of this aperture during our observations. If the target is polarised and the contaminating stars are not then the flux from them will dampen the LP measurements.

From the background-subtracted photometry I obtained the Stokes parameters $q$ and $u$ (equations \ref{polar-eqn1} and \ref{polar-eqn2}), and the variability of the total (o + e) flux across the four prism angles. Instrumental polarisation is documented to be less than 0.1\% for FORS1. To calibrate the PA, the standard small FORS1 corrections of $PA_{\rm REAL} = PA_{\rm MEASURED} + c$, where $c=-1.80^\circ$, $+1.19^\circ$ and $+2.89^\circ$ were made for the $V$, $R$ and $I$ filters, respectively. For Hiltner 652, only two prism angles were obtained so $q$ and $u$ were calculated from these two angles alone (and no variability is quoted). I measure LP$=6.2$\% for this polarised standard in $V$, consistent with that quoted by the FORS consortium based on Commissioning data taken with FORS1. Unfortunately Hiltner 652 was slightly saturated in $R$ and $I$, and the values of LP are lower than expected. The measured PA in all three filters is at most $\sim 3^\circ$ different from the documented values.

No flux standard stars were observed but the magnitude of GX 339--4 was estimated using four stars in the FOV that are listed in the NOMAD and USNO-B Catalogues \citep{zachet04,moneet03}. The magnitudes of these field stars range from $V=14.13$, $I=13.43$ to $V=17.74$, $I=16.39$. These flux measurements do not require the high level of sensitivity needed for polarimetry, so the magnitude of GX 339--4 alone was estimated (not including the contaminating stars B and C) using a circular aperture of fixed radius centred on GX 339--4. The aperture excludes most of the light from B and C. The 1$\sigma$ errors in the estimated magnitudes of GX 339--4 are inferred from the range of magnitudes implied by each of the four field stars.

GRO J1655--40 was observed on 14 and 28 October 2005 under VLT ISAAC DDT programme 275.D-5062 (Tables \ref{polar-obslog} and \ref{polar-vlt}). Short wavelength (SW) imaging polarimetry was carried out with $J$ (1.25 $\mu$m), $H$ (1.65 $\mu$m) and $Ks$ (2.16 $\mu$m) filters. The seeing was 0.71--1.58 arcsec and conditions were photometric on both nights. The unpolarised standard star WD 2359--434 was also observed on 28 October. Data reduction was performed in \small IRAF\normalsize; dark frames of equal exposure time to the science frames were subtracted to remove sky background. The science images were then flat-fielded and bad pixels were accounted for.

\begin{figure}
\centering
\includegraphics[width=8cm,angle=270]{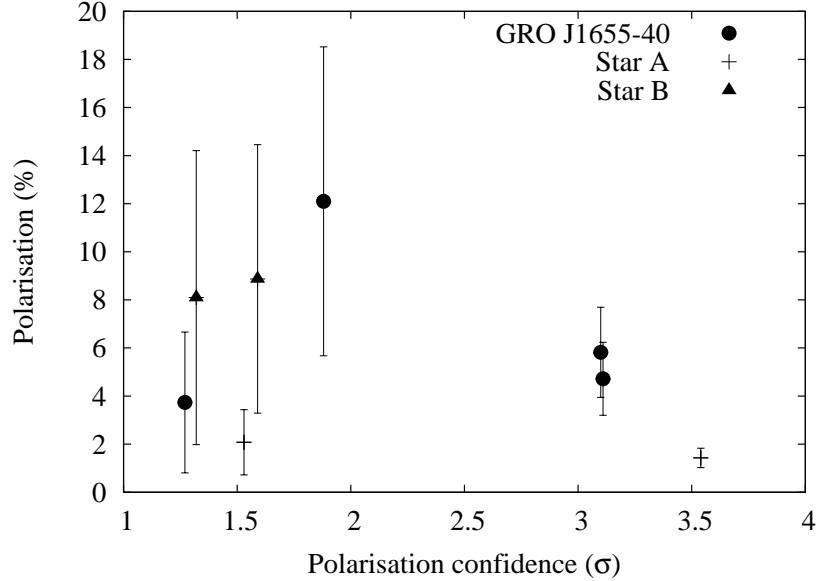}
\caption[Confidence in measured polarisation for GRO J1655--40]{Confidence in measured polarisation for GRO J1655--40 and two stars in the same field of view.}
\label{polar-j1655conf}
\end{figure}

The Stokes parameters (and the variability of the total flux) were calculated for GRO J1655--40 and two nearby field stars of similar brightness. In Fig. \ref{polar-j1655conf}, the measured LP is plotted as a function of the confidence of the measurement. The field stars are not polarised (2$\sigma$) except one observation of 1--2\% LP, which is probably interstellar in origin. Where the S/N is high enough (Table \ref{polar-vlt}), LP is detected in GRO J1655--40 at a level of 5--6\%, with 3$\sigma$ confidence. The absence of this polarisation in the field stars implies it is intrinsic to GRO J1655--40 and not due to instrumental polarisation. It is also too high to be of interstellar origin because the extinction is too low. Serkowski's law; $A_{\rm V}\geq P(\%)/3$ \citep*{serket75} indicates that the maximum LP caused by interstellar dust in the optical is 12.0\% for GRO J1655--40, adopting the extinction upper limit of $A_{\rm V} < 4.0$ (Table \ref{polar-Av}). In the $K$-band, $A_{\rm K }\sim 0.1 A_{\rm V}$, so $LP_{\rm K}\leq 1.2$\% -- much lower than is observed. In addition \cite{dubuch06} measured $LP_{\rm K}\leq 1.3$\% during quiescence, so the measured LP must be transient.

An unpolarised standard, WD 2359--434 was also observed. No LP is detected in WD 2359--434 except a low level ($\sim 2$\%) in the $J$-band, which is likely due to instrumental polarisation. The $J$-band values of $q$ and $u$ for WD 2359--434 were then subtracted from those of GRO J1655--40 to correct for this. No additional systematic correction to the values of PA are required for ISAAC. WD 2359--434 ($J=12.597$, $H=12.427$, $K=12.445$) was also used to calibrate the flux of GRO J1655--40. Atmospheric extinction was accounted for because the standard star and X-ray binary were observed at different airmasses.

\begin{table*}
\caption[List of VLT polarimetry results]{List of VLT results.}
\label{polar-vlt}
\small
\begin{tabular}{lllllllllll}
\hline
Source&MJD&Fil-&S/N &app.&$F_{\nu}$&LP &PA ($^\circ$)&Conf.		&Variabi-\\
      &   &ter	 &    &mag.&(mJy)    &(\%)	    &&($\sigma$)  &lity (\%)\\
(I)&(II)  &(III) &(IV)&(V) &(VI)     &(VII) &(VIII)             &(IX)                   &(X)\\
\hline
&\multicolumn{9}{c}{\emph{VLT + FORS1 observations$^1$:}}&\\
GX339  &53604.1&$V$&846  &18.80(06) &4.19(24)&2.15(17) &27.4(4.5)   &12.8&3.2(3)\\
GX339  &53604.1&$R$&592  &18.70(42)&2.05(79)&2.45(24) &36.4(5.6)   &10.3&1.5(4)\\
GX339  &53604.1&$I$&1630 &17.65(21)&2.07(40)&1.78(9) &34.1(2.8)   &20.5&1.9(1)\\
GX339  &53606.0&$V$&127  &18.61(17)&5.00(77)&$<5.03$	  &	    &	 &5.6(1.8)\\
GX339  &53606.0&$R$&80   &18.52(40)&2.38(88)&$<7.42$	  &	    &	 &26(3)\\
GX339  &53606.1&$I$&95   &17.46(11)&2.47(26)&$<6.59$	  &	    &	 &29(2)\\
GX339  &53613.0&$V$&237  &18.68(17)&4.66(73)&2.98(60) &24(12)  &5.0 &6.8(9)\\
GX339  &53613.0&$R$&104  &18.67(40)&2.07(76)&$<6.62$	  &	    &	 &$<11$\\
GX339  &53613.0&$I$&73   &17.41(12)&2.58(27)&$<8.79$	  &	    &	 &$<14$\\
GX339  &53625.0&$V$&454  &18.56(11)&5.21(54)&2.02(31)&34.3(8.9)   &6.5 &1.6(5)\\
GX339  &53625.0&$R$&750  &18.49(40)&2.46(89)&2.86(19)&34.3(3.8)   &15.2&2.3(3)\\
GX339  &53625.0&$I$&432  &17.54(15)&2.31(31)&2.33(33)&38.9(8.0)   &7.1 &4.1(5)\\
GX339  &53632.0&$V$&659  &18.93(17)&3.71(58)&2.10(21)&24.9(5.9)   &9.8 &2.2(3)\\
GX339  &53632.0&$R$&429  &18.87(40)&1.72(64)&2.90(33)&31.3(6.5)   &8.8 &2.7(5)\\
GX339  &53632.0&$I$&355  &17.68(12)&2.02(22)&2.57(40) &27.1(8.9)   &6.5 &2.5(6)\\
H652	&53632.0&$V$&\multicolumn{2}{l}{21000}&&6.19(1)&176.2(1)&924 &-  \\
H652$^2$&53632.0&$R$&\multicolumn{2}{l}{32000}&&4.79(1)&176.9(1)&1088&-\\
H652$^2$&53632.0&$I$&\multicolumn{2}{l}{26000}&&3.54(1)&178.6(1)&644 &-\\
\hline
&\multicolumn{9}{c}{\emph{VLT + ISAAC observations:}}&\\
J1655&53657.0&$H$&93&12.95(12)&11.8(1.3)&4.72(1.52)&44(18)&3.1&$<11$\\
J1655&53671.0&$J$&48&13.76(13)&12.4(1.5)&$< 11.7$&&&$<22$\\
J1655&53671.0&$H$&22&13.01(16)&11.2(1.6)&$< 31.4$&&&$<42$\\
J1655&53671.0&$Ks$&75 &12.62(12)&8.04(91)&5.82(1.88)&157(18)&3.1&6.9(3.0)\\
2359&53671.0&$J$ &210&&&2.04(69)&153(19)&3.0&$<4.4$\\
2359&53671.0&$H$ &72&&&$< 5.6$&&&$<11$\\
2359&53671.0&$Ks$&56&&&$< 5.4$&&&$<16$\\
\hline
\end{tabular}
\normalsize
$^1$Polarisation measurements of GX 339--4 include the contaminating stars B and C \citep{shahet01} in the aperture;
$^2$Standard star is slightly saturated, which has reduced the apparent LP level detected.
Columns give:
(I) Source name (see Table \ref{polar-obslog} for full names);
(II) MJD;
(III) filter (waveband);
(IV) S/N detection of the source;
(V) apparent magnitude;
(VI) de-reddened flux density;
(VII) measured level of polarisation and $1\sigma$ error, or $3\sigma$ upper limit if the detection of polarisation is $< 2\sigma$;
(VIII) polarisation angle and $1\sigma$ error;
(IX) confidence of polarisation detected;
(X) the standard deviation of the total (sum of o- and e-beam) source intensity as a percentage ($\Delta F_{\nu}$/$F_{\nu}$) and $1\sigma$ error, or $3\sigma$ upper limit if the variability detection is $< 2\sigma$.
\normalsize
\end{table*}

\begin{table*}
\caption[List of UKIRT polarimetry results]{List of UKIRT results$^1$.}
\label{polar-ukirt}
\small
\begin{tabular}{lllllllllll}
\hline
Source&MJD&Fil-&S/N &app.&$F_{\nu}$&LP &PA ($^\circ$)&Conf.		&Variabi-\\
      &   &ter	 &    &mag.&(mJy)    &(\%)	    &&($\sigma$)  &lity (\%)\\
(I)&(II)  &(III) &(IV)&(V) &(VI)     &(VII) &(VIII)             &(IX)                   &(X)\\
\hline
J0422 &53966.6&$K$&3.5&17.42(42)&0.072(28)&            &&   &    \\
J0422 &53966.6&$J$&44&18.36(16)&0.083(12)&$< 11.6$     &&   &$<24$\\
0614   &54016.7&$K$&37&16.64(19)&0.158(28)&$< 16.0$     &&   &$<20$\\
J1118&53781.5&$J$&91&&&$< 6.36$     &&   &$<9.6$\\
J1118&53781.6&$K$&66&&&$< 7.42$     &&   &$<15$\\
J1118&53783.5&$J$&28&&&$< 20.7$     &&   &$<28$\\
ScoX1	  &53781.6&$J$&\multicolumn{2}{l}{3000--3900}&&0.34--0.53&22--87&9.3--11.5&0.87--1.3\\
ScoX1	  &53781.7&$H$&\multicolumn{2}{l}{3000--4500}&&0.30--0.57&27--88&6.4--18.3&1.0--1.9\\
ScoX1	  &53781.6&$K$&\multicolumn{2}{l}{1600--2900}&&0.13--0.52&71--86&$<$2--10.7&1.5--4.5\\
ScoX1	  &53782.6&$J$&2000	 &&&0.36(7)&37(12)&5.0       &0.53(11)\\
ScoX1	  &53782.6&$H$&2000	 &&&0.18(7)&49(23)&2.5       &$<0.54$\\
ScoX1	  &53782.6&$K$&1600	 &&&0.23(9)&41(23)&2.5       &$<0.50$\\
ScoX1	  &53801.6&$J$&\multicolumn{2}{l}{1100--2000}&&0.30--0.65&36--95&2.4--6.9&0.44--1.4\\
ScoX1	  &53801.6&$H$&\multicolumn{2}{l}{1200--2000}&&0.07--0.54&36--49&$<$2--7.7&0.45--0.57\\
ScoX1	  &53801.6&$K$&\multicolumn{2}{l}{1000--2000}&&0.35--0.91&96--111&2.6--5.0&0.33--0.41\\
ScoX1	  &53802.6&$J$&1600	 &&&0.40(9)&46(13)&4.4     &0.44(14)\\
ScoX1	  &53802.6&$H$&700	 &&&$< 1.01$  &&	&$<1.5$\\
ScoX1     &53802.6&$K$&600	 &&&$< 1.08$	 &&	   &$<1.5$\\
AqlX1	  &53967.3&$J$&88     &16.30(15)&0.63(9)&$< 5.44$   &&	&$<8.4$\\
AqlX1	  &53967.4&$K$&41     &15.47(16)&0.46(7)&$< 11.6$   &&	&$<18$\\
AqlX1	  &54016.4&$K$&28     &15.56(22)&0.42(9)&$< 16.2$   &&	&$<26$\\
AqlX1	  &54019.2&$J$&19     &16.65(29)&0.46(12)&$< 24.5$  &&	&$<37$\\
AqlX1     &54019.2&$H$&27	 &16.06(25)&0.45(10)&$< 16.9$	 &&	   &$<26$\\
SS433       &53801.6&$J$&4600   &9.06(8)&1950(140)&0.75(3)&71(2)&24.4        &0.27(5)\\
SS433       &53966.5&$J$&4000   &8.89(19)&2290(400)&0.50(4)&76(4)&14.1        &0.47(6)\\
1344  &53783.5&$J$&500    &&&$< 1.08$     &&        &$<1.8$\\
HD38    &53802.3&$J$&2100   &&&5.56(7)&        &84.2        &$<0.47$\\
1615 &53802.6&$J$&370    &&&$< 1.79$     &&        &$<2.4$\\
\hline
\end{tabular}
\normalsize
$^1$The very close contaminating star \citep{chevet99} is included in the aperture of Aql X--1.
$^2$The position angle may have systematic errors because it is calibrated using one standard, HD 38563C.
Columns: see caption of Table \ref{polar-vlt}.
\normalsize
\end{table*}

\subsubsection{UKIRT observations and reduction}

The UKIRT targets were observed (see Table \ref{polar-obslog} for observations and Table \ref{polar-ukirt} for results) between 15 February and 11 October 2007 by the UKIRT Service Observing Programme (UKIRTSERV). $JHK$ polarimetric observations were made using UIST + IRPOL2. The conditions were mostly clear during the observations (and the seeing was 0.4--1.0 arcsec) but thin cirrus was present on some dates. The POL\_ANGLE\_JITTER standard recipe was adopted except flat field images were not taken. After dark current and bias subtraction, the flux from the target was measured in a fixed aperture (the same pixels in all images for each source). Since we are dividing the fluxes in these fixed apertures (e.g. $F_0^o/F_{45}^o$; equation \ref{polar-eqn1}) the flat fields would cancel out, so skyflats were not required.

The instrumental polarisation is known to be low ($<1$\%), which is confirmed by the non-detection of LP from the two unpolarised standards WD 1344+106 and WD 1615-154 (Table \ref{polar-ukirt}). The PA was calibrated using the polarised standard HD 38563C; I caution that there may be systematic errors in PA due to just the one calibrator being available. No flux standards were observed, however relative photometry of two stars in the FOV was used (sum of o+e beams) that are listed in the 2MASS catalogue, for each target. For XTE J1118+480 and Sco X--1, no field stars exist in the images that have 2MASS magnitudes, so no flux calibration could be achieved. The polarisation standards could not be used for flux calibration as conditions were not photometric in all observations.

\subsection{Results \& Discussion}

A significant level (3$\sigma$) of LP is detected in four sources: GX 339--4, GRO J1655--40, Sco X--1 and SS 433. Two of these four (GRO J1655--40 and Sco X--1) have intrinsic LP in the NIR which is most likely caused by optically thin synchrotron emission from the inner regions of the jets. SS 433 is polarised in the NIR due to local scattering and GX 339--4 is polarised in the optical most likely due to interstellar dust along its line of sight. The other four sources are not polarised, with 3$\sigma$ upper limits of $\sim$5--15\% (it is not possible to detect lower levels of LP in most quiescent systems due to the low S/N). Here, I discuss the results for each individual source.

\subsubsection{Individual sources}

\begin{figure}
\centering
\includegraphics[width=6.5cm,angle=270]{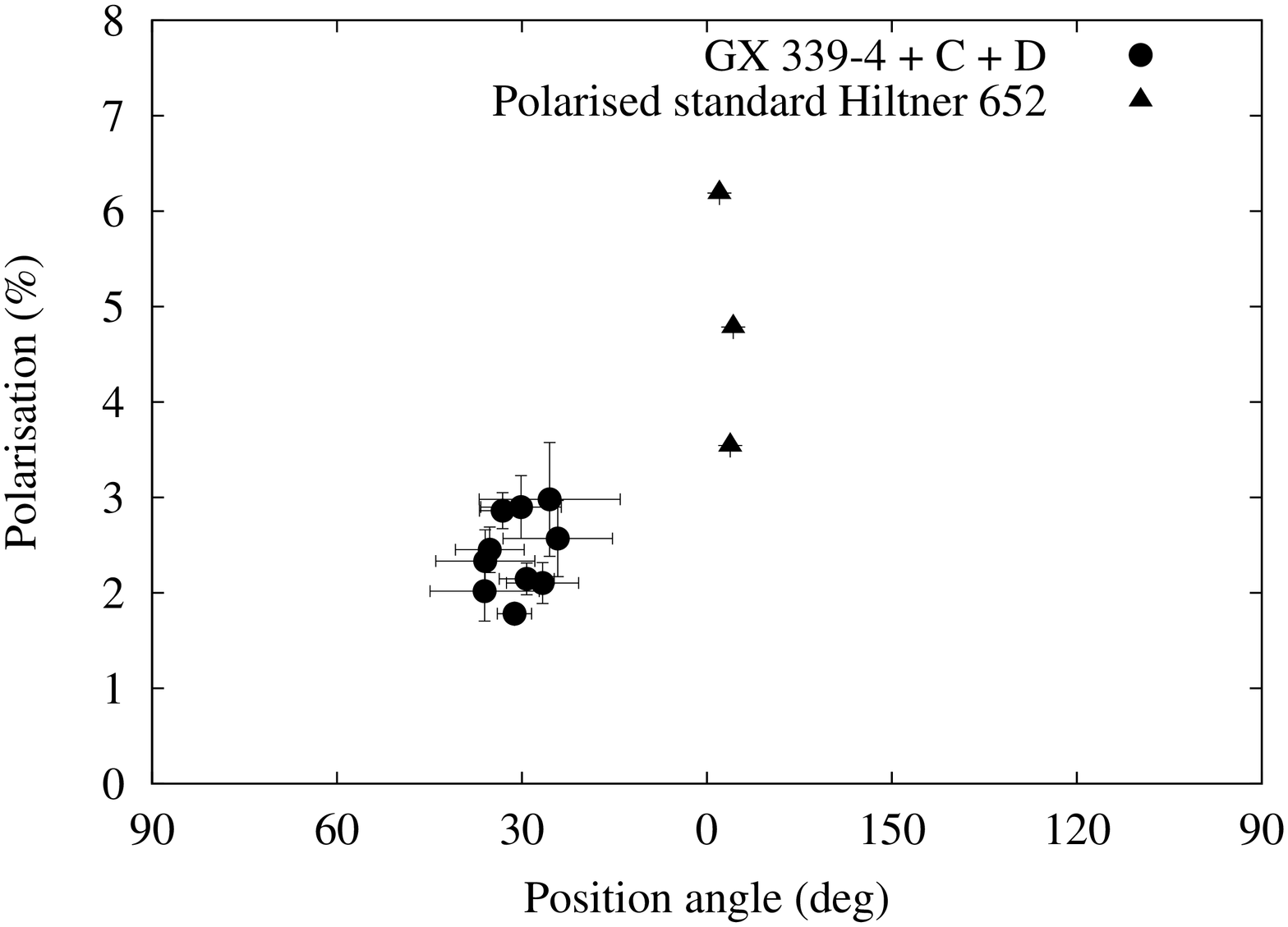}
\includegraphics[width=6.5cm,angle=270]{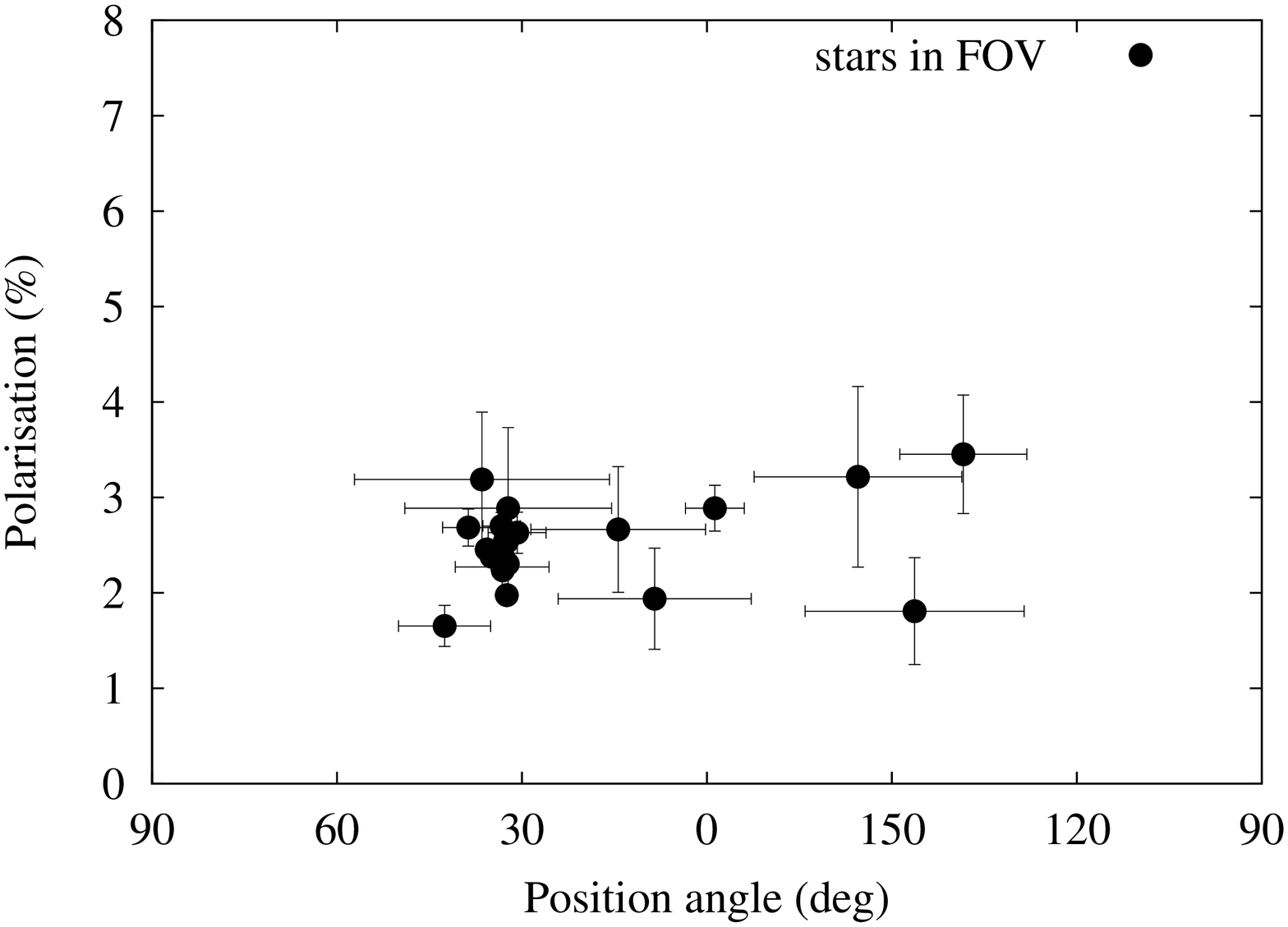}
\caption[Optical polarisation versus angle for GX 339--4]{Optical $V$, $R$ and $I$-band LP (when it is detected at the $> 3\sigma$ level) versus PA for GX 339--4 (including the two contaminating stars C and D) and the polarised standard Hiltner 652 in the upper panel and three stars in the field of view in the lower panel. A point is plotted for each observation in each filter. The polarisation of GX 339--4 is similar in level and PA to most of the data of the three field stars, and is likely to be caused by interstellar dust.}
\label{polar-gx339lp}
\end{figure}

\textbf{GX 339--4}:
\newline
This source was observed on five occasions spanning 28 days during a low luminosity state soon after its 2004--5 outburst. LP is detected at a level of 2--3\% in GX 339--4 (which includes the two close contaminating stars, see Section 5.2.2) in all three optical filters in all data with a high S/N (Table \ref{polar-vlt}). From all the $VRI$ polarisation measurements, an average LP of 2.4\% in detected. In Fig. \ref{polar-gx339lp}, LP is plotted as a function of PA for GX 339--4 (including the contaminating stars; upper panel) and three other stars in the FOV (lower panel). Both LP and PA for GX 339--4 are similar to those measured from the field stars, with the values of PA differing for a few field star observations. This is a strong indication that the polarisation in all sources (including the contaminating stars B and C) is caused by the same process, interstellar dust. Moreover, significant variability of LP is not detected. The extinction towards GX 339--4 has been measured at $A_{\rm V}=3.9\pm 0.5$ \citep{jonket04} which is high enough to explain the observed polarisation. Indeed, the extinction can be constrained from the measured level of LP using Serkowski's law, $A_{\rm V}\geq P(\%)/3$ \citep{serket75}. We find that $A_{\rm V}\geq 0.81$, which does not refine the currently estimated value of $3.9\pm 0.5$. The axis of the resolved radio jet of GX 339--4 \citep{gallet04} is at 116$^\circ$, approximately perpendicular to the PA of LP in GX 339--4, however there is no indication that the two are linked, for the above reasons.

\begin{figure}
\centering
\includegraphics[width=7cm,angle=270]{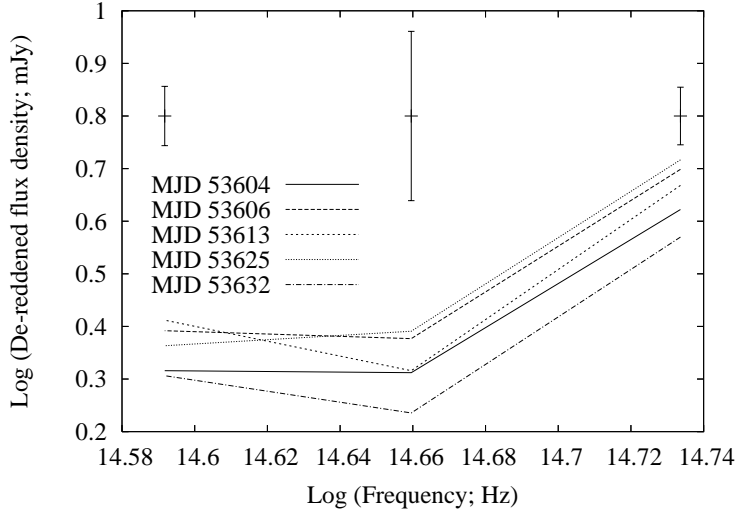}
\caption[The de-reddened optical SED of GX 339--4 from polarimetry data]{The de-reddened $IRV$ spectrum of GX 339--4 (not including the two contaminating stars C and D). The error bars represent the mean error in flux density in each filter. The $I$-to-$V$ spectral index is $\alpha = 2.1\pm 0.3$.}
\label{polar-gx339spec}
\end{figure}

\begin{figure}
\centering
\includegraphics[width=12cm,angle=0]{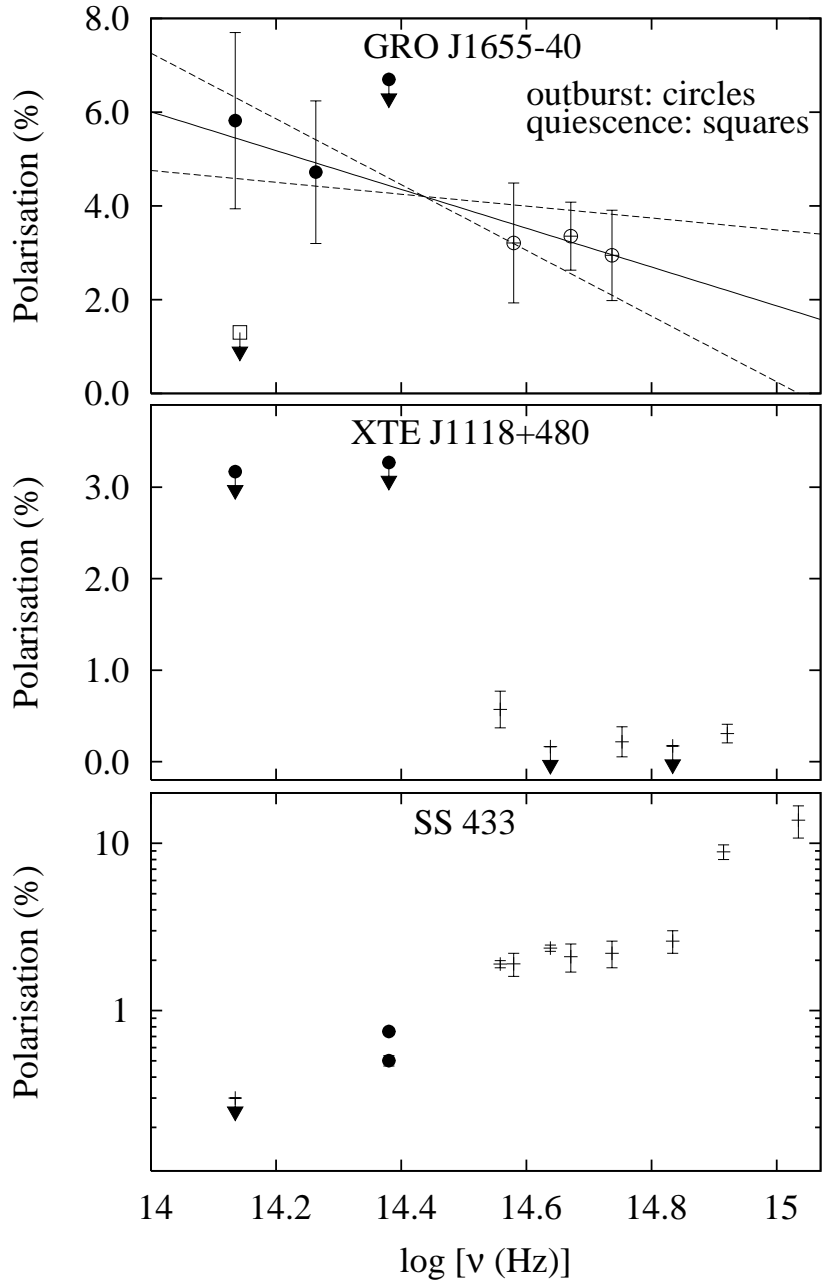}
\caption[Linear polarisation as a function of frequency for three sources]{LP as a function of frequency for three sources. Filled symbols are our results (upper limits are 1$\sigma$), other symbols are from the literature. $VRI$-band data of GRO J1655--40 are from \citet{glioet98} and a $K$-band upper limit in quiescence is from \citet{dubuch06}. The lines in the top panel show the fit to the outburst data and uncertainty in the slope. $UBVRI$ data of XTE J1118+480 are from \citet{schuet04}. $UV,UBVRI$ data of SS 433 are from \citeauthor{dolaet97} (1997; the contribution from interstellar LP has been subtracted from these data) with additional $RI$ data from \citet{schuet04} and a $K$ upper limit from \citet{thomet79}.}
\label{polar-lp3}
\end{figure}

The magnitudes obtained of GX 339--4 alone ($V\sim 18.7$, $R\sim 18.6$, $V\sim 17.5$; not including the contaminating stars) are similar to the $V$ and $I$ magnitudes measured a month or so earlier (C. Bailyn, private communication) but are $\sim$1.5 magnitudes brighter than the lowest level of $r=20.1$ \citep{shahet01}. Even at this lowest level the star in the GX 339--4 system contributes $\simlt 30$\% of the $r$-band emission \citep{shahet01}. The de-reddened optical SED of GX 339--4 for the five observations are shown in Fig. \ref{polar-gx339spec}. The spectrum is blue and fairly steep between $R$ and $V$ ($\alpha = 2.1\pm 0.3$ between $I$ and $V$) indicating thermal emission, probably from the X-ray heated disc. However there is tentative evidence for a flattening towards the $I$-band. This could be the same $I$-band flattening as seen during outburst, which is the optically thin jet spectrum beginning to dominate \citep{corbfe02,homaet05a}. If so, the detection of the jet in the $I$-band at these low luminosities suggests the break frequency in the jet spectrum (between optically thick and thin emission) does not appear to change substantially with mass accretion rate. The $R$-band flux errors are large however, and there may be no flattening. We may expect to see short timescale flux variability if the jet does play a role; GX 339--4 varies by $\simlt 5$\% in the high S/N observations (column X of Table \ref{polar-vlt}), and the variability is not stronger in the $I$-band. We cannot therefore make any conclusions as to whether the jet component is polarised or not; the jet contribution to the $I$-band at these low flux levels is likely to be low. It is interesting to note that the optically thin radio jet ejecta seen from this source are polarised at a level of 4--9\% \citep{gallet04}. A polarisation spectrum (LP versus $\nu$) is not plotted for GX 339--4 because we find no relation between P and $\nu$ for GX 339--4 or for any of the field stars, within the errors of each measurement.
\newline\newline
\textbf{GRO J1655--40}:
\newline
$\sim 5$\% LP is detected in this source at the 3$\sigma$ level in $H$ and $Ks$, when it was in a hard state at the end of its 2005 outburst. In the upper panel of Fig. \ref{polar-lp3} the polarisation spectrum is plotted, including the optical measurements of \cite{glioet98} during a soft state in 1997. The optical polarisation, which varies as a function of orbital phase (taken in outburst during a soft X-ray state), is caused by local scattering and should decrease at lower frequencies. A statistical increase is seen in LP in the NIR (shown by the fit to the optical--NIR data) which must therefore have a different origin to the optical LP. The $Ks$-band LP of 5.8$\pm$1.9\% is not consistent with the 95\% confidence upper limit of LP$<$1.3\% in $Ks$ seen during quiescence \citep{dubuch06}. The NIR LP must be intrinsic and transient, and we interpret it as originating from the optically thin region of the jets.

The PA of the optical LP (120--130$^\circ$) is also different from that measured both in $H$ (44$\pm 18^\circ$) and $Ks$ (157$\pm18^\circ$). Rotation of PA with time is seen in some jets \citep[e.g.][]{hannet00,fendet02,gallet04} indicating an overall rotation of the local magnetic field. The $Ks$ LP detection is 14 days after the $H$, implying the magnetic field may be rotating during this time. The $H$-band PA is consistent with the direction of the jet on the plane of the sky; 47$^\circ$ \citep{hjelru95}, implying that the electric field vector is parallel to the jets at this time and the magnetic field is perpendicular to the jets. Two weeks later the field orientation had appeared to change by $\sim 70^\circ$. Interestingly, \cite{hannet00} found from well-sampled radio data that the field orientation of discrete ejection events (as opposed to the steady, compact jet) changed only near the end of the 1994 outburst; here a change is also seen in the compact jet at the end of its 2005 outburst.

The de-reddened $JHKs$ flux densities of GRO J1655--40 indicate a blue spectrum, and are similar to those measured by \cite{miglet07} between two and five weeks earlier. In their Fig. 6, the broadband SED shows the companion star dominating the optical and NIR, with the jet dominating only below $\nu \approx 5\times 10^{13}$ Hz. The $JHKs$ observations here are therefore also dominated by the star. According to the model in Fig. 6 of \citeauthor{miglet07} (upper panel), the jet contributes $\sim 40$\% and $\sim 30$\% of the emission in $Ks$ and $H$, respectively. If the jet is the source of the polarisation and the star and any other contributions are not polarised, this implies the jet itself is $15\pm 5$\% polarised in $Ks$ and $16\pm 5$\% polarised in $H$. We would not expect this level of LP from the optically thick jet, so the jet spectrum must turn over to become optically thin ($\alpha \approx -0.6$) at some frequency above $\sim 5\times 10^{13}$ Hz \citep[contrary to the models of][]{miglet07}. If we take the faintest possible NIR jet scenario, one where the turnover is at $5\times 10^{13}$ Hz and the optically thin spectrum is steep ($\alpha = -0.8$), the jet would contribute 14\% of the light at $Ks$ and 9\% at $H$. This would result in a jet LP of $42\pm 13$\% in $Ks$ and $52\pm 17$\% in $H$. The likely level of jet contribution is between the two above estimates, so a jet LP of 15--42\% in $Ks$ and 16--52\% in $H$ can explain the data. These results indicate a fairly ordered magnetic field, with $f = 0.41\pm 0.19$ from the $Ks$-band result (The $H$-band is less constraining).

\begin{figure}
\centering
\includegraphics[width=12cm,angle=0]{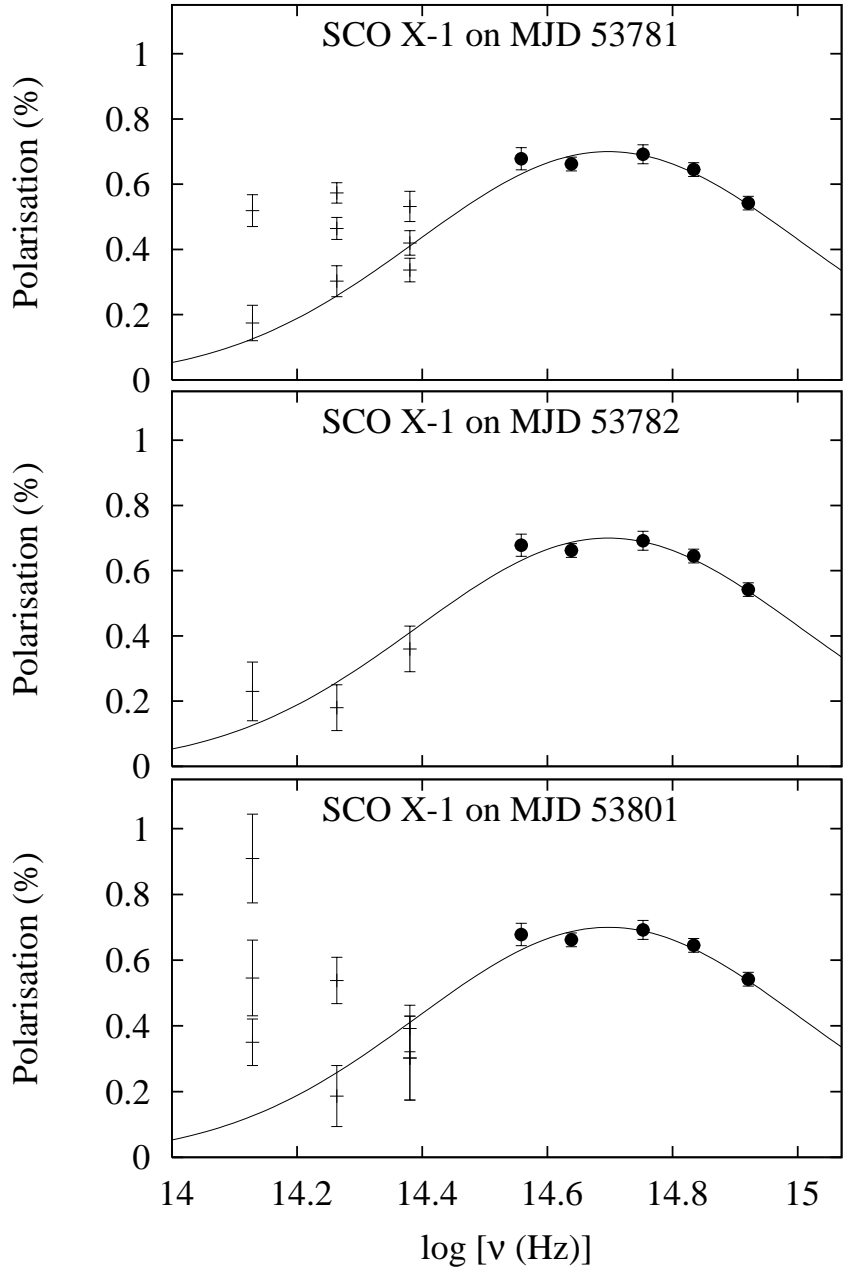}
\caption[Linear polarisation as a function of frequency for Sco X--1 in three epochs]{LP as a function of frequency for Sco X--1 in three epochs. The optical data (filled circles) and the fit to this data are included for an interstellar origin to the polarisation (solid curve) from \citet{schuet04}. The crosses on the left are our $KHJ$-band results (2$\sigma$ detections of LP); a clear variable LP component above that expected from interstellar dust is seen at the lowest frequencies.}
\label{polar-scox1lp}
\end{figure}

In addition, the flux from GRO J1655--40 is marginally variable in $Ks$ (7$\pm$3\% standard deviation; column X of Table \ref{polar-vlt}) on short timescales ($\sim 20$ sec time resolution). If the star is not variable then this suggests the jet component is varying by 9--43\%. Large-amplitude variability has been seen associated with NIR optically thin synchrotron emission from the jets in XTE J1118+480 \citep{hyneet06b}.
\newline\newline
\textbf{GRO J0422+32}:
\newline
The apparent magnitudes and $J-K$ colour of this source are found to be consistent with the quiescent values measured by \cite{geliha03}. The S/N is low; we are able to place a 3$\sigma$ upper limit on the $J$-band polarisation; LP$<$11.6\%. It has been shown that the star likely dominates the NIR light \citep{geliha03} although recent observations detect a strong flickering component which may come from the disc \citep{reynet07}.
\newline\newline
\textbf{4U 0614+09}:
\newline
This NSXB is detected with a $K$-band magnitude $\sim 0.5\pm 0.2$ brighter than reported from observations made in 2002. The system is an ultra-compact NSXB whereby the white dwarf donor has not directly been detected. \cite{miglet06} showed that the mid-IR spectrum is dominated by optically thin synchrotron emission from the jets (with $\alpha = -0.6$), and this joins the disc component ($\alpha \approx +2$) around the $H$-band. Therefore our $K$-band measurement is likely to be jet-dominated. Despite this, LP is not detected, with a 3$\sigma$ upper limit of 16\%. According to the spectrum in \cite{miglet06}, the jet contributes around twice as much light as the disc in the $K$-band. Assuming that the accretion disc contributes less than 40\% of the $K$-band light, the jet component must be less than 27\% polarised. The corresponding magnetic field ordering is $f < 0.38$.
\newline\newline
\textbf{XTE J1118+480}:
\newline
The polarisation spectrum of this source, including optical data from \cite{schuet04} is shown in the middle panel of Fig. \ref{polar-lp3}. \citeauthor{schuet04} claim the LP is variable (with a positive detection in some optical bands but not in others) and their strongest detection (2.6$\sigma$) is in the $I$-band; LP = 0.57$\pm$0.22\%. There is no apparent increase at higher frequencies, with a $B$-band upper limit of 0.38\% (3$\sigma$) so the positive detection is unlikely to be due to interstellar dust (the extinction is very low; see Table \ref{polar-Av}). The upper limits are on the order of 7\% LP in $J$ and $K$. According to a recent study of the broadband quiescent spectrum of XTE J1118+480 \citep[using for the first time optical, NIR and mid-IR data;][]{gallet07}, the jet could make a significant contribution to the mid-IR--optical spectrum. In their model (middle right panel of their Fig. 3), the jet contributes approximately 35\%, 25\% and 28\% of the flux in $K$, $J$ and $I$, respectively. From the polarisation measurements, this would correspond to a jet LP of $< 21$\% ($K$), $< 25$\% ($J$) and 2.0$\pm$0.8\% ($I$). This is by no means a solid result, but the $I$-band detection suggests the jet does indeed contribute. The accretion disc is also likely to make a contribution to at least the optical wavebands \citep{mikoet05} but other origins of LP such as scattering are likely to be stronger at higher frequencies, contrary to observations \citep{schuet04}.
\newline\newline
\textbf{Sco X--1}:
\newline
The optical light of Sco X--1 is polarised at a low level due to interstellar dust \citep*{landan72,schuet04} but there is one claim that it is variable and therefore intrinsic \citep{shakef75}. Recent NIR spectropolarimetry has revealed intrinsic LP in the $H$ and $K$ region of the spectrum, with a clear increase in LP below 1.4$\times 10^{14}$ Hz that is interpreted as coming from the (optically thin) jet \citep{shahet07}. Here, we report $JHK$ polarimetry over four epochs that shows a clear variability of LP which is stronger at lower frequencies (Fig. \ref{polar-scox1lp}). The optical data and interstellar model fit of \cite{schuet04} is included \citep[much like Fig. 3 of][]{shahet07}. The NIR LP cannot be explained by the interstellar model because it is significantly greater than that expected, and variable. The $J$-band data are all consistent with the model, whereas only two of the six $K$-band measurements agree.

\begin{figure}
\centering
\includegraphics[width=8.9cm,angle=0]{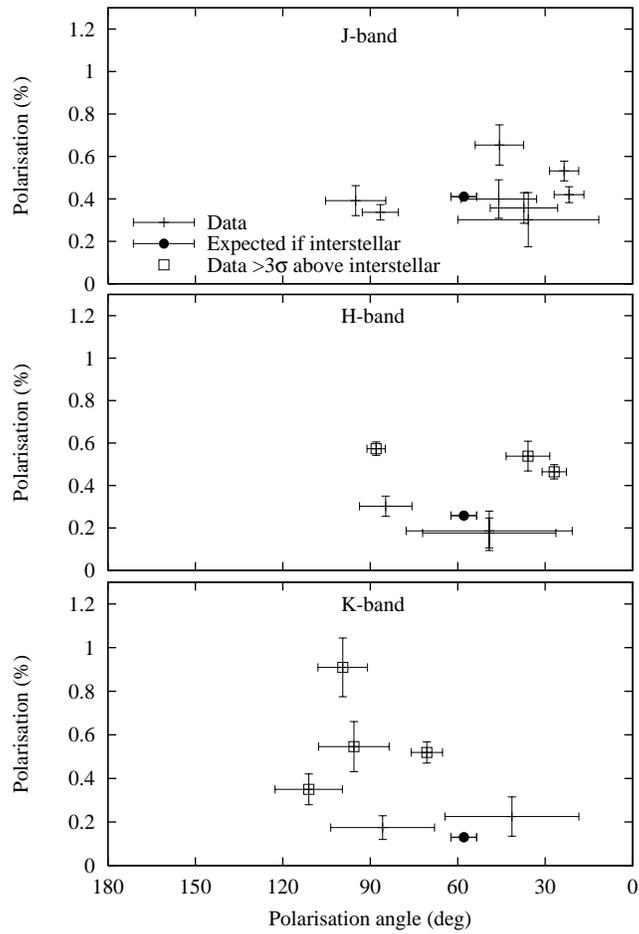}
\caption[The level of linear polarisation in Sco X--1 versus the position angle]{The level of LP in Sco X--1 versus the position angle. Square symbols indicate data with LP level $>3\sigma$ above that expected from interstellar dust \citep{schuet04}. The axis of the resolved jet of Sco X--1 is at PA = $54^\circ$ \citep{fomaet01}.}
\label{polar-scox1theta}
\end{figure}

Z-sources like Sco X--1 have the signature of OIR colours and correlated (with X-ray) variability typical of an irradiated accretion disc (e.g. \citealt{mcgoet03}; Chapters 2 and 3). They also have a radio counterpart, likely from the jet \citep[e.g.][]{miglfe06}\footnote{Some interpretations prefer an origin located in the neutron star magnetosphere \citep[e.g.][]{paizet06}, but in Sco X--1 the jet is clearly resolved \citep{fomaet01}.}, the spectrum of which may extend to the NIR, but is generally fainter than the disc in this regime (Chapters 2 and 3). The variable $K$ and $H$-band LP strongly suggest the jet is present. The low level (maximum LP $\sim$ 1.0\%) does not rule out optically thick emission, which can produce this level. The LP levels here largely agree with those of \citeauthor{shahet07}; at the $K$-band central frequency they measure LP = 0.5--1.3\% whereas our levels range from LP = 0.13--0.91\%, and at $H$ they measure LP = 0.2--0.7\% and we have LP levels between 0.07\% and 0.57\%. \citeauthor{shahet07} show the LP could increase to up to $\sim 30$\% in the mid-IR. Here, it is shown that the LP is highly variable and sometimes absent, suggesting either the conditions in the inner regions of the jet are variable, or the jet luminosity (and hence LP) is changing with time, perhaps correlated with the Z-track.

Interestingly, the PA of the most significant LP detections is $\sim 90$-$100^\circ$, differing from a PA of $\sim 60^\circ$ when the level is consistent with the interstellar model (Fig. \ref{polar-scox1theta}). $60^\circ$ is also the PA of the interstellar polarisation measured in the optical, further confirming the origin of the NIR LP is likely to be interstellar when its level is consistent with the model. The resolved radio jets \citep{fomaet01} are in a direction (which is constant over many years) on the plane of the sky $\sim 40^\circ$ different from the PA of the polarised NIR jet. There may be systematic errors in the measurements of PA (Section 5.2.3) but the fact that the low-LP measurements of this source are consistent with the interstellar model suggest these errors are low. \cite{shahet07} measured a slightly different PA to us; 116--147$^\circ$ which is perpendicular to the jet axis, indicating a magnetic field aligned with the jet.

The total flux of Sco X--1 (which is uncalibrated) is generally variable by $< 1$\%. Any jet component is likely to be variable so this suggests a low jet contribution. If this is the case, the jet must be $>> 1$\% polarised, and so it must be coming from the optically thin region. It would be interesting to use simultaneous X-ray data in a future study, to test whether the LP (and flux) is correlated with the position on the Z-track.
\newline\newline
\textbf{Aql X--1}:
\newline
An optical polarimetric study of this source was carried out by \cite{charet80} who found LP $\sim$ 1.5\%, which they interpret to be interstellar in origin. However they also detected marginal variability, with one measurement of LP = 2.3$\pm$0.5\%. The S/N was low in our NIR observations and we can place 3$\sigma$ upper limits of LP $<$ 5\%, 17\% and 12\% in $J$, $H$ and $K$, respectively \citep[the aperture includes a very close contaminating star which appears to contribute about half the flux in $K$; star $a$ in][]{chevet99}.

The observations here span three epochs, the first of which was at the end of a small outburst \citep{bailet06}. Indeed, the derived $JHK$ magnitudes are consistent with those found in quiescence by \cite{garcet99} except in the first epoch (20 Aug 2006), where $J$ and $K$ are 0.20$\pm$0.15 and 0.27$\pm$0.16 magnitudes brighter than the mean quiescent values. On this date, the $J$--$K$ de-reddened SED is blue (column VI of Table \ref{polar-ukirt}) implying the emission is from the disc, although the contaminating star will affect this measurement. The disc and donor star are expected to dominate the OIR light in quiescence, but during outburst the contributions vary. A dramatic reddening of the optical SED was seen at the peak of the bright 1978 outburst of Aql X--1 \citep{charet80} which could be explained by the domination of the jet component (Chapters 2 and 3). However the jet is likely to contribute very little to the OIR light in quiescence.
\newline\newline
\textbf{SS 433}:
\newline
High levels of LP have been detected in this HMXB in the optical and UV, which can be explained by a combination of Thompson and Rayleigh scattering \citep{dolaet97}. In the lower panel of Fig. \ref{polar-lp3} the polarisation spectrum of SS 433 in shown with the two $J$-band measurements from this work, the optical/UV data and a $K$-band upper limit from \cite{thomet79}. Our data follow the general trend of decreasing LP with decreasing frequency, so we expect the origin is the same as in the optical. However, the PAs differ between the $J$-band ($70-80^\circ$ during two epochs) and the optical ($\sim 90-140^\circ$), which could be explained by an interstellar contribution at these low levels of LP (the extinction is high; see Table \ref{polar-Av}).

\subsection{Summary and conclusions}

Linear polarimetric observations are presented of seven X-ray binaries in the NIR regime and one in the optical. A significant level (3$\sigma$) of linear polarisation is detected in four sources. The polarisation is found to be intrinsic (at the $> 3\sigma$ level) in two sources; GRO J1655--40 ($\sim 4$--7\% in $H$ and $Ks$-bands during an outburst) and Sco X--1 ($\sim 0.1$--0.9\% in $H$ and $K$), which is stronger at lower frequencies. This is likely to be the signature of optically thin synchrotron emission from the collimated jets in these systems, whose presence indicates a partially-ordered magnetic field is present at the inner regions of the jets. In Sco X--1 the intrinsic polarisation is at a comparable level to that measured recently by \cite{shahet07}, but it is also highly variable in $H$ and $K$-bands; sometimes it is absent. The variability could be due to changes in either the jet magnetic field structure or the jet power and luminosity. In the $J$-band (i.e. at shorter wavelengths) the polarisation is not significantly variable and is consistent with an interstellar origin. The optical light from GX 339--4 is also polarised, but at a level and position angle consistent with scattering by interstellar dust. The other polarised source is SS 433, which has a low level (0.5--0.8\%) of $J$-band polarisation, likely due to local scattering. The NIR counterparts of GRO J0422+32, XTE J1118+480, 4U 0614+09 and Aql X--1 (which were all in or near quiescence) have a linear polarisation level of $< 16$\% (3$\sigma$ upper limit, some are $< 6$\%).

\begin{table}
\small
\caption[The OIR emission processes that can describe the polarimetry results]{The OIR emission processes that can describe the polarimetry results.}
\label{polar-emproc}
\small
\begin{tabular}{lccccc}
\hline
Sample&X-ray       &Viscous&Optically&Optically&Intrinsic\\
      &reprocessing&disc   &thin jet &thick jet&companion\\
\hline
BHXBs; OPT; low-lum. hard state&$\surd$&$\surd$&$\times$&$\times$&$\times$\\
BHXBs; NIR; low-lum. hard state&$\times$&$\times$&$\surd$&$\times$&$\surd$\\
NSXB Z-sources; NIR&$\times$&$\times$&$\surd$&$\surd$&$\times$\\
\hline
\end{tabular}
\normalsize
\end{table}

Some results in this Chapter have constrained the dominating optical or NIR emission mechanisms in these sources (Table \ref{polar-emproc}). The optical SED of GX 339--4 is blue during a low-luminosity hard state, which can be explained by disc emission but not jet (whose spectrum is flat or red) or companion star \citep[which has never been detected even at its lowest flux levels;][]{shahet01} emission. The NIR LP of BHXB GRO J1655--40 in a low-luminosity hard state is higher than predicted from all emission processes except optically thin synchrotron emission from the jets, but its NIR SED indicates the star is the main emission contributor \citep[see][]{miglet07}. Finally, the variable polarisation seen in the Z-source Sco X--1 can be explained by either optically thick or optically thin synchrotron emission from its jets.

As far as the local conditions in the region of the inner jets is concerned, $f$, the dimensionless parameter representing the ordering of the magnetic field has been constrained in two sources. In the BHXB GRO J1655--40, $f = 0.41\pm 0.19$ whereas in the NSXB 4U 0614+09, $f < 0.38$. The BHXB result is consistent with those found from radio studies of the transient jet ejections in BHXBs \citep[e.g.][]{fendet02,brocet07} and multi-wavelength studies of jets in AGN \citep[e.g.][]{saiksa88,jorset07}. The inner regions of steady jets from BHXBs may have a similar magnetic field configuration to the brighter, transient jets. A dramatic change in PA in GRO J1655--40 at the end of its outbursts (also seen in the radio) suggests the magnetic field structure may change at low jet powers, and/or the polarisation may originate from a different region of the jet at these later times. For NSXBs, this is the first time a constraint on $f$ has been made (none exist from radio observations). The non-detection of polarisation in any of the low-luminosity sources suggests either the jet contribution is insignificant or the magnetic field is not well-ordered (although the S/N is low). This Chapter demonstrates that polarimetry is a powerful tool to constrain the conditions in the inner regions of the jets in X-ray binaries.

\newpage

\begin{center}
{\section{The Cygnus X--1 jet-powered nebula}}
\end{center}

\subsection{Introduction}

It was explained in Section 1.2.5 that XBs can indirectly produce OIR emission originating from extended matter in proximity to the system. In particular, emission which is essentially energised by the jets of XBs can be used to constrain the properties of the jets; a common technique used for AGN jets energising their surrounding media. In Chapters 2 and 3 the broadband spectrum of the steady, hard state jet was constrained, providing constraints on the total power of these jets. Here, optical observations of the extended structure (nebula) associated with the jet of Cyg X--1 are used to constrain the properties of the emitting gas and the power of the jet which powers the structure.

Wide field imaging observations were performed with the INT of the region of the Cyg X--1 nebula in H$\alpha$ and \o3 (5007 \AA) filters in order to (a) constrain the velocity and temperature of the shocked gas and hence the time-averaged power of the X-ray binary jet of Cyg X--1, (b) obtain for the first time flux-calibrated optical emission line measurements from a nebula powered by an X-ray binary jet, and (c) search for any jet--ISM interactions associated with the counter jet supposedly to the south of Cyg X--1.

The observations and data reduction procedures in this Chapter were performed by myself \citep[except the 2004 observations used in the initial paper;][which were taken by Dimitris Mislis]{gallet05}; the co-authors Rob Fender, Elena Gallo and Christian Kaiser made contributions towards the text, analysis and discussions. In addition, Elena Gallo provided the wide-field radio map of the region used for Section 6.4.4. This work is published in \emph{MNRAS} (Russell, Fender, Gallo \& Kaiser 2007; \citeauthor{paper3}, and also summarised in a published conference proceedings; \citeauthor{paper2}. The 2004 H$\alpha$ observations and analyses were used in the initial Cyg X--1 nebula paper published in \emph{Nature} \citep{gallet05}. Fraser Lewis provided the H$\alpha$ images of the Cyg X--1 nebula taken with the Faulkes Telescope North (which are not currently published).

\subsection{Observations and Data Reduction}

Cyg X--1 was observed using the Wide Field Camera (WFC) on the 2.5m INT in the H$\alpha$ filter on 2004-04-05 during a Service Night (bright time; full moon). Conditions were photometric with a seeing of $< 1.5$ arcsec. These observations were used for the first paper; \cite{gallet05}. After this publication, the field of Cyg X--1 and two Landolt standard stars were observed with the same telescope and camera in narrowband and broadband filters on 2005-10-10 as part of a Galactic survey of emission line nebulae powered by X-ray binary jets (Russell et al., in preparation; see Chapter 7). The conditions were cloudless with a seeing of $\sim$1--2 arcsec. The moon was set, making it possible to achieve a higher S/N than the 2004 images. The WFC consists of 4 CCDs (charge-coupled devices) of 2048$\times$4100 imaging pixels, each of scale 0.333 arcsec pixel$^{-1}$. The field of view is 34$\times$34 arcmin. In addition, the Cyg X--1 region was imaged in H$\alpha$ using HawkCam on the Faulkes Telescope North in 2007. Table \ref{tab-neb1-1} lists the observations used for this work.

\begin{table*}
\begin{center}
\small
\caption[List of Cyg X--1 nebula observations]{List of Cyg X--1 nebula observations.}
\label{tab-neb1-1}
\begin{tabular}{lllllll}
\hline
Run&MJD&Target&RA \& Dec. (centre of &Air-&Filter&Exp.\\
no.&   &      &CCD 4 for WFC; J2000)         &mass&      &time(sec)\\
\hline
\multicolumn{7}{c}{\emph{2004 INT data:}}\\
393072&53100.14&Cyg X--1&19 58 21.67 +35 12 05.7&1.91&H$\alpha$ 6568\AA&1200\\
393073&53100.16&Cyg X--1&19 58 21.67 +35 12 05.7&1.67&H$\alpha$ 6568\AA&1200\\
393074&53100.17&Cyg X--1&19 58 21.68 +35 12 05.7&1.53&H$\alpha$ 6568\AA&1200\\
393075&53100.19&Cyg X--1&19 58 21.68 +35 12 05.8&1.41&H$\alpha$ 6568\AA&1200\\
393076&53100.20&Cyg X--1&19 58 21.68 +35 12 05.7&1.32&H$\alpha$ 6568\AA&1200\\
\hline
\multicolumn{7}{c}{\emph{2005 INT data:}}\\
473434&53653.92&Cyg X--1 north&19 59 38.49 +35 18 20.0&1.17&[O III] 5007\AA&1200\\
473435&53653.94&Cyg X--1 north&19 59 39.31 +35 18 29.6&1.23&[O III] 5007\AA&1200\\
473436&53653.96&Cyg X--1 north&19 59 38.50 +35 18 20.0&1.31&Harris V  &600 \\
473437&53653.96&Cyg X--1 north&19 59 38.49 +35 18 20.0&1.38&H$\alpha$ 6568\AA&1200\\
473438&53653.98&Cyg X--1 north&19 59 39.31 +35 18 30.0&1.49&H$\alpha$ 6568\AA&1200\\
473439&53654.00&Cyg X--1 north&19 59 37.68 +35 18 10.0&1.62&Harris R  &600 \\
473441&53654.01&Cyg X--1 south&19 59 07.99 +34 56 59.9&1.86&[O III] 5007\AA&900 \\
473442&53654.03&Cyg X--1 south&19 59 08.80 +34 57 09.8&2.06&[O III] 5007\AA&900 \\
473452&53654.11&Landolt 94--171&02 53 38.80 +00 17 17.9&1.14&Harris R  &10\\
473454&53654.11&Landolt 94--171&02 53 38.79 +00 17 18.0&1.14&Harris V  &10\\
473457&53654.11&Landolt 94--702&02 58 13.39 +01 10 54.0&1.13&Harris V  &10\\
473458&53654.11&Landolt 94--702&02 58 13.39 +01 10 53.9&1.13&Harris R  &10\\
\hline
\multicolumn{7}{c}{\emph{2007 Faulkes Telescope data:}}\\
--&54183.55&Cyg X--1 P1&19:58:02.00 +35:20:20.0&2.45&H$\alpha$ 6568\AA&4$\times$200\\
--&54183.60&Cyg X--1 P2&19:58:02.00 +35:16:00.0&1.63&H$\alpha$ 6568\AA&4$\times$200\\
--&54183.61&Cyg X--1 P3&19:58:22.00 +35:20:20.0&1.55&H$\alpha$ 6568\AA&4$\times$200\\
--&54183.62&Cyg X--1 P4&19:58:37.00 +34:59:10.0&1.44&H$\alpha$ 6568\AA&4$\times$200\\
\hline
\end{tabular}
\normalsize
\end{center}
The positions refer to the centre of CCD 4 of the WFC (which is the middle of the three eastern CCDs; Fig. \ref{neb1-fig1}), or the four positions used to make the mosaic from the Faulkes Telescope in the right panel of Fig. \ref{neb1-naturefaulkes}. 10 bias frames and 5$\times$H$\alpha$, 7$\times$[O III] 5007 \AA, 6$\times$Harris $R$-band and 5$\times$Harris $V$-band sky flats were taken in the same observing night for the 2005 data.
\end{table*}

\begin{figure}
\centering
\includegraphics[width=11.5cm,angle=0]{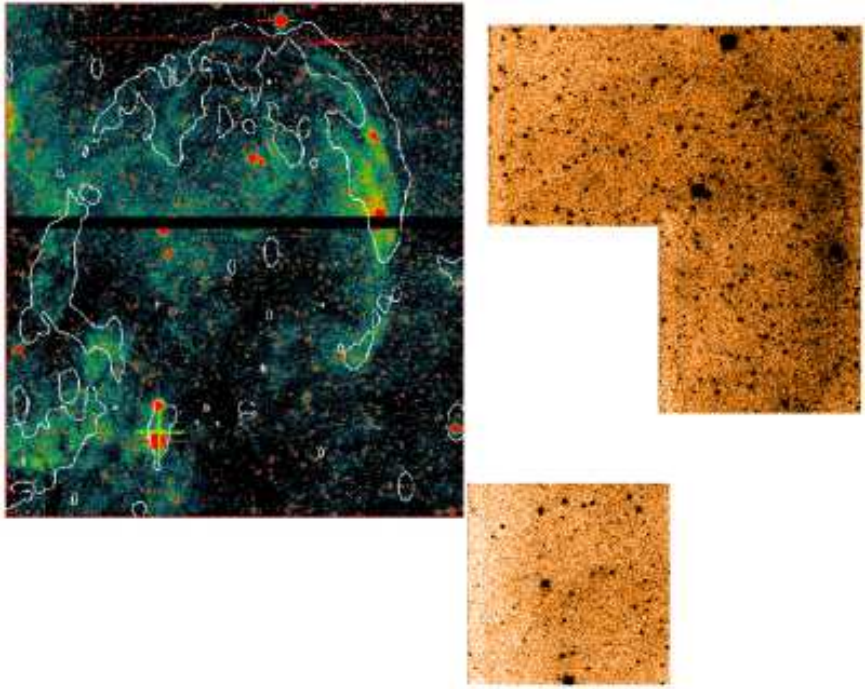}
\caption[INT image from 2004 and Faulkes Telescope image from 2007 in H$\alpha$ of Cyg X--1 and its nebula]{INT combined image from 2004 with radio contours (left panel; the colours correspond to flux intensity) and Faulkes Telescope image from 2007 (right) in H$\alpha$ of Cyg X--1 (green cross in left panel) and its nebula. In the Faulkes image, a region to the south is also imaged (see Section 6.4.4). North is up and east is to the left.}
\label{neb1-naturefaulkes}
\end{figure}

For the 2004 data, \small IRAF \normalsize was used for data reduction. The combined average bias frame for each of the four CCDs was subtracted from each image from the corresponding CCD (including the flat fields) and the science frames were divided by the normalised average flat field frame before being combined into one image for each CCD. A nearby bright star just to the west of Cyg X--1 caused a ghost on the image, which was removed by manually subtracting a constant value from the affected region of the image. The Faulkes data from 2007 are automatically de-biased and flat-fielded by the pipeline. The images were combined in \small IRAF. \normalsize The resulting images are presented in Fig. \ref{neb1-naturefaulkes}. We show radio contours from \cite{gallet05} overplotted on the INT image (left panel). Since the Faulkes image is not as deep as the INT ones (especially the 2005 data), we do not use these data in any following analyses except in Section 6.4.4, where a region to the south is used which was not imaged by the INT.

\begin{figure}
\centering
\includegraphics[width=15cm,angle=0]{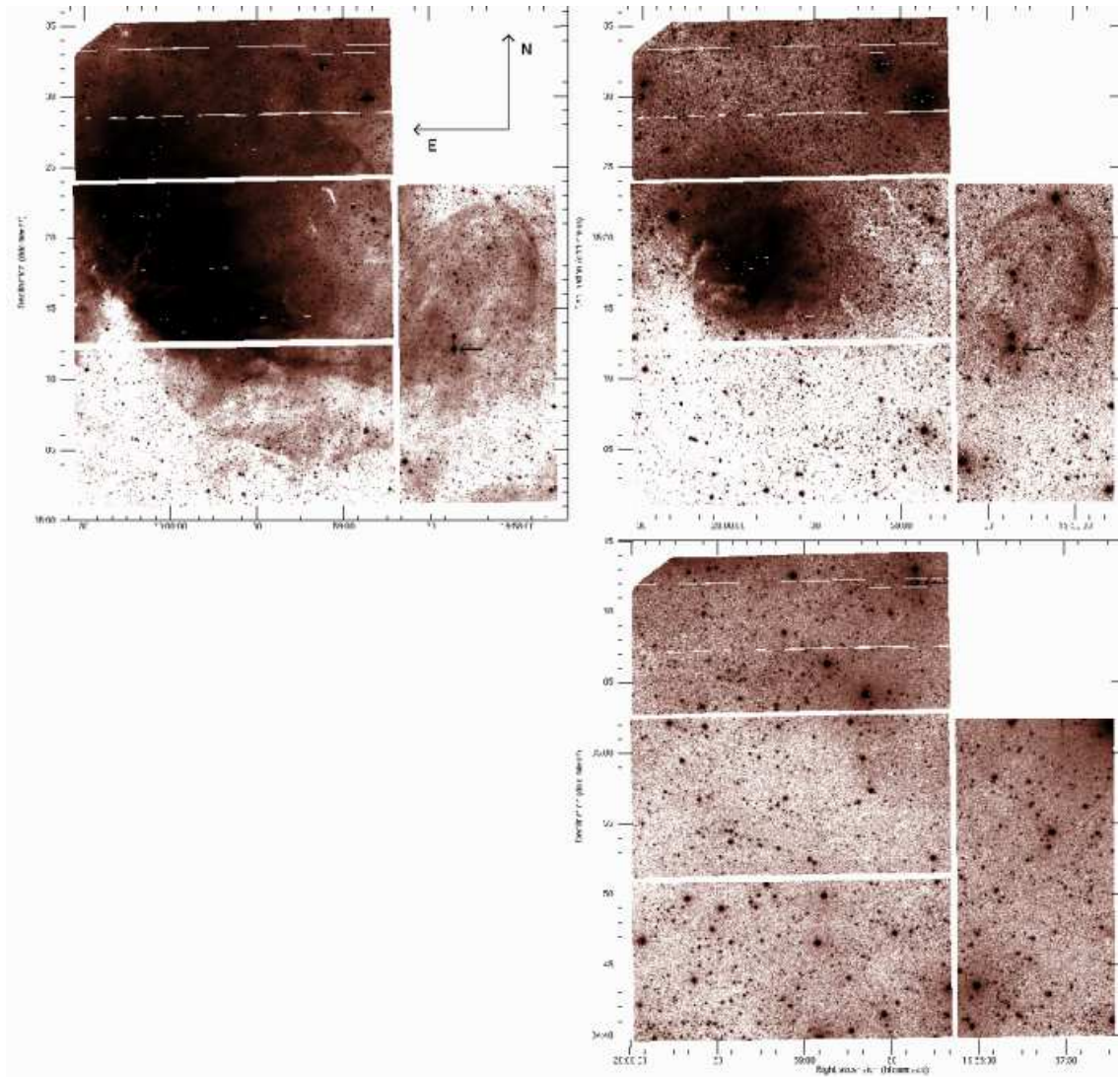}
\caption[Stacked WFC images of the field of Cyg X--1]{Stacked WFC images of the field of Cyg X--1 from the 2005 data. Top left: 2400 sec in H$\alpha$ (6568\AA); Top right: 2400 sec in [O III] (5007\AA); Lower: 1800 sec [O III] (5007\AA) of the field south of Cyg X--1. Cyg X--1 is indicated by an arrow in each panel. In the top and middle panels, the H~\small II \normalsize region Sh2--101 is prominent on the eastern CCD chips and the Cyg X--1 shell nebula is visible to the north west of the system.}
\label{neb1-fig1}
\end{figure}

Data reduction was performed using the pipeline package \small THELI \normalsize \citep[details in][]{erbeet05} for the 2005 data. After manually separating the data into type (bias/flat/science) and filter, \small THELI \normalsize separated each CCD, created a master bias and de-biased normalised master flats for each CCD before subtracting the master bias from the science frames and dividing them by the master flats. The package read the exposure time and RA \& Dec from the \small FITS \normalsize file headers (it is currently configured for reducing data from $\sim30$ instruments including the INT WFC) and matched several hundred stars in each exposure with those in the online MAST Guide Star Catalog (release 2.2). \small THELI \normalsize then aligned and stacked (for each position and filter) the images, and position-calibrated mosaics were created, with flux in counts s$^{-1}$pixel$^{-1}$. The resulting H$\alpha$ and \o3 mosaics of the Cyg X--1 field are displayed in Fig. \ref{neb1-fig1}.

\subsection{Flux calibration}

No standard star was observed that could be used to calibrate the 2004 data, however it is still possible to estimate a conservative lower limit to the flux density of the nebula. According to the exposure time calculator for the INT, given the conditions during the observations (seeing, moon phase, airmass), the sky brightness would be 18.3 mag arcsec$^{-2}$ which equals 0.138 mJy arcsec$^{-2}$ (see conversion method in Section 2.2.2). The Cyg X--1 nebula is clearly seen in not only the combined H$\alpha$ image (left panel of Fig. \ref{neb1-naturefaulkes}) but also in the individual images of 1200 sec each. For the nebula to be visible, the S/N must exceed one. The exposure time calculator predicts an extended source of magnitude 23.1 mag arcsec$^{-2}$ would have a S/N = 1.0 in 1200 sec in H$\alpha$, which equals $1.66\times 10^{-3}$ mJy arcsec$^{-2}$. Adopting the known interstellar extinction towards Cyg X--1 \citep[$A_{\rm V}=2.95\pm0.21$;][]{wuet82}, the intrinsic flux density would be 0.0219 mJy arcsec$^{-2}$. Therefore, since the nebula is visible in the 1200 sec image (and hence has S/N $\geq$ 1), it must have a flux density of $> 0.02$ mJy arcsec$^{-2}$. The measured radio flux density of the nebula is $\leq 0.05$ times this value \citep{gallet05}, so the radio--optical spectrum of the nebula cannot be explained by optically thin synchrotron emission, which requires $\alpha < 0$. The more likely scenario is bremsstrahlung emission from a thermal plasma producing the radio, with H$\alpha$ line emission from recombination in the shock. These results were used in \cite{gallet05}.

For the 2005 data, the Landolt standard stars 94--171 ($V=12.659$; $R=12.179$) and 94--702 ($V=11.594$; $R=10.838$) were observed at low airmass (see Table \ref{tab-neb1-1}). Aperture photometry was performed on the standards in the $V$ and $R$-band reduced images with \small PHOTOM \normalsize and the airmass--dependent atmospheric extinction was accounted for according to \cite{king85}. The resulting conversion between intrinsic flux density (at airmass $=0$) and counts s$^{-1}$ differed between the two standards by a factor 0.5 and 6.5 percent for $V$ and $R$, respectively. The H$\alpha$ filter is located within the passband of the Harris $R$ filter, and \o3 is located within the Harris $V$ filter. For a flat spectrum source, the WFC $R$-band filter collects 15.742 times more flux than the H$\alpha$ filter, and likewise $V$ collects 11.127 times more flux than \o3. By multiplying by these factors we obtain $7.140\times 10^{-3}$ mJy (intrinsic) per counts s$^{-1}$ for H$\alpha$ and $5.637\times 10^{-3}$ mJy (intrinsic) per counts s$^{-1}$ for \o3.

As a calibration check, the fluxes of 10 stars in our reduced Cyg X--1 H$\alpha$ image were measured, that are listed in the INT Photometric H$\alpha$ Survey of the Northern Galactic Plane \citep[IPHAS;][]{drewet05} catalogue (which uses the same telescope, instrument and filter). The ratio of the H$\alpha$ flux of each star measured here to that listed in the catalogue is 1.06$\pm$0.39 ($1\sigma$). We can therefore assume our flux calibration is accurate, and we adopt a $1\sigma$ error for each flux measurement $F$ (H$\alpha$ and \o3), of 0.14 dex (i.e. $log~F \pm 0.14$).

Since the 2005 data are the deepest, have the most number of filters used and are the only set to be flux calibrated, we use only this data in the following analyses.

\begin{figure}
\centering
\includegraphics[width=15cm,angle=0]{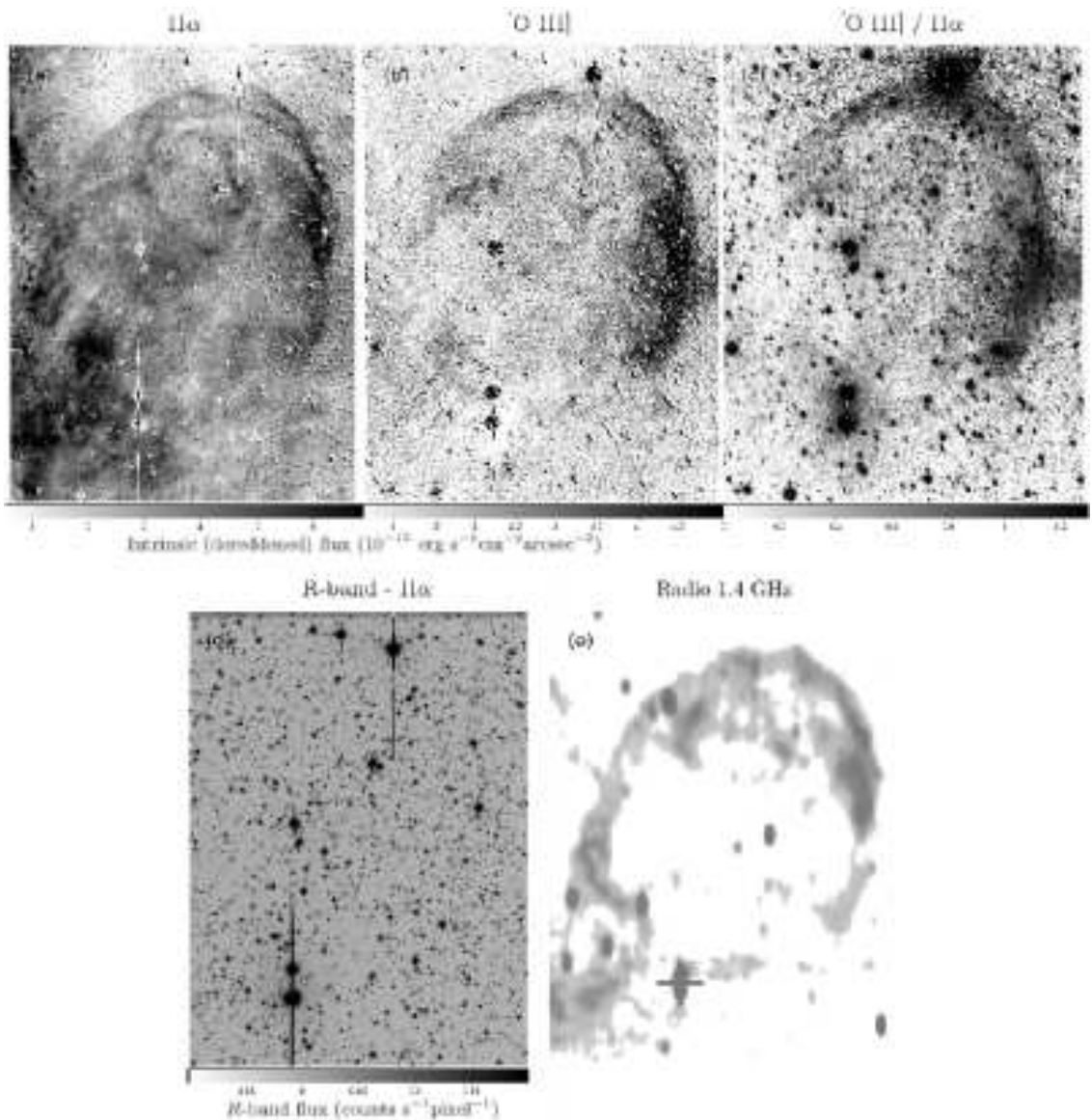}
\caption[Images of the Cyg X--1 ring nebula in five wavebands/ratios]{Images of the Cyg X--1 ring nebula in: (a) H$\alpha$; (b) [O III]; (c) [O III] / H$\alpha$; (d) $R$-band - H$\alpha$ and (e) radio 1.4 GHz \citep[from][]{gallet05}. The H$\alpha$ and [O III] images are de-reddened using $A_{\rm V}$ = 2.95 measured for Cyg X--1 \citep{wuet82}, flux calibrated and continuum-subtracted (i.e. H$\alpha-R$ and [O III]$-V$). The [O III] / H$\alpha$ flux ratio image is created from the non-continuum-subtracted [O III] and H$\alpha$ frames. The H$\alpha$ contribution has been subtracted from the $R$-band image, in which the nebula is not visible. Cyg X--1 is the lower of the two bright stars near the bottom of each image and is marked by a cross in the radio image.}
\label{neb1-fig2}
\end{figure}

\begin{figure}
\centering
\includegraphics[width=15cm,angle=0]{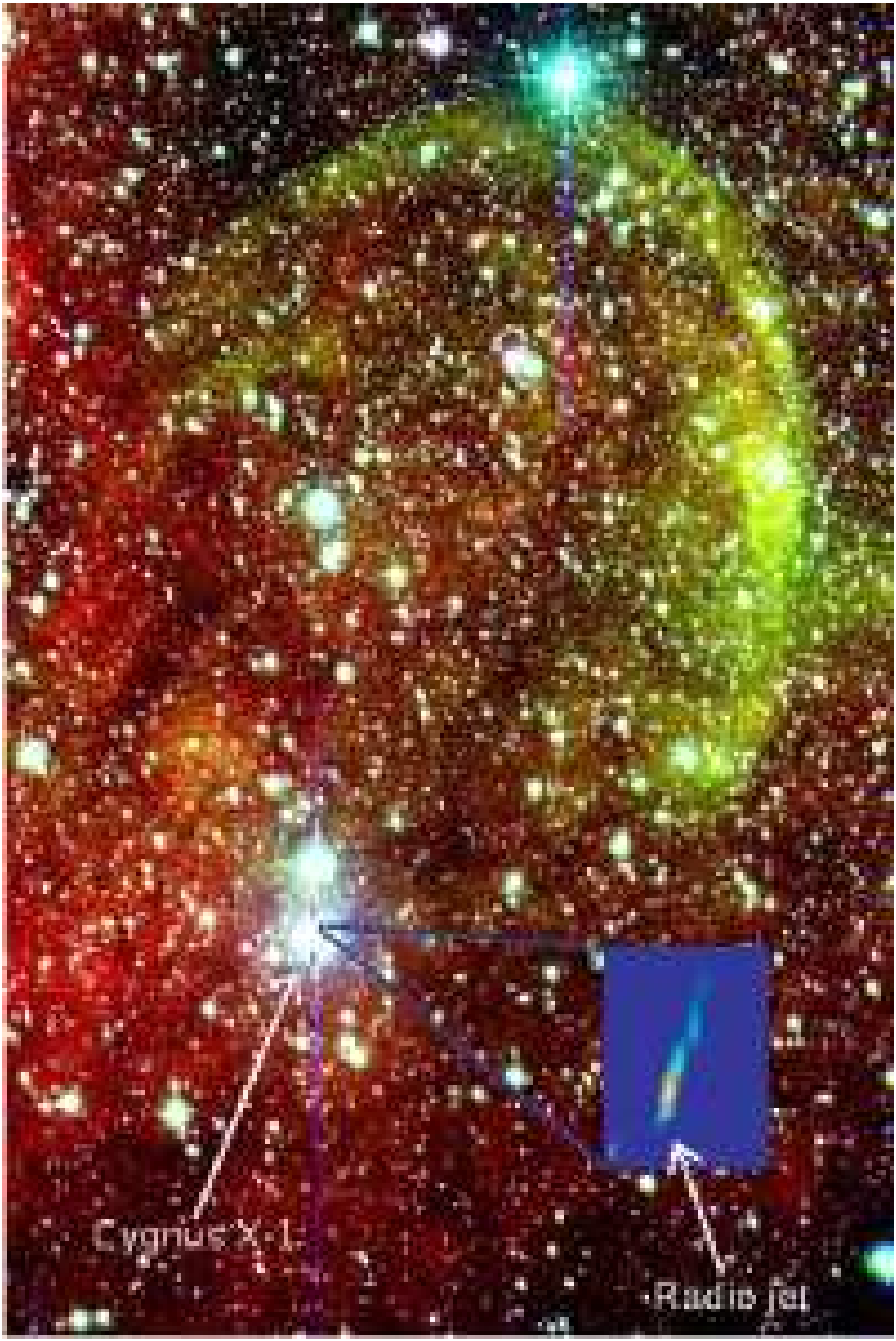}
\caption[False-colour image of the Cyg X--1 nebula]{False-colour image of the Cyg X--1 nebula with the compact radio jet of \cite{stiret01} shown. The image is combined thus: H$\alpha$ in red, [O III] in green and $V$-band in blue.}
\label{neb1-colour}
\end{figure}

\begin{figure}
\centering
\includegraphics[height=11cm,angle=270]{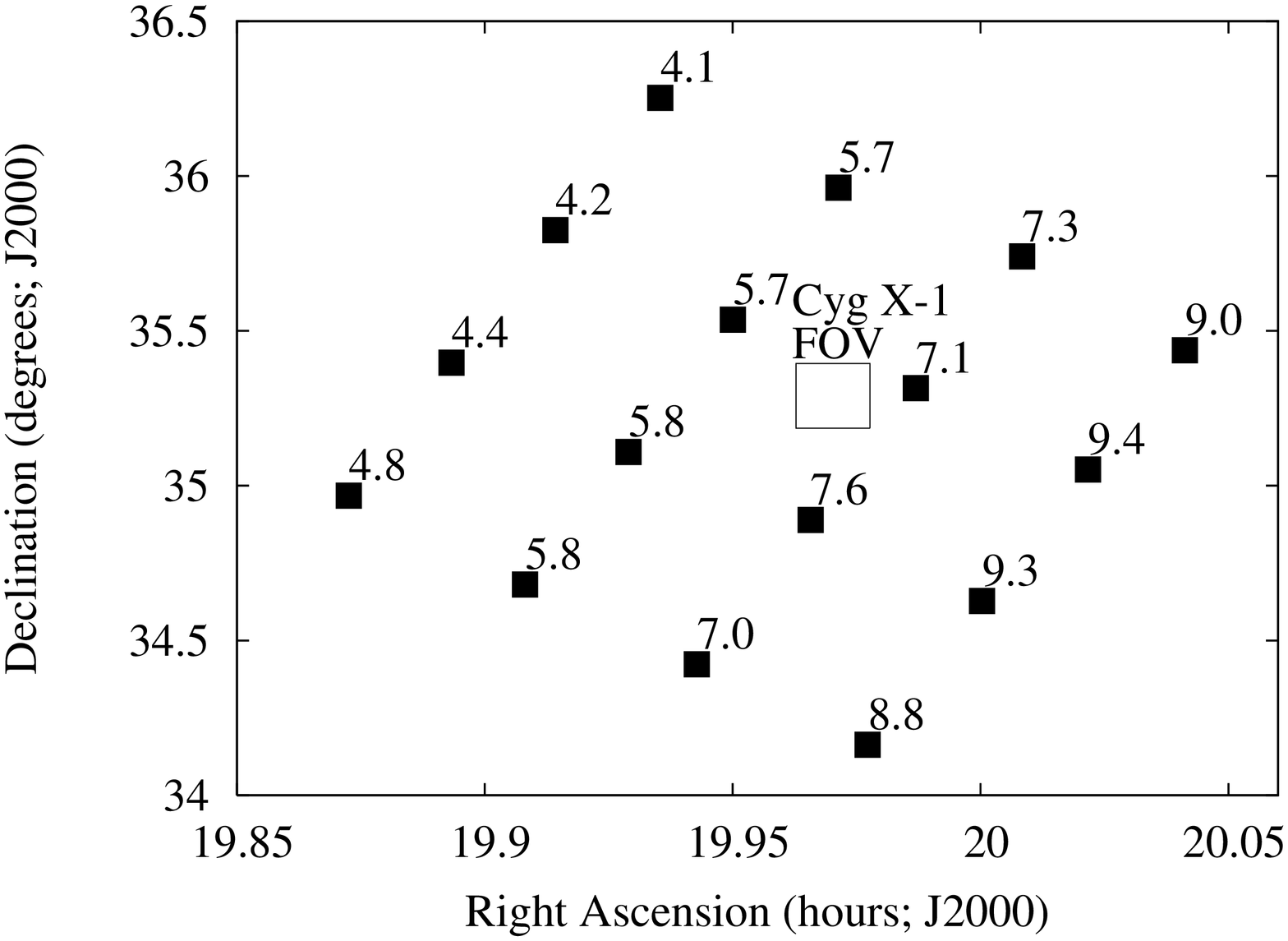}
\caption[The Galactic neutral hydrogen column density at positions close to Cyg X--1]{Total Galactic neutral hydrogen column density $N_{\rm H}$ in units of $10^{21}$ cm$^{-2}$ at positions close to Cyg X--1 \citep{hartbu97}. The box indicates the FOV containing Cyg X--1 and its nebula.}
\label{neb1-extinc}
\end{figure}

Flux-calibrated continuum-subtracted H$\alpha$ and \o3 images of the region of Cyg X--1 were created (Fig. \ref{neb1-fig2}a, \ref{neb1-fig2}b) using \small IMUTIL \normalsize and \small IMMATCH \normalsize in \small IRAF \normalsize. The resulting images are corrected for interstellar extinction towards Cyg X--1 adopting $A_{\rm V}=2.95\pm0.21$ \citep{wuet82}. The extinction towards Cyg X--1 and its nebula may vary over the field of view. The total neutral hydrogen column density $N_{\rm H}$ through the Galaxy at various positions in a $\sim4$ deg$^2$ area around Cyg X--1 \citep{hartbu97} is indicated in Fig. \ref{neb1-extinc}. $N_{\rm H}$ appears to vary smoothly over the field, and the four closest points to Cyg X--1 yield $N_{\rm H}=(6.55\pm 0.94)\times 10^{21}$ cm$^{-2}$ in the region containing Cyg X--1, or $A_{\rm V}=3.66\pm 0.53$ adopting $N_{\rm H} = 1.79 \times 10^{21} cm^{-2}A_{\rm V}$ \citep{predet95}. If the extinction $A_{\rm V}$ towards Cyg X--1 (i.e. not through the whole Galaxy in this direction) also varies by $\pm$0.53 over the much smaller field of view of the nebula (a conservative case), an error is introduced of 0.18 dex and 0.25 dex to the H$\alpha$ and \o3 flux measurements, respectively (no distance estimate is required). The measurements of \o3 / H$\alpha$ however only suffer a 0.06 dex uncertainty due to the different extinctions suffered at the two wavelengths (\o3 at 5007 \AA ~and H$\alpha$ at 6568\AA; this concerns only the contribution to the error budget of the extinction).

Propagating the above two sources of error, we arrive at:
\begin{eqnarray}
  \label{neb1-eqn1}
  \Delta (log_{10} F_{\rm H\alpha})=0.23~log_{10} F_{\rm H\alpha}\\
  \Delta (log_{10} F_{\rm [O~III]})=0.29~log_{10} F_{\rm [O~III]}\\
  \label{neb1-eqn2}
  \Delta (log_{10} (F_{\rm [O~III]}/F_{\rm H\alpha}))=0.15~log_{10} (F_{\rm [O~III]}/F_{\rm H\alpha})
  \label{neb1-eqn3}
  \end{eqnarray}
\vspace{0mm}

\o3 / H$\alpha$ and $R$-band $-$ H$\alpha$ images were also created (Fig. \ref{neb1-fig2}c, \ref{neb1-fig2}d). The \o3 / H$\alpha$ image represents the ratio of the \o3 and H$\alpha$ fluxes (in de-reddened erg s$^{-1}$cm$^{-2}$arcsec$^{-2}$) and the H$\alpha$ contribution is subtracted from the continuum in the $R$-band $-$ H$\alpha$ image. The 1.4 GHz radio image of \cite{gallet05} is shown in Fig. \ref{neb1-fig2}e for comparison. Finally, a colour image is presented in Fig. \ref{neb1-colour} which emphasises the difference in $F_{\rm [O~III]}$ / $F_{\rm H\alpha}$ ratio between the Cyg X--1 nebula and the surrounding gas, as discussed below.

\subsection{Results}

\begin{figure}
\centering
\includegraphics[width=12cm,angle=0]{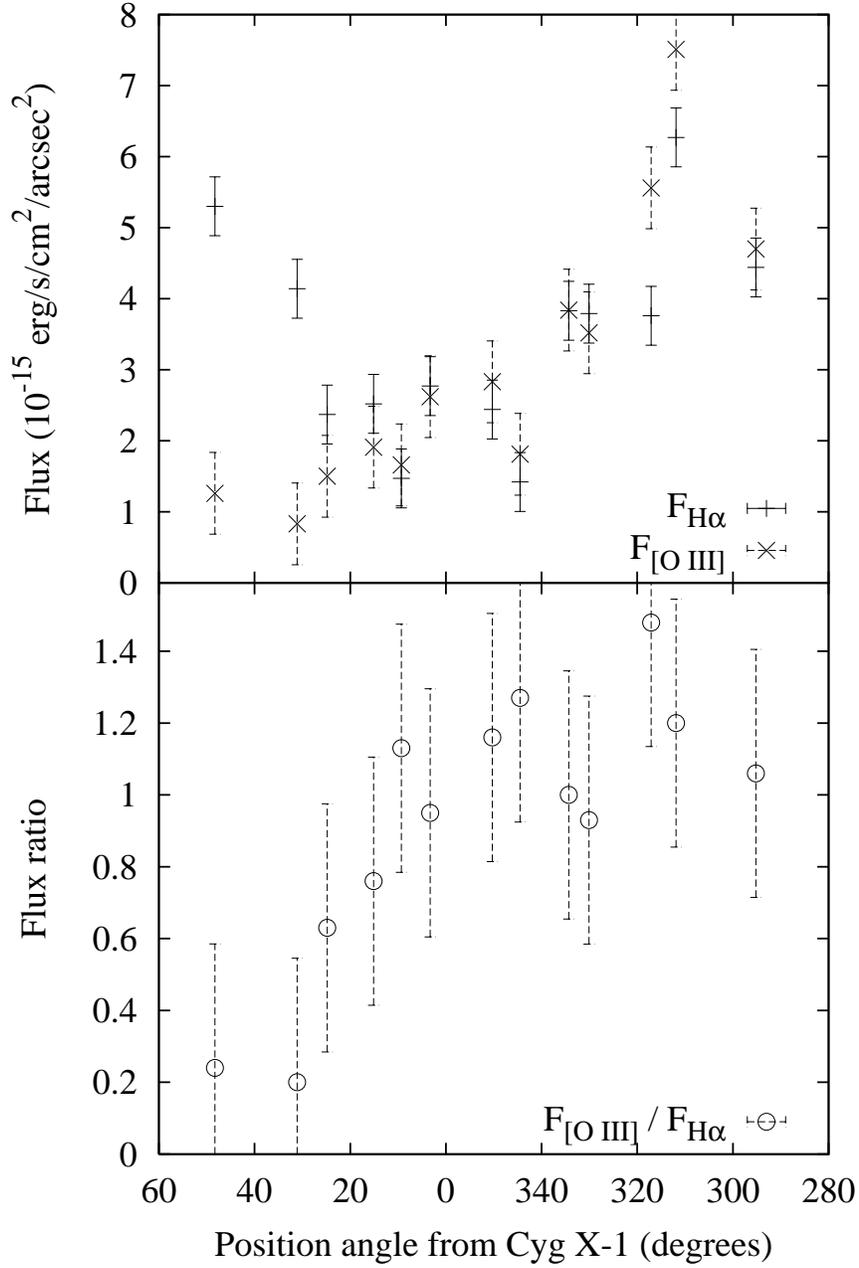}
\caption[Emission line fluxes at various positions around the Cyg X--1 shell]{Intrinsic de-reddened continuum-subtracted H$\alpha$ and [O III] fluxes at various positions around the shell, versus the position angle from Cyg X--1. See equations \ref{neb1-eqn1}--\ref{neb1-eqn3} for the absolute errors associated with each flux measurement; plotted are the relative errors. At each position, the mean flux in a 164 arcsec$^2$ circular aperture containing no visible stars is plotted. The data on the left are from the eastern side of the nebula and the data on the right are from the western side.}
\label{neb1-PA}
\end{figure}

The H$\alpha$ and \o3 fluxes measured in a number of positions in the shell of the nebula are plotted in Fig. \ref{neb1-PA}. It seems that the \o3 / H$\alpha$ ratio increases with position going clockwise around the shell, from values of $F_{\rm [O~III]}$ / $F_{\rm H\alpha}\sim 0.2$ on the eastern side to $\sim 1.2$ on the western side of the nebula. The fluxes of both emission lines are also higher on the western side. The line ratio is approximately constant in the filaments inside the nebula, with values of $F_{\rm [O~III]}$ / $F_{\rm H\alpha}\sim 0.7$. The H~\small II \normalsize region Sh2--101 and the `stream' of diffuse emission joining the region to the Cyg X--1 nebula, possess ratios of order $F_{\rm [O~III]}$ / $F_{\rm H\alpha}\sim 0$--0.2, with one area (RA 19 59 45, Dec +35 18 48) having a higher value, $F_{\rm [O~III]}$ / $F_{\rm H\alpha}\sim 0.6$. This area resembles a shell-like feature close to a bright star HD 227018 at the centre of the H~\small II \normalsize region, and could be a bow shock associated with the motion of this star \citep[which is 12 milli-arcsec per year in the direction of the observed shell in the H~\small II \normalsize region;][]{perret97}. The flux in the H~\small II \normalsize region is $F_{\rm H\alpha}\sim 5$--$15\times 10^{-14}$ erg s$^{-1}$cm$^{-2}$arcsec$^{-2}$; $F_{\rm [O~III]}\sim 0.5$--$5\times 10^{-14}$ erg s$^{-1}$cm$^{-2}$arcsec$^{-2}$.

The large quantity of stars in the line of sight of the shell prevent us from accurately calculating the total H$\alpha$ and \o3 luminosity of the structure. However, the area of the emitting gas can be conservatively estimated to be between 3 arcmin$^2$ (approximating the nebula to a $15\times 700$ arcsec$^2$ area) and 46 arcmin$^2$ (the area of a circle with radius 4 arcmin); the mean flux of the nebula (Fig. \ref{neb1-PA}) can then be used to constrain its luminosity. At a distance of 2.1 kpc, the H$\alpha$ and \o3 luminosities of the nebula are $1.8\times 10^{34}\leq L_{\rm H\alpha}\leq 2.8\times 10^{35}$ erg s$^{-1}$ and $1.3\times 10^{34}\leq L_{\rm [O~III]}\leq 2.1\times 10^{35}$ erg s$^{-1}$, respectively.

In ten regions (of $\sim 5$ arcsec$^2$ areas) inside and outside the H$\alpha$-emitting nebula, the $R$-band de-reddened H$\alpha$-subtracted flux is consistent with zero, and the mean of the standard deviations in each region is $2.57\times 10^{-15}$ erg s$^{-1}$cm$^{-2}$arcsec$^{-2}$. We therefore have a $3\sigma$ upper limit to the $R$-band flux of $7.7\times 10^{-15}$ erg s$^{-1}$cm$^{-2}$arcsec$^{-2}$, or a flux density of 0.0068 mJy arcsec$^{-2}$. Similarly, we obtain a $3\sigma$ upper limit to the $V$-band flux (from ten $\sim 5$ arcsec$^{-2}$ regions) of $2.0\times 10^{-14}$ erg s$^{-1}$cm$^{-2}$arcsec$^{-2}$, or 0.0196 mJy arcsec$^{-2}$. From the $R$-band upper limit, we conclude that the optical continuum flux density is $\leq 7$ times the radio flux density \citep{gallet05}, yielding a radio--optical spectral index $\alpha \leq 0.15$.

The broadband spectrum is consistent with bremsstrahlung emission with a flat radio--optical continuum and optical emission lines which are at least one order of magnitude more luminous than the continuum. We cannot however definitely state that the radio emission is thermal in origin, as mixed thermal/non-thermal plasma is seen in some environments \citep[e.g. the arcsec-scale jets of SS 433;][]{miglet02}. A future measurement of the radio spectral index, which is difficult to achieve, is required to test whether the radio through optical spectrum is consistent with a thermal plasma or requires two (or more) components. Currently the morphology of the radio emission, being spatially coincident with the rather thin optical shell, does seem to favour the single thermal plasma model.

\subsubsection{Morphology and nature of the nebula}

The higher S/N in the H$\alpha$ image compared to that in \cite{gallet05} is striking from the observed structure within the nebula (Fig. \ref{neb1-fig2}a). One would expect the most luminous areas of the nebula to occur in regions of the shock front where the line of sight is tangential to its surface and therefore should exist only at the apparent edge of the nebula if the structure is spherical. Filaments of gas are seen within the nebula and are unlikely to originate in unassociated line-of-sight photoionised gas because of their filamental morphology; the nebulosity is less diffuse than the gas in the photoionised H~\small II \normalsize region. The \o3 / H$\alpha$ ratios of $\sim 0.7$ in the filaments (see above) and morphology suggest either multiple shock fronts or an uneven non-spherical nebula. The latter is expected for a varying preshock gas density, which is clearly the case on inspection of the variation in H$\alpha$ nebulosity close to the nebula.

The \o3 / H$\alpha$ image reveals two striking features: firstly the H$\alpha$ nebulosity (which extends from the H~\small II \normalsize region Sh2--101 to the east of the nebula) possesses a low \o3 / H$\alpha$ flux ratio compared to the Cyg X--1 nebula, and secondly there exists a thin outer shell to the Cyg X--1 nebula with a higher \o3 / H$\alpha$ ratio. H~\small II \normalsize regions and photoionised gas in general can possess a wide range of \o3 / H$\alpha$ ratios \citep*[e.g.][]{john53,blaiet82} but this morphology of an outer \o3-emitting shell is only expected from shock-excited gas. These two observations confirm that the gas in the Cyg X--1 nebula must be ionised by a different source to the surrounding nebulosity. 

\begin{figure}
\centering
\includegraphics[width=9cm,angle=0]{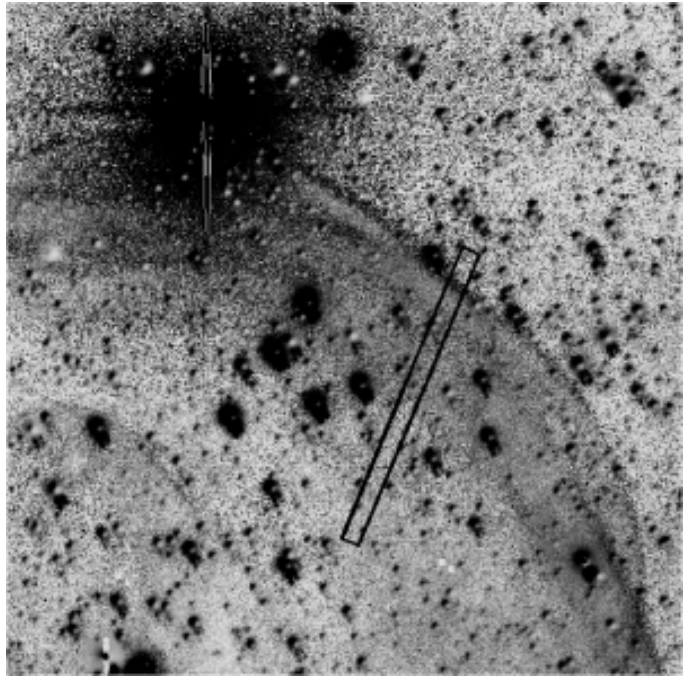}
\caption[A blow-up of part of the Cyg X--1 ring nebula]{A blow-up of part of Fig. \ref{neb1-fig2}c: [O III] / H$\alpha$ image of the Cyg X--1 ring nebula. We interpret the thin outer shell with a high [O III] / H$\alpha$ flux ratio as possibly originating in the ionised atoms close to the front of the bow shock. North is up, east to the left. The emission line fluxes and [O III] / H$\alpha$ ratio along the slice indicated in the image are shown in Fig. \ref{neb1-fig-slice}.}
\label{neb1-slice-pic}
\end{figure}

\begin{figure}
\centering
\includegraphics[width=12cm,angle=0]{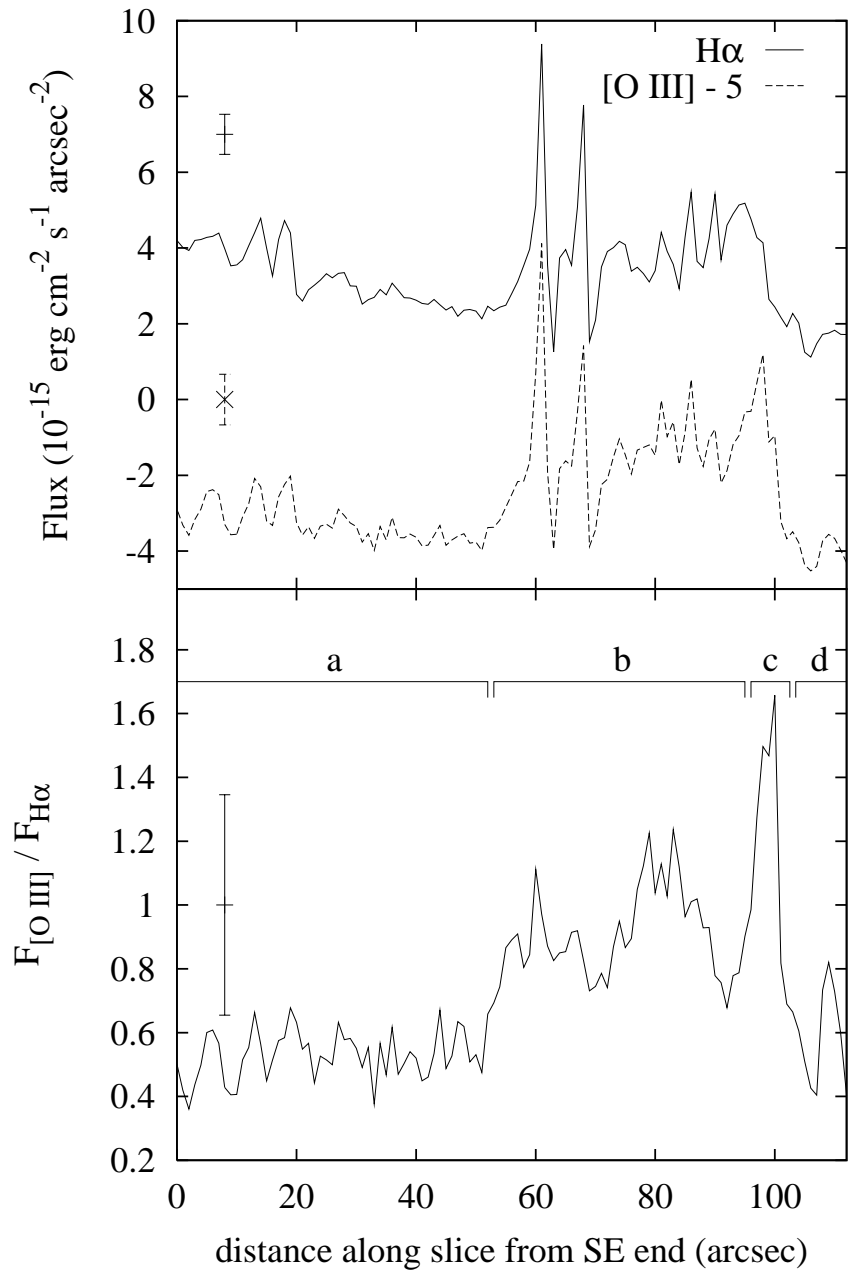}
\caption[Emission line fluxes and line ratios along a slice through the nebula]{The emission line flux (top panel) and the [O III] / H$\alpha$ ratio (lower panel) along a slice through the nebula indicated in Fig. \ref{neb1-slice-pic}. The [O III] flux has been offset by $-5\times 10^{-15}$ erg s$^{-1}$cm$^{-2}$arcsec$^{-2}$ for clarity. In the lower panel, region a = inside the nebula; b = the ring nebula; c = the thin outer shell luminous in [O III] / H$\alpha$; and d = outside the nebula. The mean relative error associated with each data set (derived from equations \ref{neb1-eqn1}--\ref{neb1-eqn3}) is indicated on the left in the panels. The spikes in the upper panel are due to stars in the slice.}
\label{neb1-fig-slice}
\end{figure}

A blow-up of the north-west area of the nebula in \o3 / H$\alpha$ is shown in Fig. \ref{neb1-slice-pic}; the figure clearly shows this thin shell of enhanced \o3 / H$\alpha$ ratio on the outer side of the ring. A 3.3~arcsec-wide slice is taken through the image (indicated in Fig. \ref{neb1-slice-pic}) and the emission line flux and $F_{\rm [O~III]}$ / $F_{\rm H\alpha}$ flux ratio along this slice is plotted in Fig. \ref{neb1-fig-slice}. The upper panel of Fig. \ref{neb1-fig-slice} shows that there is a sharp outer edge to the ring, and this edge is further out in \o3 than in H$\alpha$. This results in the thin outer shell of enhanced \o3 / H$\alpha$ ratio that is visible in the lower panel of Fig. \ref{neb1-fig-slice}. This outer shell is $\sim 6$ arcsec thick and the rest of the ring is $\sim 40$ arcsec thick, at least in the region of the slice. We note that the thin shell of enhanced \o3 / H$\alpha$ ratio is not due to inaccurate alignment of the \o3 and H$\alpha$ images as the stars are aligned and the thin shell exists on both sides of the nebula.

\subsubsection{Constraining the shock velocity and jet power}

The $F_{\rm [O~III]}$ / $F_{\rm H\alpha}$ flux ratio is measured in seven `slices' (typical thickness 12 pixels $= 4$ arcsec) of the \o3-emitting shock front, and in four larger regions behind it. In the thin outer shell, $F_{\rm [O~III]}$ / $F_{\rm H\alpha}=1.36\pm 0.28$ and behind the shell, $F_{\rm [O~III]}$ / $F_{\rm H\alpha}=0.89\pm 0.46$. Taking into account the error from equation \ref{neb1-eqn3}, $F_{\rm [O~III]}$ / $F_{\rm H\alpha}=1.36^{+0.65}_{-0.46}$ in the shock front and $F_{\rm [O~III]}$ / $F_{\rm H\alpha}=0.89^{+0.64}_{-0.49}$ in the nebula behind the shock front.

Both radiative and non-radiative shock waves are expected to produce a mix of line (radiative recombination) and continuum (due to bremsstrahlung emission) output \citep[e.g.][]{krol99}. Non-radiative shock waves (defined as when the cooling time of the shocked gas exceeds the age of the shock) produce Balmer optical emission lines only, because the heavier atoms are not excited \citep[e.g.][]{mckeho80}, so the existence of the strong \o3 line confirms the radiative nature of the Cyg X--1 nebula shock. In a radiative bow shock, the ambient ISM gas is first perturbed by the approaching radiation field. The ionisation of preshock hydrogen becomes progressively more significant in the velocity range $90\leq v_{\rm s} \leq 120$ km s$^{-1}$; below 90 km s$^{-1}$ the preshock gas is essentially neutral and above 120 km s$^{-1}$ it is entirely ionised \citep[e.g.][]{shulmc79}. In addition, at velocities $\sim 100$ km s$^{-1}$ the ionising He~\small II\normalsize~(304\AA) photons from the cooling region can doubly ionise the oxygen increasing the intensity of the \o3 lines \citep{mckeho80}. This leads to the high excitation line \o3 becoming negligible at low velocities in all shock-wave models (\citealt*{raym79,shulmc79,dopiet84,bineet85,coxra85,hartet87,dopsu95}; 1996). The outer shell of the Cyg X--1 nebula, bright in \o3 / H$\alpha$, may originate in the ionised atoms close to the front of the bow shock, and its presence confirms its association with the jet of Cyg X--1.

Radiative shock models predict that relative optical line strengths of the shocked gas are highly sensitive to the temperature and velocity of the shock wave, and vary only weakly with the conditions in the preshock gas such as the elemental abundances \citep[e.g.][]{coxra85}. Temperatures and velocities of shocked matter, mainly in SNRs, are commonly inferred by comparing their observed optical or ultraviolet (UV) line ratios to these models \citep[e.g.][]{blaiet91,vancet92,leveet95,boccet00,mavret01}. The bright \o3 / H$\alpha$ shock front of the Cyg X--1 nebula appears to be thicker on the western side of the nebula, and the H$\alpha$ and \o3 surface brightness is also higher in this region (Fig. \ref{neb1-PA}). This implies the shock may be faster here, possibly due to a lower preshock density; it is the furthest area of the nebula from the H~\small II \normalsize region (or due to the proper motion of Cyg X--1; see below).

\begin{figure}
\centering
\includegraphics[height=12cm,angle=270]{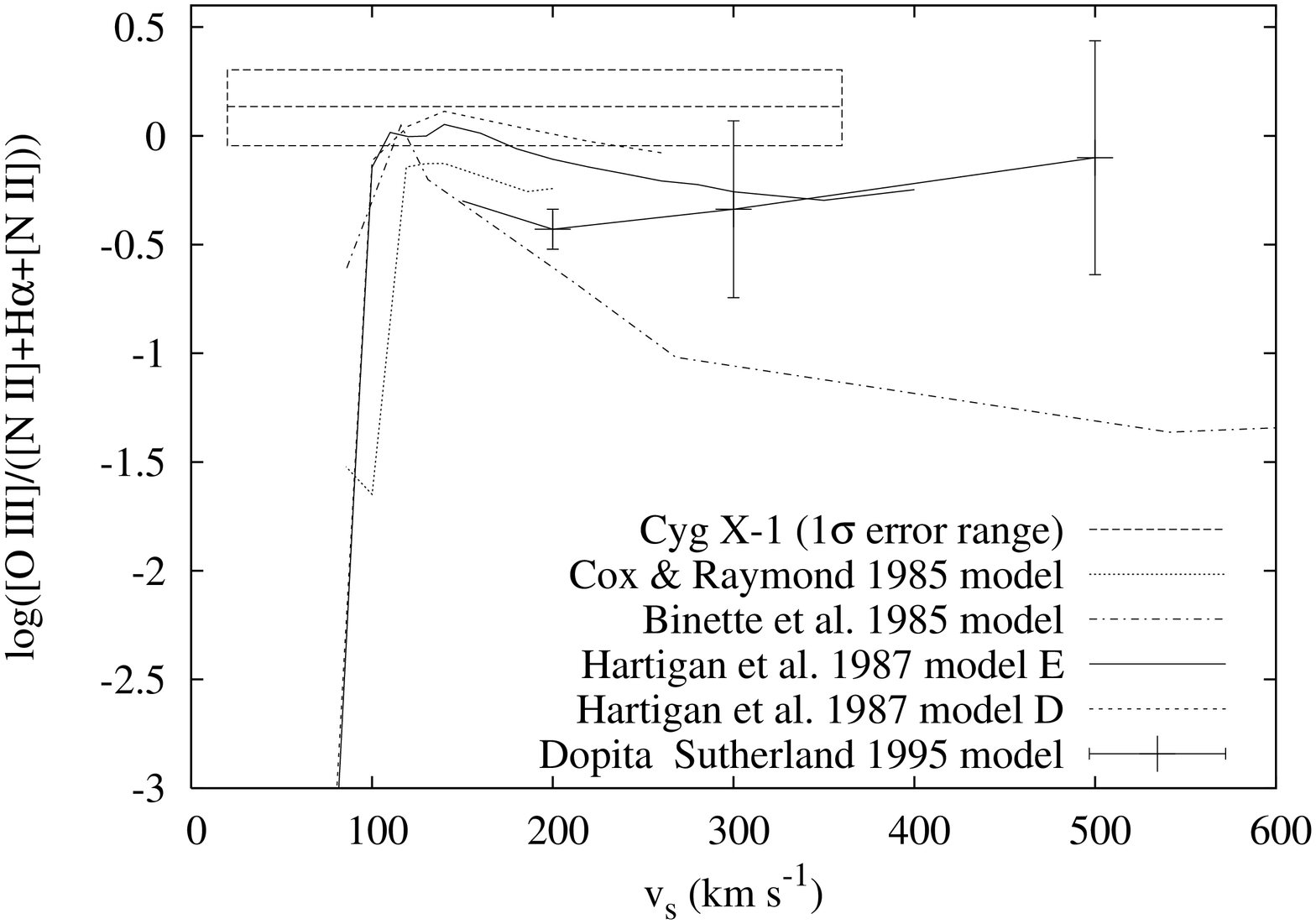}
\caption[Line ratio as a function of shock velocity predicted from five radiative shock models]{The [O III] / H$\alpha$ ratio as a function of shock velocity predicted from five radiative shock models. The dashed rectangle represents the constrained range of values measured for the Cyg X--1 shock front \citep[velocity range from][]{gallet05}. The INT [O III] filter includes both lines of the doublet; 4958.9\AA~  and 5006.9\AA, and the H$\alpha$ filter includes the two [N~II] lines at 6548.1\AA~ and 6583.4\AA~ in addition to 6562.8\AA~ H$\alpha$, within its bandpass. We therefore plot log($F_{\rm [O~III]+[O~III]}$/$F_{\rm [N~II]+H\alpha +[N~II]}$) versus velocity.}
\label{neb1-models}
\end{figure}

The relation between the shock velocity and \o3 / H$\alpha$ ratio in the shell is described for most of the published models of steady-flow radiative shocks with self-consistent pre-ionisation \citep{coxra85,bineet85,hartet87,dopsu95}.  The reliability of these models can be assessed from their level of agreement, unless there are systematic uncertainties common to all models \citep[which seems not to be the case as the velocities predicted by the models are consistent with velocities observed using other methods; e.g.][]{michet00,punet02}. In Fig. \ref{neb1-models}, the aforementioned relation predicted from five of the most advanced radiative shock models is plotted. The empirical \o3 / H$\alpha$ ratio found for the Cyg X--1 shock front is overplotted. Bearing in mind the previously estimated range of shock velocities, 20 $< v_{\rm s} < 360$ km s$^{-1}$ \citep{gallet05}, outside the range $100\leq v_{\rm s} \leq 360$ km s$^{-1}$, no models successfully predict the observed \o3 / H$\alpha$ measurement to within 1$\sigma$. Subsequently, a temperature of the gas in the shocked shell \citep{gallet05} of $2.4\times 10^5 \leq T_{\rm s} \leq 3.1\times 10^6$ K is inferred. The models in Fig. \ref{neb1-models} assume typical Galactic abundances and preshock particle densities within the range inferred for the Cyg X--1 nebula; $1\leq n_0 \leq 300$ cm$^{-3}$ \citep{gallet05}.

By constraining the velocity and temperature of the gas in the shock front, it is possible to refine a number of parameters such as the particle density in the shell and power in the jet \citep{gallet05}. Tables \ref{tab-neb1-2a} and \ref{tab-neb1-2b} list the unchanged and revised parameters inferred respectively, for the nebula and jet. The same methodology is adopted to calculate the parameters as used in \cite{gallet05}. The temperature-dependent ionisation fraction of the gas $x$ is taken from \cite{spit78}. At the temperatures inferred, the hydrogen in the shocked shell is fully ionised. The accuracy of the shock velocity, the time-averaged jet power and the jet lifetime have improved from $> 1$ dex to $\sim 0.5$ dex. Assuming the power originates in the hard state jet which is switched on for $\sim 90$ percent of the lifetime of Cyg X--1 \citep[e.g.][]{gallet05}, the total power of both these jets is between 30 and 100 percent of the bolometric 0.1--200 keV X-ray luminosity \citep[$L_{\rm X}\approx 3\times 10^{37}$ erg s$^{-1}$;][]{disaet01}.

Assuming the shock velocity has been approximately constant over the lifetime of the jet, the jet lifetime $t\sim 0.02$--0.04 Myr is much shorter than the estimated age of the progenitor of the black hole, $\sim 5$--7 Myr \citep{miraro03}. However, given that the proper motion of Cyg X--1 is $\sim 8.7$ milli-arcsec year$^{-1}$ in a direction $\sim 130$ degrees from the jet direction \citep{stiret01,lestet99,miraro03}, the system must have travelled between 2.5 and 9.1 arcmins during the lifetime of the jet. We speculate that the jet may have been present for the lifetime of Cyg X--1, but only when the system moved into the ISM close to the H~\small II \normalsize region did the local density become large enough for a bow shock to form. In fact, the age of the nebula (if it is powered by the jet) cannot exceed $\sim 0.04$ Myr because the position of Cyg X--1 at the birth of the jet would be \emph{in front of} the bow shock. This scenario may also explain the apparent higher \o3 / H$\alpha$ ratio and velocity of the gas on the western side of the nebula compared with the eastern side: the western side is further from the jet origin as the Cyg X--1 system moves.

\begin{table}
\begin{center}
\small
\caption[Unchanged Cyg X--1 nebula and jet parameters]{Unchanged nebula and jet parameters ($1\sigma$ confidence limits where quoted).}
\label{tab-neb1-2a}
\begin{tabular}{|ll|}
\hline
Parameter&Value\\
\hline
Monochromatic radio luminosity&$L_{\rm \nu, 1.4GHz}\approx 10^{18}$ erg s$^{-1}$Hz$^{-1}$\\
Cyg X--1 -- shell separation&$L\approx 3.3\times 10^{19}$ cm\\
Jet inclination angle to line of sight&$\theta \approx 35^o$\\
Thickness of shell&$\Delta R\approx 1.6\times 10^{18}$ cm\\
Source unit volume&$V(L,\Delta R)\approx 4\times 10^{53}$ cm$^3$\\
Bremsstrahlung emissivity for hydrogen gas&$\epsilon_{\nu}(V,L_{\rm \nu, 1.4GHz})\approx 2.5\times 10^{-36}$ erg s$^{-1}$Hz$^{-1}$ cm$^{-3}$\\
Gaunt factor&$g\approx 6$\\
Number density of preshock gas&$1\leq n_0 \leq 300$ cm$^{-3}$\\
\hline
\end{tabular}
\normalsize
\end{center}
\end{table}

\begin{table}
\begin{center}
\small
\caption[Refined Cyg X--1 nebula and jet parameters]{Refined nebula and jet parameters ($1\sigma$ confidence limits where quoted).}
\label{tab-neb1-2b}
\begin{tabular}{|ll|}
\hline
Parameter&Value\\
\hline
H$\alpha$ flux of shell&$F_{\rm H\alpha}\approx 3.6\times 10^{-15}$ erg s$^{-1}$cm$^{-2}$arcsec$^{-2}$\\
H$\alpha$ luminosity of nebula&$1.8\times 10^{34}\leq L_{\rm H\alpha}\leq 2.8\times 10^{35}$ erg s$^{-1}$\\
\ofig flux of shell&$F_{\rm [O~III]}\approx 2.8\times 10^{-15}$ erg s$^{-1}$cm$^{-2}$arcsec$^{-2}$\\
\ofig luminosity of nebula&$1.3\times 10^{34}\leq L_{\rm [O~III]}\leq 2.1\times 10^{35}$ erg s$^{-1}$\\
\ofig / H$\alpha$ ratio in shock front&$F_{\rm [O~III]}$ / $F_{\rm H\alpha}=1.36^{+0.65}_{-0.46}$\\
\ofig / H$\alpha$ ratio behind shock front&$F_{\rm [O~III]}$ / $F_{\rm H\alpha}=0.89^{+0.64}_{-0.49}$\\
Optical continuum flux density of shell&$F_{\rm \nu,OPT}\leq 6.8\mu$Jy arcsec$^{-2}$ (3$\sigma$)\\
Shock velocity&$100 \leq v_{\rm s} \leq 360$ km s$^{-1}$\\
Shocked gas temperature&$2.4\times 10^5 \leq T_{\rm s} \leq 3.1\times 10^6$ K\\
Number density of ionised particles in shell&$56\leq n_{\rm e}\leq 103$ cm$^{-3}$\\
Ionisation fraction&$x\approx 1$\\
Number density of total particles in the shell&$56\leq n_{\rm p}\leq 103$ cm$^{-3}$\\
Jet lifetime&$0.017\leq t\leq 0.063$ Myr\\
Time-averaged jet power&$P_{\rm Jet} = (4$--$14) \times 10^{36}$ erg s$^{-1}$\\
Hard state jet power including both jets&$P_{\rm Jets} = (9$--$30) \times 10^{36}$ erg s$^{-1}$\\
Outflow power / X-ray luminosity&$0.3\leq f_{\rm Jet/X}\leq 1.0$\\
\hline
\end{tabular}
\normalsize
\end{center}
\end{table}

\subsubsection{Ruling out alternative origins of the shell}

It has been shown (Section 6.4.1) that the morphology of the Cyg X--1 nebula indicates the gas is shock-excited and not photoionised; the structure is therefore not a planetary nebula or an H~\small II \normalsize region. The nebula also cannot be a SNR associated with Cyg X--1 itself \citep{gallet05}. The proximity of the nebula to Cyg X--1, its alignment with the Cyg X--1 jet and the agreement of its age inferred from its velocity and from the proper motion of Cyg X--1 (Section 4.2) all point towards the jet-powered scenario. The one remaining alternative is a field SNR not associated with Cyg X--1; this possibility is explored here.

\cite*{xuet05a} plot the diameter of 185 Galactic SNRs against their 1 GHz surface brightness (Fig. 3 of that paper). The radio surface brightness of the Cyg X--1 nebula is $\Sigma_{\rm 1~GHz}\approx 3\times 10^{-23}$ W m$^{-2}$ Hz$^{-1}$ sr$^{-1}$ \citep[lower than any SNR in the sample of][]{xuet05a}, and would have a diameter of $\simgt 100$ pc if it were to lie close to the diameter--surface brightness relation for SNRs. The nebula has an angular diameter of $\sim 7$ arcmin, suggesting a distance to the source of $d \simgt 50$ kpc if it is a SNR. However, the authors note that this surface brightness--diameter relation could be severely biased by selection effects \citep*{xuet05b}. Large, low surface-brightness SNRs may not have been detected in surveys due to confusion with strong foreground and background sources \citep[e.g.][]{gree04}. We therefore cannot completely rule out a SNR origin to the nebula, but its alignment with the relativistic jet of Cyg X--1 (for example) strongly suggests it is related to the system.

\begin{figure*}
\centering
\includegraphics[width=19cm,angle=90]{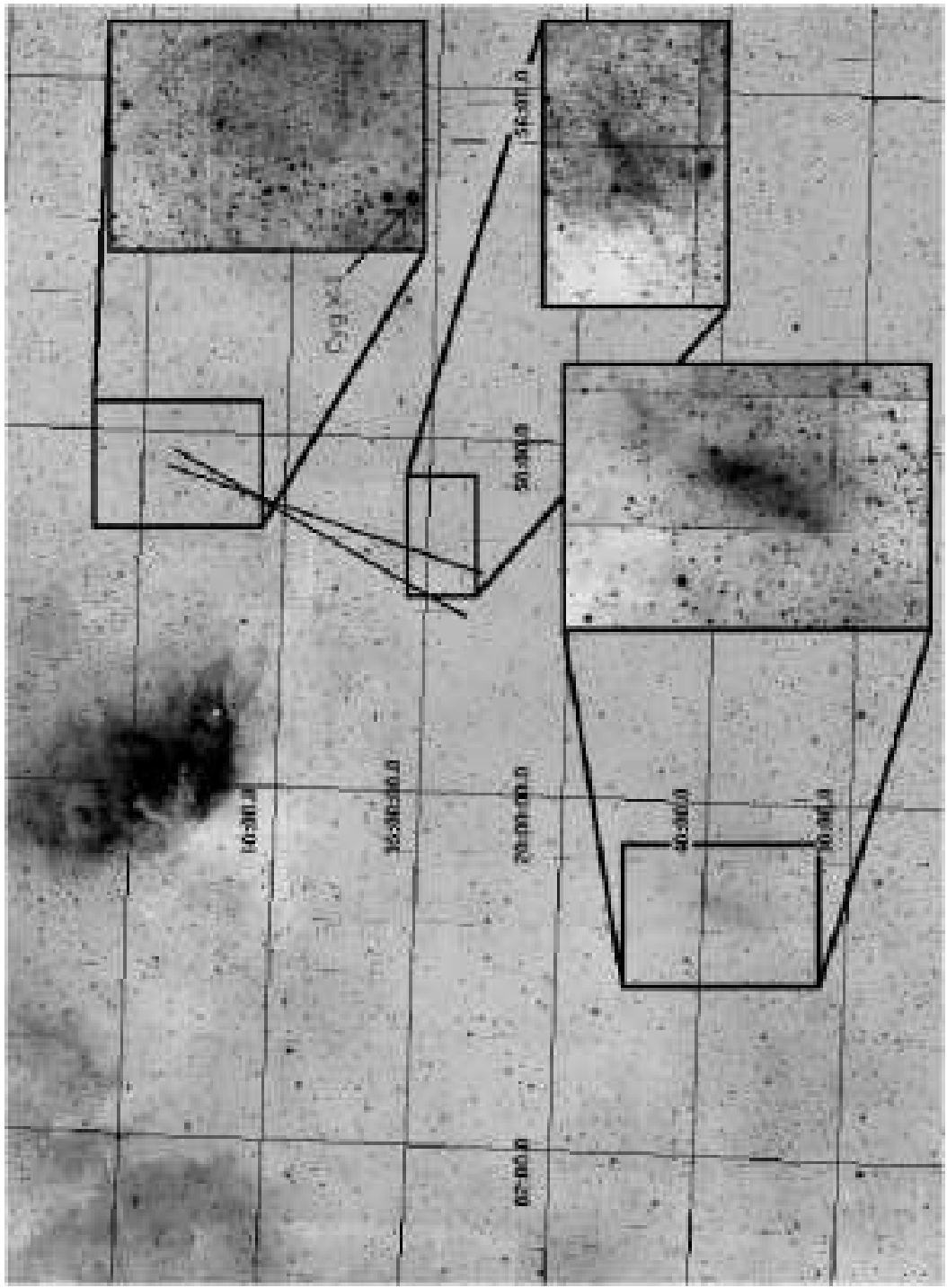}
\caption[Mosaic from the IPHAS survey in the region of Cyg X--1]{A $\sim 1.5$ degree$^2$ mosaic from the IPHAS survey. Each CCD frame is a 120 sec H$\alpha$ exposure. North is to the left of the page (top of the image), east to the bottom of the page (left in the image). RA and Dec are epoch J2000. The most extreme opening position angles of the resolved radio jet (both the steady hard state and transient jets) north of Cyg X--1; $17^o\leq \theta \leq 28^o$ \citep[][also projected to the south assuming the jets are anti-parallel]{stiret01,fendet06} are shown. The north nebula and two areas of candidate ISM interactions with the southern jet of Cyg X--1 are expanded.}
\label{neb1-fig-iphas}
\end{figure*}

\subsubsection{The search for ISM interactions with the southern jet}

A $\sim 0.25$ deg$^2$ field in \o3 south of Cyg X--1 was searched for a shell powered by the southern jet (lower panel of Fig. \ref{neb1-fig1}). No shell was observed in the 1800 sec image, nor any visual nebulosity detected in the \o3 images that is consistent with the southern jet of Cyg X--1 interacting with the ISM. The mass density of ISM may be much less to the south; the jet should travel essentially ballistically and undetected until being decelerated by a denser region. \cite{hein02} found that the rarity of jet-powered lobes in microquasars compared to AGN is likely due to the lower density environment. In addition, it is interesting to note that the X-ray-bright shocked edges surrounding AGN radio lobes are sometimes one-sided \citep{krafet03}. A $\sim1$ deg$^2$ H$\alpha$ mosaic from the IPHAS survey \citep[][see Fig. \ref{neb1-fig-iphas}]{drewet05,withet06} was searched. Two candidate `hot spots' analogous to those associated with AGN jets were identified, and are expanded in the figure.

Radio emission exists in the region close to the northern hot spot according to \cite{martet96}, but it is uncertain whether it originates in point sources or extended emission. We inspected the southern region of the 1.4 GHz radio image obtained by \cite{gallet05} to search for diffuse emission. In a 1600$\times$1600 arcsec region close to the more northern H$\alpha$ hot spot (the southern hot spot is not within the radio field of view), the mean r.m.s. flux density is $4\times 10^{-4}$ Jy beam$^{-1}$, yielding a 3$\sigma$ upper limit to the (putative) radio hot spot in that region of 1.2 mJy beam$^{-1}$. Deeper radio observations are needed to test whether the candidate H$\alpha$ hot spots have radio counterparts, as would be expected if they are jet--ISM interaction sites.

A follow-up observation was made of the region of the more northern of the two hot spots, in H$\alpha$ with the Faulkes Telescope North (see Section 6.2). The image is presented in the right panel of Fig. \ref{neb1-naturefaulkes} (the lower, southern CCD). The hot spot is visible in the image, but its S/N is no higher than the H$\alpha$ image from IPHAS.

\subsection{Conclusions}

H$\alpha$ and \o3 (5007 \AA) images of the nebula powered by the jet of the black hole candidate and microquasar Cygnus X--1 have been presented. The nebula and surrounding regions were observed with the 2.5 m INT and the 2.0 m Faulkes Telescope North. The ring-like structure is luminous in \o3 and there exists a thin outer shell with a high \o3 / H$\alpha$ flux ratio.  The morphology and line ratios indicate that this outer shell probably originates in the collisionally excited atoms close to the front of the bow shock. Its presence indicates that the gas is shock excited as opposed to photoionised, supporting the jet-powered scenario. The shock velocity was previously constrained at 20 $< v_{\rm s} < 360$ km s$^{-1}$; here it is shown that $v_{\rm s} \geq 100$ km s$^{-1}$ (1$\sigma$ confidence) based on a comparison of the observed \o3 / H$\alpha$ ratio in the bow shock with a number of radiative shock models. The H$\alpha$ flux behind the shock front is typically $4\times 10^{-15}$ erg s$^{-1}$ cm$^{-2}$ arcsec$^{-2}$, and an upper limit of $\sim 8\times 10^{-15}$ erg s$^{-1}$cm$^{-2}$arcsec$^{-2}$ (3$\sigma$) to the optical ($R$-band) continuum flux of the nebula is also estimated.

Alternative interpretations of the structure such as photoionised gas are ruled out, but an unrelated SNR cannot completely be ignored as a possible origin. However, the morphology of the nebula and its alignment with the jet of Cyg X--1 strongly suggest that it is indeed powered by this jet. We have probably isolated for the first time the thin, hot, compressed gas close to the bow shock front of a nebula powered by an X-ray binary jet.

The time-averaged power of the jet must be $P_{\rm Jet} = (4$--$14) \times 10^{36}$ erg s$^{-1}$ to power a shock wave of the inferred size, luminosity, velocity and temperature. If the steady hard state jet of Cyg X--1 (that is active for $\sim90$ percent of the time) is providing this power \citep[rather than the transient jet;][]{fendet06}, the total energy of both jets in the hard state is 0.3--1.0 times the bolometric X-ray luminosity. This provides the strongest evidence so far that at an accretion rate of $\sim 2$ percent Eddington, Cyg X--1 is in a state of approximate equipartition between radiative and kinetic output. Lower accretion rates should therefore be jet-dominated (\citealt{fendet03}; \citealt{kordet06}).

The inferred age of the structure is similar to the time Cyg X--1 has been close to a bright H~\small II\normalsize~region (due to the proper motion of the binary), indicating a dense local medium is required to form the shock wave. In addition, a $> 1$ degree$^2$ field of view to the south of Cyg X--1 in H$\alpha$ (provided by the IPHAS survey) is searched for evidence of the counter jet interacting with the surrounding medium. Two candidate regions are identified, whose possible association with the jet could be confirmed with follow-up optical emission line and radio observations.

\newpage

\begin{center}
{\section{Conclusions \& Future Work}}
\end{center}

The majority of this work has concerned to some extent the dominating OIR continuum emission processes in LMXBs. From Chapters 2--5 a general picture seems to emerge, in which the detected OIR emission originates from different components depending on the luminosity and spectral state of the source at the time and the nature of the compact object; i.e. essentially the rate and nature of the mass accretion. Four methods were adopted to constrain these emission processes: correlation analysis (C), SEDs (the OIR spectrum; S), changes in the OIR luminosity during state transitions (T), and linear polarimetry (P). In Table \ref{emproc-final} I summarise which emission processes can explain the OIR data, using the four methods. No processes were tested using all four methods, so the number of tests (methods) used for each emission process is shown in the second column. The processes that can explain all of the data when tested are shown in bold.

It is clear that not all of the sources in a given luminosity/state have the same dominating optical and NIR emission processes, probably due to differing source parameters (see Section 1.1.1) like the size of the accretion disc. However, many results in each of the Chapters 2--5 agree, and by summarising these results in Table \ref{emproc-final} it is possible to identify the emission processes that dominate \emph{most of the time} at a given luminosity/state and to see how these differ between BHXBs and NSXBs.

Quiescent LMXBs (both black hole and neutron star systems) are dominated by the companion star most of the time, but the jet and disc may play a role in BHXBs in quiescence. Comparing these results to the literature, there is evidence for a disc contribution in quiescence in many BHXBs \citep{mcclet95,oroset96,zuriet03}; this is generally thought to be the viscous disc as opposed to the X-ray heated disc (as the X-rays are probably too faint in quiescence to heat the disc). A jet contribution is as yet fairly unexplored \citep[recent works have begun to explore this;][]{gallet07}. The BHXBs with the faintest discs and stars are likely to have a larger fraction of OIR jet emission \citep[see][]{hyneet06b,reynet07,gallet07}. The OIR emission of NSXBs, like the BHXBs are mostly dominated by the companion in quiescence but at least some probably have a disc contribution \citep[Table \ref{emproc-final} and e.g.][]{campet02,casaet02}. There is no empirical evidence for a jet contribution; this agrees with what is expected from correlation analysis of NSXBs (see Sections 1.2.3; 2.4.2).

\begin{table*}[h]
\small
\caption[The OIR emission processes that can describe the data in this thesis]{The OIR emission processes that can describe the data in this thesis.}
\label{emproc-final}
\small
\begin{center}
\begin{tabular}{|lc|ccccc|}
\hline
Waveband / Luminosity   &Tests&\multicolumn{2}{c}{--- Accretion Disc ---}&\multicolumn{2}{c}{----- Jet emission -----}&Comp-\\
/ X-ray state&&X-ray &Viscous&Optically&Optically&anion\\
             &&reprocessing&disc&thin  &thick    &\\
\hline
\multicolumn{7}{|c|}{--------------- \emph{Black hole X-ray binaries} ---------------}\\
OPT; Quiescence&2&\bf{C,S}&S&C&C&\bf{C,S}\\
NIR; Quiescence&1&\bf{S}&\bf{S}&&&\bf{S}\\
OPT; Low lum. hard state &3&C,S&S&C&C&\\
NIR; Low lum. hard state &3&S&S&\bf{C,S,P}&C&P\\
OPT; High lum. hard state&3&C,S&S&C,T&C,T&\\
NIR; High lum. hard state&3&S,T&S&\bf{C,S,T}&C,T&\\
OPT; Soft state  	&2&\bf{S,T}&\bf{S,T}&&&\\
NIR; Soft state  	&3&S,T&\bf{C,S,T}&&&\\
\hline
\multicolumn{7}{|c|}{--------------- \emph{Neutron star X-ray binaries} ---------------}\\
OPT; Quiescence&2&\bf{C,S}&\bf{C,S}&&&\bf{C,S}\\
NIR; Quiescence&2&S&S&&&\bf{C,S}\\
OPT; Low lum; atolls/MSXPs&2&\bf{C,S}&\bf{C,S}&&&\\
NIR; Low lum; atolls/MSXPs&2&S&S&&&\\
OPT; High lum; atolls/MSXPs&2&C&C&\bf{C,S}&C&\\
NIR; High lum; atolls/MSXPs&2&&&\bf{C,S}&C&\\
OPT; High lum; Z-sources  &2&\bf{C,S}&S&C&\bf{C,S}&\\
NIR; High lum; Z-sources  &3&S&S&C,P&\bf{C,S,P}&\\
\hline
\multicolumn{7}{|c|}{--------------- \emph{High mass X-ray binaries} ---------------}\\
OIR; All&1&&&&&\bf{C}\\
\hline
\end{tabular}
Low lum. means outbursting sources with $L_{\rm X}\simlt 10^{36-37}$ erg s$^{-1}$; High lum. means outbursting sources with $L_{\rm X}\simgt 10^{36-37}$ erg s$^{-1}$. C = correlation analysis; S = SEDs (the OIR spectrum); T = changes in the OIR luminosity during state transitions; P = linear polarimetry.
\end{center}
\normalsize
\end{table*}

BHXBs in a hard X-ray state in outburst appear to be dominated by the X-ray heated disc and the optically thin region of the jet spectrum in the optical and NIR, respectively (both at high and low luminosities). The OIR seems to be the regime in which the two components have comparable luminosities, and the `V'-shape formed by the two components is seen in the SEDs of some BHXBs (Chapter 3). The position of the `V' changes between sources (likely due to different positions of the turnover in the jet spectrum or different sizes of accretion disc); for example XTE J1859+226 has a blue OIR spectrum in the hard state but \cite{hyneet02a} showed that a non-thermal component is present that reddens the NIR slightly (hence the `V' is further into the NIR here). In GX 339--4 and 4U 1543--47, the jet component is more dominant at high luminosity than at low luminosity in the hard state \citep{buxtet04,homaet05a}, as is expected from the models in Section 2.4.2 whereby the jet component drops off with $L_{\rm X}$ slightly more steeply than the X-ray heated disc does. This suggests the disc may dominate the OIR in quiescence (in the absence of a star) and the jet may be more luminous than the disc only redward of the NIR here. On the other hand, the SEDs of BHXBs are often redder at lower luminosities in the hard state (Chapter 3); this could be explained by the jet dominating more (the opposite to above) or the disc blackbody being cooler. The only direct measurement of the level of jet contribution is made at the highest luminosity in the hard state (from the level of soft state quenching), where the jet contributes $\sim 90$\% of the NIR light and $< 76$\% of the optical.

The jet and disc may not always dominate the NIR and optical emission (respectively) of BHXBs in the hard state; in fact most emission processes considered can sometimes explain the emission (Table \ref{emproc-final}). Further studies of BHXBs in outburst may reveal relations between the dominating processes and some parameters of the binary like the size of the disc. For BHXBs in the soft state the jet is quenched and there is empirical evidence for the viscous disc to dominate the OIR \citep[shown in Chapter 4 in particular; see also][]{kuul98,homaet05a} but many results are also consistent with X-ray reprocessing; it is likely that both play a role and their level of contribution depends on the source parameters.

NSXBs in outburst at low luminosities seem to be always dominated by the accretion disc in the OIR, which largely agrees with the literature \citep[e.g.][]{mcgoet03,hyneet06a}. At high luminosities the optically thin jet component becomes more dominant (moreso in the NIR than the optical); this is seen more prominently in sources with smaller accretion discs, i.e. the MSXPs \citep{wanget01,krauet05,torret07}. The NSXBs with the highest luminosities and largest discs, the Z-sources, are dominated by the (probably X-ray reprocessing) disc but most likely have a jet contribution. The position of Z-sources in their colour--colour diagrams or HIDs changes on timescales of hours--days and the jet seems to be quenched during episodes in the flaring branch \citep[e.g.][]{miglfe06}. The linear polarisation of Sco X--1 (which most likely originates in the jet) seems to flicker (Chapter 5); further NIR or mid-IR polarimetric observations of Z-sources, correlated with X-ray hardness may reveal the X-ray state-dependent NIR contribution of the jet in Z-sources (the level of jet OIR contribution in Z-sources is currently not certain, although it will be more prominent in the NIR than the optical).

In general there seems to be less ambiguity as to the dominating OIR processes in NSXBs compared to BHXBs; few emission processes can explain the OIR behaviour of NSXBs except those mentioned above. This is probably at least partly due to the very different relations between $L_{\rm OIR}$ and $L_{\rm X}$ for the emission processes of NSXBs; in BHXBs, the OIR light from X-ray reprocessing and the jet have roughly the same dependency on $L_{\rm X}$, whereas in NSXBs the relations are not so similar (see Section 2.4.2). This work has demonstrated that it is important not to make assumptions about the dominating OIR emission, especially in BHXBs during outburst. However since the contributions have been constrained and separated fairly satisfactorily in some cases, this work has made it possible to make such assumptions in these cases, allowing analysis of the different components. For example, in Chapter 4 the origin of the hysteresis affect could not have been discussed in depth without knowing the the NIR is dominated by the jet in hard state BHXBs.

Promising future methods to constrain the dominating OIR emission mechanisms include well-sampled photometry (i.e. tracking the OIR colours through an outburst) and linear polarimetry; optically thin synchrotron emission is expected to be highly polarised if the magnetic field is ordered. NIR linear polarimetry during times in which the jet is known to dominate (e.g. BHXBs in bright hard states) will constrain the conditions in the inner regions of the jets close to where they are launched. Future polarimetric studies may reveal changes in these conditions as a function of jet power (i.e. jet luminosity) and differences between object types (black hole or neutron star systems). In particular, the BHXB GX 339--4 undergoes quasi-regular outbursts with a bright NIR jet \citep[e.g.][]{homaet05a}. In quiescence, a positive polarisation detection from the jet will indicate its contribution is significant, supporting jet-dominated scenarios. In addition, circular polarimetry could provide clues towards the orientation of the magnetic field and/or the conditions in the local medium.

A method not adopted in this work is fast timing analysis. Variability is expected and seen from many of the OIR emission processes \citep[e.g.][]{zuriet03,einket07}. A future work could be to perform fast timing photometry of the hard state NIR optically thin jet component, when it is known to dominate. Comparing the features in the resulting power spectrum to X-ray power spectra could constrain the contribution of the synchrotron jet component in the X-ray regime (which is currently debated). One source so far has shown dips in optical light curves in advance of X-rays, which are likely to be a signature of jet emission \citep{kanbet01,malzet04}. In addition, if the behaviour of the OIR variability differs between disc and jet components for example (like above), this analysis could help to identify their respective levels of contribution, e.g. in quiescence. Monitoring campaigns of LMXBs like ours using the Faulkes Telescopes (see Section 4.5) will reveal the long-term variability of quiescent systems; differences between BHXBs and NSXBs could be interesting because the jet component is expected to be present in the BHXBs but not in the NSXBs at these low luminosities.

It has been found (Chapter 2) that at a given X-ray luminosity, a BHXB is typically $\sim 20$ times more luminous in OIR than an atoll or MSXP NSXB (Z-sources can be as OIR-luminous as BHXBs or NSXBs). The reasons for this are unclear but have direct implications for the nature and physical location of the illuminating X-ray source; a hot topic of debate.

It was shown in Chapter 2 that the data of BHXBs (in hard and soft states), NSXBs (atolls/MSXPs and Z-sources) and HMXBs occupy different areas of the $L_{\rm X}$--$L_{\rm OIR}$ diagram, with small areas of overlap. Therefore this diagram is a useful tool to identify the nature of a new X-ray binary, if its distance and reddening are known. For most Galactic XBs the distance and reddening of new transients are not known, but this tool may instead be useful for extragalactic XBs; many XBs with X-ray and optical counterparts have now been discovered in nearby galaxies such as M31 \citep[e.g.][]{willet05}.

The correlation analysis method used in Chapter 2 could be applied to OIR--X-ray data not used so far, such as BHXBs in intermediate/very high states (to identify when exactly the dominating processes change during state transitions), NSXBs with high magnetic fields that have faint or no jets (good for a test sample to compare to), ULXs and (incorporating a mass term) SMBHs. The empirical hysteresis effect found in Chapter 4 has direct implications for the behaviour of many parameters such as the disc viscosity parameter $\alpha$ and the power and spectrum of the jets; future works on these topics should take into account this effect.

It has been known for a number of years that the the broadband optically thick synchrotron spectrum of the jet extends to the IR in some sources \citep[e.g.][]{corbfe02}, but much of the work here has shown that this is probably common to all LMXBs. The power in the jets is higher than thought previously under the assumption that the spectrum turns over in the radio or mm-regime. It is difficult to estimate quantitatively the power of the jets without constraining their radiative efficiency, which is also hard to constrain. However the results here lead us one step closer to estimating the jet power. The jet power of one BHXB, Cyg X--1, is constrained in Chapter 6 by a completely different method which is independent of the jet radiative efficiency -- from the interaction of the jet with the surrounding ISM. If eventually the jet power can be constrained in a number of systems using this method, and the jet spectrum can be constrained for the same sources, the jet radiative efficiency can be estimated and compared between sources.

I (with Rob Fender, Elena Gallo, Christian Kaiser, James Miller-Jones and Sebastian Heinz) am gathering observations of potential X-ray binary jet-powered nebulae to test the ubiquity of the jet power from X-ray binaries and constrain the matter and energy they input into the ISM. LMXBs may have less powerful jets than Cyg X--1 when time-averaged, due to their long phases in quiescence, but they are longer lived and so the [jet power $\times$ lifetime] product may be comparable and nebulae may be detected. Any nebulae discovered associated with the jet of an LMXB will provide constraints on these speculations.

\begin{figure*}
\centering
\includegraphics[width=15cm,angle=0]{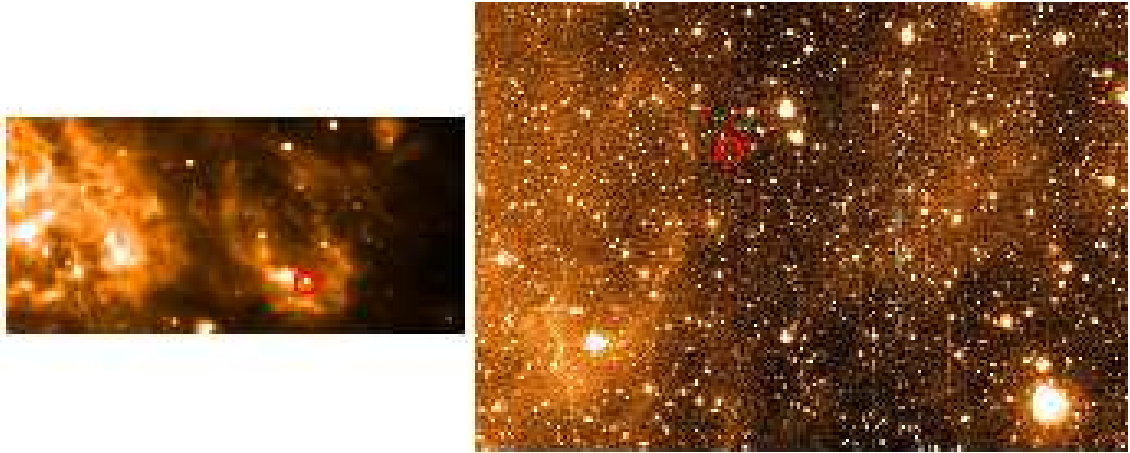}
\caption[H$\alpha$ images of the regions local to LMC X--1 and GRS 1009--45]{H$\alpha$ images of the regions local to LMC X--1 (left) and GRS 1009--45 (right). North is up and east is to the left. The X-ray binaries are marked in red. An XIN surrounds LMC X--1 \citep{pakuan86} and a shell-like structure lies to the south-east of GRS 1009--45. Follow-up observations are required to identify the nature of this latter nebula. The jets of LMC X--1 may be interacting with the XIN; see \cite{cooket07}.}
\label{conc-lmcx1grs1009}
\end{figure*}

So far, wide-field imaging of the fields around 30 X-ray binaries have been obtained in H$\alpha$ (and some in \o3) using the INT and the ESO/MPI 2.2 m telescopes (for some preliminary results see also \citeauthor{paper2}). In most fields no shell-like nebulae are discovered, however a few candidates were found. The analyses of these data are ongoing, and follow-up observations of some candidate nebulae are required to confirm their nature. A couple of candidates are presented in Fig. \ref{conc-lmcx1grs1009}. A known X-ray-ionised nebula (XIN) surrounds LMC X--1 which is photoionised by the UV and X-ray photons from the XB \citep{pakuan86} and the jets are probably interacting with this nebula \citep{cooket07}. For GRS 1009--45, an oval-shaped nebula to the east of the X-ray binary may be shock-excited from the jets, and is consistent with the morphology of some AGN radio lobes. Follow-up narrowband observations in e.g. [S~\small II\normalsize ] (as this is much stronger in shocked gas than photoionised gas) are required to test this.

\begin{figure}
\centering
\includegraphics[width=15cm,angle=0]{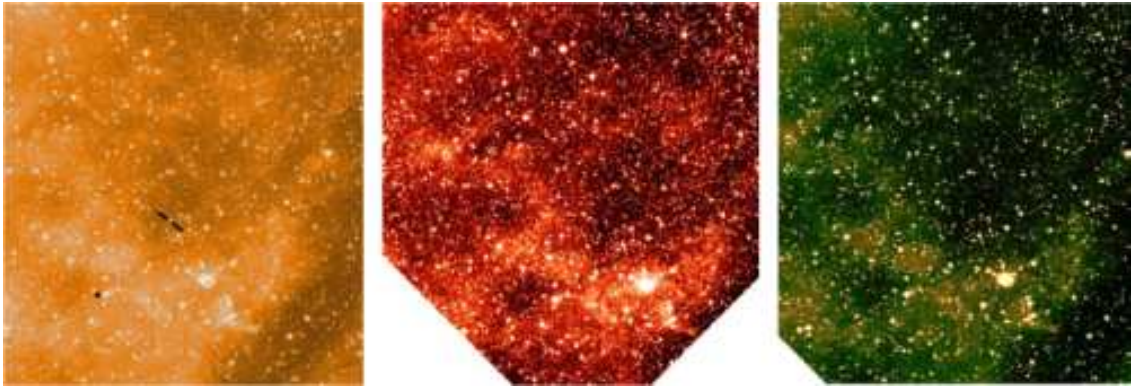}
\caption[GRO J1655--40 and its candidate jet-powered nebula]{GRO J1655--40 \citep[marked black in the left panel; the direction is that of the resolved radio jets;][]{hjelru95} and its candidate jet-powered nebula to the lower right in each panel. North is up and east is to the left. From left to right: H$\alpha$ image taken with the ESO/MPI 2.2~m; [S II] image taken with the Danish 1.5~m; [S II] / H$\alpha$ ratio image (H$\alpha$ is green, [S II] is red).}
\label{conc-groj1655}
\end{figure}

An apparent H$\alpha$ shell to the south west of GRO J1655--40 is aligned with the resolved radio jet \citep{hjelru95}. Follow-up imaging in [S~\small II\normalsize ] (performed with the Danish 1.5m telescope) revealed the shell to be clearly visible in [S~\small II\normalsize ] line emission, implying the gas is shocked-excited as opposed to photoionised (Fig. \ref{conc-groj1655}). There is also extended radio emission in the region of this nebula \citep{combet01}. Flux calibration of the optical images and overlaying of archival radio data may confirm this structure to be associated with the jet.

The properties of the gas in the Cyg X--1 nebula (any any other jet-powered structures discovered), and hence the properties of the jet that powers it, may be constrained with optical and UV \citep[e.g.][]{blaiet91} spectroscopic observations. Many of these properties are related to optical emission line ratios. 
In addition, observations in [S~\small II\normalsize ] would test whether the H$\alpha$-emitting gas in the hot spots to the south of Cyg X--1 are shock-excited or photoionised, and deeper \o3 observations may constrain the velocity of any shocked gas.

Imaging in infrared emission lines may also help to identify nebulae hidden behind many magnitudes of interstellar dust extinction. In general, few Galactic jet-powered nebulae are discovered so far, probably because the conditions required are rare: a high local mass density is required for a bow shock to form \citep{hein02}, however low Galactic extinction towards the source is also needed to see the faint structures in the optical regime, and a low space velocity of the system may be needed so that the power is not dissipated over too large a volume.

Evidence for the existence of synchrotron-emitting jets in BHXBs is solid; spectral evidence for the existence of these jets in NSXBs presented in this work \citep[and showed for the first time by][]{miglet06}, along with radio detections which are sometimes resolved \citep[see][]{fomaet01,miglfe06}, have direct implications for the local conditions and accretion processes in NSXBs.

To conclude, \emph{`we know more than we ever have, but every answer unveils a whole new set of questions'} (my father Edwin Russell, 2007).

\end{document}